\documentclass{ws-rv961x669}
\usepackage{subfigure}   
\usepackage{ws-rv-thm}   
\usepackage{ws-rv-van}   
\usepackage[subpreambles=true]{standalone}
\usepackage{comment}
\usepackage[dvipsnames]{xcolor}
\usepackage{hyperref}
\hypersetup{
    colorlinks=true,
    linkcolor=BrickRed,
    filecolor=magenta,      
    urlcolor=purple,
    citecolor=BrickRed,
}
\usepackage{titlesec}
\makeindex
\begin{document}

\title{Neutrino Physics and Astrophysics}

\chapter[IceCube Neutrinos]{IceCube and High-Energy Cosmic Neutrinos}

\author[F. Halzen and A. Kheirandish]{Francis Halzen\footnote{halzen@icecube.wisc.edu}
}

\address{Dept. of Physics and Wisconsin IceCube Particle Physics Center\\
Univeristy of Wisconsin{\textendash}Madison, Madison, WI 53706 USA}

\author{Ali Kheirandish\footnote{kheirandish@psu.edu}}
\address{Dept. of Physics, Department of Astronomy and Astrophysics, and Center for Multimessenger Astrophysics, Institute for Gravitation and the Cosmos\\ The Pennsylvania State University, University Park, PA 16802, USA}
\vspace{1cm}

\begin{abstract}
The IceCube experiment discovered PeV-energy neutrinos originating beyond our Galaxy with an energy flux that is comparable to that of TeV-energy gamma rays and EeV-energy cosmic rays. Neutrinos provide the only unobstructed view of the cosmic accelerators that power the highest energy radiation reaching us from the universe. We will review the rationale for building kilometer-scale neutrino detectors that led to the IceCube project, which transformed a cubic kilometer of deep transparent natural Antarctic ice into a neutrino telescope of such a scale. We will summarize the results from the first decade of operations: the status of the observations of cosmic neutrinos and of their first identified source, the supermassive black hole TXS 0506+056. Subsequently, we will introduce the phenomenology associated with cosmic accelerators in some detail. Besides the search for the sources of Galactic and extragalactic cosmic rays, the scientific missions of IceCube and similar instruments under construction in the Mediterranean Sea and Lake Baikal include the observation of Galactic supernova explosions, the search for dark matter, and the study of neutrinos themselves. This review resulted from notes created for summer school lectures and should be accessible to nonexperts.\footnote{To be published in Neutrino Physics and Astrophysics, edited by F. W. Stecker, in the Encyclopedia of Cosmology II, edited by G. G. Fazio, World Scientific Publishing Company, Singapore, 2022.}
\end{abstract}

\markboth{IceCube Neutrinos}{F. Halzen \& A. Kheirandish}
\newpage
\tableofcontents

\newpage

\setcounter{footnote}{0}
\section{Neutrino Astronomy: A Brief History}
\vspace{.2cm}

Soon after the 1956 observation of the neutrino~\cite{Reines1956}, the idea
emerged that it represented the ideal astronomical messenger. Neutrinos travel
from the edge of the Universe without absorption and with no
deflection by magnetic fields. Having essentially no mass and no electric
charge, the neutrino is similar to the photon, except for one important
attribute: its interactions with matter are extremely feeble. So, high-energy
neutrinos may reach us unscathed from cosmic distances: from the inner
neighborhood of black holes and from the nuclear furnaces where
cosmic rays are born. But, their weak interactions also make cosmic neutrinos very
difficult to detect. Immense particle detectors are required to collect cosmic
neutrinos in statistically significant numbers~\cite{Halzen:2008zz}. By the
1970s, it was clear that a cubic-kilometer detector was needed to observe cosmic
neutrinos produced in the interactions of cosmic rays with background microwave
photons~\cite{Roberts:1992re}. Subsequent estimates for observing potential cosmic
accelerators such as Galactic supernova remnants and gamma-ray bursts
unfortunately pointed to the same exigent requirement~\cite{Gaisser1995,Learned:2000sw,Halzen:2002pg}. 
Building a neutrino telescope has been a daunting technical challenge.

Given the detector's required size, early efforts concentrated on transforming
large volumes of natural water into Cherenkov detectors that collect the light
produced when neutrinos interact with nuclei in or near the
detector~\cite{Markov1960}. After a two-decade-long effort, building the
Deep Underwater Muon and Neutrino Detector (DUMAND) in the sea off the main
island of Hawaii unfortunately failed~\cite{Babson:1989yy}. However, DUMAND
paved the way for later efforts by pioneering many of the detector technologies
in use today, and by inspiring the deployment of a smaller instrument in Lake
Baikal~\cite{Balkanov:2000cf} as well as efforts to commission neutrino
telescopes in the Mediterranean~\cite{Aggouras:2005bg,Aguilar:2006rm,Migneco:2008zz}. 
These efforts in turn have led towards the construction of KM3NeT. But the first telescope
on the scale envisaged by the DUMAND collaboration was realized instead by
transforming a large volume of transparent natural Antarctic
ice into a particle detector, the Antarctic Muon and Neutrino Detector Array
(AMANDA). In operation beginning in 2000, it represented a proof of concept for the
kilometer-scale neutrino observatory,
IceCube~\cite{ICPDD2001,Ahrens:2003ix}.

\begin{figure}[t!]
\begin{center}
\includegraphics[width=0.9\columnwidth]{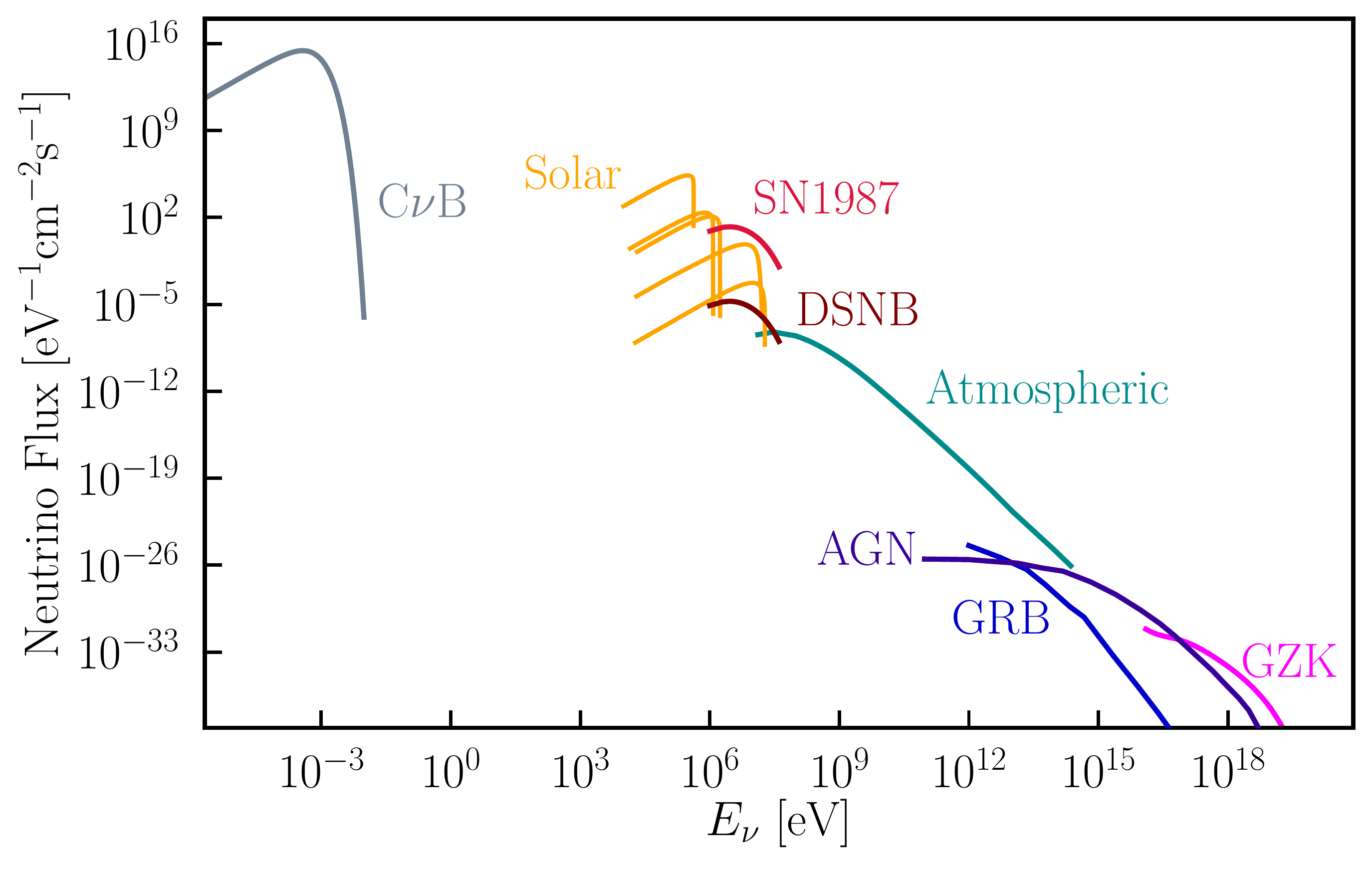}
\caption{The cosmic-neutrino spectrum. Sources are the Big Bang (C$\nu$B), the Sun,
supernovae (SN), atmospheric neutrinos, active galactic nuclei (AGN) galaxies,
and GZK neutrinos. 
}
\label{fig-XX.1}
\end{center}
\end{figure}

Neutrino astronomy has achieved spectacular successes in the past: neutrino
detectors have ``seen" the Sun and detected a supernova in the Large Magellanic
Cloud in 1987. Both observations were of tremendous importance; the former
showed that neutrinos have mass, opening the first crack in the Standard
Model of particle physics, and the latter confirmed the basic nuclear physics of
the death of stars. Fig.~\ref{fig-XX.1} illustrates the neutrino energy
spectrum covering an enormous range, from microwave energies ($10^{-12}$\,eV) to
$10^{20}$\,eV~\cite{Becker:2007sv}. The figure is a mixture of observations
and theoretical predictions. At low energy, the neutrino sky is dominated by
neutrinos produced in the Big Bang. Nuclear fusion in the sun generates neutrinos with keV energy and a spectrum that extends to MeV. At MeV energy, neutrinos are produced by
supernova explosions; the flux from the 1987 event is shown.  At yet higher energies, the figure displays
the atmospheric-neutrino flux, up to energies of 100 TeV which were measured at the Frejus underground
laboratory~\cite{Rhode:1996es} AMANDA experiment~\cite{Achterberg:2007qp}. Atmospheric neutrinos are a main player in our story, because they are a dominant background for extraterrestrial searches. 
The flux of atmospheric neutrinos falls with increasing energy;
events above 100 TeV are rare, leaving eventually a clear field of view for
extraterrestrial sources at the highest energies.

The highest energy neutrinos in Fig.~\ref{fig-XX.1} are the decay products of
pions produced by the interactions of cosmic rays with microwave
photons~\cite{Engel:2001hd}. Above a threshold of $\sim 4\times10^{19}$\,eV,
cosmic rays interact with the microwave background introducing an absorption
feature in the cosmic-ray flux, the Greisen-Zatsepin-Kuzmin (GZK) cutoff. As a
consequence, the mean free path of extragalactic cosmic rays propagating in the
microwave background is limited to roughly 75 megaparsecs, and, therefore, the
secondary neutrinos are the only probe of the still enigmatic sources at longer
distances. What they will reveal is a matter of speculation. The calculation of
the neutrino flux associated with the observed flux of extragalactic cosmic rays
is straightforward and yields one event per year in a kilometer-scale detector. 
 The flux, labeled GZK in Fig. \ref{fig-XX.1}, shares the high-energy neutrino sky with neutrinos anticipated from gamma-ray bursts and active galactic
nuclei (AGN)~\cite{PhysRevLett.66.2697, Gaisser1995,Learned:2000sw,Halzen:2002pg}.

A population of extragalactic cosmic neutrinos with energies of 60\,TeV--1\,PeV was revealed by the first two years of IceCube data. We will review the present status of the observations, the identification of the sources by multimessenger astronomy, as well as attempts to decipher the phenomenology of the heavenly beam dumps producing neutrinos. Subsequently, we will describe the status of the search for Galactic neutrino sources and conclude with brief discussions of other uses of neutrino telescopes.

\section{Rationale for the Construction of Kilometer-Scale Neutrino Detectors}
\vspace{.2cm}

\begin{figure}[t]
 \includegraphics[width=\columnwidth]{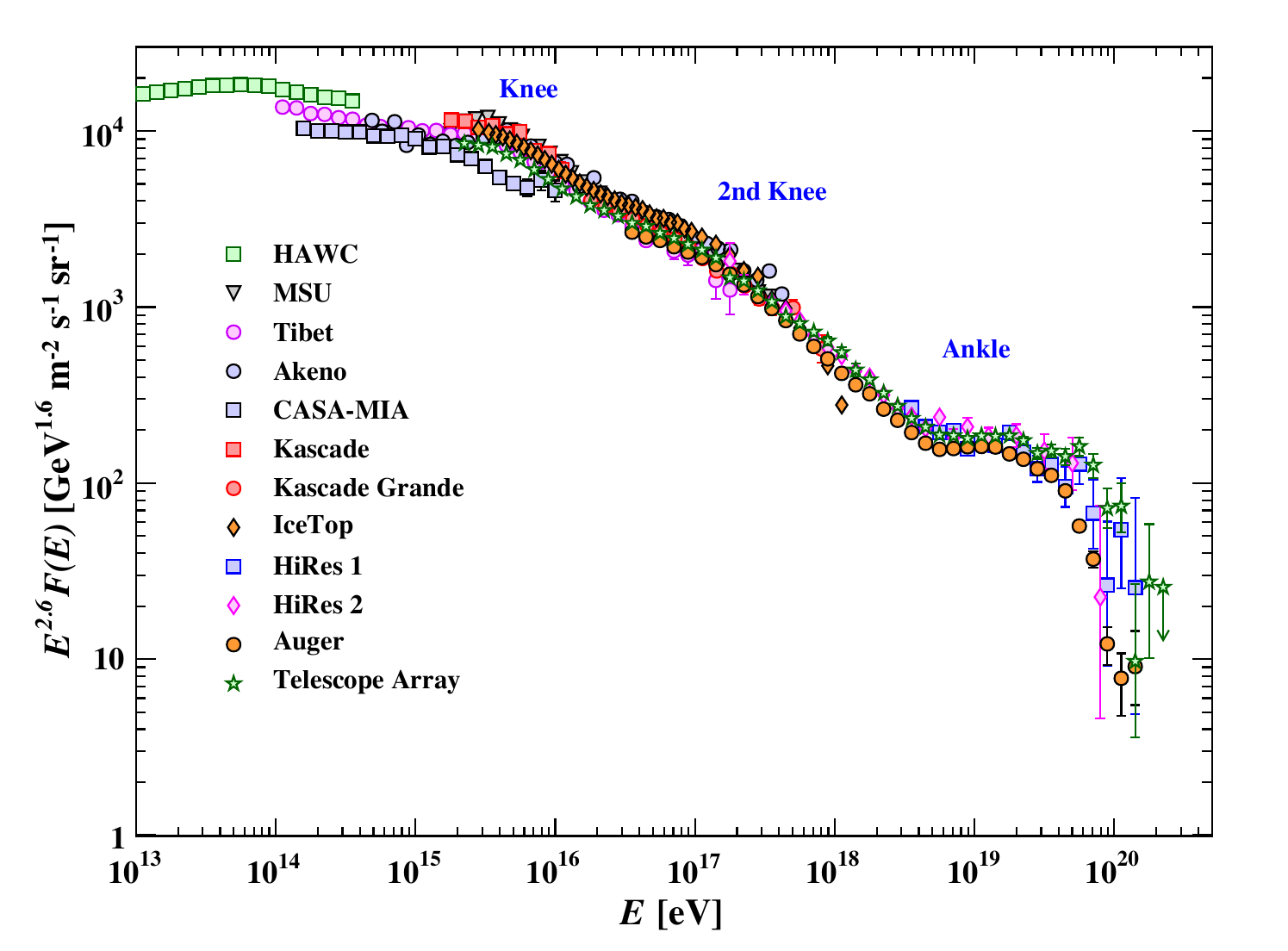}
 \caption{At the energies of interest here, the cosmic-ray spectrum
follows a sequence of three power laws. The first two are separated by the
``knee,'' the second and third by the ``ankle.'' Cosmic
rays beyond the ankle are a new population of particles produced in
extragalactic sources. Note that the spectrum $F(E)(=dN/dE)$ has been multiplied by a power $E^{2.7}$ in order to visually enhance the structure in the spectrum. Figure from Particle Data Group \citep{Tanabashi:2018oca}.} 
 \label{fig:crspectrum}
 \end{figure}

The construction of kilometer-scale neutrino detectors was primarily motivated by the prospect of detecting neutrinos associated with the sources of high-energy cosmic rays.
Cosmic accelerators produce particles with energies in excess of $100$\,EeV; we still do not know where or how~\cite{Sommers:2012px}; see Fig.~\ref{fig:crspectrum}\footnote{We will use energy units TeV, PeV and EeV, increasing by factors of 1000 from GeV energy.}. The bulk of the cosmic rays are Galactic in origin. Any association with our Galaxy presumably disappears at EeV energy when the gyroradius of a proton in the Galactic magnetic field exceeds its size. The cosmic-ray spectrum exhibits a rich structure above an energy of a few PeV, the so-called ``knee" in the spectrum, but where exactly the transition to extragalactic cosmic rays occurs is a matter of debate.

\subsection{Cosmic-Ray Accelerators}

The detailed blueprint for a cosmic-ray accelerator must meet two challenges: the highest-energy particles in the beam must reach energies beyond $10^3$\,TeV ($10^8$\,TeV) for Galactic (extragalactic) sources and their luminosities must accommodate the observed flux. Both requirements represent severe constraints that have guided theoretical speculations. Acceleration of protons (or nuclei) to TeV energy and above requires massive bulk flows of relativistic charged particles. Instead of accelerating protons, cosmic accelerators can boost really large masses to relativistic velocities. The radio emission reveals that the plasma in the jets of active galaxies flows with velocities of $0.99\,c$. A fraction of a solar mass per year can be accelerated to relativistic Lorentz factors of order 10 leading to luminosities of $10^{46}$\,erg/s close to the Eddington limit. In the collapse of very massive stars $10^{51} \sim 10^{52}$\,erg/s is released in a fireball that expands with velocities of $0.999\,c$.

\begin{figure}
\centering
\includegraphics[width=0.9\columnwidth]{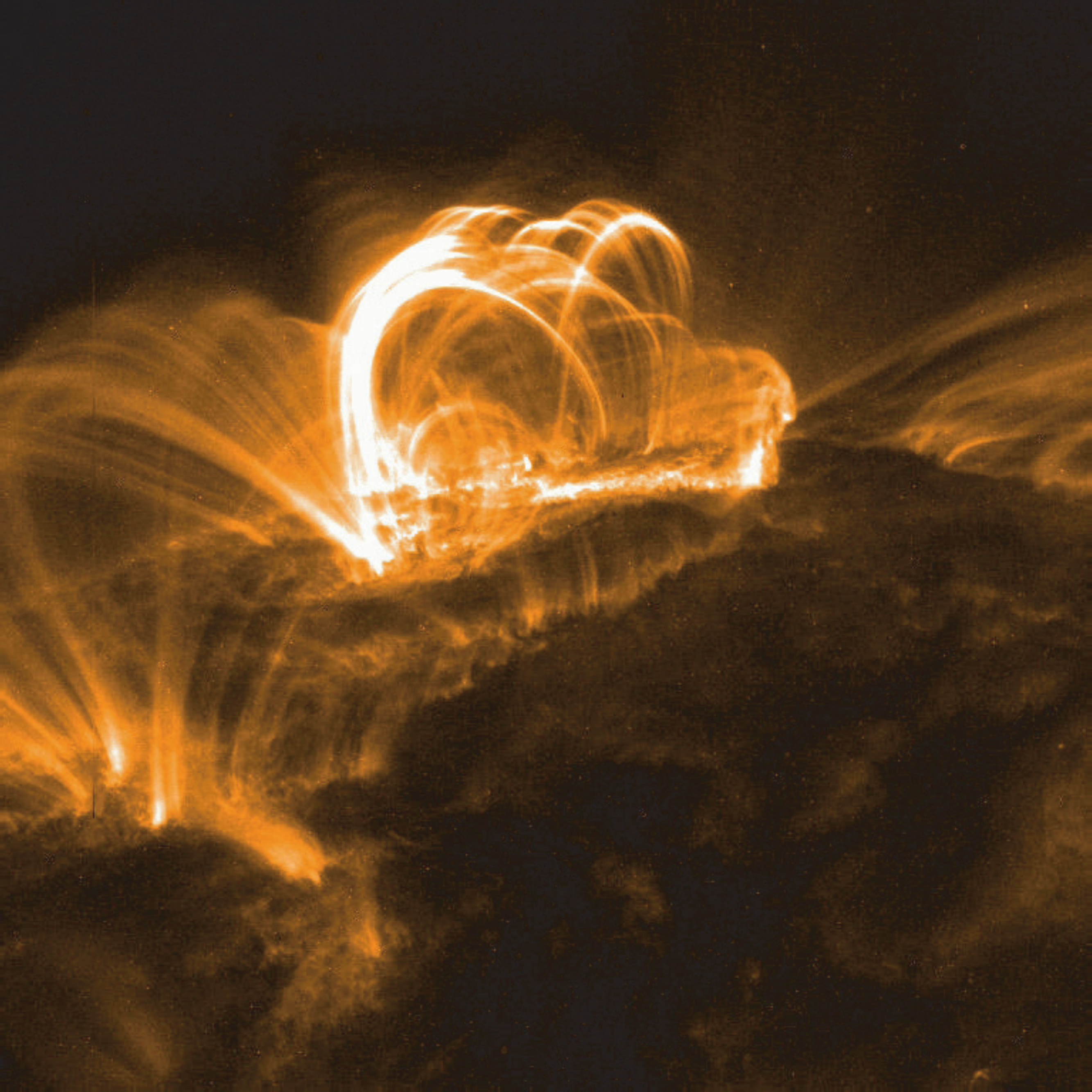}
\caption{Opportunities exist near intense charged particle flows, seen as filaments in this X-ray picture of a solar flare, for solar particles to accelerate to GeV energy.}
\label{fig:solar}
\end{figure}

The blueprint of the accelerator can be copied from solar flares where particles are accelerated to GeV energy by shocks and, possibly, magnetic reconnection; see Fig.~\ref{fig:solar}. Requiring that the gyroradius of the accelerated particle be contained within the accelerating B-field region, $E/ZevB \leq R$, leads to an upper limit on the energy of the particle, the Hillas \cite{hillas1984origin} formula
    
\begin{equation}
E \leq Ze\,v\,B\,R\,.
\end{equation}
Reaching energies much above 10\,GeV in solar flares is dimensionally impossible. In a solar flare, the extent $R$ of the accelerating region and the magnitude of the magnetic fields $B$ are not large enough to accelerate particles of charge $Ze$ to energies beyond GeV even if their velocity is taken to be the speed of light, $c$. Another way to view the dimensional argument is by estimating the energy of a particle from the Lorentz force

\begin{equation}
E = -e \frac{d\Phi}{dt} = e \, \pi R^2 \, \frac{dB}{dt}  \simeq 1\rm{GeV} \big(\frac{R}{10^4 \rm \, km} \big)^2 \big(\frac{\Delta B}{10^3 \, \rm G}) \big(\frac{10^5 \rm \, s}{\Delta t} \big),
\end{equation}
\\
where $\Phi$ is the magnetic flux set up in the loops of gyrating particles in Fig.~\ref{fig:solar} that can reach several thousand Gauss in a time of order one day. The result follows from the loops have radii of more than $10^4$\,km. In the spirit of dimensional analysis the two estimates above are the same by identifying the velocity $v$ with $dR/dt$.

While it is not a challenge to find astronomical sources with larger B and R, the other challenge is that the luminosity of the cosmic ray sources is large as well, and here a central idea for accommodating the high luminosities of the Galactic and extragalactic cosmic rays observed is that a fraction of the gravitational energy released in a stellar collapse is converted into particle acceleration, presumably by shocks.

\begin{figure}[ht!]
\centering
\includegraphics[trim =0 400 2 50, clip,width=0.9\columnwidth]{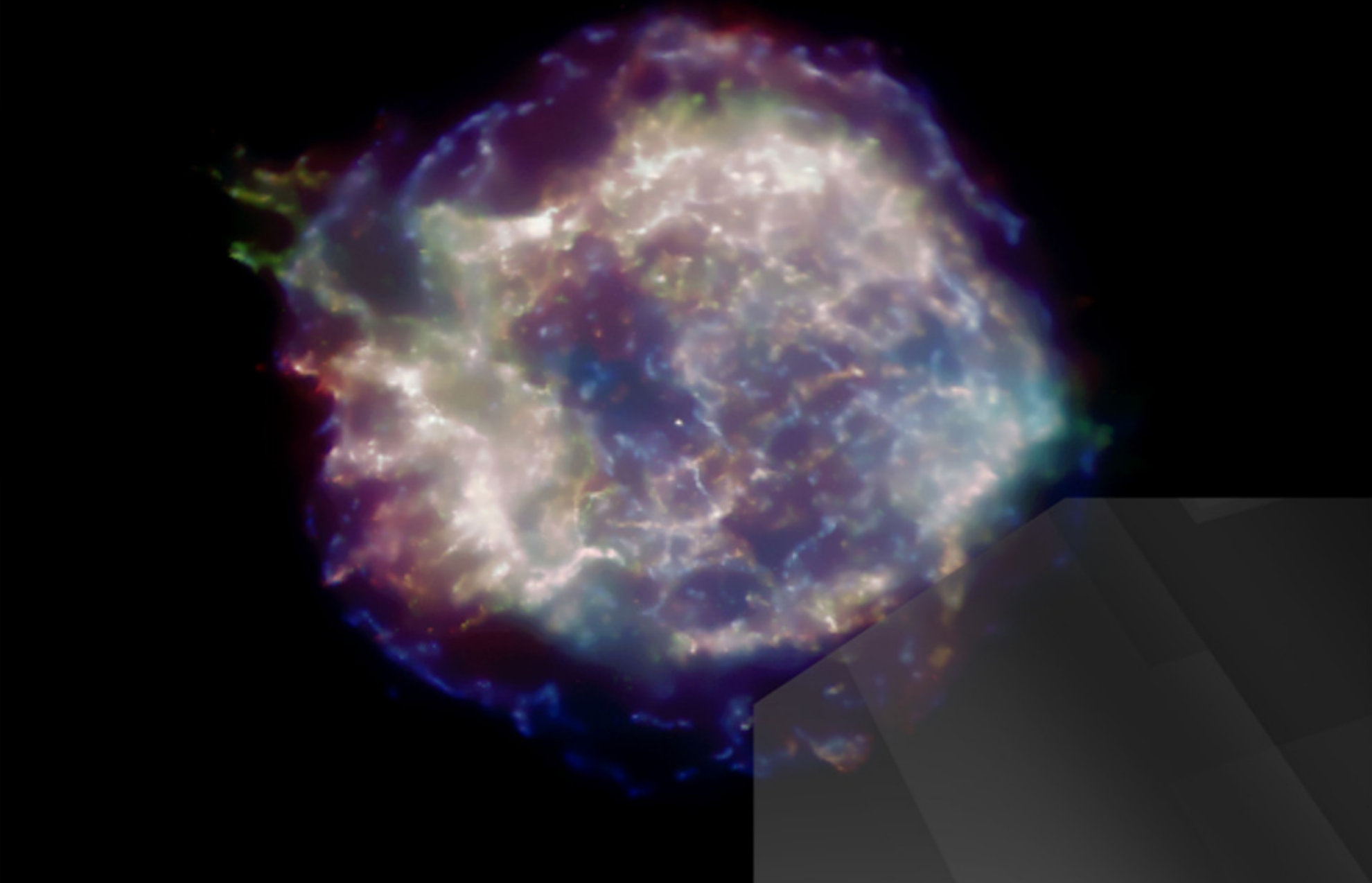}
\caption{This X-ray picture of the supernova remnant CasA reveals strong particle flows near its periphery. We believe they are the site for accelerating Galactic cosmic rays to energies reaching the ``knee" in the spectrum.}
\label{fig:casA}
\end{figure}

\begin{figure}
\centering
\includegraphics[trim =2 395 2 2, clip,width=0.9\columnwidth]{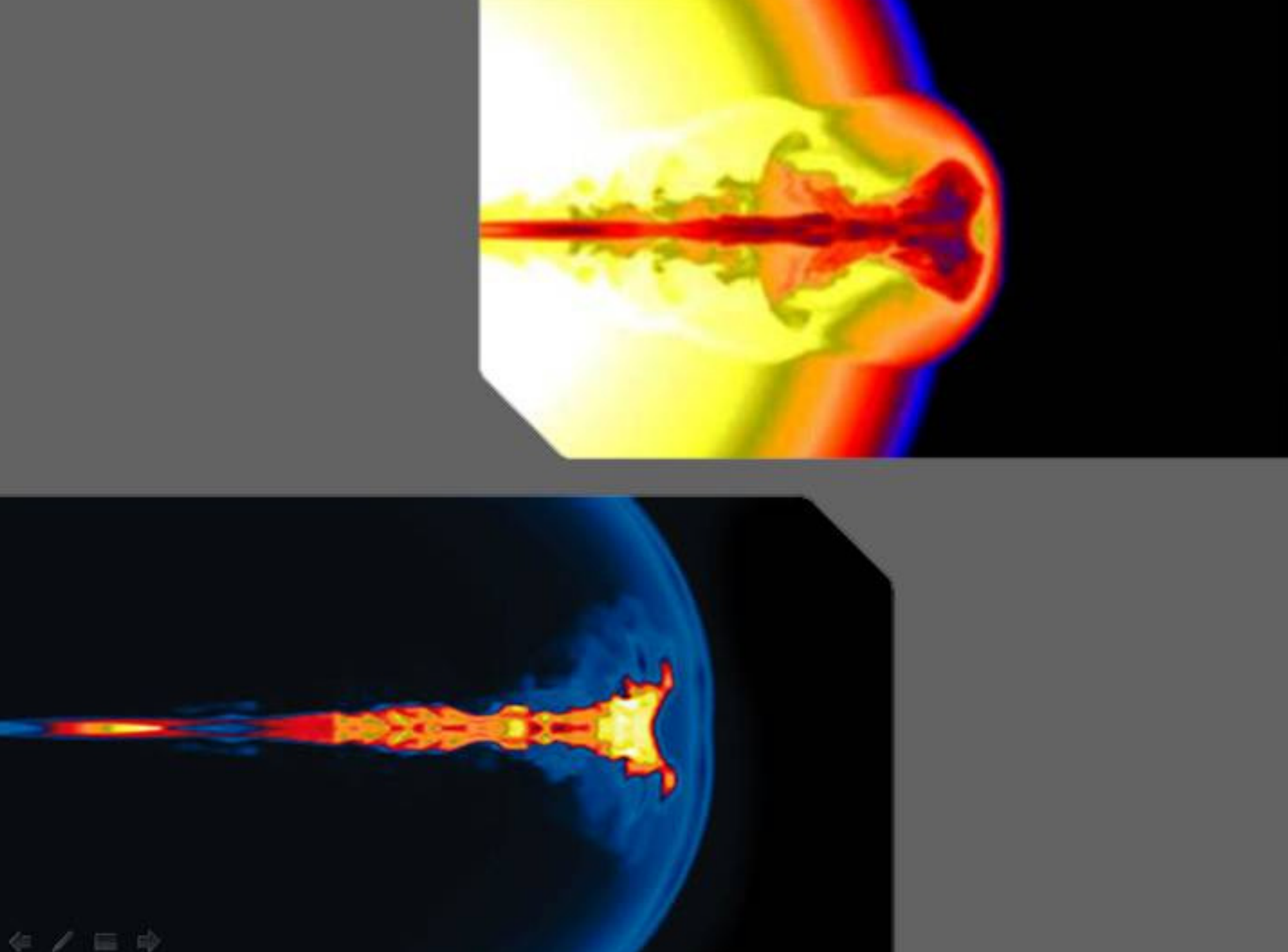}
\caption{Colliding shocks in the simulation of a gamma-ray burst (GRB) fireball may accelerate cosmic rays to the highest energies observed. The filaments in the particle flow are directed along the rotation axis of the black hole. Animated view at http://www.nasa.gov/centers/goddard/news/topstory/2003/0618rosettaburst.html.}
\label{fig:GRBsim}
\end{figure}

Baade and Zwicky~\cite{Baade1934} suggested as early as 1934 that supernova remnants could be sources of the Galactic cosmic rays. It is assumed that, after the collapse, $\sim\!\! 10^{51}$\,erg of energy is transformed into particle acceleration by diffusive shocks associated with young ($\sim\!\! 1000$ year old) supernova remnants expanding into the interstellar medium. Like a snowplow, the shock sweeps up the $\sim\!\! \,1\,\rm proton/cm^3$ density of hydrogen in the Galactic plane. The accumulation of dense filaments of particles in the outer reaches of the shock, clearly visible as sources of intense X-ray emission, are the sites of high magnetic fields; see Fig.~\ref{fig:casA}. It is theorized that particles crossing these structures multiple times can be accelerated to high energies following an approximate power-law spectrum $dN/dE \sim\!\!  E^{-2}$. The mechanism copies solar flares where filaments of high magnetic fields, visible in Fig.~\ref{fig:solar}, are the sites for accelerating nuclear particles to tens of GeV. The higher energies reached in supernova remnants are the consequence of particle flows of much larger intensity powered by the gravitational energy released in the stellar collapse.

This idea has been widely accepted despite the fact that to date no source has been conclusively identified, neither by cosmic rays nor by accompanying gamma rays and neutrinos produced when the cosmic rays interact with Galactic hydrogen. Galactic cosmic rays reach energies of at least several PeV, the ``knee" in the spectrum; therefore, their interactions should generate gamma rays and neutrinos from the decay of secondary pions reaching hundreds of TeV. Such sources, referred to as PeVatrons, have not been found; see, however, reference~\citen{pevatron}. Nevertheless, Zwicky's suggestion has become the stuff of textbooks, and the reason is energetics: three Galactic supernova explosions per century converting a reasonable fraction of a solar mass into particle acceleration can accommodate the steady flux of cosmic rays in the Galaxy. It is interesting to note that Zwicky originally assumed that the sources were extragalactic since the most recent supernova in the Milky Way was in 1572. After diffusion in the interstellar medium was understood, supernova explosions
in the Milky Way became the source of choice for the origin of Galactic cosmic 
rays~\cite{GinzburgSyrovatskii}, although after more than 50 years the issue is still debated~\cite{Butt2009}.

Energetics also guides speculations on the origin of extragalactic cosmic rays.
By integrating the cosmic-ray spectrum above the ankle at $\sim4$\,EeV, it is possible
to estimate~\cite{Gaisser2001} the energy density in extragalactic cosmic rays as $\sim 3 \times 10^{-19}\rm\,erg\ cm^{-3}$.  This value is rather uncertain because of our ignorance of the energy where the transition from Galactic to extragalactic sources occurs. The power required for a population of sources to generate this energy density over the Hubble time of $10^{10}$\,years is $2 \times 10^{37}\,\rm\,erg\ s^{-1}$ per Mpc$^3$. Long-duration gamma-ray bursts have been associated with the collapse of massive stars to black holes, and not to neutron stars, as is the case in a collapse powering a supernova remnant. A gamma-ray-burst fireball converts a fraction of a solar mass into the acceleration of electrons, seen as synchrotron photons. The observed energy in extragalactic cosmic rays can be accommodated with the reasonable assumption that shocks in the expanding gamma-ray burst (GRB) fireball convert roughly equal energy into the acceleration of electrons and cosmic rays~\cite{Waxman1995}; see Fig.~\ref{fig:GRBsim}. It so happens that $2 \times 10^{51}$\,erg per GRB will yield the observed energy density in cosmic rays after $10^{10}$ years, given that their rate is on the order of 300 per $\textrm{Gpc}^{3}$ per year. Hundreds of bursts per year over a Hubble time produce the observed cosmic-ray density, just as three supernovae per century accommodate the steady flux in the Galaxy.

\begin{figure}[t]
\begin{center}
\includegraphics[width=\columnwidth]{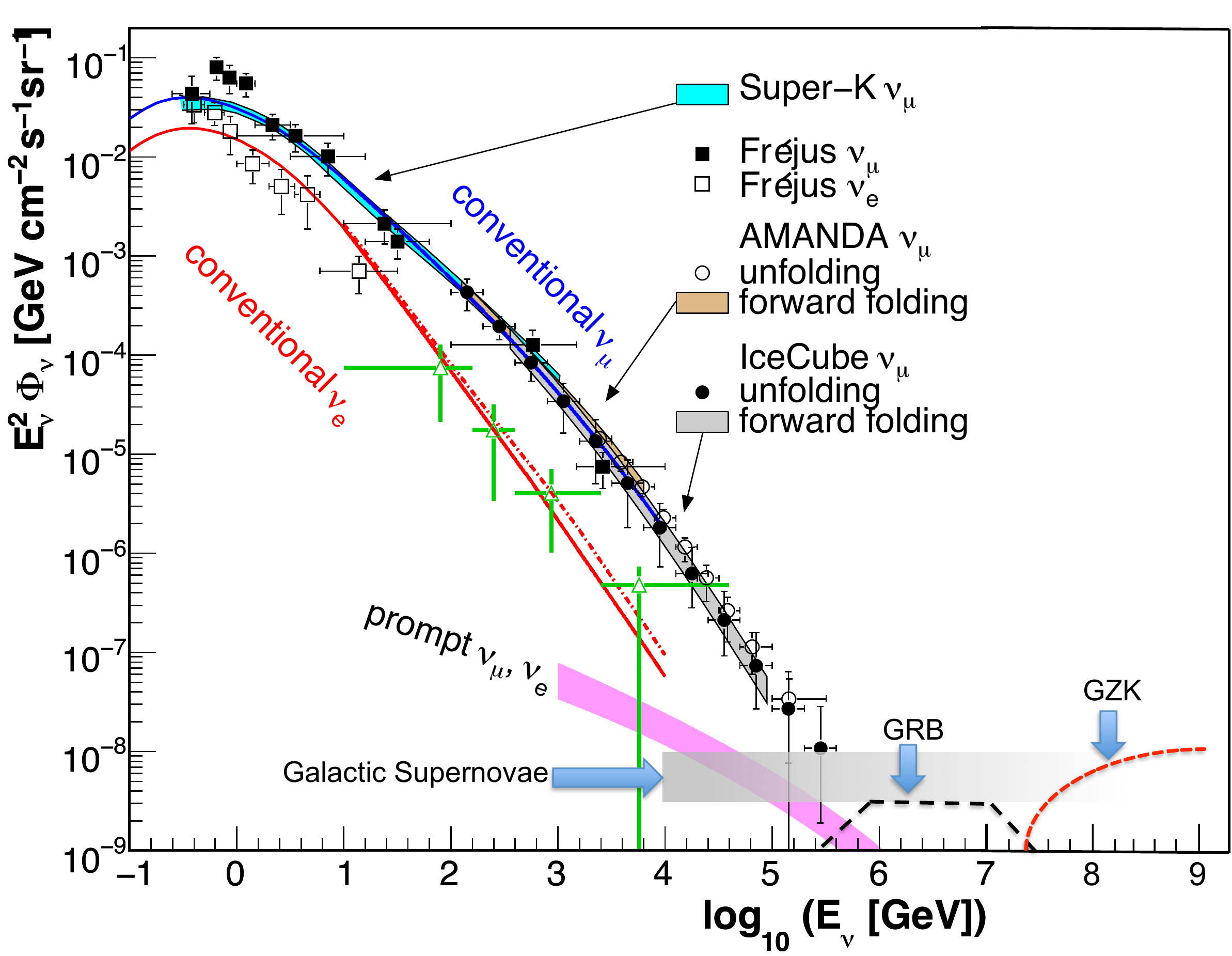}
\caption{Anticipated astrophysical neutrino fluxes compared with measured and calculated fluxes of atmospheric neutrinos.  Measurements of $\nu_\mu$ from Super-K~\cite{GonzalezGarcia:2006ay}, Frejus~\cite{Daum:1994bf}, AMANDA,~\cite{Abbasi:2009nfa,Abbasi:2010qv} and IceCube~\cite{Abbasi:2010ie,Abbasi:2011jx} are shown along with the electron-neutrino spectrum at high energy from reference~\cite{Aartsen:2012uu} (green open triangles). Calculations
of conventional $\nu_e$ (red line) and $\nu_\mu$ (blue line) from Honda et al.~\cite{Honda:2006qj}, $\nu_e$ (red dotted line) from Bartol,~\cite{Barr:2004br} and charm-induced neutrinos (magenta band)~\cite{Enberg:2008te} are also shown. }
\label{discovery}
\end{center}
\end{figure}

Problem solved? Not really: it turns out that the same result can be achieved assuming that active galactic nuclei convert, on average, $2 \times 10^{44}\,\rm\,erg\ s^{-1}$ each into particle acceleration~\cite{Becker:2007sv}. This is an amount that matches their output in electromagnetic radiation.

\begin{figure}
\centering
\includegraphics[width=\columnwidth]{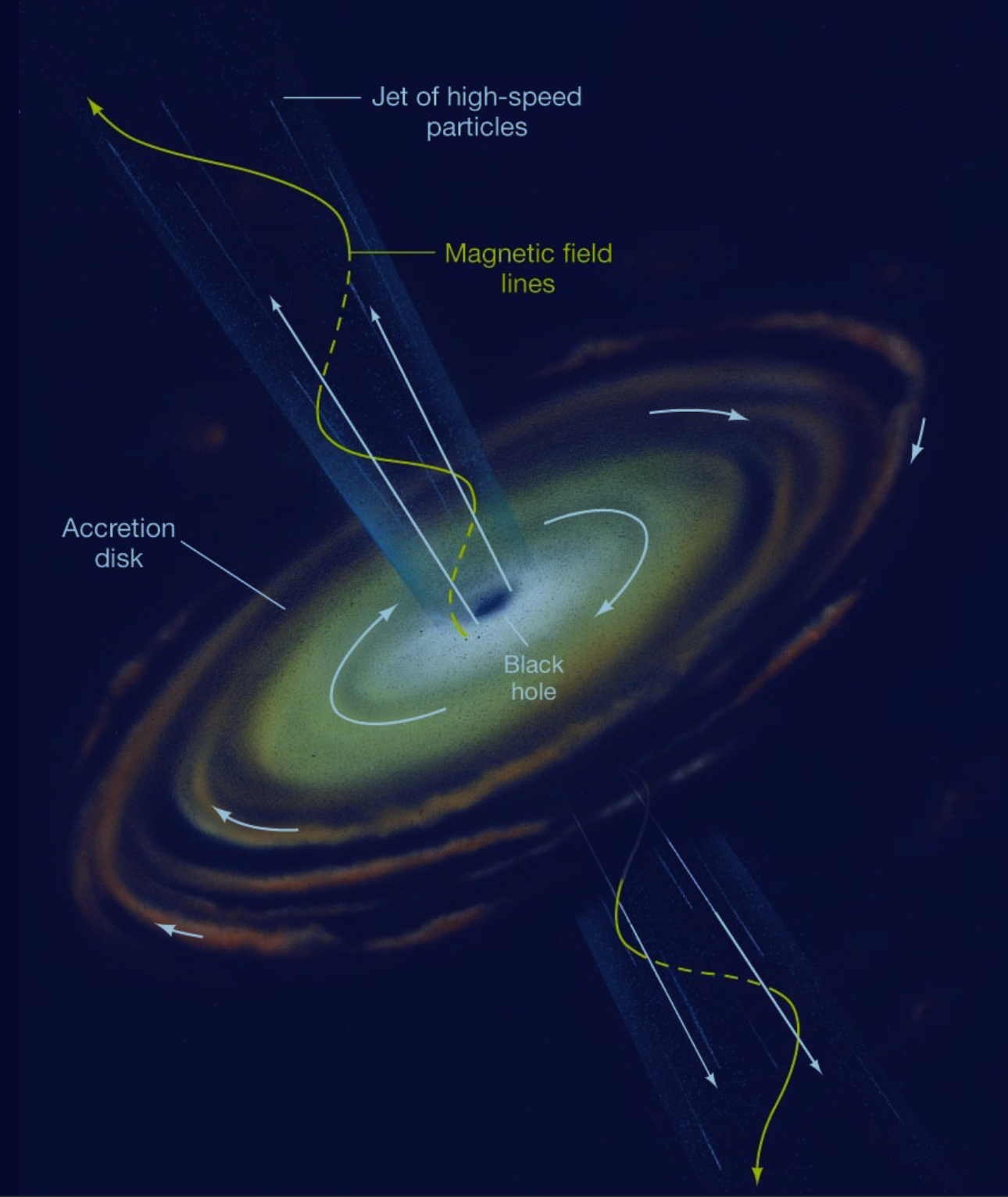}
\caption{The accretion disk meets the spinning black holes that winds up the disk's magnetic field lines. Credit: http://www.physics.ucc.ie/gabuzda/Nadia.html. \label{fig:windingB}} 
\end{figure}

In contrast with our own Galaxy where the black hole is mostly dormant, in an active galaxy the supermassive black hole is absorbing the matter in its host galaxy at a very high rate. An active galactic nucleus (AGN) hosts a rotating supermassive black hole. Fast spinning matter falling onto it swirls around the black hole in an accretion disk, like the water approaching the drain of your bath tub. When the accretion disk comes in contact with the rotating black hole space-time drags on the magnetic field winding it into a tight cone around the rotation axis into a jet of particles; see Fig.~\ref{fig:windingB}. Not just particles but huge ``blobs'' of plasma from the accretion disk are flung out along these field lines. It is not clear whether it is the rotation energy of the black hole or the magnetic energy in the rotating plasma that powers the accelerator. When this jet runs into a target material, for instance the ubiquitous 10\,eV ultraviolet photons in some galaxies, neutrinos can be produced.

Active galaxies are actually complex systems with many possible sites for accelerating cosmic rays and for targets to produce neutrinos. Acceleration of particles may occur at the spectacular termination shocks of the jets in intergalactic space~\cite{Berezhko} at distances of hundreds of Mpc from the center where there would be little target material available and few neutrinos produced. In contrast, production of neutrinos near the black hole~\cite{Tjus:2014dna}, or in collisions with interstellar matter of the accelerated particles diffusing in the magnetic field of the galaxy hosting the black hole~\cite{Hooper:2016jls}, could yield fluxes at the level observed. We will work through these examples further on.
\subsection{Neutrinos and Gamma Rays Associated with Cosmic Rays}

Neutrinos must be produced at some level in association with the cosmic-ray beam.  Cosmic rays accelerated in regions of high magnetic fields near black holes or neutron stars inevitably interact with the matter or radiation surrounding them. Thus, cosmic-ray accelerators are also part of a ``beam dump'', like the ones producing neutrino beams in accelerator laboratories: the beam is dumped is a target where it produces pions and kaons that decay into neutrinos. Dense targets absorb all secondary particles except for the neutrinos. This is typically not the case for a cosmic beam dump where neutrinos are expected to be accompanied by other stable particles: protons, neutrons and photons. For example, cosmic rays accelerated in supernova shocks interact with gas in the Galactic disk, producing equal numbers of pions of all three charges that decay into pionic photons and neutrinos. A larger source of secondaries is likely to be produced by the interaction of accelerated particles with the gas near the sources, for example cosmic rays interacting with high-density molecular clouds that are ubiquitous in the star-forming regions where supernovae are more likely to explode. For extragalactic sources, the neutrino-producing target may be electromagnetic, for instance photons radiated by the accretion disk of an AGN, or synchrotron photons that coexist with protons in the expanding fireball producing a GRB.

How many neutrinos and, inevitably, gamma rays are produced in association with the
cosmic-ray beam? A Galactic supernova shock is an example of a hadronic beam dump. Cosmic rays interact with the hydrogen in the Galactic disk, producing equal numbers
of pions of all three charges in hadronic collisions $p+p \rightarrow
n_\pi\,[\,\pi^{0}+\pi^{+} +\pi^{-}]+X$; $n_\pi$ is the pion multiplicity. In the case of a photon target, neutral and
charged pion secondaries are produced by the photoproduction processes
\begin{equation}
p + \gamma \rightarrow \Delta^+ \rightarrow \pi^0 + p
\mbox{ \ and \ }
p + \gamma \rightarrow \Delta^+ \rightarrow \pi^+ + n.
\label{eq:delta}
\end{equation}
Only the neutrinos and neutrons will escape the source. While secondary protons may remain trapped and loose energy in the high magnetic fields of the accelerator, neutrons will decay and the decay products escape the dump with high energy. The energy escaping the source is therefore distributed among cosmic rays, gamma rays and neutrinos, particles produced by the decay of neutrons, neutral pions and charged pions, respectively. Photoproduction produces charged
and neutral pions according to Eq.~\ref{eq:delta}, with probabilities of 2/3 and
1/3, respectively. Subsequently, the pions decay into gamma rays and neutrinos
that carry, on average, 1/2 and 1/4 of the energy of the parent pion. It is a good approximation to assume that, on average, the four leptons in the decay $\pi^{+}
\rightarrow \nu_{\mu} + \mu^+ \rightarrow \nu_{\mu} + \left (e^+ + {\nu}_{e} +
\bar\nu_{\mu}\right)$ equally share the charged pion's energy. The energy of the
pionic leptons relative to the proton is:
\begin{equation}
x_{\nu} = \frac {E_{\nu}}{E_{p}} = \frac {1}{4} x_\pi \; \simeq \,\frac {1}{20}
\label{eq:xnu}
\end{equation}
and
\begin{equation}
x_{\gamma} = \frac {E_{\gamma}}{E_{p}} = \frac {1}{2} x_\pi \;\simeq\, \frac {1}{10}.
\end{equation}
Here,
\begin{equation}
x_\pi \, = \, \langle 
\frac{E_{\pi}}{E_{p}}\rangle \;\simeq \,0.2
\end{equation}
is the average energy transferred from the proton to the pion that produces the neutrino  \cite{Stecker:1978ah}.

Interestingly, these relations are approximately valid whether the pions are produced in $pp$ or $p\gamma$ interactions. Pion production is usually described in terms of their average multiplicity $n_\pi$ and a pion's average energy $E_\pi$ in the final state. After interaction, the initial state energy is distributed  between the proton and the production of pions. The fraction of energy going into the production of pions is referred to as the {\em inelasticity} $\kappa$ which is $0.5(0.2)$ for $pp(p\gamma)$ interactions. The fraction of energy going into a single pion $x_\pi$ is
\begin{equation}
x_\pi \, = \,\frac{\kappa}{n_\pi} \simeq 0.2.
\end{equation}
It turns out that $x_\pi \simeq 0.2$ yields the pion energy for both $pp$ and $p\gamma$ interactions, despite the very different particle processes. For photoproduction, the charged pion takes 0.2 of the initial energy and the pion multiplicity is 1.

While both gamma-ray and neutrino fluxes can be calculated knowing the luminosity of the accelerated protons and the density of the target material, their relative flux is independent of the details of the production mechanism. Their production rates ${\rm d}N/{\rm d}E{\rm d}t$ of neutrinos and gamma rays are related by known particle physics. The above discussion can be summarized as:
\begin{equation}\label{eq:Qgamma}
\frac{1}{3}\sum_{\nu_\alpha}E_\nu \frac{{\rm d}N_\nu}{{\rm d}E_\nu {\rm d}t}(E_\nu) \simeq \frac{K_\pi}{2}E_\gamma \frac{{\rm d}N_\gamma}{{\rm d}E_\gamma {\rm d}t}(E_\gamma)\,.
\end{equation}
Here, $N$ and $E$ denote the number and energy of neutrinos and gamma rays and $\nu_\alpha$ stands for the neutrino flavor. Note that this relation is solid and depends only on the charged-to-neutral secondary pion ratio, with $K_\pi=1(2)$ for $p\gamma$($pp$) neutrino-producing interactions. In deriving the relative number of neutrinos and gamma rays, one must be aware of the fact that the neutrino flux represents the sum of the neutrinos and antineutrinos, which cannot be separated by current experiments: in short, a $\pi^0$ produces two $\gamma$ rays for every charged pion producing a $\nu_\mu + \bar\nu_\mu$ pair. A more detailed discussion of this relation will follow in the context of multimessenger astronomy; see Section 7.

The production rate of gamma rays at their origin described by Eq.~\ref{eq:Qgamma} is not necessarily the emission rate observed. For instance, in cosmic accelerators that efficiently produce neutrinos via $p\gamma$ interactions, the target photon field can also efficiently reduce the energy of the pionic gamma rays produced via pair production. Gamma rays with energies above the threshold for pair production will lose energy in the source. Their maximum energy, in the comoving frame, is determined by the energy of the target photons:

\begin{equation}\label{eq:pairprod}
    E_{\gamma}^{\prime \rm max} E^{\prime}_{\rm ph} < m^2_e.
\end{equation}

This is a calorimetric process that will, however, conserve the total energy of hadronic gamma rays. The production of photons in association with cosmic neutrinos is inevitable but unlike neutrinos, photons may reach Earth with reduced energy after losing energy in the target and after propagation in the universal microwave and infrared photon backgrounds. However, their cascaded energy must appear in some electromagnetic wave band because of energy conservation. Furthermore, one must be aware of the fact that inverse-Compton scattering and synchrotron emission by accelerated electrons in magnetic fields in the source have the potential to produce gamma rays; not every high-energy gamma ray is pionic.

A couple of decades of modeling potential neutrino sources yielded generic predictions shown in Fig.~\ref{discovery}, estimates of astrophysical neutrino fluxes are compared with measurements of atmospheric neutrinos. While the models varied, the result were typically dictated by the relation between the gamma ray and neutrino fluxes discussed above. The shaded band indicates the level of model-dependent expectations for high-energy 
neutrinos of astrophysical origin. The estimates that we will discuss in more detail further on predicted a neutrino flux at a level of
\begin{equation}
E_\nu^2\,{\rm d}N_\nu/{\rm d}E_\nu\,{\simeq}\,10^{-8}\,{\rm GeV\,cm}^{-2}{\rm s}^{-1}{\rm sr}^{-1}
\label{energyflux}
\end{equation}
per flavor. The figure illustrates the rationale for building a kilometer-scale detector because it yields about 100 neutrino events per year in a cubic kilometer detector. We now know that this is indeed the magnitude of the cosmic component of the neutrino spectrum above 100\,TeV revealed by IceCube's data although we have found no evidence for the specific sources shown in the picture! The ongoing search by IceCube for neutrinos in coincidence with and in the direction of GRB alerts issued by astronomical telescopes has limited the GRB neutrino flux to less that 1\% of the diffuse cosmic neutrino flux actually observed by the experiment~\cite{Aartsen:2016qcr,Adrian-Martinez:2013gel}. However, this may not conclusively rule out GRBs as a source of cosmic rays; the events that produce the spectacular photon displays catalogued by astronomers as GRBs may not be the stellar collapses that are sources of high-energy neutrinos. We will return to this point further on when we discuss acceleration of cosmic rays in GRB fireballs.

\begin{figure}[htb]
\begin{center}
\includegraphics[width=1.0\columnwidth]{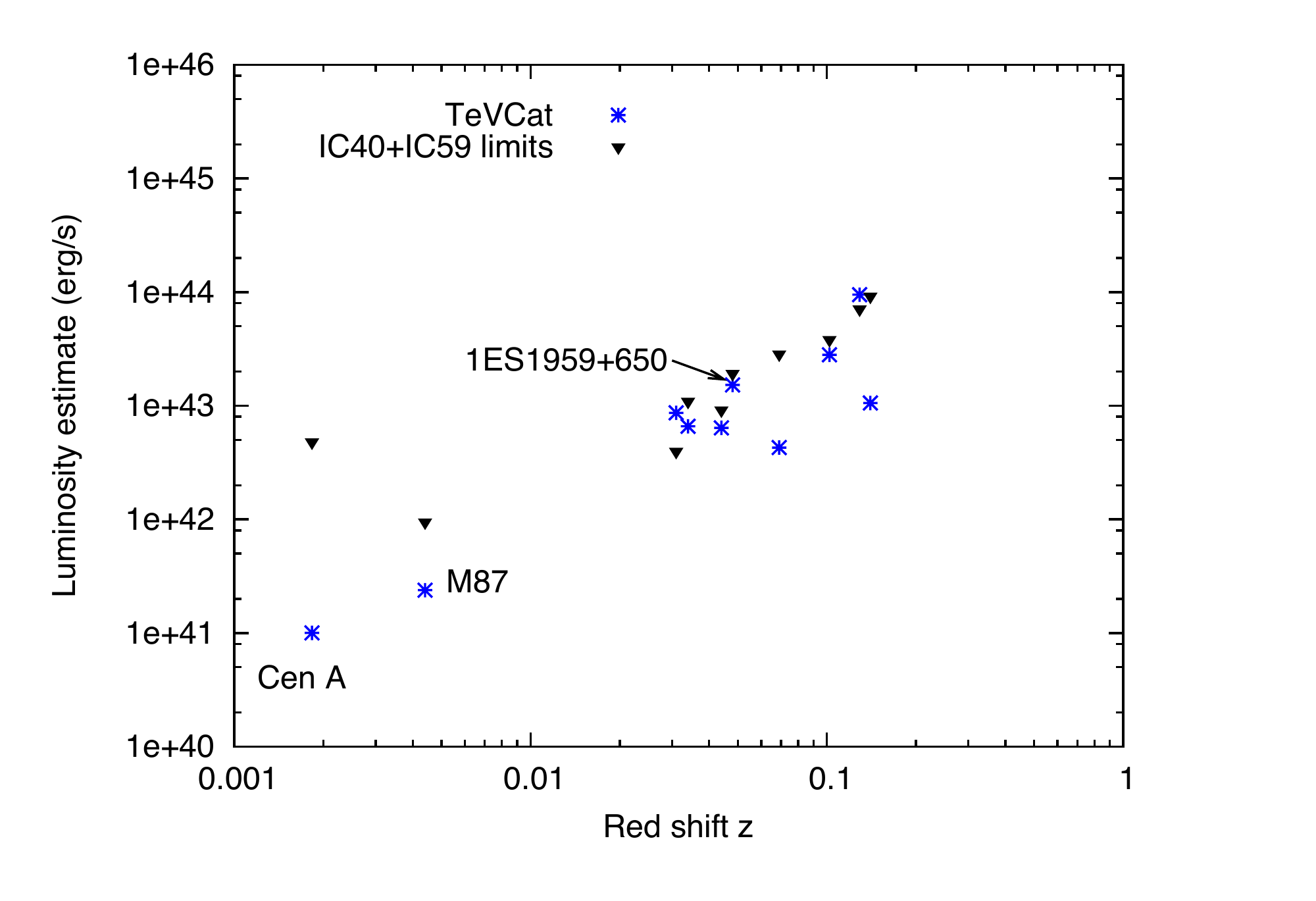}
\caption{Limits on the neutrino flux from selected active galaxies derived from IceCube data taken during construction, when the instrument was operating with 40 and 59 strings of the total 86 instrumented strings of DOMs~\cite{Abbasi:2010rd}. These are compared with the TeV photon flux for nearby AGNs. Note that energy units are in ergs, not TeV.}
\label{agnem}
\end{center}
\end{figure}

The failure of IceCube to observe neutrinos from GRBs has lately promoted AGNs as the best-bet source of the cosmic neutrinos observed. Here again Gaisser~\cite{Gaisser1997,Ahlers2005} has emphasized the relation between IceCube limits and the electromagnetic energy of the sources; see Fig.~\ref{agnem}~\cite{GaisserCERN}. In this context, we introduce Fig.~\ref{agnem}~\cite{GaisserCERN} showing IceCube upper limits~\cite{Abbasi:2010rd} on the neutrino flux from nearby AGNs as a function of their distance.  The sources at red shifts between 0.03 and 0.2 are Northern Hemisphere blazars
for which distances and intensities are listed in TeVCat~\cite{TeVCat} and for which IceCube also has upper limits.
In several cases, the muon-neutrino limits have reached the level of the TeV photon flux. One can sum the sources shown in the figure into a diffuse flux. The result, after accounting
for the distances and luminosities, is $3 \times 10^{-9}\,\rm GeV\,cm^{-2}\,s^{-1}\,sr^{-1}$, or approximately $10^{-8}\,\rm GeV\,cm^{-2}\,s^{-1}\,sr^{-1}$ for all neutrino flavors. This is at the level of the generic astrophysical neutrino flux of Eq.~\ref{energyflux}.
At this intensity, neutrinos from theorized cosmic-ray accelerators will cross the steeply falling atmospheric neutrino flux above an energy of $\sim300$\,TeV; see Fig.~\ref{discovery}. 
The level of events observed in a cubic-kilometer neutrino detector is $10\sim100$ 
$\nu_\mu$-induced events per year. Such estimates reinforce the logic for building a cubic kilometer neutrino detector~\cite{Halzen:2013bta}.

\section{IceCube}

\subsection{Detecting Very High Energy Neutrinos}

Cosmic rays have been studied for more than a century. They reach energies in excess of $10^8$\,TeV, populating the extreme universe that is opaque to photons because they interact with the background radiation fields, mostly microwave photons, before reaching Earth. We don't yet know where or how cosmic rays are accelerated to these extreme energies, and with the recent observation of a the rotating supermassive black hole TXS 0506+056 in coincidence with the direction and time of a very high energy muon neutrino, neutrino astronomy might have taken a first step in solving this puzzle~\cite{Kotera:2011cp,Ahlers:2015lln}. The rationale is however simple: near neutron stars and black holes, gravitational energy released in the accretion of matter or binary mergers can power the acceleration of protons or heavier nuclei that subsequently interact with gas (``$pp$'') or ambient radiation (``$p\gamma$''). Neutrinos are produced by cosmic-ray interactions at various epochs: in their sources during their acceleration, in the source environment after their release, and while propagating through universal radiation backgrounds from the source to Earth.

Because of their weak interactions, high-energy neutrinos will reach our detectors without deflection or absorption. They essentially act like photons; their small mass is negligible relative to the TeV to EeV energies targeted by neutrino telescopes. They do however oscillate over cosmic distances. For instance, for an initial neutrino flavor ratio of  $\nu_e:\nu_\mu:\nu_\tau \simeq 1:2:0$ from the decay of pions and muons, the oscillation-averaged composition arriving at the detector is approximately an equal mix of electron, muon, and tau neutrino flavors, $\nu_e:\nu_\mu:\nu_\tau \simeq 1:1:1$~\cite{Farzan:2008eg}.

\begin{figure}[t]
  \centering
    \includegraphics[width=1.0\linewidth]{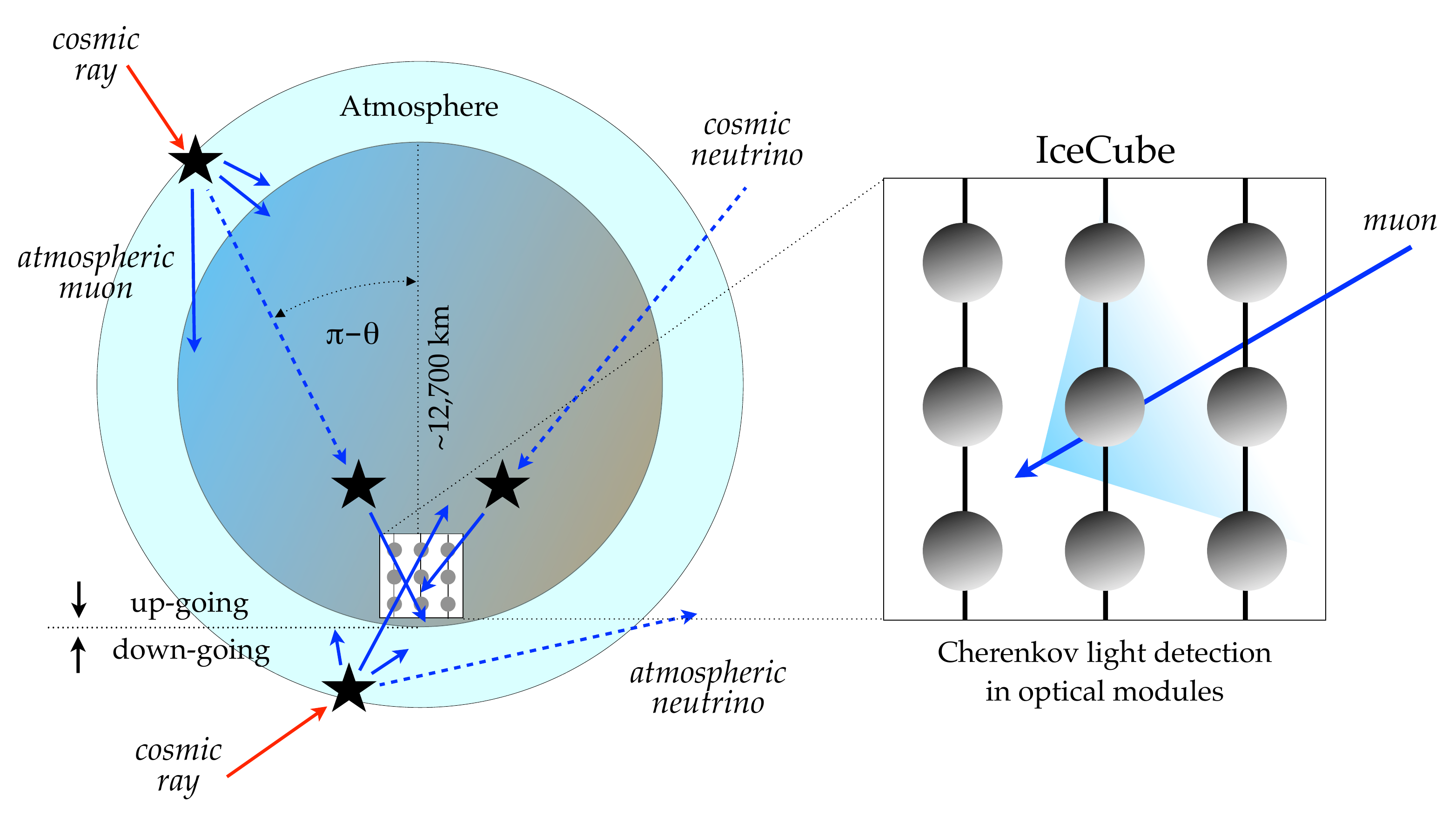}
  \caption{The concept of a neutrino telescope from the point of view of IceCube located at the South Pole. Muon neutrinos interact with the nuclei of atoms in a transparent medium like water or ice and produce a secondary muon track that is detected by the wake of Cherenkov photons it leaves inside the detector. The background of high-energy muons (solid blue arrows) produced in the atmosphere can be reduced by placing the detector underground. The surviving fraction of muons is further reduced by looking for upgoing muon tracks that originate from muon neutrinos that have traversed the Earth (dashed blue arrows) interacting close to the detector. This still leaves the contribution of muons generated by atmospheric muon neutrino interactions. This contribution can be separated from the diffuse cosmic neutrino emission by an analysis of the combined neutrino spectrum.}
 \label{earth}
\end{figure}

High-energy neutrinos interact predominantly with matter via deep inelastic scattering off nucleons: the neutrino scatters off quarks in the target nucleus by the exchange of a $Z$ or $W$ weak boson, referred to as {\it neutral current} (NC) and {\it charged current} (CC) interactions, respectively. Whereas the NC interaction leaves the neutrino state intact, in a CC interaction a charged lepton is produced that shares the initial neutrino flavor. The average relative energy fraction transferred from the neutrino to the lepton is at the level of $80$\% at high energies. The inelastic CC cross section on protons is at the level of $10^{-33}~{\rm cm}^{2}$ at a neutrino energy of $10^3$~TeV and grows with neutrino energy as $\sigma_{\rm tot}\propto E_\nu^{0.36}$~\cite{Gandhi:1995tf,CooperSarkar:2011pa}. The struck nucleus does not remain intact and its high-energy fragments typically initiate hadronic showers in the target medium.

Immense particle detectors are required to collect cosmic neutrinos in statistically significant numbers. Already by the 1970s, it had been understood~\cite{Roberts:1992re} that a kilometer-scale detector was needed to observe the cosmogenic neutrinos produced in the interactions of cosmic rays with background microwave photons~\cite{Beresinsky:1969qj, Stecker:1973sy}. A variety of methods are used to detect the high-energy secondary particles created in CC and NC neutrino interactions. One particularly effective method observes the radiation of optical Cherenkov light radiated by secondary charged particles produced in CC and NC interactions that travel faster than the speed of light in the medium.

\begin{figure}[t]
  \centering
   \includegraphics[width=1.0\linewidth]{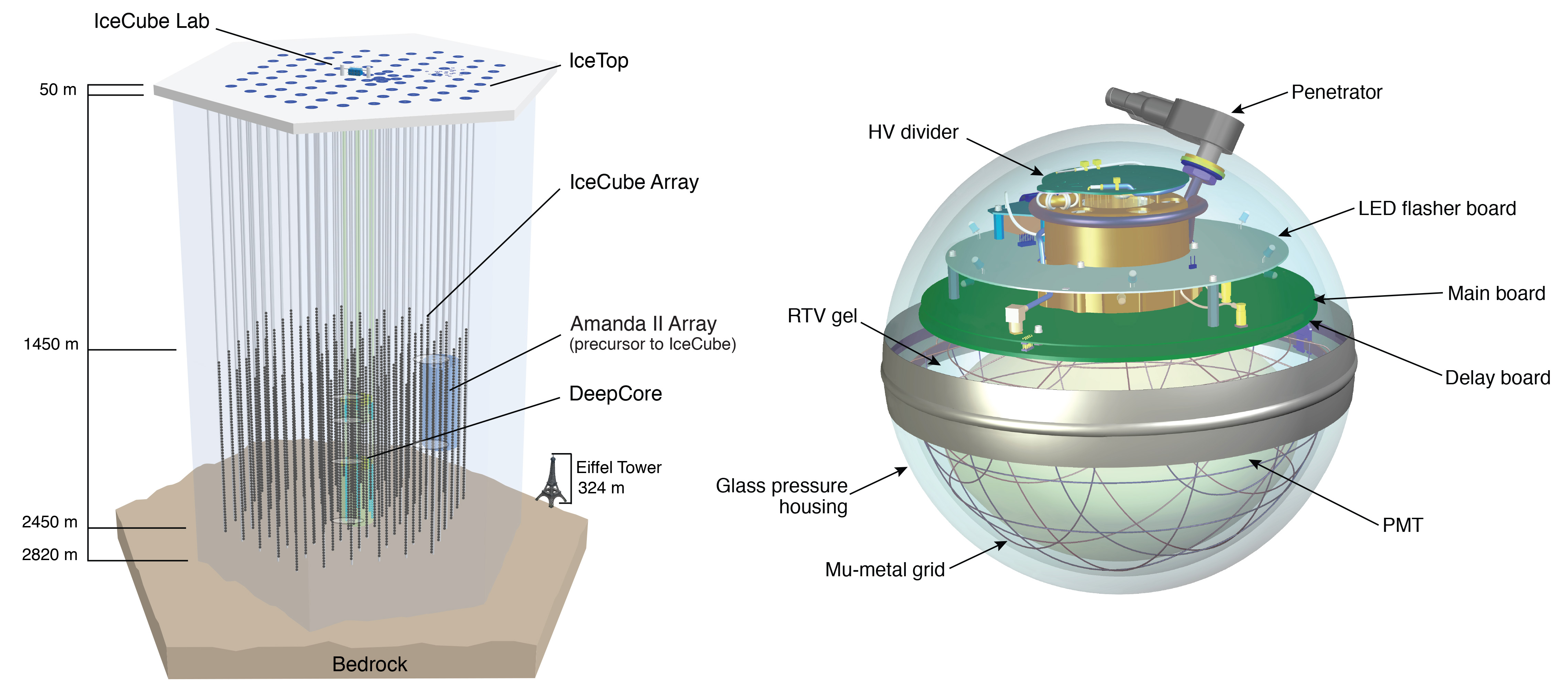}
  \caption{Sketch of the IceCube observatory (left) and the digital optical module (right).}
  \label{fig:detector_DOM}
\end{figure}

The detection concept is that of a conventional Cherenkov detector, a transparent medium is instrumented with photomultipliers that transform the Cherenkov light into electrical signals by the photoelectric effect; see Figs.~\ref{earth} and \ref{fig:detector_DOM}. IceCube consists of 80 strings, each instrumented with 60 10-inch photomultipliers spaced by 17\,m over a total length of 1 kilometer. The deepest
module is located at a depth of 2.450\,km so that the instrument is shielded
from the large background of cosmic rays at the surface by approximately 1.5\,km
of ice. Strings are arranged at apexes of equilateral triangles that are 125\,m
on a side. The instrumented detector volume is a cubic kilometer of dark, highly
transparent and sterile Antarctic ice. The radioactive background in the detector is dominated by the instrumentation deployed into this natural ice.

\begin{figure}[htb]
\begin{center}
\includegraphics[width=10cm,height=5cm,angle=0]{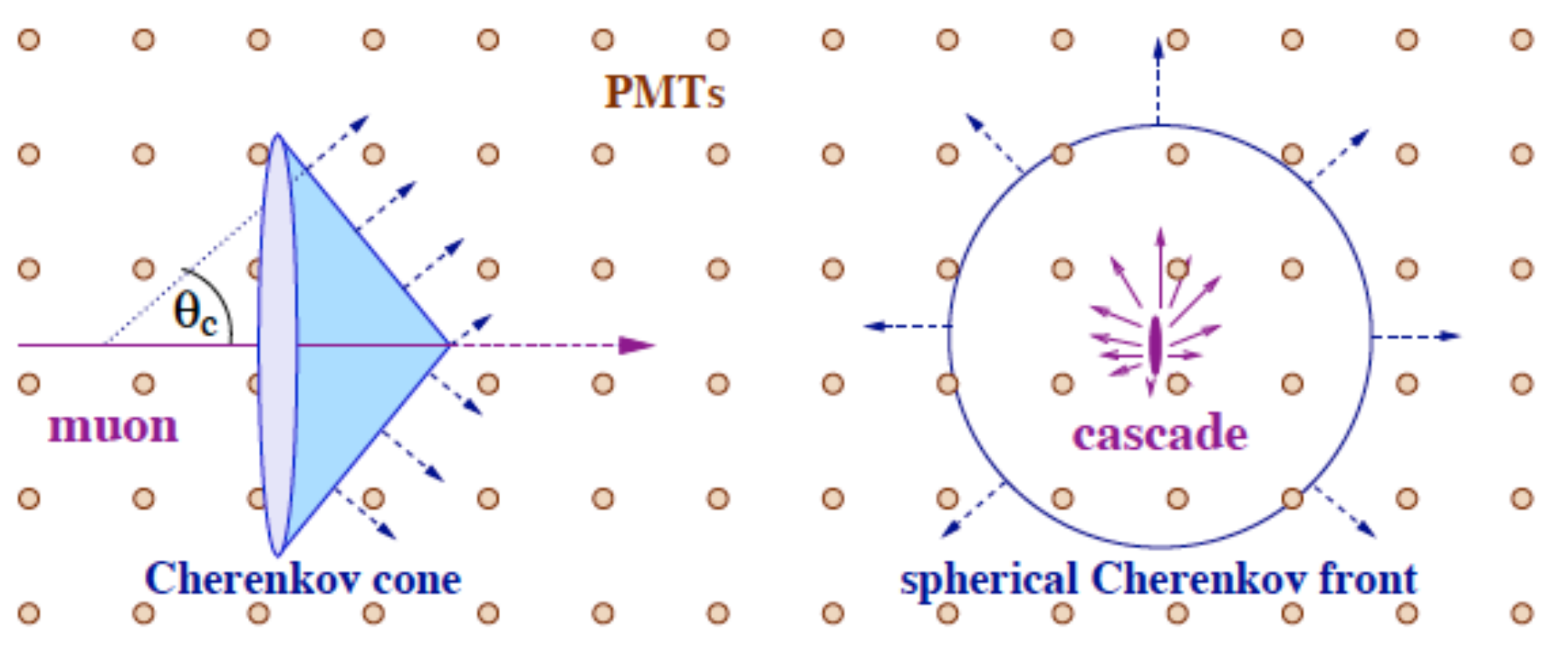}
\caption{Contrasting Cherenkov light patterns produced by muons (left) and by showers
initiated by electron and tau neutrinos (right) and by neutral current
interactions. The patterns are referred to as tracks and cascades (or
showers). Cascades are produced by the radiation of particle showers, whose 
dimensions are less than 10\,m, i.e., essentially a point source of light relative to the dimensions of the detector.}
\label{fig-XX.8}
\end{center}
\end{figure}

Each optical sensor consists of a glass sphere containing the photomultiplier
and the electronics board that captures and digitizes the signals locally using an on-board computer. The digitized signals are given a global time stamp with residuals
accurate to less than 3\,ns and are subsequently transmitted to the surface. 
Processors at the surface continuously collect the time-stamped signals from
the optical modules, each of which functions independently. The digital messages are sent
to a string processor and a global event trigger. They are subsequently sorted
into the Cherenkov patterns emitted by secondary muon tracks, or particle showers for electron and tau neutrinos, that reveal the flavor, energy and direction of the incident neutrino~\cite{Halzen:2006mq}.
There are two principle classes of Cherenkov events that must be separated by the detector, ``tracks'' and ``cascades''. The two basic topologies are illustrated in Fig.~\ref{fig-XX.8}: tracks initiated by $\nu_{\mu}$ neutrinos and cascades from $\nu_e$, $\nu_{\tau}$ neutrinos as well as the neutral current interactions from all flavors. On the scale of IceCube, PeV cascades, with a length of less than 10\,m, are therefore essentially point sources of Cherenkov light in a detector of kilometer size.
The term ``tracks'' refers to the Cherenkov emission of long-lived muons produced in CC interactions of muon neutrinos interacting inside or in the vicinity of the detector. Energetic electrons and taus produced in CC interactions of electron and tau neutrino interactions, respectively, will not produce elongated tracks: electrons will initiate electromagnetic showers in the ice and, because of the relatively short lifetime, taus will decay and also produce a shower in the ice. Because of the large background of muons produced by cosmic ray interactions in the atmosphere, the observation of muon neutrinos is typically limited to upgoing muon tracks that are produced in interactions inside or close to the detector by neutrinos that have passed through the Earth as illustrated in Fig.~\ref{earth}. The remaining background consists of atmospheric neutrinos, which are indistinguishable from cosmic neutrinos on an event-by-event basis. However, the steeply falling spectrum ($\propto E^{-3.7}$) of atmospheric neutrinos allows identifying diffuse astrophysical neutrino by a spectral analysis as we will highlight in the following sections. Above $\sim 300$\,TeV atmospheric neutrinos are relatively rare, even in a cubic kilometer detector, and every neutrino exceeding this energy is likely to be of cosmic origin. The atmospheric background is further reduced when looking in the specific direction of a point source, in particular transient neutrino sources like GRB.

The hadronic particle shower generated by the target struck by a neutrino also radiates Cherenkov photons. Because of the large multiplicity of secondary particles at these energies and the repeated scattering of the Cherenkov photons in the ice, the light pattern is essential spherical tens of meters from its point of origin. The light patterns produced by the particle showers initiated by the electron or tau produced in CC interactions of electron or tau neutrinos, respectively, will be superimposed on the hadronic cascade. The two are not typically not separated. The direction of the initial neutrino can only be reconstructed from the Cherenkov emission of secondary particles produced close to the neutrino interaction point, and the angular resolution is inferior to that for track events. 

In contrast, the energy resolution of the neutrino is superior for cascades than for tracks. For both, the observable energy of the secondaries can be estimated from the total number of Cherenkov photons after accounting for kinematic effects and detection efficiencies. The Cherenkov light observed in cascades is proportional to the energy transferred to the cascade and is often fully contained in the instrumented volume. It is actually sufficient that the cascade is partially contained, as long as one can reconstruct its actuals size. In contrast, muons produced by CC muon neutrino interactions lose energy gradually by ionization, bremsstrahlung, pair production, and photo-nuclear interactions while passing through the detector. The secondary charged particles produced in each energy-loss interaction radiate Cherenkov photons that allow for a measurement of the total energy lost by the muon in the detector. The energy deposited by the muon in the detector represents a lower limit on the initial neutrino energy. 

Catastrophically losing energy by the process mentioned above, muon tracks range out, over kilometers at TeV energy to tens of kilometers at EeV energy. Because the energy of the muon thus degrades along its track, the energy of the secondary showers decreases, which reduces the distance from the
track over which the associated Cherenkov light can trigger a PMT. The geometry of the light pool surrounding the muon track is therefore
a kilometer-long cone with a gradually decreasing radius. On average, in its first kilometer, a high-energy muon loses energy in a couple of showers with one-tenth of the muon's initial energy. So the initial radius of the cone is the radius of a shower with 10\% of the muon energy. At lower energies of hundreds of GeV and less, the muon becomes minimum-ionizing.

Because of the stochastic nature of the muon's energy loss, the relationship
between the observed energy loss inside the detector and the muon energy varies from
muon to muon. Additionally, only the muon energy lost in the detector can be determined; we do not know the energy lost before entering the instrumented volume, nor how much energy it carries out upon exiting.  An unfolding process is required to derive the
neutrino energy from the energy lost inside the detector; fortunately, it is described by well-understood Standard Model physics. One derives a probability distribution for the energy of the initial neutrino that determines its most probable value. The neutrino energy may only be determined within a factor of 2 or thereabout, depending on the energy, but the uncertainties drop out when measuring a neutrino spectrum involving multiple events. In contrast, for $\nu_e$
and $\nu_{\tau}$, the detector is a total energy calorimeter capturing all or most of the Cherenkov light produced, and the
determination of their energy is superior.

The different topologies each have advantages and disadvantages. For 
$\nu_{\mu}$ charged-current interactions, the long lever arm of muon tracks allows for a measurement of the muon direction with an angular resolution of better than $0.4^{\circ}$. Superior angular resolution can be reached for
selected high-energy events. At the highest energies the neutrino is aligned with the muon within the angular resolution and the sensitivity to point sources thus maximized. The disadvantages are a large background of atmospheric neutrinos below 100\,TeV, and of cosmic-ray muons at all energies, and the indirect
determination of the neutrino energy that must be inferred from sampling the
energy loss of the muon when it transits the detector.

Observation of $\nu_e$ and $\nu_{\tau}$ flavors represents significant
advantages. They are detected from both Northern and Southern
Hemispheres. (This is also true for $\nu_{\mu}$ with energy in excess of several hundred TeV,
where the background from the steeply falling atmospheric spectrum becomes
negligible.) At TeV energies and above, the background of atmospheric $\nu_e$ is
lower by over an order of magnitude because long-lived pions, the source of atmospheric $\nu_e$,
no longer decay, and relatively rare K-decays become the dominant source of background $\nu_e$. Atmospheric $\nu_{\tau}$, produced by oscillations, are rare above an energy of $100$\,GeV. High-energy  $\nu_{\tau}$ are therefore of cosmic origin; one such event with an energy of $100$\,TeV has been identified and represents an independent discovery of cosmic neutrinos~\cite{Stachurska:2019wfb}. Furthermore, one can establish the cosmic origin of a single single cascade event by demonstrating that the energy cannot be reached by muons and neutrinos of atmospheric origin.

Finally, $\nu_{\tau}$ are
not absorbed by the Earth~\cite{Halzen:1998be}: $\nu_{\tau}$ interacting
in the Earth produce a secondary $\nu_{\tau}$ of lower energy, either directly in
a neutral current interaction or via the decay of a secondary tau lepton
produced in a charged-current interaction. High-energy $\nu_{\tau}$ will thus
cascade down to energies of hundred of TeV where the Earth becomes transparent. 
In other words, they are detected with a reduced energy, but not absorbed. By this mechanism GZK $\nu_{\tau}$ with EeV energies produce a signal at PeV energy~\cite{Safa:2019ege}. 

Although cascades are nearly point-like sources of Cherenkov light and, in practice, spatially isotropic, the pattern of arrival times of the photons at individual optical modules reveals the direction of the secondary lepton. While a fraction of cascade events may be reconstructed to within a degree~\cite{Middell:2011nca}, the precision is inferior to 
that reached for $\nu_{\mu}$ events, typically $8^{\circ}$ using the present techniques. A campaign is underway to better characterize the optical properties of the ice and the calibration of the detector in order to improve this resolution. 

At energies above about $100$\,PeV, electromagnetic showers begin to elongate because of the Landau-Pomeranchuk-Migdal effect~\cite{Klein:2004er}. An extended length scale, associated with the abundant radiation of soft photons, results in the interaction of the secondary shower particles with two atoms. Negative interference of this process relative to interactions with a single atom results in a reduction of the energy loss. 

\subsection{The Neutrino Detector's Telescope Area}

Cosmic neutrinos must be separated from the large backgrounds of atmospheric neutrinos and atmospheric cosmic-ray muons. Two principal methods have been developed: isolating neutrinos that interact inside the instrumented volume (``starting events"), and specializing to events where a muon enters the detector from below, created by a neutrino that has traversed the Earth (``throughgoing events"), thus pointing back to its origin. In the latter case, the Earth is used as a filter for cosmic-ray muons.

For starting events, neutrinos are detected provided they interact within the detector volume, i.e., within the instrumented volume of one cubic kilometer. That probability is
\begin{equation}
P (E_\nu) = 1 - \exp[-l/\lambda_{\nu} (E_\nu)] \simeq l/\lambda_{\nu}(E_\nu) \,,
\label{eq:prob_int}
\end{equation}
where $l$($\theta$) is the path length traversed by a neutrino with zenith angle $\theta$ within the detector volume and $\lambda_{\nu} (E_\nu) = [\rho_{\rm ice} \ N_A \ \sigma_{\nu N} (E_\nu)
]^{-1}$ is the mean free path in ice for a neutrino of energy $E_{\nu}$. Here,
$\rho_{\rm ice} = 0.9~{\rm g}{\rm cm}^{-3}$ is the density of the ice, $N_A =
6.022 \times 10^{23}$ is Avogadro's number, and $\sigma_{\nu N}(E_\nu)$ is the
neutrino-nucleon cross section. The path length is determined by the detector's geometry and is typically much shorter than the neutrino mean free path $\lambda_{\nu}$.

A neutrino flux $dN/dE_\nu$ (with typical units neutrinos per GeV or erg per cm$^2$ per second) crossing a detector with energy threshold $E_\nu^{\rm th}$ and cross sectional area A($E_{\nu}$) facing the incident neutrino beam will produce
\begin{equation}
N_{ev} = T \, \int_{E_\nu^{\rm th}} 
A(E_\nu) \, P (E_\nu) \, \frac{dN}{dE_\nu}\,\, dE_\nu
\end{equation}
events after a time $T$. The ``effective" detector area A($E_{\nu}$) is a function of neutrino energy and of the zenith angle $\theta$. It isn't strictly equal to the geometric cross section of the instrumented volume facing the incoming neutrino, because
even neutrinos interacting outside the instrumented volume may produce enough
light inside the detector to be detected. In practice, A($E_{\nu}$) is
determined as a function of the incident neutrino direction and zenith angle by
a full-detector simulation, including the trigger and the cuts that are used to isolate ``starting'' events.

In contrast, throughgoing muon neutrinos will be detected provided the secondary muon reaches the detector with sufficient energy to trigger it. Because the muon travels kilometers at TeV energy and tens of kilometers at PeV energy, neutrinos are detected outside
the instrumented volume with a probability
\begin{equation}
P = \lambda_{\mu}/\lambda_{\nu}
\end{equation}
obtained by the substitution
\begin{equation}
l \rightarrow \lambda_{\mu} \,,
\end{equation}
in Eq.~\ref{eq:prob_int}.
Here, $ \lambda_{\mu}$ is the range of the muon determined by its energy losses. Values for the neutrino nucleon cross section and the range of the muon can be found in reference~\citen{PDG}.
 
At energies above tens of TeV, one has to account for the fact that the neutrinos may be absorbed in the Earth before reaching the detector. The reduced flux of $\nu_\mu$-induced muons reaching the detector is given by \cite{Gaisser1995,Learned:2000sw,Halzen:2002pg}:
\begin{equation}
\phi_\mu(E_\mu^{\rm min},\theta) =
\int_{E_\mu^{\rm min}}\,P(E_\nu,E_\mu^{\rm min})\,
\exp[-\sigma_{\rm tot}(E_\nu)\,N_A\,X(\theta)]\,\phi(E_\nu,\theta) dE_{\nu}.
\label{N_mu}
\end{equation}
The additional exponential factor accounts for the absorption of neutrinos along
a chord through the Earth of length $X(\theta)$ at zenith angle $\theta$. 

For back-of-the-envelope calculations, the $P$-function describing the probability that a neutrino is detected can be approximated by
\begin{eqnarray}
P \simeq 1.3 \times 10^{-6} E^{2.2}  && \mbox{for $E = 10^{-3}$--1 TeV} \,, \\
    \simeq 1.3 \times 10^{-6} E^{0.8}  && \mbox{for $E= 1$--$10^3$ TeV} \,.
\end{eqnarray}
At EeV energy, the increase is reduced to only $E^{0.4}$. The parametrization describes how neutrinos of higher energy are more likely to be detected because of the increase with energy of both the cross section and muon range. But at neutrino energies of tens of TeV and above, this gain is partially mitigated by absorption in the Earth. The highest energy neutrinos reach the detector from zenith angles near the horizon.

Tau neutrinos interacting outside the detector can be observed provided the secondary tau lepton reaches the
instrumented volume within its lifetime. In Eq.~\ref{eq:prob_int}, $l$ is replaced by
\begin{equation}
l \rightarrow \Gamma c \tau = \frac{E}{m} c \tau\,,
\end{equation}
where $m$, $\tau$, and $E$ are the mass, lifetime, and energy of the tau,
respectively. The tau's decay length $\lambda_{\tau} = \Gamma c \tau \approx
50~{\rm m} \times (E_\tau/\rm{PeV})$ grows linearly with energy and exceeds the range of the muon near 1\,EeV. At yet higher energies, the
tau eventually ranges out by catastrophic interactions, just like the muon,
despite the reduction of the leading energy-loss cross sections by a factor of $(m_{\mu}/m_{\tau})^2$.

At sub-PeV energies, tracks and showers produced by tau neutrinos are
difficult to distinguish from those initiated by muon and electron neutrinos, respectively. Only at PeV energies it is possible to detect both the initial neutrino interaction and the subsequent tau decay that are separated by tens of meters. Additionally, both must be contained within the detector volume; for a cubic-kilometer detector, this can be  realized for neutrinos with energies from a few hundreds of TeV to a few tens of PeV~\cite{Learned:1994wg}.

For an in-depth discussion of neutrino detection, energy measurement, and flavor
separation, and for detailed references, see the IceCube Preliminary Design
Document~\cite{ICPDD2001} and reference~\citen{Halzen:2008zz}.

\subsection{Atmospheric Neutrinos: Calibration and Background}

The 3-kHz trigger rate of the IceCube detector is dominated by atmospheric
muons from the decay of pions and kaons produced 
in the atmosphere above the detector.  Their distribution peaks near the zenith and decreases with increasing angle because the muon energy required to reach the deep detector increases. Most atmospheric muons are identified as tracks entering the detector from above and are rejected because the Earth shields the detector from atmospheric muons in the Northern Hemisphere. Even after the removal of cosmic ray muons, the neutrinos from decay of mesons produced by cosmic-ray interactions in the atmosphere are a residual background in the search for neutrinos of extraterrestrial origin. Because of the large ratio of atmospheric muons to neutrinos misreconstructed atmospheric muons remain an important source of background for most searches.

Measurement of the relatively well-established spectrum of atmospheric neutrinos is an important benchmark and a useful calibration tool for a neutrino telescope. IceCube detects an atmospheric neutrino every few minutes, more than one hundred thousand per year. The spectrum of atmospheric $\nu_\mu$ has been measured by unfolding the measured rate and energy deposition of neutrino-induced muons entering the detector from below the horizon~\cite{Abbasi:2010ie}, as shown in Fig.~\ref{discovery}. More challenging is the measurement of the flux of atmospheric electron neutrinos.  This has been achieved by making use of DeepCore, the more densely instrumented subarray in the deep center of IceCube, to identify contained shower events.
The measured spectrum of $\nu_\mu$ is used to calculate the contribution of neutral current interactions to the observed rate of showers. Subtracting the neutral current contribution leads to the measurement of the spectrum of atmospheric electron neutrinos from 100 GeV to 10 TeV~\cite{Aartsen:2012uu}; see Fig.~\ref{discovery}.

In general, atmospheric neutrinos are indistinguishable
from astrophysical neutrinos.  An important exception is when muon neutrinos reaching the detector from above can be tagged as atmospheric by detecting the muon produced in the same decay as the neutrino. The neutrino energy must be sufficiently high and the zenith angle sufficiently small that this muon to reach the detector~\cite{Schonert:2008is}. Monte Carlo simulation are performed to evaluate the rejection rate or, equivalently, the atmospheric neutrino passing rate. Also high-energy muons other than the muon associated directly with the neutrino and produced in
the same cosmic-ray shower as the neutrino are included in the {\em veto}.  In this way, the method can be extended to electron neutrinos.  In practice, the passing rate is significantly
reduced for $E_\nu>100$~TeV and zenith angles $\theta < 70^\circ$ where the ice overburden is not too large.

The spectrum of atmospheric neutrinos becomes one power steeper than the
spectrum of primary nucleons at high energy because the competition between
interaction in the atmosphere and the decay of pions and kaons increasingly suppresses their decay.
For the kaon channel, dominant at high energies, the characteristic energy for the steepening
is $E_\nu \sim 1~{\rm TeV}/\cos\theta$.  A further steepening occurs above
$100$~TeV as a consequence of the knee in the primary cosmic ray spectrum.  
In contrast, astrophysical neutrinos should reflect the cosmic-ray spectrum of the cosmic accelerator expected
to be significantly harder spectrum than atmospheric neutrinos.  Establishing
an astrophysical signal above the steep atmospheric background requires
an understanding of the atmospheric neutrino spectrum at energies of $100$~TeV and above.

Although there is some uncertainty associated with the composition through
the knee region~\cite{Gaisser:2013bla}, the major uncertainty in the spectrum
of atmospheric neutrinos at high energy is the level of charm production.  The short-lived
charmed hadrons preferentially decay, up to a characteristic energy of $10^7$~GeV,
producing ``prompt" muons and neutrinos with the same spectrum as their
parent cosmic rays.  This prompt flux of leptons has yet to be measured. 
Existing limits~\cite{Aglietta:1999ic,Schukraft:2013ya} 
allow a factor of two or three around the level predicted
by model calculations~\cite{Enberg:2008te}. For reasonable assumptions, the charm contribution is expected to dominate
the conventional spectrum above $\sim 10$~TeV for $\nu_e$, above $\sim 100$~TeV
for $\nu_\mu$, and above $\sim 1$~PeV for muons~\cite{Gaisser:2013ira}.

The expected hardening in the spectrum of atmospheric neutrinos due to prompt neutrinos, is partially degenerate with a hard astrophysical component.
However, the spectrum of astrophysical neutrinos should reflect the
spectrum of cosmic rays at their sources, which is expected to be harder than the spectrum of cosmic rays at Earth.  
It should eventually be possible with IceCube to measure the charm contribution
by requiring a consistent interpretation of neutrino flavors and cosmic-ray muons for which
there is no astrophysical component.  An additional signature of atmospheric charm
is the absence of seasonal variations for this component~\cite{Desiati:2010wt}.

As we will discuss further on, with a good understanding of the energy and zenith angle dependence of the atmospheric neutrino spectrum, supplemented by the {\em veto} technique and the shielding of the muons by the Earth, IceCube has been successful in separating a diffuse flux of cosmic neutrinos from the atmospheric backgrounds. The next step is to find the origin of this flux by identifying individual sources.

The strategy of searching for neutrino sources is to look for spatial clustering in the arrival direction of neutrinos to find any excess over the expected isotropic distribution of background. 
The technique used by IceCube to search for point sources is described in reference~\citen{Braun:2008bg}. In this method, an unbinned maximum likelihood is constructed to search for spatial clustering of the events. Significances are estimated by repeating each hypothesis test on data sets that are randomized in right ascension and dominated by background. This provides robust p-values that are largely independent of detector systematic uncertainties.

The unbinned maximum likelihood ratio method used to look for a localized, statistically significant excess of events above the background allows full use of spatial and spectral information from the data. The data are hypothesized to be a mixture of events from signal and background. 

For an event with reconstructed direction $\Vec{x}_i = (\alpha_i, \delta_i)$, the probability of originating from the source at $x_s$ is modeled as a circular two-dimensional Gaussian. The signal probability distribution function (PDF)  $\mathcal{S}_i$ incorporates directional information for each individual event and its angular uncertainty, $\sigma_i$, and the angular difference between the reconstructed direction of the event and the source:
\begin{equation}\label{SPDF}
\mathcal{S}_i ={S}_i(|\Vec{x}_i - \Vec{x}_s|, \sigma_i)\, \mathcal{E}_i(\delta_i, E_i, \gamma),
\end{equation}

where the spatial distribution is modeled as a two-dimensional Gaussian 
\begin{equation}
S_i(|\Vec{x}_i - \Vec{x}_s|, \sigma_i) = \frac{1}{2\pi \sigma_i^2} \exp \Big( {-\frac{|\Vec{x}_i - \Vec{x}_s|^2}{2\sigma_i^2}}\Big)
\end{equation}

The background PDF, $\mathcal{B}_i$, contains similar terms that describe the angular and energy distributions of background events.
The likelihood function for a point source is defined as 
\begin{equation}
\mathcal{L}(n_s,x_s, \gamma) = \prod_i^{events}\bigg( \frac{n_s}{N} \mathcal{S}_i(|x_i - x_s|,\sigma_i, E_i, \gamma) + \frac{N - n_s}{N} \mathcal{B}_i(\delta_i, E_i) \bigg)
\end{equation}

The likelihood ratio test statistic (TS) is used to perform statistical tests. The results of these tests can be clearly defined in the context of testing between two hypotheses: the null hypothesis $H_0$ and the alternative hypothesis $H_1$ that signal events exceed the background. $H_0$ represents the case of background only ($n_s = 0$).

After maximizing and determining the best fit number of signal events $\hat n_s$ and their spectral index $\hat \gamma$, the test statistic (TS) is defined as the log likelihood ratio between the null and signal hypothesis. In this case, the null hypothesis is that all events are generated from the isotropic background distribution, i.e., $n_s = 0.$ The alternative hypothesis is that $n_s$  neutrinos originate from the source. As is the case for the diffuse flux, a harder spectrum can indicate a signal. The TS is calculated as:
\begin{equation}
{\rm TS} = 2 \log \big[ \frac{\mathcal{L}(\hat n_s,\hat \gamma)}{\mathcal{L}(n_s=0)} \big]
\end{equation}

The significance of an observation is determined by comparing the observed TS to the TS distribution from data sets randomized in right ascension. The TS distribution for randomized data sets represents the probability that a given observation could occur by random chance within the data set. 
For large sample sizes, this distribution approximately follows a chi-squared distribution, where the number of degrees of freedom corresponds to the difference in the number of free parameters between the null hypothesis and the alternate hypothesis.

In addition to triggered and untriggered searches for neutrino sources, stacking a collection of candidate sources could be an effective way to enhance the discovery potential. in stacking searches, the correlation to a catalog of sources is tested instead of searching in the direction of a single source. 
The stacking likelihood is defined as

\begin{equation}
\mathcal{L}(n_s,\gamma) = \prod_i^{events} \sum_j^M  \frac{n_s}{N} \frac{w_j}{M} \mathcal{S}_i^j(|x_i - x_j|,\sigma_i, E_i, \gamma) + \frac{N - n_s}{N} \mathcal{B}_i(\delta_i, E_i)\,, 
\end{equation}

where $\mathcal{B}_i$ represents the isotropic background PDF, and $\mathcal{S}_i$ the signal PDF, for each event. $M$ is the number of sources in the catalog and $w_j$ the normalized theoretical weight for each source. This weight could be associated with properties of the individual sources in the catalog such as distance or the magnitude of their flux at some wavelength.

Either the point source search or the stacking search can be modified in order to search for the extended sources of neutrinos. For this purpose, the angular uncertainty in the spatial PDF has to be modified using the extension of the source. In this case the effective angular uncertainty for an extended source is given by 
\begin{equation}
\sigma_{\rm eff} =  \sqrt{\sigma_i^2+\sigma_{\rm ext}^2}
\end{equation}

The technique described above can be extended to a time-dependent search for transient neutrino sources \cite{Aartsen:2013uuv}. A time component is added to the signal function in Eq.~(\ref{SPDF}). This component will take into account the temporal structure of the neutrino emission. This temporal dependency can be modeled as a Heaviside function or a Gaussian, for example. 

Time-dependent searches are generally more sensitive than time-integrated searches because they accumulate lower background rates. These searches, therefore, offer an alternative opportunity to pinpoint the origin of cosmic neutrinos.

\section{The Discovery of Cosmic Neutrinos}
\label{sec:cosmicnu}

For neutrino astronomy, the first challenge is to select a pure sample of neutrinos, more than 100,000 per year above a threshold of 0.1\,TeV for IceCube, in a background of ten billion cosmic-ray muons, while the second is to identify the small fraction of these neutrinos that is astrophysical in origin, expected to be at the level of ten to hundred events per year according to the expectations of Fig.~\ref{discovery}. Atmospheric neutrinos are a background for cosmic neutrinos, at least at neutrino energies below $\sim300$\,TeV. Above this energy, the atmospheric neutrino flux reduces to less than one event per year, even in a kilometer-scale detector, and thus events in that energy range are predominantly cosmic in origin.

Searching for high-energy neutrinos of cosmic origin, IceCube continuously monitors the whole sky collecting very high statistics data sets of atmospheric neutrinos. Neutrino energies cover more than six orders of magnitude, from 5~GeV in the highly instrumented inner core (DeepCore) to beyond 10~PeV. Soon after the completion of the detector, with two years of data, IceCube discovered an extragalactic flux of cosmic neutrinos with an energy flux in the local universe that is, surprisingly, similar to that in gamma rays; see references~\citen{Ackermann:2014usa} and~\citen{Fang:2017zjf}.

\begin{figure}[t]\centering
\includegraphics[width=0.47\linewidth,viewport=20 0 200 170,clip=true]{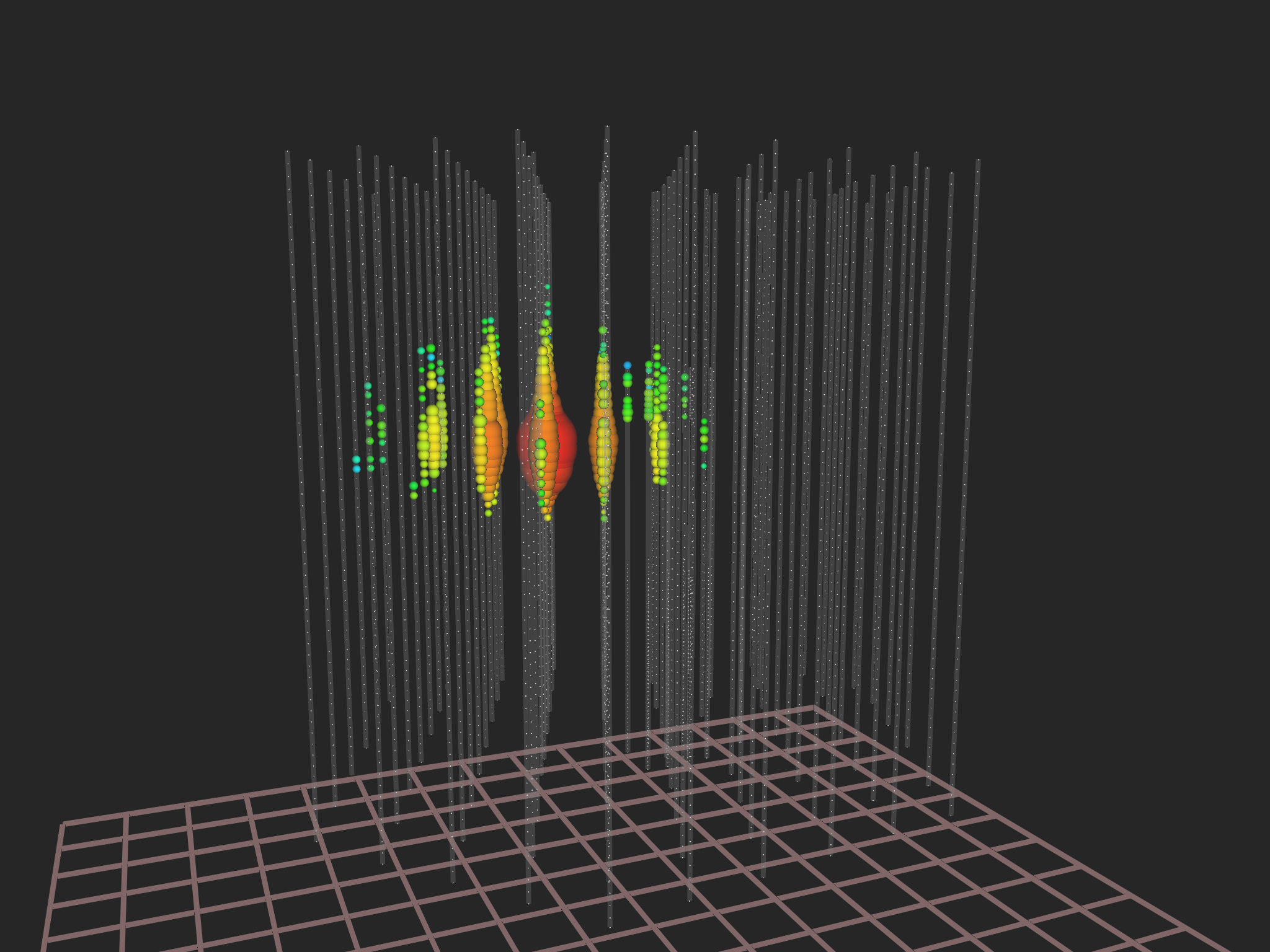}\hfill\includegraphics[width=0.47\linewidth,viewport=20 0 200 170,clip=true]{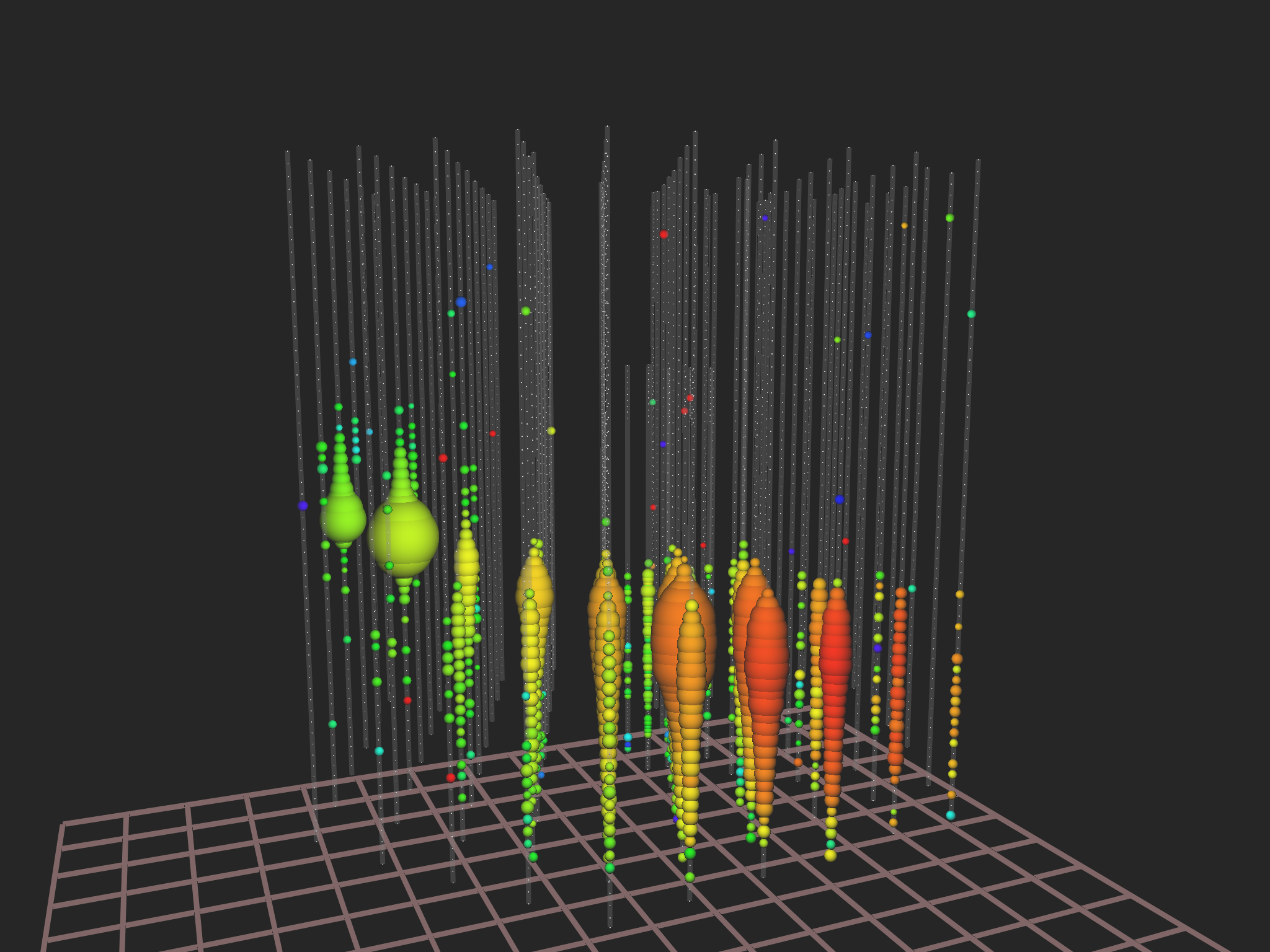}
\caption{{\bf Left Panel:}  Light pool produced in IceCube by a shower initiated by an electron or tau neutrino of $1.14$ PeV, which represents a lower limit on the energy of the neutrino that initiated the shower. White dots represent sensors with no signal. For the colored dots, the color indicates arrival time, from red (early) to purple (late) following the rainbow, and size reflects the number of photons detected. {\bf Right Panel:}  A muon track coming up through the Earth, traverses the detector at an angle of $11^\circ$ below the horizon. The deposited energy, i.e., the energy equivalent of the total Cherenkov light of all charged secondary particles inside the detector, is 2.6\,PeV.}
\label{fig:erniekloppo}
\end{figure}

Two principal methods are used to identify cosmic neutrinos. The first method reconstructs upgoing muon tracks initiated by muon neutrinos and the second identifies neutrinos of all flavors interacting inside the instrumented volume of the detector. For illustration, the Cherenkov patterns initiated by an electron (or tau) neutrino of 1\,PeV energy and a neutrino-induced muon losing 2.6\,PeV energy while traversing the detector are contrasted in Fig.~\ref{fig:erniekloppo}.

\begin{figure}[t]\centering
\includegraphics[width=0.95\textwidth]{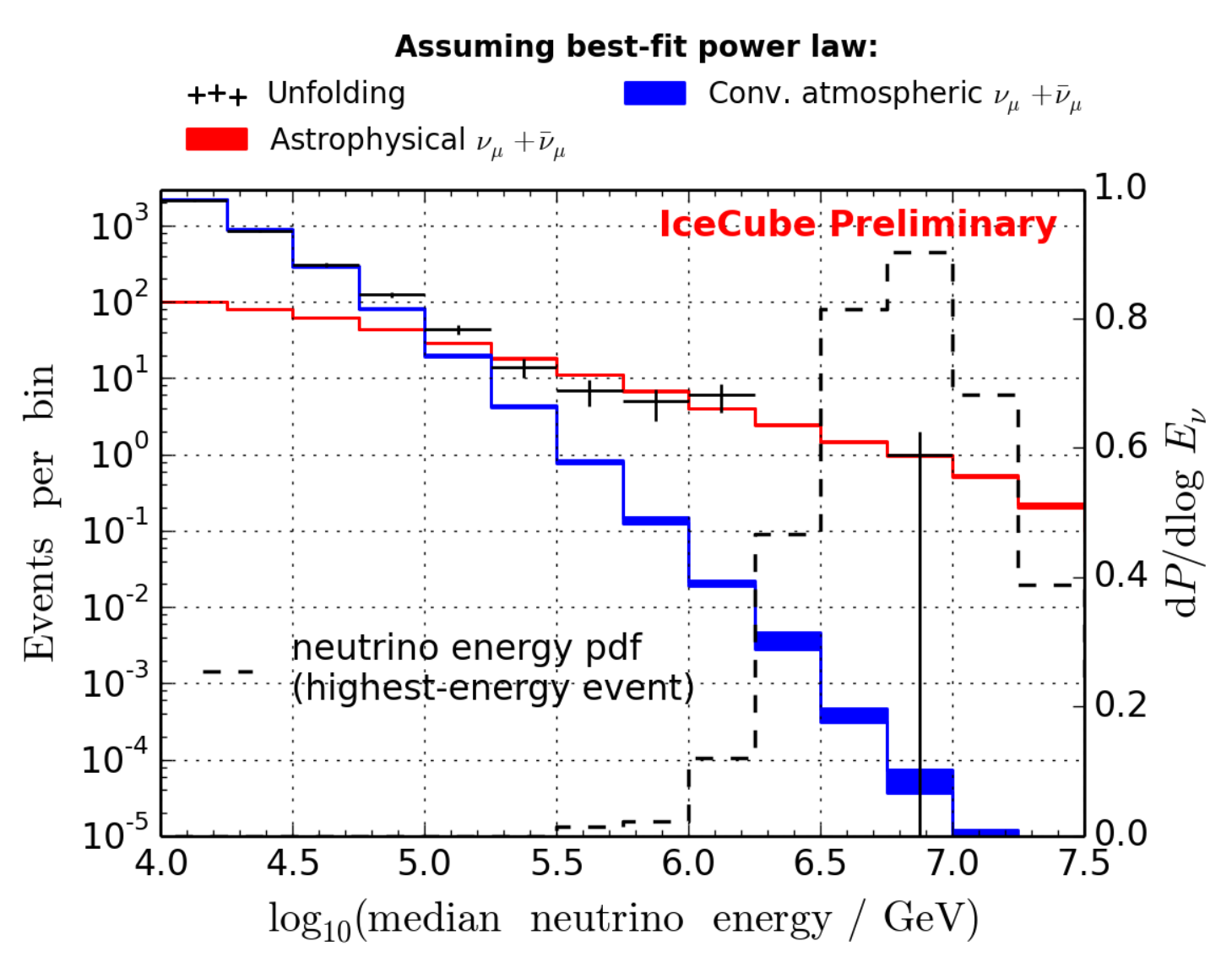}
\caption{Distribution of muon tracks arriving from the Northern Hemisphere, i.e., with declination greater than -5$^\circ$, as a function of median neutrino energy. The black crosses correspond to observed data. The blue band represents the expected conventional atmospheric neutrino. The red lines shows the high-energy cosmic neutrino flux with the best fit spectral index of 2.13. The black dashed line demonstrates the probability density function for the highest energy neutrino event when the best fit spectrum is assumed.
}
\label{fig:diffusenumu}
\end{figure}

The kilometer-long muon range makes it possible to identify neutrinos that interact outside the detector and to separate them from the the background of atmospheric muons using Earth as a filter. Using this method, IceCube has measured the background atmospheric neutrino flux over more than five orders of magnitude in energy with a result that is consistent with theoretical calculations. However, with eight years of data, IceCube has observed an excess of neutrino events at energies beyond 100\,TeV~\cite{Aartsen:2015rwa,Aartsen:2016xlq,Aartsen:2017mau} that cannot be accounted for by the atmospheric flux; see Fig.~\ref{fig:diffusenumu}. Although the detector only records the energy of the secondary muon inside the detector, from Standard Model physics we can infer the energy spectrum of the parent neutrino. The high-energy cosmic muon neutrino flux is well described by a power law with a spectral index of $2.28\pm0.10$ and a normalization at 100\,TeV neutrino energy of $(1.44^{+0.25}_{-0.24})\,\times10^{-18}\,\rm GeV^{-1}\rm cm^{-2} \rm sr^{-1}$~\cite{Stettner:2019tok}.

\begin{figure*}[t!]
\centering
\includegraphics[trim=100 0 100 0,clip,width=0.50
\columnwidth,angle=-90]{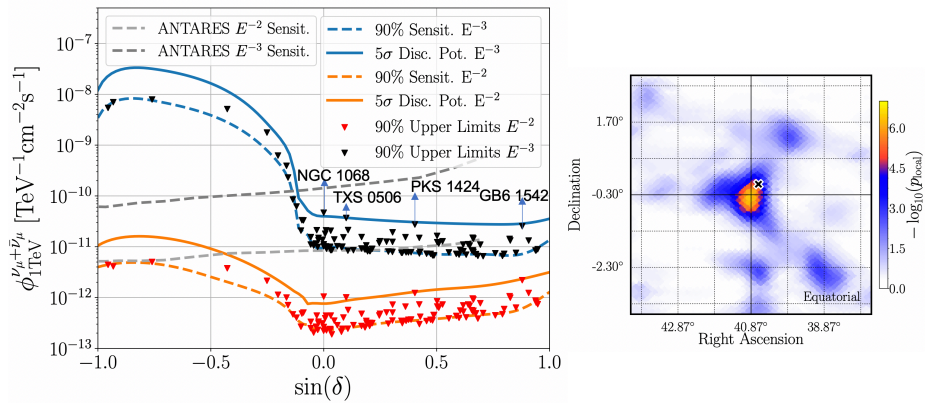}
\caption{Upper limits on the flux from candidate point sources of neutrinos in 10 years of IceCube data assuming two spectral indices of the flux. Four sources exceed the $4\sigma$ level (pretrial) and collectively result into a $3\sigma$ anisotropy of the sky map.}
\label{fig:10year}
\end{figure*}

The arrival directions of the muon tracks have been analyzed by a range of statistical methods, yielding a first surprise: there is no evidence for any correlation to sources in the Galactic plane. IceCube is recording a diffuse flux of extragalactic sources. Only after analyzing 10 years of data~\cite{Aartsen:2019fau} has evidence emerged at the $3\sigma$ level that the neutrino sky map might not be isotropic. The anisotropy results from four sources--TXS 0506+056 among them (you will hear more about that source later on), that emerge as point sources at the $4\sigma$ level (pretrial); see Fig.~\ref{fig:10year}. The strongest of these sources is the nearby active galaxy NGC 1068, also known as Messier 77. 

NGC 1068 is one of the best-studied Seyfert 2 galaxies that motivated the AGN unification model. Its measured infrared luminosity implies a high level of starburst activity. In addition, NGC 1068 has a heavily obscured nucleus and the column density of the gas surrounding it reaches to  $\simeq 10^{25} \rm \,  cm^{- 2}$ \cite{Marinucci:2015fqo}.  Therefore, the high-energy electromagnetic emission is absorbed in the Compton thick molecular gas, which even makes measuring its intrinsic X-ray emission challenging \cite{Janssen_2015, Marinucci:2015fqo}. As such, the optically thick environment at the core of NGC 1068 provides a favorable environment for the efficient production of high-energy neutrinos. The $\sim$51 signal neutrinos identified in the direction of NGC 1068~\cite{Aartsen:2019fau} implies a neutrino flux that considerably exceeds the gamma-ray emission measured by the Fermi satellite, implying a very efficient neutrino emission concurrent with suppression of the very high energy gamma rays. 

Modeling of the high-energy neutrino emission from the corona of AGN \cite{Murase:2019vdl,Inoue:2019yfs} can accommodate these conditions. In Seyfert galaxies and quasars, a magnetized corona is formed above the accretion disk on the central black hole; see Miller \& Stone (2000)~\cite{Miller:1999ix} for details. The hot, turbulent, and highly magnetized corona of AGN facilitates particle acceleration. Combined with the high density and abundance of target gas and radiation in the vicinity of the AGN core, a sizable neutrino flux can be expected \cite{Kheirandish:2021wkm}. The accompanying pionic photons will cascade down due to the high opacity of the target. The efficient production of neutrinos in the vicinity of supermassive black holes by this mechanism can accommodate the measured flux of high-energy neutrinos. We will discuss this further in Sec.~\ref{TXS} where we highlight implications of astrophysical beam dumps for neutrino emission, especially in the presence of major accretion activity onto a supermassive black hole.

The second method for separating cosmic from atmospheric neutrinos exclusively identifies high-energy neutrinos interacting inside the detector, so-called high-energy starting events. It divides the instrumented volume of ice into an outer veto shield and a $\sim500$-megaton inner fiducial volume. The advantage of focusing on neutrinos interacting inside the instrumented volume of ice is that the detector functions as a total absorption calorimeter~\cite{Aartsen:2013vja} allowing for a good energy measurement that separates cosmic from lower-energy atmospheric neutrinos. With this method, neutrinos from all directions in the sky and of all flavors can be identified, which includes both muon tracks as well as secondary showers produced by charged-current interactions of electron and tau neutrinos and neutral current interactions of neutrinos of all flavors. A sample event with a light pool of roughly one hundred thousand photoelectrons extending over more than 500 meters is shown in the left panel of Fig.~\ref{fig:erniekloppo}. The starting event sample revealed the first evidence for neutrinos of cosmic origin~\cite{Aartsen:2013bka,Aartsen:2013jdh}. Events with PeV energies, and no trace of accompanying muons from an atmospheric shower, are highly unlikely to be of atmospheric origin. The present seven-year data set contains a total of 60 neutrino events with deposited energies ranging from 60\,TeV to 10\,PeV that are likely to be of cosmic origin.

\begin{figure}[t]\centering
\includegraphics[width=0.95\textwidth]{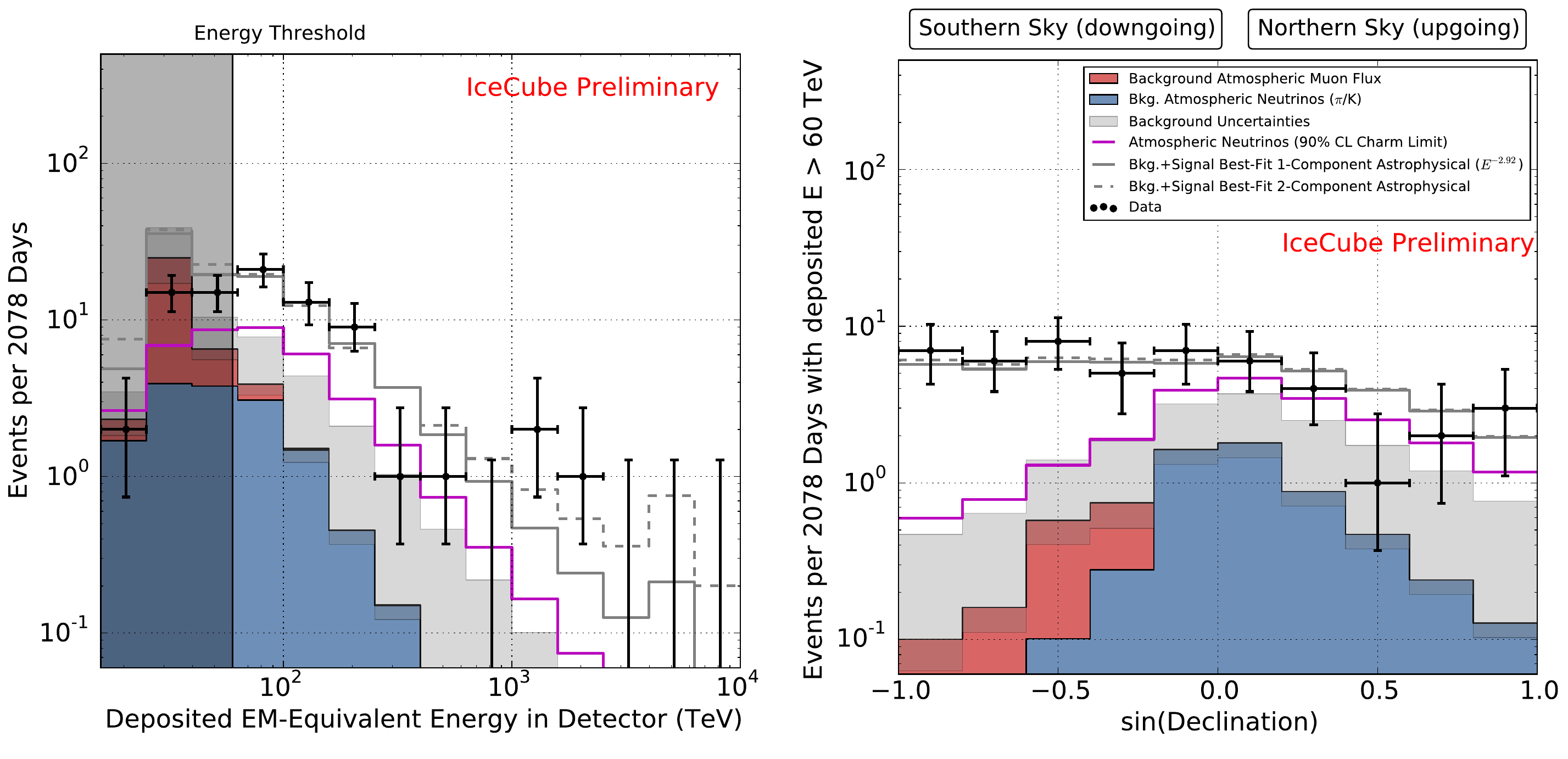}
\caption[]{{\bf Left Panel:} Deposited energies, by neutrinos interacting inside IceCube, observed in six years of data~\cite{Aartsen:2017mau}. The grey region shows uncertainties on the sum of all backgrounds. The atmospheric muon flux (blue) and its uncertainty is computed from simulation to overcome statistical limitations in our background measurement and scaled to match the total measured background rate. The atmospheric neutrino flux is derived from previous measurements of both the $\pi, K$, and charm components of the atmospheric spectrum \protect\cite{Aartsen:2013vca}. Also shown are two fits to the spectrum, assuming a simple power-law (solid gray) and a broken power-law (dashed gray). {\bf Right Panel:} The same data and models, but now showing the distribution of events with deposited energy above 60~TeV in declination. At the South Pole, the declination angle $\delta$ is equivalent to the distribution in zenith angle $\theta$ related by the identity, $\delta = \theta-\pi/2$. It is clearly visible that the data is flat in the Southern Hemisphere, as expected from the contribution of an isotropic astrophysical flux.}
\label{hese_energy}
\end{figure}

\begin{figure}[t]\centering
\includegraphics[width=0.7\linewidth]{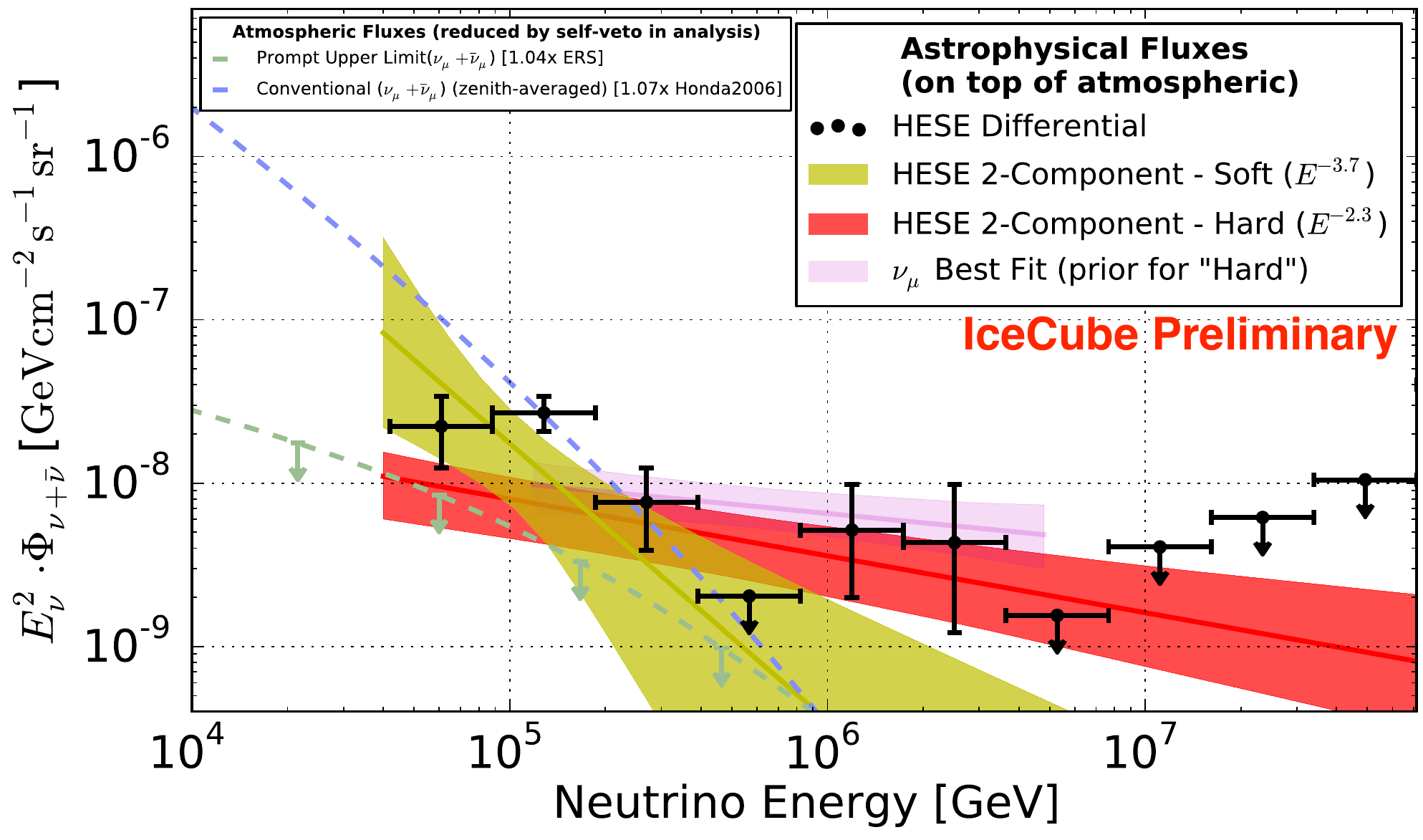}

\caption[]{Unfolded spectrum for six years of HESE neutrino events starting inside the detector. The yellow and red bands show the $1\sigma$ uncertainties on the result of a two-power-law fit. Superimposed is the best fit to eight years of the upgoing muon neutrino data (pink). Note the consistency of the red and pink bands. Figure from reference~\citen{Aartsen:2017mau}.}
\label{figs:two_component}
\end{figure}

The deposited energy and zenith dependence of the high-energy starting events~\cite{Aartsen:2017mau} is compared to the atmospheric background in Fig.~\ref{hese_energy}. The expected number of events for the best-fit astrophysical neutrino spectrum following a two-component power-law fit is shown as dashed lines in the two panels. The corresponding neutrino spectrum is also shown in Fig.~\ref{figs:two_component}. It is, above an energy of $200$\,TeV, consistent with a power-law flux of muon neutrinos penetrating the Earth inferred by the data shown in Fig.~\ref{fig:diffusenumu}. A purely atmospheric explanation of the observation is excluded at $8\sigma$.

\begin{figure*}[t!]
\centering
\includegraphics[width=0.95\linewidth]{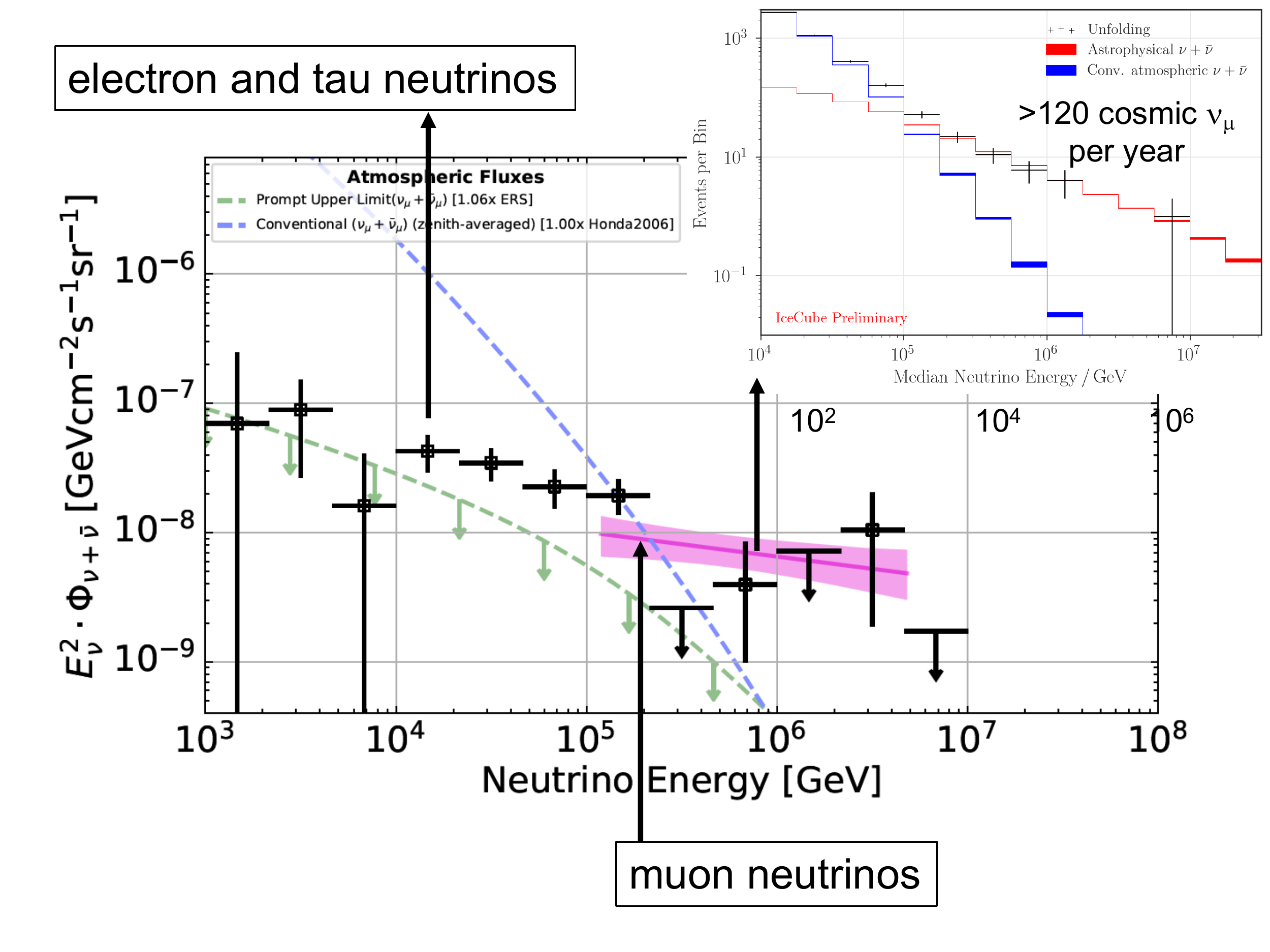}
\caption{The flux of cosmic muon neutrinos~\cite{Aartsen:2017mau} inferred from the eight-year upgoing-muon track analysis (red solid line) with $1\sigma$ uncertainty range (shaded range; from fit shown in upper-right inset) is compared with the flux of showers initiated by electron and tau neutrinos~\cite{Aartsen:2020aqd}. The measurements are consistent assuming that each flavor contributes the same flux to the diffuse spectrum.}
\label{fig:showerstracks-2}
\end{figure*}

Both measurements of the cosmic neutrino flux using cascades \cite{Aartsen:2020aqd} and throughgoing muons yield consistent determinations of the cosmic neutrino flux; see Fig.~\ref{fig:showerstracks-2}. The data are consistent with an astrophysical component with a spectrum close to $E^{-2.2}$ above an energy of $\sim 100$\,TeV~\cite{Aartsen:2017mau}.

\begin{figure*}[ht]
\includegraphics[width=\textwidth]{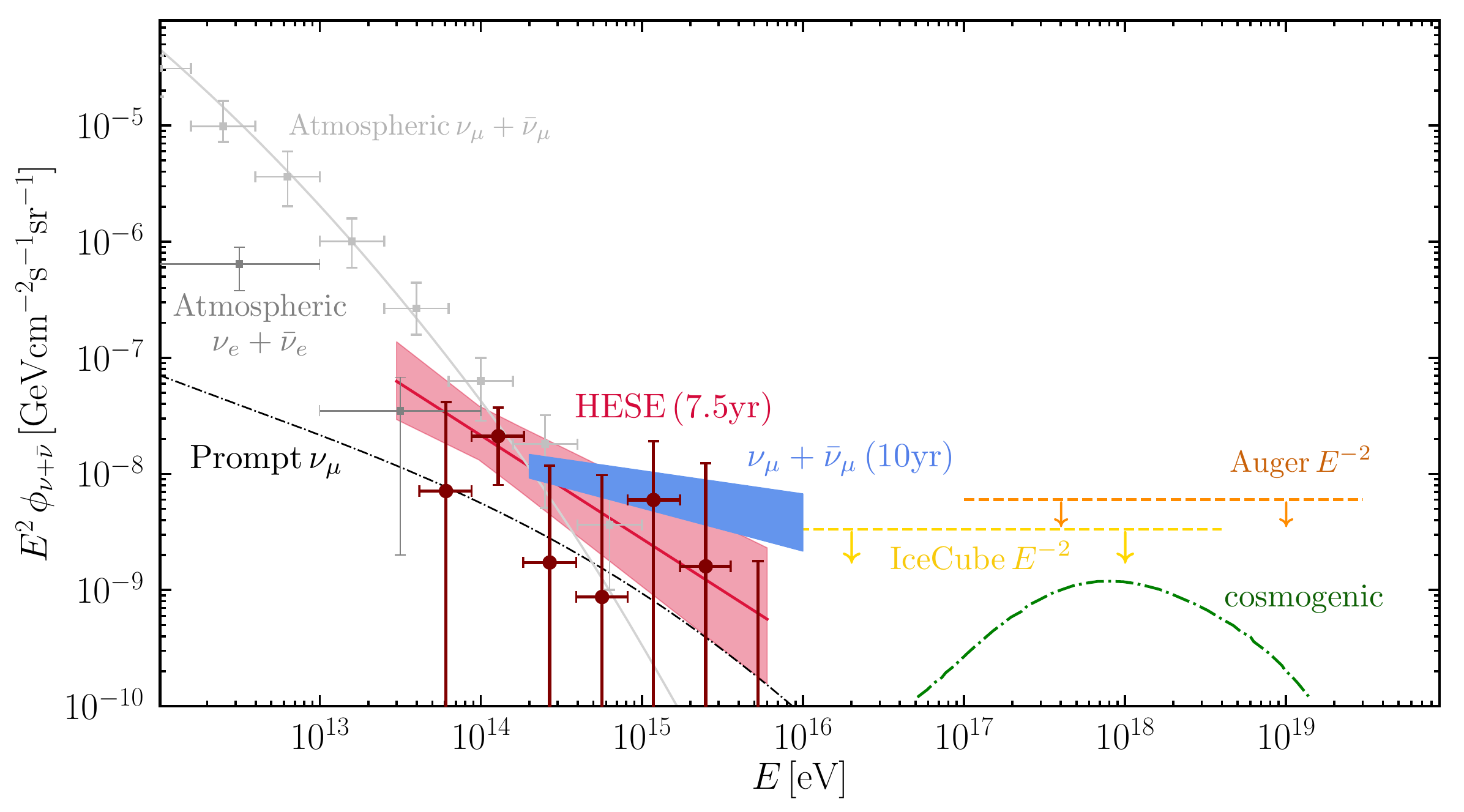}
\caption{The flux of high-energy neutrinos (per flavor) above 1 TeV.  At low energies, IceCube's measured atmospheric muon and electron neutrinos \citep{Aartsen:2014qna,Aartsen:2015xup} are shown in light and dark grey, respectively. The modeled prompt flux \citep{Enberg:2008te} from atmospheric charm production in cosmic ray showers is shown in black. The red line shows the fitted spectra for the HESE analysis covering 7.5 years of data and the shaded region shows the uncertainties \citep{Schneider:2019ayi}. The dark red points show segmented flux data points in the HESE 7.5-year analysis. The results for the 10-year muon neutrino flux measurement is shown as the blue shaded area \citep{Stettner:2019tok}. At higher energies: the predicted flux of cosmogenic neutrinos \citep{Ahlers:2012rz} shown in green, as well as the upper limit on the extremely high energy flux of neutrinos from IceCube \citep{Aartsen:2016ngq} and Pierre Auger Observatory \citep{Aab:2015kma}.}
\label{fig:nu-spec}
\end{figure*}

In summary, IceCube has observed cosmic neutrinos using both methods for rejecting background. Based on different methods for reconstruction and energy measurement, their results agree, pointing at extragalactic sources whose flux has equilibrated in the three flavors after propagation over cosmic distances~\cite{Aartsen:2015ivb} with $\nu_e:\nu_\mu:\nu_\tau \sim 1:1:1$.

Figure \ref{fig:nu-spec} summarizes the measurements of the cosmic neutrino flux using starting events and throughgoing muons from the northern sky. An extrapolation of this high-energy flux to lower energy may suggest an excess of events in the $10-100$\,TeV energy range over and above a single power-law fit. This conclusion is however not statistically compelling~\cite{Aartsen:2016tpb}.

\begin{figure}[t]\centering
\includegraphics[width=0.95\linewidth,viewport=5 30 645 360,clip=true]{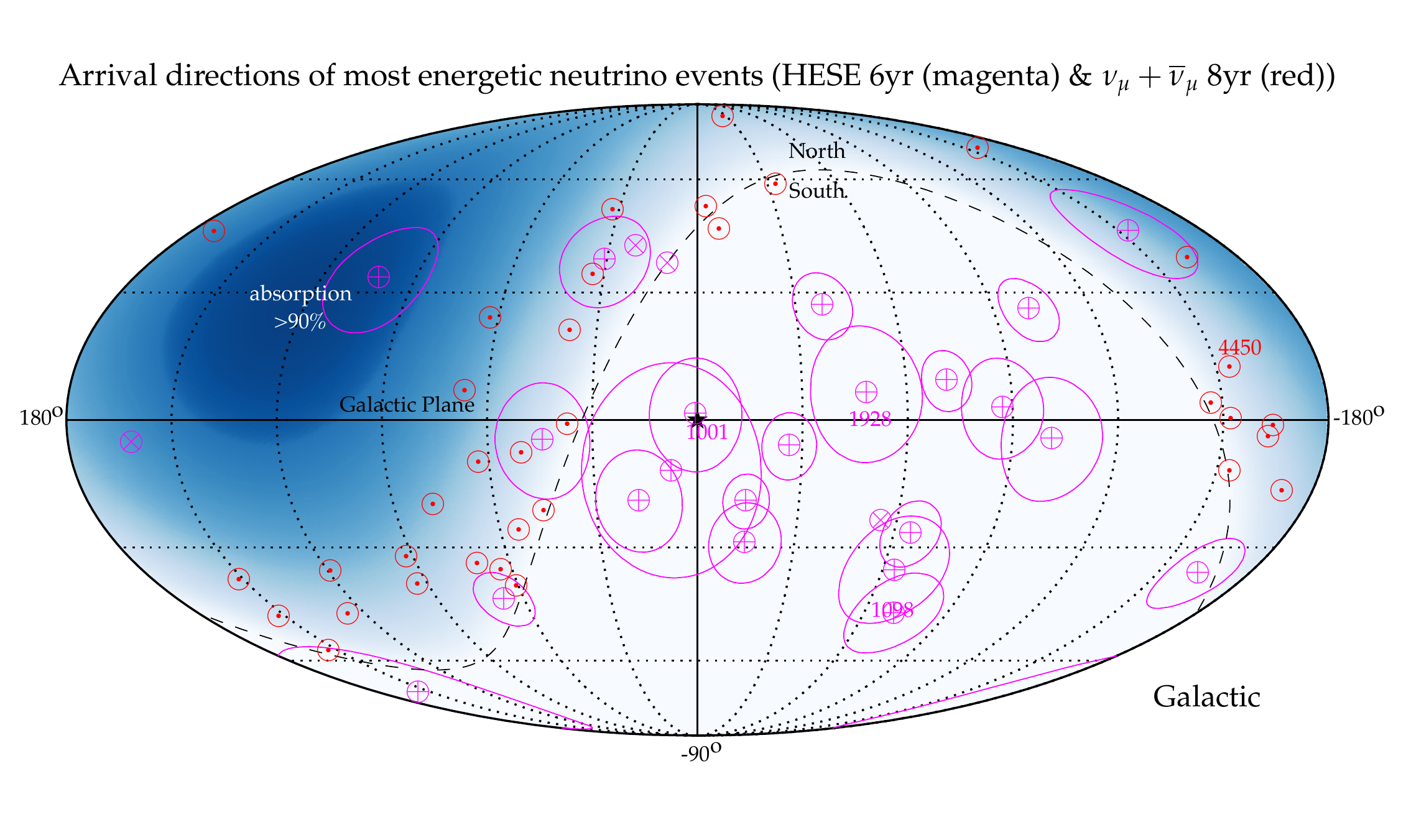}
\caption[]{Mollweide projection in Galactic coordinates of the arrival direction of neutrino events. We show the results of the eight-year upgoing track analysis~\cite{Aartsen:2017mau} with reconstructed muon energy $E_\mu\gtrsim200$~TeV ({$\odot$}). The events of the six-year high-energy starting event (HESE) analysis with deposited energy larger than 100\,TeV (tracks {$\otimes$} and cascades {$\oplus$}) are also shown~\cite{Aartsen:2014gkd,Aartsen:2015zva,Aartsen:2017mau}. The thin circles indicate the median angular resolution of the cascade events ({$\oplus$}). The blue-shaded region indicates the zenith-dependent range where Earth absorption of 100~TeV neutrinos becomes important, reaching more than 90\% close to the nadir. 
The dashed line indicates the horizon and the star ({\scriptsize $\star$}) the Galactic Center. We highlight the four most energetic events in both analyses by their deposited energy (magenta numbers) and reconstructed muon energy (red number).}
\label{NeutrinoMap}
\end{figure}

In Fig.~\ref{NeutrinoMap} we show the arrival directions of the most energetic events in the eight-year upgoing $\nu_\mu+\bar\nu_\mu$ analysis ($\odot$) and the six-year HESE data sets. The HESE data are separated into tracks ($\otimes$) and cascades ($\oplus$). The median angular resolution of the cascade events is indicated by thin circles around the best-fit position. The most energetic muons with energy $E_\nu>200$~TeV in the upgoing $\nu_\mu+\bar\nu_\mu$ data set accumulate near the horizon in the Northern Hemisphere. Elsewhere, muon neutrinos are increasingly absorbed in the Earth before reaching the vicinity of the detector because of their relatively large high-energy cross sections. This causes the apparent anisotropy of the events in the Northern Hemisphere. Also HESE events with deposited energy of $E_{\rm dep}>100$~TeV suffer from absorption in the Earth and are therefore mostly detected when originating in the Southern Hemisphere. After correcting for absorption by the Earth, the arrival directions of cosmic neutrinos are isotropic, suggesting extragalactic sources. In fact, no correlation of the arrival directions of the highest energy events, shown in Fig.~\ref{NeutrinoMap}, with potential sources or source classes has reached the level of $3\sigma$~\cite{Aartsen:2016tpb}.

\begin{figure}[t]
\centering
\includegraphics[width=0.95\linewidth]{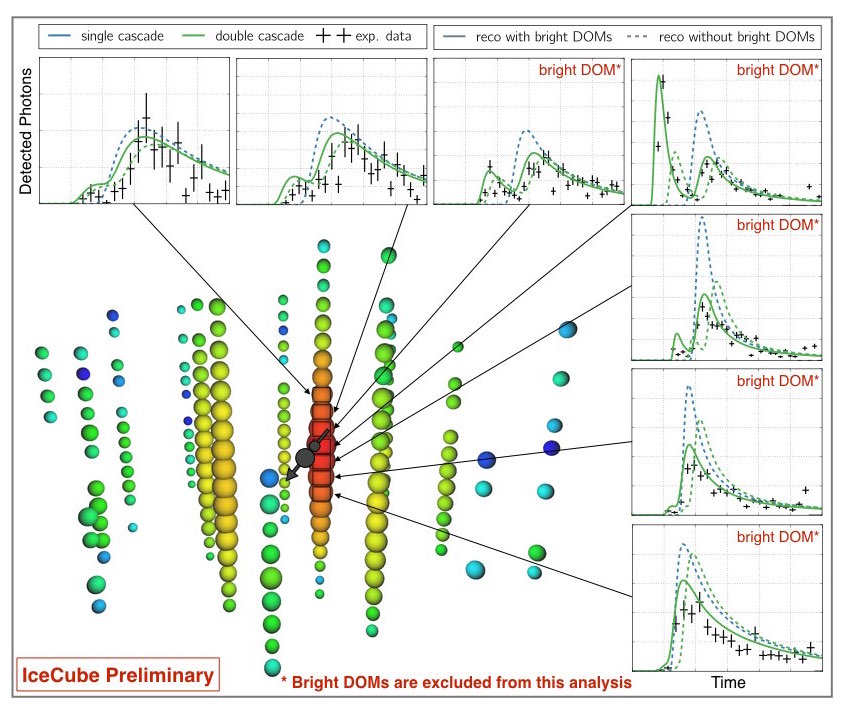}
\caption[]{Event view of a tau neutrino. The Cherenkov photons associated with the production and subsequent decay of the tau neutrino are identified by the double-peaked photon count as a function of time for the bright DOMs, for instance, the one shown in the top-right corner. The best fit (solid line) corresponding to a 17\,m decay length and is far superior to fits assuming a single electromagnetic or hadronic shower (dashed lines).}
\label{fig:double bang}
\end{figure}

We should comment at this point that there is yet another method to conclusively identify cosmic neutrinos: the observation of very high energy tau neutrinos. Below an energy of 100\,GeV,  tau neutrinos are abundantly produced in the atmosphere by the oscillation of muon into tau neutrinos. Above that energy, they must be of cosmic origin, produced in cosmic accelerators whose neutrino flux has approximately equilibrated between the three flavors after propagating over cosmic distances. Tau neutrinos produce two spatially separated showers in the detector, one from the interaction of the tau neutrino and the second one from its decay; the mean tau lepton decay length is $\lambda_{\tau} = \Gamma c \tau \approx 50~{\rm m} \times (E_\tau/\rm{PeV})$, where $m$, $\tau$, and $E$ are the mass, lifetime, and energy of the tau, respectively. Two such candidate events have been identified~\cite{ignationeutrino2018}. An event with a decay length of 17\,m and a probability of 98\% of being produced by a tau neutrino is shown in Fig.~\ref{fig:double bang}.

\begin{figure}[t]
\centering
\includegraphics[width=0.95\linewidth]{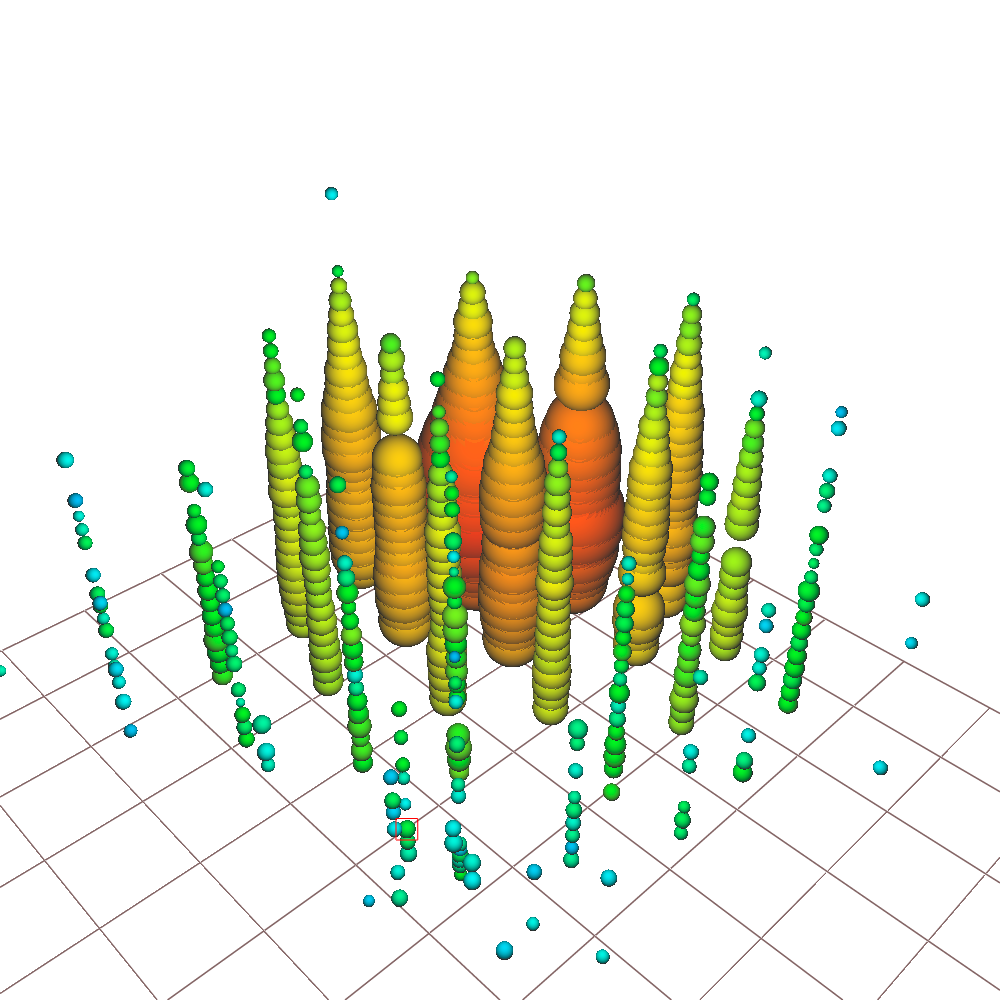}
\caption[]{Particle shower created at the Glashow resonance. The energy, measured via Cherenkov radiation in the Antarctic ice sheet, is reconstructed at the resonant energy for the production of a weak intermediate boson $W^-$ in the interaction of an antielectron neutrino with an atomic electron in the ice. Features consistent with the production of secondary muons in the particle shower indicate the hadronic decay of a resonant $W^-$ boson.}
\label{fig:glashow}
\end{figure}

Yet another independent confirmation of the observation of neutrinos of cosmic origin appeared in the form of a Glashow resonance event shown in Fig.~\ref{fig:glashow}. The event was identified in a search for partially contained events in a 4.5 year data set: an antielectron neutrino interacting with an atomic electron produced an event compatible with an incident neutrino energy of 6.3\,PeV, characteristic of the resonant production of a weak intermediate $W^-$ with a mass of $M_W=80.38\, \rm GeV$~\cite{IceCube:2021rpz}.

Given its energy and direction, the event is classified as an astrophysical neutrino at the $5\sigma$ level. Furthermore, data collected by the sensors closest to the interaction point, as well as the measured energy, are consistent with the hadronic decay of a $W^-$ produced on the Glashow resonance. In the observer frame, where the electron mass ($m_e=0.511 \, \rm MeV$) is at rest, the resonance energy is given by $E_R = M_W^2/2m_{e} = 6.32\, \rm PeV$ for $M_W=80.38 \, \rm GeV$. The measured energy of $6.05\pm 0.72$ PeV translates into a neutrino energy of 6.3 PeV after correcting the visible energy produced by the hadronic decay of the W for shower particles that do not radiate. Taking into account the detector's energy resolution, the probability that the event is produced off resonance by deep inelastic scattering is only 0.01 assuming a spectrum with a spectral index of $\gamma = -2.5$. Its observation extends the measured astrophysical flux to 6.3 PeV. Assuming the Standard Model resonant cross section, we expect 1.55 events in the sample assuming an antineutrino:neutrino ratio of 1:1 characteristic of a cosmic beam dump producing an equal number of pions of all three electric charges.

The observation of a Glashow resonant event heralds the presence of electron antineutrinos in the cosmic neutrino flux. Its unique signature illustrates a method to disentangle neutrinos from antineutrinos, thus opening a path to distinguish astronomical accelerators that produce neutrinos via hadronuclear or photohadronic interactions with or without strong magnetic fields ~\cite{IceCube:2021rpz}. As such, knowledge of both the flavor and charge of the incident neutrino will add a new tool for doing neutrino astronomy.

\section{Multimessenger Astronomy: General Considerations}
\label{sec:multimessenger}

\begin{figure}[t]
\centering
\includegraphics[width=0.95\linewidth]{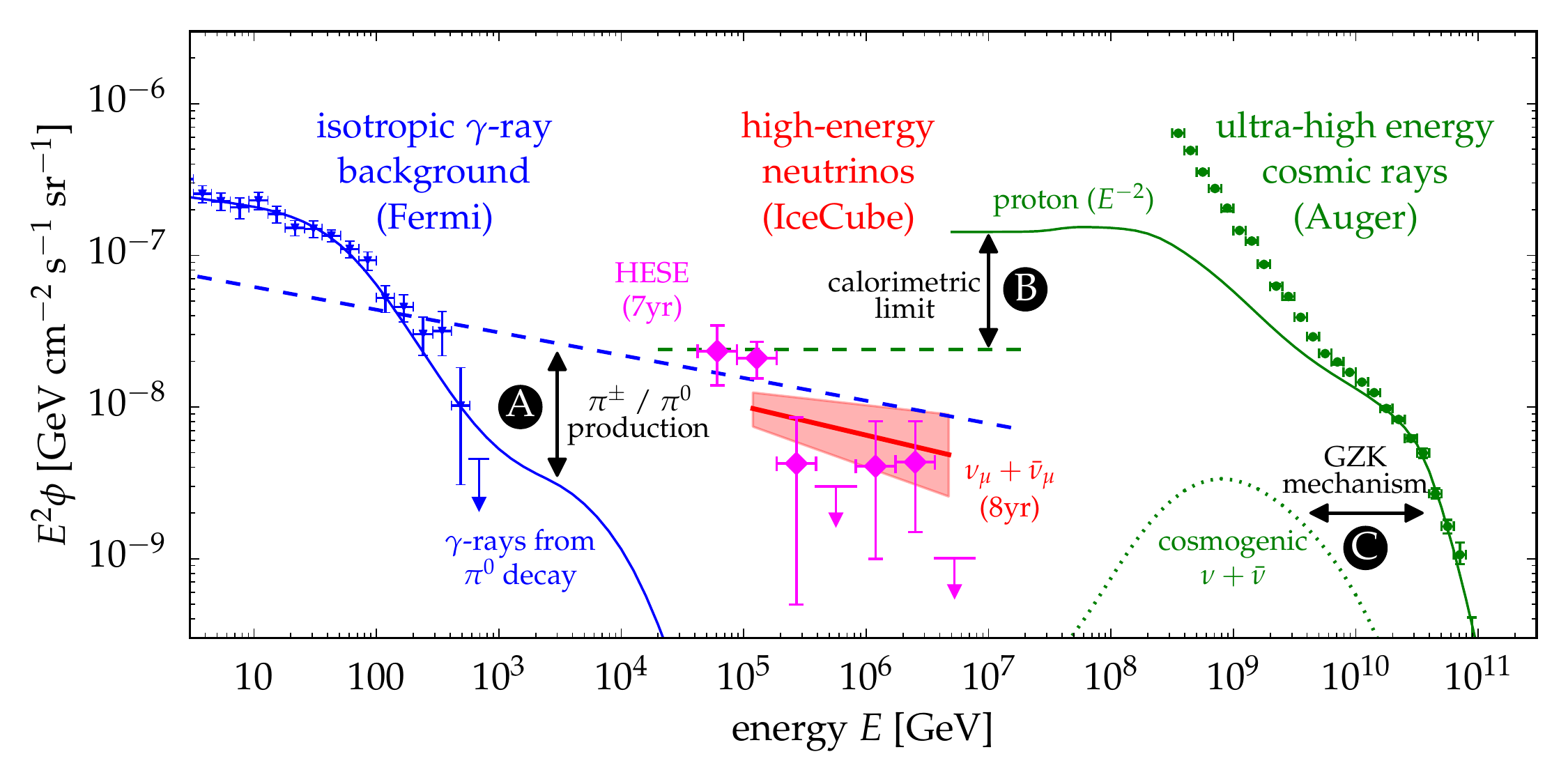}
\caption[]{The spectral flux ($\phi$) of neutrinos inferred from the eight-year upgoing track analysis (red fit) and the six-year HESE analysis (magenta fit) compared to the flux of unresolved extragalactic gamma-ray sources~\cite{Ackermann:2014usa} (blue data) and ultra-high-energy cosmic rays~\cite{Aab:2015bza} (green data). The neutrino spectra are indicated by the best-fit power-law (solid line) and $1 \sigma$ uncertainty range (shaded range). We highlight the various multimessenger interfaces: {A:} The joined {production of charged pions ($\pi^\pm$) and neutral pions ($\pi^0$)} in cosmic-ray interactions leads to the emission of neutrinos (dashed blue) and gamma-rays (solid blue), respectively. {B:} Cosmic ray emission models (solid green) of the most energetic cosmic rays imply a maximal flux ({calorimetric limit}) of neutrinos from the same sources (green dashed). {C:} The same cosmic ray model predicts the emission of cosmogenic neutrinos from the collision with cosmic background photons ({GZK mechanism}).}
\label{fig:panorama}
\end{figure}

The most important message emerging from the IceCube measurements may not be apparent yet: the prominent and surprisingly important role of neutrinos relative to photons in the nonthermal universe. To illustrate this point, we show in Fig.~\ref{fig:panorama} the observed energy flux of neutrinos, $E^2\phi$. One can see that the cosmic energy density of high-energy neutrinos is comparable to that of gamma-rays observed with the Fermi satellite~\cite{Ackermann:2014usa} and to that of the ultra-high-energy (UHE) cosmic rays above $10^{9}$~GeV, observed, by the Auger observatory~\cite{Aab:2015bza}. This may indicate a common origin and, in any case, provides an excellent opportunity for multi-messenger studies.

The pionic gamma rays accompanying the neutrino flux is shown as the solid blue. It has been derived from Eq.~\ref{energyflux} but, for extragalactic neutrinos, this gamma-ray emission is not directly observed because of strong absorption of photons by $e^+e^-$ pair production in the extragalactic background light (EBL) and CMB. 
The high-energy photons initiate electromagnetic showers by repeated pair production and inverse-Compton scattering, mostly on CMB photons, that eventually yield the lower energy photons observed by Fermi in the GeV-TeV range.

The extragalactic gamma-ray background observed by Fermi~\cite{Ackermann:2014usa} has contributions from  identified point-like sources on top of an isotropic gamma-ray background (IGRB) shown in Fig.~\ref{fig:panorama}. This IGRB is expected to consist mostly of emission from the same class of gamma-ray sources that are individually below Fermi's point-source detection threshold (see, e.g., reference~\citen{DiMauro:2015tfa}). The significant contribution of gamma-rays associated with IceCube's neutrino observation has the somewhat surprising implication that indeed many extragalactic gamma-ray sources are also neutrino emitters, while none has been detected so far.

Another intriguing observation is that the high-energy neutrinos observed at IceCube could originate in the sources of the highest energy cosmic rays. These could, for instance, be embedded in environments that act as ``storage rooms'' for cosmic rays with energies below the ``ankle'' ($E_{\rm CR}\ll1$EeV). This energy-dependent trapping can be achieved via cosmic ray diffusion in magnetic fields. While these cosmic rays are trapped, they can produce gamma-rays and neutrinos via collisions with gas. If the conditions are right, this mechanism can be so efficient that the total energy stored in low-energy cosmic rays is converted to that of gamma rays and neutrinos. These ``calorimetric'' conditions can be achieved in starburst galaxies~\cite{Loeb:2006tw} or galaxy clusters~\cite{Berezinsky:1996wx}. We will discuss these multimessenger relations in more detail next.

\subsection{IceCube Neutrinos and Fermi Photons}

Recall that photons are produced in association with neutrinos when accelerated cosmic rays produce neutral and charged pions in interactions with photons or nuclei. Targets include strong radiation fields that may be associated with the accelerator as well as concentrations of matter or molecular clouds in their vicinity. Additionally, pions can be produced in the interaction of cosmic rays with the EBL when propagating through the interstellar or intergalactic background medium from their source to Earth. A high-energy flux of neutrinos is produced in the subsequent decay of charged pions via $\pi^+\to\mu^++\nu_\mu$ followed by $\mu^+ \to e^++\nu_e+\bar\nu_\mu$ and the charge--conjugate processes. High-energy gamma rays result from the decay of neutral pions, $\pi^0\to\gamma+\gamma$. Pionic gamma rays and neutrinos carry, on average, 1/2 and 1/4 of the energy of the parent pion, respectively. With these approximations, the neutrino production rate $Q_{\nu_\alpha}$ (units of ${\rm GeV}^{-1} {\rm s}^{-1}$) can be related to the one for charged pions as
\begin{equation}
\sum_{\alpha}E_\nu Q_{\nu_\alpha}(E_\nu) \simeq 3\left[E_\pi Q_{\pi^\pm}(E_\pi)\right]_{E_\pi \simeq 4E_\nu}\,.
\end{equation}
Similarly, the production rate of pionic gamma-rays is related to the one for neutral pions as
\begin{equation}
E_\gamma Q_{\gamma}(E_\gamma) \simeq 2\left[E_\pi Q_{\pi^0}(E_\pi)\right]_{E_\pi \simeq 2E_\gamma}\,.
\label{eq:PIONtoNU}
\end{equation}

Note, that the relative production rate of pionic gamma rays and neutrinos only depends on the ratio of charged-to-neutral pions produced in cosmic-ray interactions, denoted by $K_\pi = N_{\pi^\pm}/N_{\pi^0}$. Pion production of cosmic rays in interactions with photons can proceed resonantly in the processes $p + \gamma \rightarrow \Delta^+ \rightarrow \pi^0 + p$ and $p + \gamma \rightarrow \Delta^+ \rightarrow \pi^+ + n$. These channels produce charged and neutral pions with probabilities 2/3 and 1/3, respectively. However, the additional contribution of nonresonant pion production changes this ratio to approximately 1/2 and 1/2.

In contrast, cosmic rays interacting with matter, for instance with hydrogen in the Galactic disk, produce equal numbers of pions of all three charges: $p+p \rightarrow n_\pi\,[\,\pi^{0}+\pi^{+} +\pi^{-}]+X$, where $n_\pi$ is the pion multiplicity. From above arguments we have $K_\pi\simeq2$ for cosmic ray interactions with gas ($pp$) and $K_\pi\simeq1$ for interactions with photons ($p\gamma$). 

With this approximation we can combine Eqs.~(\ref{eq:Qgamma}) and (\ref{eq:PIONtoNU}) to derive a simple relation between the pionic gamma-ray and neutrino production rates:

\begin{equation}\label{eq:GAMMAtoNU}
\frac{1}{3}\sum_{\alpha}E^2_\nu Q_{\nu_\alpha}(E_\nu) \simeq \frac{K_\pi}{4}\left[E^2_\gamma Q_\gamma(E_\gamma)\right]_{E_\gamma = 2E_\nu}\,.
\end{equation}

The prefactor $1/4$ accounts for the energy ratio $\langle E_\nu\rangle/\langle E_\gamma\rangle\simeq 1/2$ and the two gamma rays produced in the neutral pion decay.  
This powerful multimessenger relation connects pionic neutrinos and gamma rays without any reference to the cosmic ray beam; it simply reflects the fact that a $\pi^0$ produces two $\gamma$ rays for every charged pion producing a $\nu_\mu +\bar\nu_\mu$ pair, which cannot be separated by current experiments.

Before applying this relation to a cosmic accelerator, we have to be aware of the fact that, unlike neutrinos, gamma rays interact with photons of the cosmic microwave background before reaching Earth. The resulting electromagnetic shower subdivides the initial photon energy, resulting in multiple photons in the GeV-TeV energy range by the time the photons reach Earth. Calculating the cascaded gamma-ray flux accompanying IceCube neutrinos is straightforward~\cite{Protheroe1993,Ahlers:2010fw}.

 An example of the relation between pionic gamma-ray (solid) and neutrino (dashed) emission is illustrated in Fig.~\ref{fig:panorama} by the blue lines. We assume that the underlying $\pi^0$ / $\pi^\pm$ production follows from cosmic-ray interactions with gas in the universe; $K_\pi\simeq2$. In this way, the initial emission spectrum of gamma-rays and neutrinos from pion decay is almost identical to the spectrum of cosmic rays, after accounting for the different normalizations and energy scales. The flux of neutrinos arriving at Earth (blue dashed line) follows this initial CR emission spectrum, assumed to be a power law, $E^{-\gamma}$. However, the observable flux of gamma-rays (blue solid lines) is strongly attenuated above 100~GeV by interactions with extragalactic background photons.

The overall normalization of the emission is chosen in a way that the model does not exceed the isotropic gamma-ray background observed by the Fermi satellite (blue data). This implies an upper limit on the neutrino flux shown as the blue dashed line. Interestingly, the neutrino data shown in Fig.~\ref{fig:panorama} saturates this limit above 100~TeV. Moreover, the HESE data that extends to lower energies is only marginally consistent with the upper bound implied by the model (blue dashed line). If the underlying assumptions hold, the neutrino spectrum cannot be harder than $E^{-1.5}$, a result somewhat challenged by the data. This example shows that multi-messenger studies of gamma-ray and neutrino data are powerful tools to study the neutrino production mechanism and to constrain neutrino source models~\cite{Murase:2013rfa}.

Above exercise relating gamma rays and neutrinos comes with more warnings than a TV commercial for drugs:
\begin{itemize}
    \item The target for producing the neutrinos may be photons. This changes the value of $K_\pi$ and the shape of the flux. Yoshida and Murase recently worked through this example in detail~\cite{Yoshida:2020div}.
    \item Electrons accelerated along with protons may contribute photons to the observed flux on top of those of pionic origin.
    \item The source itself may not be transparent to high-energy photons that will lose energy in the source event before reaching the EBL. As a result, part of the pionic flux will emerge below the threshold of the Fermi satellite, at MeV and lower energies. This will be an important consideration when we discuss the first identified source of high-energy neutrinos, the black hole TXS 0506+056. 
\end{itemize}

While the details matter, the main message should not be lost: the matching energy densities of the extragalactic gamma-ray flux detected by Fermi and the high-energy neutrino flux measured by IceCube suggest that, rather than detecting some exotic sources, it is more likely that IceCube to a large extent observes the same universe astronomers do. Clearly, an extreme universe modeled exclusively on the basis of electromagnetic processes is no longer realistic. The finding implies that a large fraction, possibly most, of the energy in the nonthermal universe originates in hadronic processes, indicating a larger role than previously thought. The high intensity of the neutrino flux below 100~TeV in comparison to the Fermi data might indicate that these sources are even more efficient neutrino than gamma-ray sources~\cite{Murase:2015xka,Bechtol:2015uqb}.

IceCube is developing methods that enable real-time multiwavelength observations in cooperation with astronomical telescopes, to identify the sources and build on the discovery of cosmic neutrinos to launch a new era in astronomy~\cite{Aartsen:2016qbu,Aartsen:2016lmt}. We will return to a coincident observation of a flaring blazar on September 22, 2017, further on.

\subsection{IceCube Neutrinos and Ultra-High-Energy Cosmic Rays}
\label{sec:nuanduhcr}
 
The charged pion production rate $Q_{\pi^\pm}$ is proportional to the density of the cosmic-ray nucleons in the beam, $Q_N$, by a ``bolometric'' proportionality factor $f_\pi$. For a target with nucleon density $n$ and extension $\ell$, the efficiency factor for producing pions is $f_\pi \simeq 1-\exp(-\kappa\ell\sigma n)$, where $\kappa$ is the inelasticity, i.e., the average relative energy loss of the leading nucleon going into the production of pions, $(\sigma n)^{-1}$ its interaction length with $\sigma$ the cross section for either $p\gamma$ or $pp$ interactions. The pion production efficiency $f_\pi$ normalizes the conversion of cosmic-ray energy into pion energy on the target as:
\begin{equation}\label{eq:CRtoPION}
E_\pi^2Q_{\pi^\pm}(E_\pi) \simeq f_\pi\, \frac{K_\pi}{1+K_\pi}\,\left[E^2_NQ_N(E_N)\right]_{E_N = E_\pi/x_\pi}\,.
\end{equation}
We already introduced the pion ratio $K_\pi$ in the previous section, with $K_\pi\simeq2$ for $pp$ and $K_\pi\simeq1$ for $p\gamma$ interactions. The factor $x_\pi$ denotes, as before, the average inelasticity {\it per pion} that depends on the average pion multiplicity $n_\pi$. For, both, $pp$ and $p\gamma$ interactions this can be approximated as $x_\pi = \kappa/n_\pi \simeq 0.2$. The average energy per pion is then $\langle E_\pi\rangle  = x_\pi E_N$ and the average energy of the pionic leptons relative to the nucleon is $\langle E_\nu\rangle \simeq \langle E_\pi\rangle /4 = (x_\pi/4)E_N \simeq 0.05 E_N$.

In general, the cosmic ray nucleon emission rate, $Q_N$, depends on the composition of the highest energy cosmic rays and can be obtained by integrating the measured spectra\footnote{Note that the integrated number of nucleons is linear to mass number, $\int {\rm d}E Q_N(E) = \sum_AA\int {\rm d}EQ_A(E)$.} of nuclei with mass number $A$ as $Q_N(E_N) = \sum_A A^2Q_A(AE_N)$. In the following we will derive a upper limit on diffuse neutrino fluxes under the assumption that the highest energy cosmic rays are mostly protons~\cite{Waxman:1998yy,Bahcall:1999yr}. The local emission rate {\it density}, $\mathcal{Q} = \rho_0Q$,  is at these energies insensitive to the luminosity evolution of sources at high redshift and can be estimated to be at the level of $\left[E_p^2\mathcal{Q}_p(E_p)\right]_{10^{19.5}{\rm eV}} \sim (0.5-2.0)\times10^{44} {\rm erg}/{\rm Mpc}^{3}/{\rm yr}$~\cite{Ahlers:2012rz,Katz:2013ooa,Waxman:2015ues}. Note, that composition measurements indicate that the mass composition above the ankle also requires a contribution of heavier nuclei. However, the estimated local power density based on proton models is a good proxy for that of cosmic ray models that include heavy nuclei, as long as the spectral index is close to $\gamma\simeq2$. For instance, a recent analysis of Auger~\cite{Aab:2016zth} provides a solution with spectral index $\gamma\simeq2.04$ and a combined nucleon density of $[E_N^2\mathcal{Q}_N(E_N)]_{10^{19.5}{\rm eV}} \sim 2.2\times10^{43}\,{\rm erg}/{\rm Mpc}^{3}/{\rm yr}$. 

We construct the diffuse neutrino flux from the contribution of individual sources. A neutrino point-source (PS) at redshift\,$z$ with spectral emission rate $Q_{\nu_\alpha}$ contributes a neutrino flux summed over flavors (in units ${\rm GeV}^{-1} {\rm cm}^{-2} {\rm s}^{-1}$)
\begin{equation}\label{eq:PS}
\phi^{\rm PS}_{\nu}(E_\nu) = \frac{(1+z)^2}{4\pi d^2_L(z)}\sum_\alpha Q_{\nu_\alpha}((1+z)E_\nu)\,,
\end{equation}
where $d_L$ is the luminosity distance
\begin{equation}
d_L(z) = (1+z)\int\limits_0^z\frac{{\rm d}z'}{H(z')}\,.
\end{equation} 
Here, the Hubble parameter $H$ has a local value of $c/H_0 \simeq 4.4$~Gpc and scales with redshift as $H^2(z) = H^2_0[(1+z)^3\Omega_{\rm m} + \Omega_\Lambda]$, with $\Omega_{\rm m} \simeq 0.3$ and $\Omega_\Lambda\simeq 0.7$ assuming the standard $\Lambda$CDM cosmological model~\cite{Olive:2016xmw}. Note that the extra factor $(1+z)^2$ in Eq.~(\ref{eq:PS}) follows from the definition of the luminosity distance and accounts for the relation of the energy flux to the differential neutrino flux $\phi$. The diffuse neutrino flux from extragalactic sources is given by the integral over co-moving volume ${\rm d}V_c = 4\pi (d_L/(1+z))^2 {\rm d}z/H(z)$. Weighting each neutrino source by its density per co-moving volume $\rho(z)$ gives~\cite{Ahlers:2014ioa}
\begin{equation}\label{eq:Jtot}
\phi_{\nu}(E_\nu) = \frac{c}{4\pi}\int_0^\infty\frac{{\rm d}z}{H(z)}\mathcal{\rho}(z)\sum_\alpha Q_{\nu_\alpha}((1+z)E_\nu)\,,
\end{equation} 
and, assuming that $Q_{\nu_\alpha}$ follows a power law $E^{-\gamma}$, %
\begin{equation}\label{eq:Lnu}
\frac{1}{3}\sum_\alpha E_\nu^2\phi_{\nu_\alpha}(E_\nu) = \frac{c}{4\pi}\frac{\xi_z}{H_0}\frac{1}{3}\sum_\alpha E_\nu^2 \mathcal{Q}_{\nu_\alpha}(E_\nu)\,,
\end{equation}
where $\mathcal{Q}_{\nu_\alpha}= \rho_0Q_{\nu_\alpha}$ is the neutrino emission rate {\it density} and
\begin{equation}\label{xi}
\xi_z = \int_0^\infty{\rm d}z\frac{(1+z)^{-\Gamma}}{\sqrt{\Omega_\Lambda+(1+z)^3\Omega_{\rm m}}}\frac{\rho(z)}{\rho_0}\,.
\end{equation}
A spectral index of $\gamma\simeq2.0$ and no source evolution in the local ($z<2$) universe, $\rho(z)=\rho_0$, yields $\xi_z\simeq0.5$. For sources following the star-formation rate, $\rho(z)=(1+z)^3$ for $z<1.5$ and $\rho(z)=(1+1.5)^3$ for $1.5<z<4$, with the same spectral index yields $\xi_z\simeq 2.6$. 

We can now derive the relation between the diffuse neutrino flux from the cosmic ray rate density by combining Eqs.~(\ref{eq:PIONtoNU}) and (\ref{eq:CRtoPION}):
 
\begin{equation}\label{eq:WBbound}
\frac{1}{3}\sum_{\alpha}E_\nu^2\phi_{\nu_\alpha}(E_\nu)
\simeq  3\times10^{-8}f_\pi\left(\frac{\xi_z}{2.6}\right)\left(\frac{[E^2_p\mathcal{Q}_p(E_p)]_{E_p=10^{19.5}{\rm eV}}}{10^{44} \,{\rm erg}/{\rm Mpc}^{3}/{\rm yr}}\right)\frac{\rm GeV}{\rm cm^2\,s\,sr}\,.
\end{equation}

Here, we have assumed $pp$ interactions with $K_\pi=2$. The calorimetric limit, previously discussed corresponds to the assumption that $f_\pi\to 1$, and is also referred to as the Waxman-Bahcall ``bound"~\cite{Waxman:1998yy,Bahcall:1999yr}. Sources with $f_\pi > 1$ are referred to as ``dark" sources; their thick target efficiently converts nucleons to neutrinos and rendering them opaque to high-energy photons. Accelerator-based beam dumps are the ultimate dark sources of neutrinos.

It is intriguing that the observed intensity of diffuse neutrinos indicates that $f_\pi \simeq 1$. The correspondence is illustrated by the green lines in Fig.~\ref{fig:panorama} showing a parametrization that accounts for the most energetic cosmic rays (green data). Note, that the cosmic ray data below $10^{10}$~GeV are not described by this fit and must be supplied by additional sources, e.g., in our own Galaxy. Assuming that the cosmic ray energy density is transformed in neutrinos we derive the maximal neutrino emission (green dashed line). Interestingly, the observed neutrino flux saturates this calorimetric limit. It is therefore possible that the highest energy cosmic rays and neutrinos have a common origin. If this is the case, the neutrino spectrum beyond 200~TeV should reflect the energy-dependent release of cosmic rays from the calorimeters. Future studies of the neutrino spectrum beyond 1~PeV can provide supporting evidence for this and, in particular, the transition to a thin environment ($f_\pi\ll1$), that is a necessary condition for the emission of the highest energy cosmic rays by the source, implies a break or cutoff in the neutrino spectrum.

Note that the proton model in Fig.~\ref{fig:panorama} also contributes to the flux of EeV neutrinos shown as a dotted green line. Ultra-high-energy cosmic rays are strongly attenuated by resonant interactions with background photons, as first pointed out by Greisen, Zatsepin and Kuzmin~\cite{Greisen:1966jv,Zatsepin:1966jv} (GZK). This GZK mechanism is responsible for the suppression of the proton flux beyond $5\times10^{10}$~GeV (``GZK cutoff'') in Fig.~\ref{fig:panorama} (green solid line) and predicts a detectable flux of cosmogenic neutrinos~\cite{Beresinsky:1969qj} (green dotted line) in a kilometer-scale detector. In case a significant fraction of the highest energy cosmic rays are heavy nuclei as indicated by the Auger data, the GZK flux is reduced accordingly because the individual nucleons fall below the threshold for producing neutrinos. This may explain the failure of IceCube to observe GZK neutrinos at the level predicted in Fig.~\ref{fig:panorama}.

\begin{figure}[t]\centering
\includegraphics[width=0.95\linewidth]{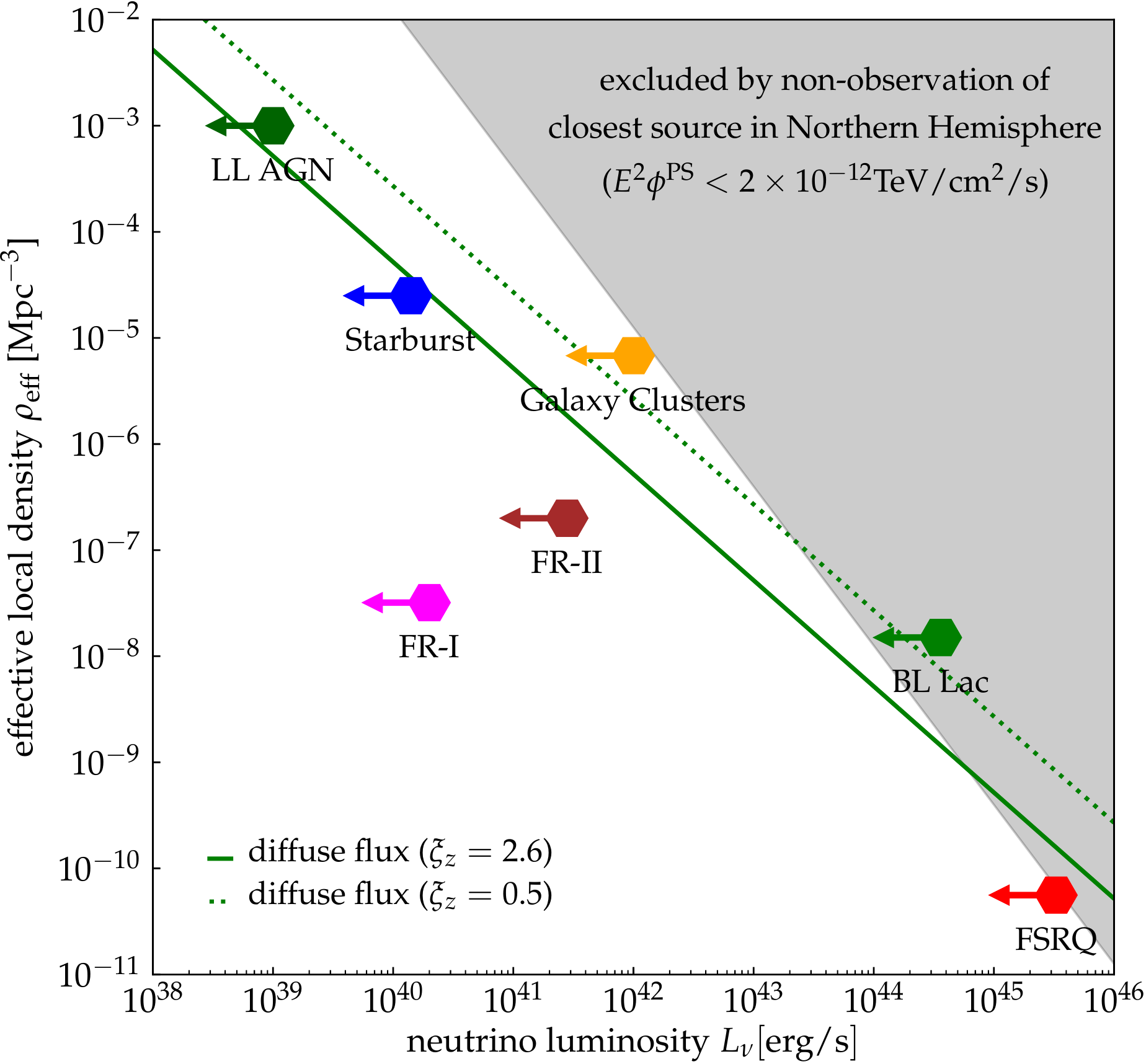}\\[0.3cm]
\caption[]{The effective local density and (maximal) neutrino luminosity of various neutrino source candidates from reference~\citen{Mertsch:2016hcd}. The green solid (green dotted) line shows the local density and luminosity of the population of sources responsible for the diffuse neutrino flux of $E^2\phi\simeq 10^{-8}~{\rm GeV}~{\rm cm}^{-2}~{\rm s}^{-1}~{\rm sr}^{-1}$ observed with IceCube, assuming source evolution following the star-formation rate ($\xi_z\simeq2.6$) or no source evolution ($\xi_z\simeq0.5$), respectively. The grey-shaded area indicates source populations that are excluded by the non-observation of point sources in the Northern Hemisphere ($f_{\rm sky} \simeq 0.5$) with discovery potential $E^2\phi^{\rm PS}\simeq 2\times10^{-12}~{\rm TeV}~{\rm cm}^{-2}~{\rm s}^{-1}$~\cite{Aartsen:2017kru}.
}\label{fig:Luminosity}
\end{figure}

\subsection{Pinpointing the Astrophysical Sources of Cosmic Neutrinos: Relating the Point Source and Diffuse Fluxes}

The diffuse flux of neutrinos measured by IceCube provides a constraint on the flux from the individual sources that it is composed of. In the sense of Olbers' paradox, one can investigate under what circumstances IceCube can detect the neutrino emission from individual, presumably nearby, point sources that contribute to the quasi-diffuse emission. 

One can relate the average luminosity of individual neutrino sources to the diffuse flux 

\begin{equation}\label{eq:Lnu1}
\frac{1}{3}\sum_\alpha E_\nu^2\phi_{\nu_\alpha}(E_\nu) \simeq \frac{c}{4\pi}\frac{\xi_z}{H_0}\rho_0\frac{1}{3}\sum_\alpha E_\nu^2 Q_{\nu_\alpha}(E_\nu)\,,
\end{equation}

where the $\xi_z$ summarizes the evolution of the sources, $H_0$ is the Hubble constant, and $\rho_0$ is the effective source density.

Equation~(\ref{eq:Lnu1}) relates the average luminosity of individual neutrino sources to the diffuse flux on the left hand side, which is measured by the experiment to be at the level of $E^2\phi_\nu \simeq 10^{-8}~{\rm GeV} {\rm cm}^{-2} {\rm s}^{-1} {\rm sr}^{-1}$ for energies in excess of $\simeq 100$\,TeV; see~Fig.~\ref{fig:nu-spec}. From the measurement, we can infer the average emission from a single source~\cite{Ahlers:2014ioa} as a function their local density,
\begin{equation}\label{eq:rate}
\frac{1}{3}\sum_\alpha E_\nu^2 Q_{\nu_\alpha}(E_\nu) \simeq 1.7\times10^{45} \left(\frac{\xi_z}{2.6}\right)^{-1}\left(\frac{\rho_0}{10^{-10}{\rm Mpc}^{-3}}\right)^{-1}{\rm erg}\,{\rm s}^{-1}\,.
\end{equation}

This relation builds in the very strong assumption that all sources corresponding to some astronomical classification are identical.

For a homogenous distribution of sources, we expect, within the partial field of view of the full sky, $f_{\rm sky}$, of the telescope, one source within a distance $d_1$ determined by
\begin{equation}
f_{\rm sky} 4\pi d_1^3/3\,\rho_0 = 1\,.
\end{equation}
In other words, $d_1$ defines the volume containing one nearby source for a homogeneous source density $\rho_0$. Defining $\phi_1$ as the flux of a source at distance $d_1$, given by Eq.~(\ref{eq:PS}) with $z\simeq0$ and $d_L = d_1$, we can write the probability distribution $p(\phi)$ of finding the {\it closest} source of the population with a flux $\phi$ as
\begin{equation}
p(\phi) = \frac{3}{2} \frac{1}{\phi}\left(\frac{\phi_1}{\phi}\right)^{\frac{3}{2}}e^{-\left(\frac{\phi_1}{\phi}\right)^{\frac{3}{2}}}\,.
\end{equation}
(This result takes into account fluctuation in the source distribution; for details see appendix of reference~\citen{Ahlers:2014ioa}). The average flux from the closest source is then $\langle \phi\rangle \simeq 2.7 \,\phi_1$, with median $\phi_{\rm med} \simeq 1.3\, \phi_1$. Applying this to the closest source introduced above, we obtain its neutrino flux:

\begin{equation}\label{eq:point_diffuse}
\frac{1}{3}\sum_\alpha E_\nu^2\phi_{\nu_\alpha} \simeq 2\times10^{-12}\left(\frac{f_{\rm sky}}{0.5}\right)^{\frac{2}{3}}\left(\frac{\xi_z}{2.6}\right)^{-1}\left(\frac{\rho_0}{10^{-10}{\rm Mpc}^{-3}}\right)^{-\frac{1}{3}} {\rm TeV}\,{\rm cm}^{-2}\,{\rm s}^{-1}\,.
\end{equation}

Interestingly, this value is not far from IceCube's average point-source discovery potential at the level of $2\times10^{-12}\,{\rm TeV}\,{\rm cm}^{-2}\,{\rm s}^{-1}$ in the Northern Hemisphere~\cite{Aartsen:2019fau}.

In Fig.~\ref{fig:Luminosity}, we show the local density and luminosity of theorized neutrino sources~\cite{Mertsch:2016hcd}. The grey-shaded region is excluded by the failure to observe these sources as individual point sources. The green lines show the combination of density and luminosity for which sources produce a diffuse neutrino flux at the level observed by IceCube. We show the result for a source densities following the star formation rate (solid line) or no evolution (dotted line). We conclude that IceCube is presently sensitive to source populations with local source densities smaller than, conservatively, $10^{-8}\,{\rm Mpc}^{-3}$. Much lower local densities, like FSRQs, are challenged by the nonobservation of individual sources. Some source classes, like Fanaroff-Riley (FR) radio galaxies, have an estimated neutrino luminosity that is likely too low for the observed flux. This simple estimate can be refined by considering not only the closest source of the population, but the combined emission of {\it known} local sources; see, e.g., reference~\citen{Ahlers:2014ioa}. 

The luminosity-density parameter space discussed above is often exploited to study sources identified in astronomical catalogs as potential contributors to the cosmic neutrino flux. However, it is critical to note that astronomical classifications are meant to explain the collective behavior of sources with regard to common spectral features and power. Thus, they may lack the necessary condition for a neutrino source: provide a beam dump with sufficient target density to produce neutrinos. Spectral features, deviating from a simplified assumption of the single power law, together with the variations of the power and target densities, modifies the allowed parameter space. In addition, note that the source luminosity distribution is neglected by assuming an average luminosity for all sources. Finally, the evolution of the sources plays a key role in the allowed parameter space. As such, the built-in assumption of a specific cosmological evolution that is common to all sources is problematic.

In addition to the strong dependence of these bands on the spectrum of the sources and their evolution, we should note that the all-sky untriggered searches for the sources of high-energy neutrinos allocate a large number of trial factors that penalizes the significance for any potential signal excess. Therefore, the presence of outlying sources in the neutrino data could result in the violation of such bands when additional information provided by other messengers distinguishes a specific source or direction in the sky.

Is it possible that the sources of the extragalactic cosmic rays are themselves neutrino sources? We previously derived the emission rate density of nucleons from the measured cosmic-ray spectrum~\cite{Ahlers:2012rz,Katz:2013ooa}
\begin{equation}
\mathcal{L}_p = \rho_0E^2_pQ_p(E_p) \simeq {(1-2)\times10^{44}\,{\rm erg}\,{\rm Mpc}^{-3}\,{\rm yr}^{-1}}\,.
\end{equation}
Combining this with Eq.\ref{eq:CRtoPION} we can derive the diffuse neutrino flux:
\begin{equation}\label{eq:WBbound}
\frac{1}{3}\sum_{\alpha}E_\nu^2\phi_{\nu_\alpha}(E_\nu) \simeq {f_\pi}{\frac{\xi_zK_\pi}{1+K_\pi}}(2-4)\times10^{-8}\,{\rm GeV}\,{\rm cm}^{-2}\,{\rm s}^{-1}\,{\rm sr}\,.
\end{equation}
The equation has been rewritten and some notation adjusted in order to accommodate both $pp$ and $\gamma p$ interactions. Counting particles we derive that
\begin{equation}
\frac{1}{3}\sum_\alpha E_\nu Q_{\nu_\alpha}(E_\nu)=E_\pi Q_\pi\,,
\end{equation}
from which the relation for energy follows by multiplying both sides with $E_\nu$; an additional factor of 1/4 multiplies the right-hand side accounting for the ratio of neutrino to pion energy within the approximations routinely used throughout.

The requirement $f_\pi\leq1$ may limit the neutrino production by the actual sources of the cosmic rays as pointed out by the seminal work by Waxman and Bahcall~\cite{Waxman:1998yy}. For optically thin sources, $f_\pi\ll1$, neutrino production is only a small by-product of the  acceleration process. The energy loss associated with pion production must not limit the sources' ability to accelerate the cosmic rays. On the other hand, optically thick sources, $f_\pi\simeq 1$, are more efficient neutrino emitters. Realistic sources of this type need different zones, one zone for the acceleration process ($f_\pi\ll1$) and a second zone for the efficient  conversion of cosmic rays to neutrinos ($f_\pi\simeq1$). An example for this scenario are sources embedded in starburst galaxies, where cosmic rays can be stored over sufficiently long timescales to yield significant neutrino production. A more relevant example is the multimessenger source TXS 0506+056 which we will discuss in detail further on.

For $\xi_z \simeq2.4$ and $K_\pi\simeq 1-2$, the upper bound resulting from Eq.~(\ref{eq:WBbound}) and $f_\pi=1$ is at the level of the neutrino flux observed by IceCube. Therefore, it is possible that the observed extragalactic cosmic rays and neutrinos have the same origin. A plausible scenario is a ``calorimeter'' in which only cosmic rays with energy below a few $10$~PeV interact efficiently. An energy dependence of the calorimetric environment can be introduced by energy--dependent diffusion. If $D(E)$ is the diffusion coefficient, then the timescale of escape from the calorimeter is given by the solution to $6D(E)t = d^2$, where $d$ is the effective size of the region. Typically, we have $D(E)\propto E^\delta$ with $\delta \simeq 0.3-0.6$. In the following, we again consider the case of protons. Taking $\sigma_{pp} \simeq 8\times10^{-26}\,{\rm cm}^2$ at $E_p=100$~PeV and the diffusion coefficient of $D(E_p) \simeq D_{\rm GeV}(E_p/1{\rm GeV})^{1/3}$, the $pp$ thickness can be expressed as $\tau_{pp} \simeq ctn_{\rm gas}
\sigma_{pp}$ or
\begin{equation}
\tau_{pp} \simeq 0.18\left(\frac{d}{100\,{\rm pc}}\right)^2\left(\frac{D_{\rm GeV}}{10^{26}\,{\rm cm}^2/{\rm s}}\right)^{-1}\left(\frac{E_p}{10~{\rm PeV}}\right)^{-1/3}\left(\frac{n}{100\,{\rm cm}^{-3}}\right)\,.
\end{equation}
Here, we have used feasible parameters of starburst galaxies~\cite{Loeb:2006tw,Murase:2013rfa}. Therefore, depending on the calorimetric environment, it is possible that the flux below a few PeV is efficiently converted to neutrinos and contributes to the TeV--PeV diffuse emission observed by IceCube.

\section{The First-Identified Cosmic-Ray Accelerator: the Rotating Supermassive Black Hole TXS 0506+056}\label{TXS}

The qualitative matching of the energy densities of photons and neutrinos, discussed in the previous section, may suggest that the unidentified neutrino sources contributing to the diffuse flux might have already been observed as strong gamma-ray emitters. Theoretical modeling~\cite{Ajello:2011zi,DiMauro:2013zfa} and recent data analyses~\cite{TheFermi-LAT:2015ykq,Zechlin:2015wdz,Lisanti:2016jub} show that Fermi's extragalactic gamma-ray flux is dominated by blazars. However, a dedicated IceCube study~\cite{Aartsen:2016lir} of Fermi-observed blazars shows no evidence of neutrino emission from these source candidates. The inferred limit on their quasi-diffuse flux leaves room for a significant contribution to IceCube's diffuse neutrino flux at the 10\% level, and increasing towards PeV. The multimessenger campaign launched by the neutrino alert IC170922 not only identified the first source of cosmic neutrinos, it will also shed light on this apparent contradiction.

Since 2016, the IceCube multimessenger program has grown from Galactic supernova alerts and early attempts to match neutrino observations with early LIGO/Virgo gravitational wave candidates to a steadily expanding set of automatic filters that selects in real time rare very high energy events that are potentially cosmic in origin. Within less than a minute of their detection in the deep Antarctic ice, the astronomical coordinates are sent to the Gamma-ray Coordinate Network for potential follow-up by astronomical telescopes.

\begin{figure}[t]
    \centering
    \includegraphics[width=\textwidth]{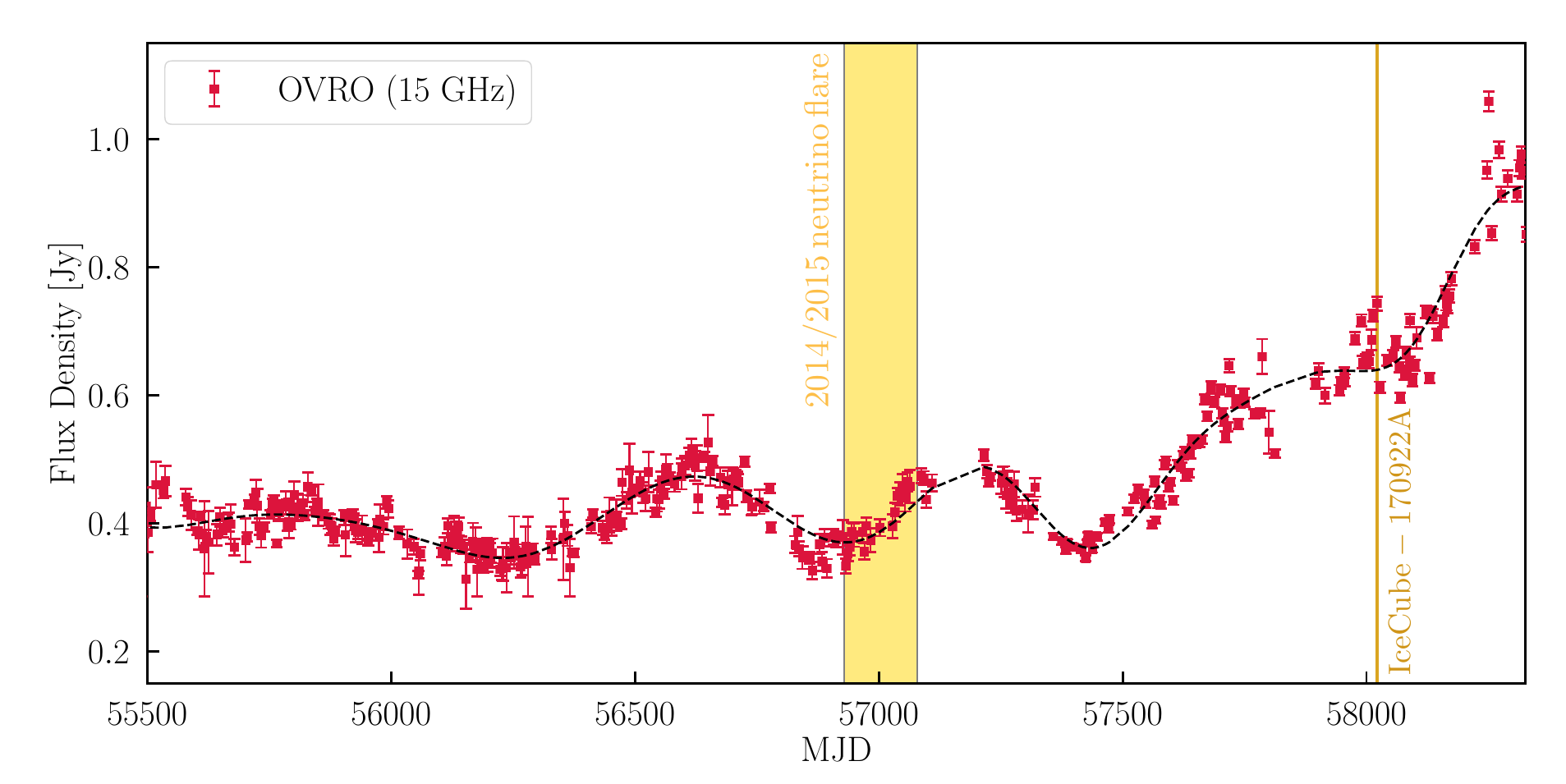}
    \caption{TXS 0506+056 radio light curve from Owen Valley Radio Observatory (OVRO) at 15 GHz (red). THe dashed line illustrates the pattern of the radio flux density. The 2014/15 110-day neutrino flare (yellow band) and the IceCube-170922A episodes are shown. Radio data indicate that the neutrinos arrive during periods of enhanced radio emission.}
    \label{fig:txs-ovro}
\end{figure}

On September 22, 2017, the tenth such alert, IceCube-170922A~\cite{2017GCN.21916....1K}, reported a well-reconstructed muon that deposited 180 TeV inside the detector, and with an estimated neutrino energy of 290\,TeV. Its arrival direction aligned with the coordinates of a known blazar, TXS 0506+056, to within $0.06^\circ$. The source was ``flaring" with a gamma ray flux that had increased by a factor 7 in recent months. A variety of estimates converge on the probability of order $10^{-3}$ that the coincidence is accidental. The identification of the neutrino with the source reaches the level of evidence, but no more. What clinches the association is a series of subsequent observations, culminating with the optical discovery of a switch of the source from an off to an on state two hours after the emission of IC190922, conclusively associating the neutrino with TXS 0506+056:

\begin{itemize}
    \item The redshift of the host galaxy was measured to be $z\simeq0.34$~\cite{Paiano:2018qeq}. It is important to realize that nearby blazars like the Markarian sources are at a redshift that is ten times smaller, and therefore TXS 0506+056, with a similar flux despite the greater distance, is one of the most luminous sources in the Universe. It must belong to a special class of sources that accelerate proton beams in dense environments, revealed by the neutrino. The source is somehow special, and this eliminates any conflict between its observation and the lack of correlation between the arrival directions of IceCube neutrinos and the bulk of the blazars observed by Fermi~\cite{Aartsen:2016lir}.
    \item Originally detected by NASA's Fermi~\cite{2017ATel10791....1T} and Swift\cite{2017ATel10792....1E} satellite telescopes, the alert was followed up by the MAGIC and other air Cherenkov telescopes~\cite{2017ATel10817....1M}. MAGIC detected the emission of gamma rays with energies exceeding 100 GeV starting five days after the observation of the neutrino, indicating that the source is a relatively rare TeV blazar given the large redshift of the source. After correcting for the absorption on the microwave background, we conclude that the source is a TeV blazar.
    \item Given where to look, IceCube searched its archival neutrino data up to and including October 2017 for evidence of neutrino emission at the location of TXS0506+056. When searching the sky for point sources of neutrinos, two analyses have routinely been performed: one that searches for steady emission of neutrinos and one that searches for flares over a variety of timescales. Evidence was found for 19 high-energy neutrino events on a background of less than 6 in a burst lasting 110 days. This burst dominates the integrated flux from the source over the last 9.5 years for which we have data, leaving the 2017 flare as a second subdominant feature. We note that this analysis applied a published prescription to data; the chances that this observation is a fluctuation are small.
    \item Radio interferometry images of the source revealed a jet that loses its tight collimation beyond 5 milliarcseconds running into target material that creates the opportunity for producing the neutrinos. The origin of this target material is still a matter of debate, and speculations include the merger with another galaxy affecting the jet from the dominant supermassive black hole. Alternatively, in a structured jet, the accelerated protons may collide with a slower moving and denser region of jetted photons. The jet may interact with the dense molecular clouds of a star-forming region, or simply with supermassive stars in the central region of the host galaxy \cite{Britzen2019,Kun:2018zin}. The VLBA observations not only identify a striking feature in the jetted material, but also reveal that the neutrino burst occurs at the peak of enhanced radio emission at 15 GHz, which started five years ago, see Fig. \ref{fig:txs-ovro}. The radio flare may be a signature of a galaxy merger; correlations of radio bursts with the process of merging supermassive black holes have been anticipated \cite{Gergely_2009}. 
    \item The robotic optical telescope MASTER network has been monitoring the source since 2005 and found the strongest time variation of the source over a period of two hours after the emission of IC170922, with a second variation following the 2014-15 burst~\cite{2020ApJ...896L..19L}. The blazar switches from the off to the on state two hours after the emission of the neutrino. The observation conclusively associates the source with the neutrino.
\end{itemize}

Additionally, it is important to note the striking fact that all high-energy spectra, for both photons and neutrinos and for both the 2014 and 2017 bursts, are consistent with a hard $E^{-2}$ spectrum, which is expected for a cosmic accelerator. In fact, the gamma-ray spectrum hints at flattening beyond that during the 110-day period of the 2014 burst \cite{Padovani:2018acg,Garrappa:2018}.

The problem with modeling TXS 05060+056 as a blazar is that it is not a blazar at the times of neutrino emission as shown by the multimessenger observation itemized above. It is evident that a source that is transparent to high-energy gamma rays is unlikely to host the target material to produce neutrinos with the opacity for $\gamma\gamma$ interactions two orders of magnitude larger than that for $p\gamma$ interactions. We will work through these arguments quantitatively further on.

\begin{figure}[t]\centering
\includegraphics[width=0.60
\columnwidth,angle=-90]{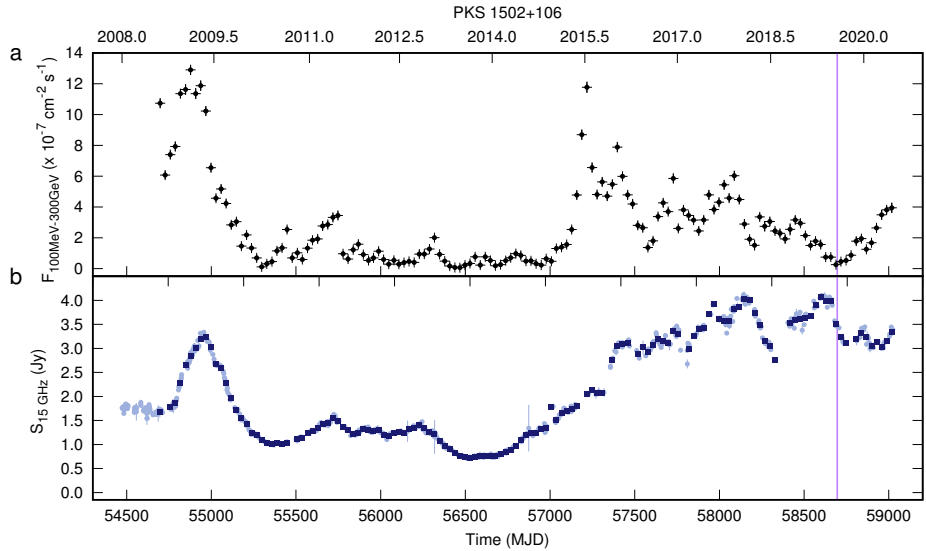}
\caption[]{Temporal variation of the $\gamma$-ray and radio brightness of PKS\,1502+106. {\bf a}: Fermi-LAT likelihood light curve integrated between 100\,MeV and 300\,GeV (marked by black dots with error bars). {\bf b}: OVRO flux density curve of PKS\,1502+106 plotted with light blue dots, that is superimposed by the radio flux density curve binned to the Fermi-LAT light curve (marked with dark blue squares). The detection time of the neutrino IC-190730A is labeled by a vertical purple line.}
\label{fig:fermiovro}
\end{figure}

In this context, an alert recorded by IceCube on July 30, 2019, provides striking support for the idea that cosmic neutrinos are produced by temporarily gamma-suppressed blazars. IC190730 and IC170922 are the two highest energy alerts recorded so far. A well-reconstructed 300-TeV muon neutrino is observed in spatial coincidence with the blazar PKS 1502+106 \cite{2019ATel12971....1L}. OVRO radio observations~\cite{2019ATel12996....1K} show that the neutrino is coincident with the highest flux density of a flare at 15 GHz that started five years ago~\cite{2016A&A...586A..60K}, matching the similar long-term radio outburst observed from TXS 0506+056. Even more intriguing is that the gamma ray flux observed by Fermi shows a clear minimum at the time that the neutrino is emitted; see Fig.~\ref{fig:fermiovro}. At the time that the target crosses the jet and produces a neutrino beam, the high-energy gamma rays are absorbed and their energy cascades to energies below the Fermi threshold, i.e., MeV or lower. For further discussion, see reference~\citen{Kun:2020njy}.

Other IceCube alerts have triggered interesting observations. Following up on a July 31, 2016, neutrino alert, the AGILE collaboration, which operates an orbiting X-ray and gamma-ray telescope, reported a day-long blazar flare in the direction of the neutrino one day before the neutrino detection~\cite{Lucarelli:2017hhh}. Even before IceCube issued automatic alerts, in April 2016, the TANAMI collaboration argued for the association of the highest energy IceCube event at the time, dubbed ``Big Bird,'' with the flaring blazar PKS B1424-418~\cite{Kadler:2016ygj}. The event was produced at a minimum of the Fermi flux~\cite{Kun:2020njy}, as was the case for PKS 1502 +106. AMANDA, IceCube's predecessor, observed three neutrinos in coincidence with a rare flare of the blazar 1ES 1959+650, detected by the Whipple telescope in 2002~\cite{Daniel:2005rv}. However, none of these identifications reach the significance of the observations triggered by IceCube-170922A.

Note that a subset of blazars, around $1 \sim10\,\%$ of all blazars, bursting once in 10 years at the levels of TXS 0506+056, can accommodate the diffuse cosmic neutrino flux observed by IceCube. The energy of the neutrino flux generated by the flaring blazars is also at the same level as the flux in extragalactic cosmic rays. In order to calculate the flux of the high-energy neutrinos from a population of sources, we use the formalism developed in the previous section~\cite{Halzen:2002pg} to relate the diffuse neutrino flux to the energy injection rate of the cosmic rays and their efficiency to transfer this energy from protons to neutrinos. For a population of sources with source density $\rho$ and neutrino luminosity $L_\nu$, the diffuse neutrino flux is
\begin{eqnarray}
E^2 \frac{d N}{dE} = \frac{1}{4\pi} \int d^3 r \frac{L_\nu}{4\pi r^2} \rho\,.
\end{eqnarray}
which can be simplified into
\begin{eqnarray}
E^2 \frac{d N}{dE} = \frac{c}{4\pi} \,t_H \,\xi \,L_\nu \,\rho\,,
\end{eqnarray}
where $\xi$ it the result of the integration over the redshift history of the sources; see Eq.~\ref{xi}. We revise the equation to account for the duration of the flares $\Delta t$, the total time of observation $T_{obs}$, and the fraction of sources $\mathcal{F}$:
\begin{eqnarray}
E^2 \frac{d N}{dE} = \frac{c}{4\pi} \,t_H \,\xi \,L_\nu \,\rho \,\frac{\Delta t}{T} \,\mathcal{F}\,,
\end{eqnarray}
corresponding to 
\begin{eqnarray}\label{diffuse_flux}
3\times 10^{-11} \, {\rm{TeV cm^{-2} s^{-1} sr^{-1}}} = \frac{ \mathcal{F}}{4\pi} \bigg(\frac{R_H}{4.3\, \rm{Gpc}} \bigg) \bigg(\frac{\xi}{0.7 \rm{}} \bigg)
\bigg(\frac{L_\nu}{1.2\times10^{47}\, \rm{erg/s}} \bigg) \nonumber\\ \times \bigg(\frac{\rho}{10^{-8}\, \rm{Mpc^{-3}}} \bigg) \bigg(\frac{\Delta t}{110 \,{\rm d}} \frac{10 {\, \rm yr}}{T_{\rm obs}}\bigg) 
\end{eqnarray}
which results in $\mathcal{F}=0.05$. In summary, a special class of BL Lac blazars with a neutrino luminosity similar to TXS 05060+056, can accommodate the diffuse flux of high-energy cosmic neutrinos observed by IceCube. These high-energy neutrino flaring sources constitute 5\% of the sources. 

The energetics in neutrino production from these sources also matches the energy flux of the highest energy cosmic rays, i.e.,
\begin{eqnarray}
E^2 \frac{d N}{dE} \simeq \frac{c}{4 \pi}\,\bigg( \frac{1}{2}
(1-e^{-f_\pi})\, \xi \,t_H \frac{dE}{dt} \bigg)\,.
\end{eqnarray}
From the observed cosmic ray injection rate of $(1-2) \times 10^{44}\, \rm erg\,Mpc^{-3}yr^{-1}$ we can determine the average pion efficiency of the neutrino sources that makes the match of Eq.~\ref{diffuse_flux} possible:
\begin{eqnarray}
\bigg(\frac{L_\nu}{1.2\times10^{47}\, \rm{erg/s}} \bigg) \bigg(\frac{\rho}{10^{-8}\, \rm{Mpc^{-3}}} \bigg) \bigg(\frac{\Delta t}{110 \,{\rm d}} \frac{10 {\, \rm yr}}{T_{\rm obs}}\bigg) \bigg(\frac{\mathcal{F}}{0.05} \bigg) \nonumber\\ \simeq  \frac{1}{2} (1-e^{-f_\pi})\,\frac{dE/dt}{(1-2) \times 10^{44}\, \rm erg\,Mpc^{-3}yr^{-1}}\,. 
\end{eqnarray}
We find that $f_\pi > 0.8$.

In the section on blazar jets, we discussed how this can be achieved with a proton beam with low boost factor interacting with the blue photons in the active galaxy. The jet producing the neutrinos is not transparent to TeV photons, only to photons with tens of GeV; this is indeed what is observed by Fermi at the time of the 2014 flare.

That TXS 0506+056 belongs to a special class of sources is reinforced by the fact that conventional blazar modeling of the multiwavelength spectrum has been unsuccessful despite the opportunity of tuning 14 free parameters. Blazar models at best meet the requirements for the 2017 burst by invoking Eddington bias \cite{Strotjohann:2018ufz} for the single neutrino produced. The answer to this puzzle could be a new subclass of high-energy gamma-ray sources, labeled as blazars by astronomers, that is responsible for producing the high-energy cosmic neutrinos and cosmic rays observed \cite{Halzen:2018iak}. In fact, if every source labelled as a blazar produced neutrinos at the level of TXS 0506+056, they would overproduce the total diffuse flux observed by IceCube by close to two orders of magnitude.

The observation that the energy content in neutrinos and very high energy cosmic rays are similar, underscores the fact that the cosmic rays must be highly efficient at producing neutrinos, requiring a large target density that renders them opaque to high-energy gamma rays. A consistent picture emerges when the source opacity exceeds a value of 0.4 \cite{Halzen:2018iak}, resulting in a gamma-ray cascade where photons lose energy in the source before cascading to even lower energies in the extragalactic background light. Most of their energy falls below the Fermi threshold by the time they reach Earth. The 2014-15 burst cannot be, and is not, accompanied by a large gamma-ray flare. This is consistent with the premise that a special class of efficient sources is responsible for producing the high-energy cosmic neutrino flux seen by IceCube.

The nature of the special class of sources has not been settled. One straightforward explanation could be a subclass of blazars selected by redshift evolution: powerful proton accelerators producing neutrinos may have been active in the past but are no longer today. This accommodates the large redshift of TXS 0506+056~\cite{Neronov:2018wuo}, which would be the closest among a set of sources that only accelerated cosmic rays at early redshifts. It is a requirement that the sources must have a sufficient target density in photons or protons in order to produce neutrinos at the high flux level observed from TXS 0506+056~\cite{Halzen:2018iak}. If neutrino sources only represent a few percent of those labelled by astronomers as blazars, attempts by IceCube and others to find correlations between the directions of high-energy neutrinos and $\it{all}$ Fermi blazars must inevitably be unsuccessful.

Merger activity in blazars in not uncommon. In merging galaxies there is plenty of material for accelerated cosmic rays to interact with the jet. This fresh material provides optically thick environments, and allows for rapid variation of the Lorentz factors. A cursory review of the literature on the production of neutrinos in galaxy mergers is sufficient to conclude that it can indeed accommodate the observations of both the individual sources discussed above and the total flux of cosmic neutrinos; see, e.g., references~\citen{Kashiyama:2014rza,Yuan:2017dle,Yuan:2018erh}.

There is another, possibly related, development associated with the search for steady neutrino sources that recently delivered a new time-integrated neutrino sky map covering ten years of IceCube data \cite{Carver:2019jcd}. Evidence at the 3$\sigma$ level suggests that the neutrino sky map might no longer be isotropic. The anisotropy results from four sources-TXS 0506+056 among them-that show evidence for clustering at the 4$\sigma$ level. The strongest of these sources is the nearby active Seyfert galaxy NGC 1068 (Messier 77) with starburst activity. There is evidence for shocks near the core and for molecular clouds with densities of more than $10^5\,\rm cm^{-3}$. Similarly to TXS 0506+056, a merger onto the black hole is observed - either with a satellite galaxy or, more likely, with a star-forming region~\cite{2014A&A...567A.125G}. This major accretion event may be the origin of the increased neutrino emission.

The assumption that the evolution of most radio galaxies is governed by mergers  \cite{2001PhDT.......173R, Gergely_2009} supports the proposal of a redshift evolution selection of a special class of sources to account for the cosmic neutrino flares observed by the IceCube Collaboration. Alternative scenarios proposed to describe the evolution of the radio galaxies based on their morphology, such as twin active galactic nuclei \cite{Lal_2019} and back-flow \cite{Capetti:2002da}, also provide a potential production mechanism for the cosmic neutrino flares. In this context, previous studies of the neutrino emission from radio galaxies are also relevant as they require a high density close to the source \cite{BeckerTjus2014, Hooper:2016jls}. Furthermore, far-infrared observations support the argument that the central activity of the galaxies, such as starburst activity, is fueled by the concentration of matter that may be preceded by a merger \cite{sanders1996}.

In a major merger scenario, when two galaxies with supermassive black holes at their centres with different spins merge, a spin-flip is expected to occur due to the spin-orbit precession and energy dissipation from gravitational radiation  \cite{Gergely_2009, Biermann:2008yv}. This is accompanied by precession of the jet, and the activity may therefore not always look like a blazar because the jet might be only briefly pointing into our direction. It has been suggested that the precessional phase of the merger of two black holes, occurring prior to the spin-flip, is visible as a superdisk in radio galaxies where the precessing jet emerges as a superwind separating the radio lobes in the final stages of the merger\cite{Gopal-Krishna2007}.

As discussed in the previous sections, the absence of a strong anisotropy of neutrino arrival directions raises the possibility that the cosmic neutrinos originate from a number of relatively weak extragalactic sources. It is indeed important to keep in mind that the interaction rate of a neutrino is so low that it travels unattenuated over cosmic distances through the tenuous matter and radiation backgrounds of the Universe. This makes the identification of individual point sources contributing to the IceCube flux challenging~\cite{Lipari:2006uw,Becker:2006gi,Silvestri:2009xb,Ahlers:2014ioa}. Even so, it is also important to realize that IceCube is capable of localizing the sources by observing multiple neutrinos originating in the same location. Not having observed strong neutrino clusters in the present data raises the question of how many events are required to make such a model-independent identification possible. The answer to this question suggests the construction of a next-generation detector that instruments a ten times larger volume of ice~\cite{Aartsen:2014njl}.

\newpage
\section{Modeling Cosmic Beam Dumps}

In this section we will introduce the generic theoretical framework used to calculate the neutrino flux associated with an astronomical source where protons are accelerated to energies exceeding the threshold for producing pions that decay into neutrinos. We will work through several examples. We start with the generic beam dump where protons accelerated near the black hole hole at the center of an active galaxy interact with the gas (protons) in their vicinity, i.e the equivalent of the $1 \simeq \rm{proton}/cm^{3}$ in our own galaxy. Next, we will discuss two typical blazar models where accelerated protons interact with photons associated with the jet, or, alternatively, with strong external radiation fields near the black hole. Finally, we turn from jets to fireballs and discuss the production of neutrinos when protons interact with photons in the relativistically expanding fireball following a stellar collapse, such as in a gamma-ray burst.

Neutrinos are produced when pions, and, at higher energies, kaons and charm particles decay. We start by calculating the number of pions produced {\it at the source} when an accelerated proton beam interacts with gas of density $n$ in the rest of the galaxy. We introduce the source function $q_\pi\left(E_\pi\right)$ defined such that $q_\pi\left(E_\pi\right)dE_\pi$ represents the rate at which pions are produced within the energy range $E_\pi$ and $E_\pi+dE_\pi$ per unit time:
\begin{equation}
q_\pi=\frac{dN_\pi}{dE_{\pi}dt}=\int dE_p\int_{0}^{\tau} d\tau'\,q_p(E_p,\tau')\,\frac{dN_\pi}{dE_\pi}\left(E_\pi\right)\,,
\label{defq}
\end{equation}
where:
\begin{equation}
q_p(E_p,\tau')= \exp(-\tau')\,dN_p/dE_p.
\end{equation}
Here $dN_p/dE_p$ is the {\it unattenuated} proton production rate at the source, in units $\rm GeV^{-1}s^{-1}$; it is often labeled  as $j_p(E_p)$ in the literature. The proton rate is absorbed on a target with total optical depth $\tau$ producing secondary pions with an energy distribution $dN_\pi/dE_\pi$, normalized to unity; it can be parametrized in terms of the multiplicity of $n_\pi$ and the average energy $<E_\pi>$ of the pions produced:
\begin{equation}
\frac{dN_\pi\left(E_\pi\right)}{dE_\pi}=n_{\pi^\pm}\,\delta\left(E_{\pi}-\left<E_{\pi}\right>\right)\,,
\end{equation}
The transmittance $T$ of the target is
\begin{equation}
T = \exp(- \int_0^l dr' \alpha\left(r'\right))\,,
\end{equation}
where $l$ is the path length of the beam in a target and $\alpha$ is the attenuation coefficient (the inverse of the mean-free path $\lambda$) which is determined by the cross section and the density $n$
\begin{equation}
\alpha=n\sigma\,,
\end{equation}
and, assuming isotropy,
\begin{equation}
\tau=nl\sigma\,=N\sigma\,,
\end{equation}
where $N$ is the column density.

Typically, one also makes the approximation that the proton-proton cross section is independent of energy, $\sigma \simeq 3\cdot 10^{-26}$~cm$^{2}$ and therefore the integrals over $\tau$ and $E_p$ separate: 
\begin{equation}
q_{\pi^{\pm}} = \left(1-\exp(-\tau)\right) \, \int dE_{\rm p}\, \frac{dN_p}{dE_p}\,n_\pi\, \delta\left(E_{\pi}-\left<E_{\pi}\right>\right)\,,
\label{dump}
\end{equation}
In most astrophysical situations, where the optical depth is small, one can replace the production efficiency $\left(1-\exp(-\tau)\right)\rightarrow \tau$, with $\tau=l\,n\,\sigma$.

The integral can be performed by rewriting the delta function as a function of $E_p$. Each time a proton interacts, it deposits $\kappa E_p$ energy into $\left<n_\pi\right>$ pions of average energy $\left<E_\pi\right>$; here, $\kappa$ is the total proton inelasticity previously introduced. Energy conservation implies that
\begin{equation}
\kappa E_p=E_\pi^{tot} = n_{\pi^\pm} \left<E_\pi\right>\,.
\label{energyconservation}
\end{equation}
As before, $x_\pi$ is the average relative energy of a pion relative to the initial proton energy
\begin{equation}
x_\pi = \frac{\left<E_\pi\right>}{E_p} = \frac{\kappa}{n_\pi}.
\end{equation}
We obtain the result that
\begin{equation}
q_{\pi^{\pm}} = \tau\,n_\pi \int dE_{\rm p} \frac{dN_p}{dE_p}\delta\left(E_{\pi}-x_\pi E_p\right)\,,
\end{equation}
or
\begin{equation}
q_{\pi^{\pm}} = N\,\sigma\,n_\pi\,\frac{1}{x_\pi}\,\frac{dN_p}{dE_p}\left(\frac{E_\pi}{x_\pi}\right)\,.
\label{eq:ptopienergy}
\end{equation}
The result is transparent: the pion source function is proportional to the intensity of the proton beam, the optical depth of the target and the cross section, and the multiplicity of the pions produced.

We are now able to compute the rate at which neutrinos are produced, with the total rate given by summing the emissivities of the muon neutrino from the pion decay, and those of the second muon neutrino and the electron neutrino from the muon decay:
\begin{equation}
q_{\nu,{\rm tot}}=q_{\nu_{\mu}}^{(1)}+q_{\nu_{\mu}}^{(2)}+q_{\nu_{e}}\,.
\end{equation}
Assuming that the total energy of the pions is distributed equally among the
four decay leptons,
\begin{equation}
q_{\nu_{i}}(E_{\nu_{i}})=q_{\pi}(4 E_{\nu_{i}})\,\frac{dE_{\pi}}{dE_{\nu_{i}}}=4 q_{\pi}(4 E_{\nu_{i}})
\end{equation}
for each neutrino,
$\nu_{i}=\overline{\nu}_{e} \rm \,or\,\nu_{e},\,\nu_{\mu},\,
\overline\nu_{\mu}$. Because IceCube does not distinguish neutrinos and antineutrinos, we will not separate them. In this approximation the number of neutrinos produced per flavor in the energy bin
$dE_{\nu}$ originate from the original pion in the energy bin
$dE_{\pi}=4 dE_{\nu}$; therefore, $q_{\nu}(E_{\nu})\,dE_{\nu}=q_{\pi}(4\,E_{\nu})dE_{\pi}$.

We next have to translate the results for neutrino production at the source to the point source flux $E_\nu\phi_{\nu_\alpha}(E_\nu)$ observed at Earth, where $\alpha$ labels the neutrino flavor.  For a single source the emissivity at Earth is $q_{\nu}(E_{\nu})/4\pi r^2$, where $r$ is the distance to the source. So far IceCube has not pinpointed such a flux, instead it discovered a diffuse flux from a, yet unidentified, source population with uniform density $\rho(r)$ in the Universe:
\begin{equation}
\phi_{\nu_\alpha}(E_\nu) = \frac{1}{4\pi} \int d^3r \,\rho \left(r\right) \frac{q_\nu\left(E_\nu\right)}{4\pi r^2}\,,
\end{equation}
where the first factor $1/4\pi$ is introduced in order to define the diffuse flux with the conventional units $GeV^{-1}cm^{-2}s^{-1}sr^{-1}$.
Therefore,
\begin{equation}
\phi_{\nu_\alpha}(E_\nu) = \frac{1}{4\pi} \int dr\,4\pi r^2 \rho \left(r\right) \frac{q_\nu\left(E_\nu\right)}{4\pi r^2}\,,
\end{equation}
or
\begin{equation}
\phi_{\nu_\alpha}(E_\nu) = \frac{1}{4\pi} \int dr \rho \left(r\right) q_\nu\left(E_\nu\right)\,.
\end{equation}
For illustration we assumed a Euclidian Universe, an approximation that is approximately valid for nearby sources. Integrating over the cosmology of the Universe is done by changing the integration from $dr \Rightarrow cdt \Rightarrow cdz(dt/dz)$, with $dz/dt = H(z)$, the Hubble scaling factor. We thus can rewrite the integral in covariant form
\begin{equation}
\phi_{\nu_\alpha}(E_\nu)= \frac{c}{4\pi} \int \frac{dz}{H\left(z\right)} \,\rho \left(z\right) q_\nu\left(\left(1+z\right) E_\nu\right)\,.
\end{equation}
For the standard $\Lambda$CDM cosmological model, the Hubble parameter scales as $H^2(z) = H^2_0[(1+z)^3\Omega_{\rm m} + \Omega_\Lambda]$, with $\Omega_{\rm m} \simeq 0.3$, $\Omega_\Lambda\simeq 0.7$, and the Hubble distance $c/H_0 \simeq 4.4$~Gpc~\cite{Olive:2016xmw}.

In the following, we will typically assume that the neutrino emission rate $q_{\nu_\alpha}$ follows a power law $E^{-\gamma}$ and the flavor-averaged neutrino energy density can then be written as
\begin{equation}\label{eq:Lnu}
\frac{1}{3}\sum_\alpha E_\nu^2\phi_{\nu_\alpha}(E_\nu) \simeq \frac{c}{4\pi}\frac{\xi_z}{H_0}\rho_0\frac{1}{3}\sum_\alpha E_\nu^2 q_{\nu_\alpha}(E_\nu)\,,
\end{equation}
where we introduce the redshift factor
\begin{equation}\label{xi}
\xi_z = \int_0^\infty{\rm d}z\frac{(1+z)^{-\gamma}}{\sqrt{\Omega_\Lambda+(1+z)^3\Omega_{\rm m}}}\frac{\rho(z)}{\rho(0)}\,.
\end{equation}
A spectral index of $\gamma\simeq2.0$ and no source evolution, $\rho(z)=\rho_0$, yields $\xi_z\simeq0.6$, whereas the same spectral index and source evolution following the star formation rate yields $\rho\simeq 2.4$. An earlier discussion on how to introduce the cosmological evolution of the sources can be found in section~\ref{sec:nuanduhcr}.

The identical procedure can be followed for the production of neutral pions that subsequently decay into two gamma rays. This will lead to Eq.~\ref{eq:Qgamma}, previously introduced:
\begin{equation}
\frac{1}{3}\sum_{\nu_\alpha}E_\nu \frac{{\rm d}N_\nu}{{\rm d}E_\nu {\rm d}t}(E_\nu) = \frac{K_\pi}{2}E_\gamma \frac{{\rm d}N_\gamma}{{\rm d}E_\gamma {\rm d}t}(E_\gamma)\,.
\end{equation}
This is a powerful relation because it only depends on the ratio of neutral to charged pions produced which is determined by isospin.

We complete this section with a discussion of the multimessenger relations between neutrinos, photons and protons. Eq.~\ref{eq:ptopienergy} can be rewritten as
\begin{equation}
q_{\pi^{\pm}} = \tau\,\frac{n_\pi}{x_\pi}\,q_p\left(\frac{E_\pi}{x_\pi}\right)\,,
\end{equation}
or,
\begin{equation}\label{eq:CRtoPION2}
E_\pi^2q_{\pi^\pm}(E_\pi) \simeq f_\pi\, \frac{K_\pi}{1+K_\pi}\,\left[E^2_pq_p(E_p)\right]_{E_p = E_\pi/x_\pi}\,,
\end{equation}
with, as before, $K_\pi = 2$ for $pp$ and $K_\pi = 1$ for $p\gamma$ interactions. The charged pion production rate $q_{\pi^\pm}$ is proportional to the rate of the protons accelerated in the cosmic accelerator, $q_p$, by the ``bolometric'' proportionality factor $\kappa f_\pi$ previously introduced, and $\kappa$ is the inelasticity factor of 0.5(0.2) for $pp(p\gamma)$ interactions. In general, the ``proton'' emission rate, $q_p$, has to be generalized to the composition of the cosmic rays that may not all be protons. Using superposition one can relate the spectra of nuclei with mass number $A$ as $q_N(E_N) = \sum_A A^2q_A(AE_N)$.

As previously pointed out, the relation between the photons and neutrinos do not depend on the initial proton beam. With the usual approximations
\begin{equation}
x_{\nu} = \frac {E_{\nu}}{E_{p}} = \frac {1}{4}\, x_\pi  \;\simeq \,\frac {1}{20}\,\,{\rm and}\,\, x_{\gamma} = \frac {E_{\gamma}}{E_{p}} = \frac {1}{2}\, x_\pi
 \;\simeq\, \frac {1}{10}\,,
\label{eq:xnu}
\end{equation}
the neutrino production rate $q_{\nu_\alpha}$ can be related to the one for charged pions:
\begin{equation}\label{eq:PIONtoNU2}
\frac{1}{3}\sum_{\alpha}E_\nu q_{\nu_\alpha}(E_\nu) \simeq \left[E_\pi q_{\pi^\pm}(E_\pi)\right]_{E_\pi \simeq 4E_\nu}\,.
\end{equation}
Using Eqs.~(\ref{eq:CRtoPION2}) and (\ref{eq:PIONtoNU2}), we arrive at the final relation for neutrino production:
\begin{equation}\label{eq:CRtonu2}
\frac{1}{3}\sum_{\alpha}E^2_\nu q_{\nu_\alpha}(E_\nu) \simeq \frac{1}{4}f_\pi \frac{K_\pi}{1+K_\pi}\left[E^2_pq_p(E_p)\right]_{E_p = 4E_\nu/x_\pi}\,.
\end{equation}
The production rate of gamma rays from the decay of neutral pions can be obtained in exactly the same way.

From the two equations for the productions of neutrinos and gamma rays, one can eliminate $q_p$ to obtain a model-independent relation that only depends on the relative contribution of charged-to-neutral pions,
\begin{equation}\label{eq:GAMMAtoNU}
\frac{1}{3}\sum_{\alpha}E^2_\nu q_{\nu_\alpha}(E_\nu) \simeq \frac{K_\pi}{4}\left[E^2_\gamma q_\gamma(E_\gamma)\right]_{E_\gamma = 2E_\nu}\,.
\end{equation}
Here, the prefactor $1/4$ accounts for the energy ratio $E_\nu/E_\gamma\simeq 1/2$ and the two gamma rays produced in the neutral pion decay. The relation simply reflects the fact that a $\pi^0$ produces two $\gamma$ rays for every charged pion producing a $\nu_\mu + \bar\nu_\mu$ pair, which cannot be separated by current experiments. This is the simple counting that was used to derive the energy density in gamma rays in the universe accompanying the flux of cosmic neutrinos observed by IceCube, shown in Fig.~\ref{fig:panorama}.

\subsection{Active Galaxies: A Worked Example}

An active galaxy presents multiple opportunities for the acceleration of cosmic rays in the inflows and outflows or jets associated with the supermassive black hole. The high-energy cosmic rays will subsequently produce neutrinos in interactions with a variety of targets such as the hydrogen and molecular clouds in the galactic disk. It is therefore useful to start by considering a generic beam dump where a beam of protons with an initial rate $dN_p/dE_p$ interacts with a target of density $n$ over a distance $l$ using the general formalism introduced above. For analytic calculations one can use a simple parameterization of the average pion multiplicity that allows for its observed increase with proton energy above the threshold $pp \rightarrow pp+\pi^+\pi^-$~\cite{Mannheim:1994sv}:
\begin{equation}
n_{\pi^\pm}=2  \left(\frac{E_{p}-E_{th}}{\rm GeV}\right)^{1/4}\,.
\end{equation}
Also the increase of their average energy can be adequately parametrized as:
\begin{equation}
\left<E_{\pi}\right>=\frac{1}{6}(E_{p}-m_p c^2)^{3/4}\,{\rm GeV}.
\end{equation}
Given the number of protons produced by the accelerator per energy and time interval
\begin{equation}
\frac{dN_{p}}{dE_{p}}=A_{p}  \left(\frac{E_{p}-m_p  c^2}{{\rm GeV}}\right)^{-\gamma},
\end{equation}
we obtain a pion rate {\it at the source} using Eq.~\ref{dump}:
\begin{equation}
q_{\pi^{\pm}}(E_\pi)\approx 26\,n\,l\,A_{p}\,\sigma \left(\frac{6  E_{\pi}}{\rm GeV}\right)^{-\frac{4}{3}(\gamma-\frac{1}{2})}\,.
\end{equation}
The final result for the total neutrino emission rate at the source is given by
\begin{equation}
q_{\nu,{\rm tot}}\approx3\times10^{2}\,n\,l\,A_{p}\, \sigma  \left(\frac{24  E_{\nu}}{{\rm GeV}}\right)^{-\frac{4}{3}\gamma+\frac{2}{3}}\,,
\label{nusource}
\end{equation}
which provides us with an estimate of the total neutrino flux at the source in terms of three key quantities: $A_{p}$ and $\gamma$, the normalization and spectral slope of the luminosity of the accelerator, and the column density of the target $N \sim l   n$. The spectral index is routinely taken to be $\gamma=2$, a value suggested by diffusive shock acceleration and, in any case, typical for the spectra of gamma rays observed for nonthermal sources.

The column density $N=l  n$ is taken from astronomical information, with $l$ being the distance that the cosmic rays travel through a target of density $n$. For instance $l$ could be the diffusion length of the cosmic rays propagating in the magnetic field of the galaxy. Diffusion allows the protons to interact with the typical density of hydrogen of $n\simeq1$\,cm$^{-3}$ over an extended path length determined by the diffusion time. The diffusion distance before escaping galaxy is given by
\begin{equation}
d_{diff}=2\,\sqrt{D(E_{\rm p})\,t_{esc}}\,,
\end{equation}
where $D$ is the diffusion coefficient and $t_{esc}$ is the escape time. In this case, the size of the target is identified with $l=c   t_{esc}$. For instance $l \simeq 10\,$\,kpc for our own Galaxy.

An interesting exercise can be performed by fixing the proton luminosity $A_{p}$ from the requirement that the aggregate diffuse proton flux from all AGN reproduces the cosmic ray flux observed at Earth; see, for instance, reference~\citen{Hooper:2016jls}. Assuming all other parameters entering in the calculation to be similar to those observed in our own Galaxy, one obtains a diffuse neutrino flux that is consistent with the one observed by IceCube. The accompanying flux of photons also accommodates the diffuse high-energy photon flux observed by Fermi. This attractive solution to the cosmic ray problem is unfortunately not supported by other aspects of the IceCube data, most prominently the lack of correlation of the arrival directions of high-energy neutrinos with Fermi sources~\cite{Aartsen:2016lir}. Also, this attractive solution to the cosmic ray problem implies that radio galaxies are responsible for the majority of Fermi photons, which may be an issue with an extragalactic diffuse flux apparently dominated by blazars. 

An alternative exercise is to determine $A_{p}$ from the radio luminosity of AGNs, $L$, that results from the synchrotron radiation of electrons accelerated along with the protons, with $L_e=\kappa  L$ and $\kappa \geq 1$. $A_p$ is obtained from the assumptions that protons and electrons are connected by a constant fraction $f_e$, with $L_{\rm e}=f_e  L_{\rm p}$, which is on the order of $0.1$~\cite{Tjus:2014dna}. With these assumptions we obtain a relation between the total cosmic ray and radio luminosities:
\begin{equation}
L_{p}=\int \frac{dN_{\rm p}}{dE_{p}}\,dE_{p} \approx \frac{\chi  L}{f_e}\,,
\end{equation}
and
\begin{equation}
A_{p}=A_{p}(L,z)= \frac{\chi L}{f_e}  \left[\ln\left(E_{\max}/E_{\min}\right)\right]^{-1}\,{\rm GeV}^{-2}\,
\label{norm}
\end{equation}
for $\gamma=2$; generalization for $\gamma \neq2$ is straightforward. Observed values of $\chi$ can be found in reference~\cite{Tjus:2014dna}.

Having related the proton flux to the radio luminosity, we obtain the neutrino for a single AGN from Eq.~\ref{nusource}. The diffuse flux that can be confronted with the IceCube data is obtained using the formalism for summing over the sources previously introduced:
\begin{equation}
\Phi_\nu=\int_{L}\int_{z}\frac{q_{\nu,tot}}{4\,\pi\,d_L(z)^{2}}  \frac{dn_{\rm AGN}}{dV\,dL}  \frac{dV}{dz}\,dz\,dL\,.
\label{diffuse}
\end{equation}
Here, $d_L$ is the luminosity distance, $dn_{\rm AGN}/dV\,dL$ is the radio luminosity function of the AGN, and $dV/dz$ is the comoving volume at a fixed redshift~$z$. The radio luminosity function is usually separated into the product of a luminosity-dependent and a redshift-dependent function, $dn_{\rm AGN}/dV\,dL = g(L) f(z)$. The result matches the IceCube observations; more details can be found in reference~\citen{Tjus:2014dna} for column densities typical for the relatively dense targets near the AGN cores, as had been suggested by reference \citen{Stecker:2005hn}.

The two calculations illustrate how two different mechanisms manage to accommodate the IceCube result. Hooper~\cite{Hooper:2016jls} normalizes the proton flux to the  cosmic ray flux and generates the neutrino flux by diffusing the protons confined in the Galaxy, while Tjus {\it et al.}~\cite{Tjus:2014dna} relate the proton flux to the radio emission of the galaxies and produce the neutrinos in the dense matter near the supermassive black hole.

High-energy neutrino emission from photohadronic ($p\gamma$) interactions, where the accelerated particles interact with dense radiation fields, can be modeled in a similar way. Following the scheme previously introduced, the optical depth of $p\gamma$ interaction is given by 
\begin{equation}
    \tau_{p\gamma} = \frac{ct_{\rm esc}}{\lambda_{p\gamma}}\,,
\end{equation}

where $ct_{\rm esc}$ is the distance traveled by the accelerated protons in the photon target, and $\lambda_{p\gamma}$ is the mean free path in that target
\begin{equation}
\lambda^{-1}_{p\gamma}
= \int ds \frac{dn_{\gamma}}{d\epsilon_\gamma}(s) \sigma_{\gamma p}(s). 
\label{lambdapg}
\end{equation}

Here, ${dn_{\gamma}}/{d\epsilon_\gamma}$ is the spectrum of the target photons and $\sigma_{p\gamma}$ is the cross section for $p\gamma$ interaction. The square of the center of mass energy $s$ in the interaction is given by 
\begin{equation}
    s = m_p^2+2m_p\epsilon^\prime,
\end{equation}

where $\epsilon^\prime$ is the photon energy in proton's rest frame. The target photon spectrum can be approximated as a power law, or it may be a monochromatic ``line", approximated by a delta function at a specific energy. By considering the photo-meson interaction kinematics, one can find the neutrino flux for a photohadronic beam dump by taking in to account the initial cosmic ray energy, the optical depth, and the kinematics of the photoproduction of pions:
\begin{equation}
    \frac{dN_\nu}{dE_{\nu}}=\int dE_{\rm p} \frac{dN_{\rm p}}{dE_{\rm p}} \xi(E_{\rm p}, E_\nu) \tau_{p\gamma},
\end{equation}

\noindent where $\xi(E_{\rm p}, E_\nu)$ encapsulates the energy distribution of the final state pions: 
\begin{equation}
\xi(E_{\rm p}, E_\nu) = \int dE_{\pi} \frac{d \sigma_{\gamma p}}{dE_{\pi}}(E_{\pi}; E_{\rm p}) \frac{dN_\pi}{dE_{\nu}}(E_{\nu}; E_{\pi}). 
\label{eq:nudist}
\end{equation}

${dN_\pi} / dE_{\nu}$ is the energy distribution of neutrinos produced in the decay of charged pions with energy $E_{\pi}$. 

As was done for the $pp$ case, Eq.~\ref{defq}, the initial proton spectrum can be modeled as a power law. The integral on $s$ in Eq. \ref{eq:nudist} can be simplified when particles energies meet the required threshold for the $\Delta$-resonance. We will use this approximation later when exploring the neutrino emission from blazars. For more details on the kinematics of the $p\gamma$ interaction, see references~\citen{1990cup..book.....G} and~\citen{Yoshida:2012gf}.

In reality, AGNs are episodic sources producing gamma rays in bursts, with episodes where the flux increases by orders of magnitude for periods from seconds to months, sometimes years. A correlation of the arrival of IceCube neutrinos in coincidence with such bursts can provide a smoking gun for their origin as illustrated by the observation of TXS 0506+056. We will this possibility next.

\subsection{Neutrino Flux Associated with the Plasma Blobs in the Jets of Active Galaxies}
Particle accelerators provide a textbook example of special relativity. So do cosmic accelerators, but they can do much better. Instead of accelerating protons, they can boost really large masses to relativistic velocities. The radio emission reveals that the plasma in the jets of active galaxies flows with velocities of $0.99\,c$. A fraction of a solar mass per year can be accelerated to relativistic Lorentz factors of order 10 leading to luminosities close to the Eddington limit. In the collapse of very massive stars $10^{51} \sim 10^{52}$\,erg/s is released in a fireball that expands with velocities of $0.999\,c$. It is therefore more convenient to introduce the Lorentz factor $\Gamma = [1-(\frac{v}{c})^2]^{-1/2}$.
\begin{figure}[ht!]
  \centering
   \includegraphics[width=0.5\linewidth]{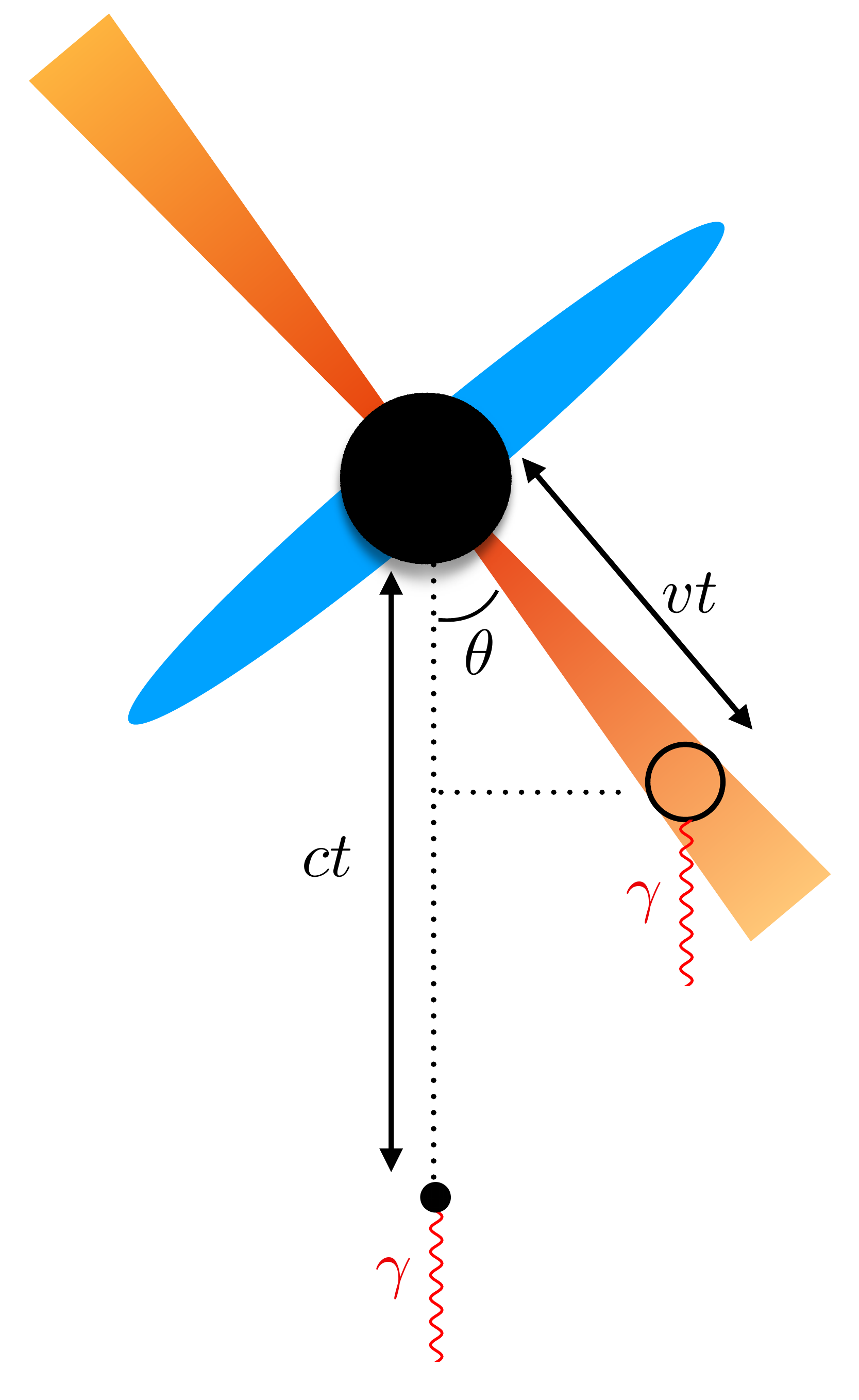}
  \caption{A blob of plasma moves with a velocity v during a time t at an angle $\theta$ relative to the line of sight of the observer.}
  \label{fig:superluminal}
\end{figure}
The plasma flow in the jets can be efficiently extracted from the black hole in terms of ``blobs" of plasma whose radiation is clearly visible in radio and X-ray maps. When observing relativistic bulk motion moving close to the speed of light, the different paths of the photons reaching us must be taken into account. Fig.~\ref{fig:superluminal} shows a blob B moving with a velocity $v$ at an angle $\theta$ relative to our line of sight to the black hole A positioned at the base of the jet. After a time $\Delta t$ the relative separation of the the two photons on the sky is $v sin\theta$ and the photon from A is delayed by a time $(c-v\, cos\theta)/c$. This leads to an apparent velocity of the photon from point A:
\begin{equation}
v_{app} = \frac{v\, sin\theta}{1 - \frac{v}{c}\, cos\theta}.
\end{equation}
For small viewing angles the apparent velocity exceeds the speed of light when $v \rightarrow c$. This is a purely geometric effect arising from the Doppler contraction of the photons. The maximal superluminal velocity $v_{app} = \Gamma v$ is reached for $\theta = \Gamma v/c$. Some important implications: 
\begin{itemize}
\item If the blob A radiates isotropically, the radiation will be strongly anisotropic in the frame of the observer with one half the photons observed in a semi-aperture $sin\theta = 1/\Gamma$. 
\item The difference $\Delta t$ between the photons will be contracted in the frame of the observer. After a travel time $t_b$ of the blob in Fig.~\ref{fig:superluminal} the photons are separated by a time
\begin{equation}
\Delta t = t_b\, (1-v\,cos\theta)=\Gamma t'_b\, (1-v\,cos\theta) \rightarrow \Gamma t'_b\,,
\end{equation}
with the last equality valid for the maximal superluminal effect. The factor $[ \Gamma t'_b\, (1-v\,cos\theta)]^{-1}$ is referred to as the boost factor $\delta$.
\item The energies of the photons are boosted because frequency $\nu$ behaves as the inverse of time.
\item The photon luminosity is actually boosted by a factor $\Gamma^2$ at maximum. This follows from the fact that the {\it number} of photons has to be conserved between the two frames. Origin of the $\Gamma^2$ dependence of the luminosity follows from the frame invariance of the number of particles in a relativistically invariant volume $d^3 x d^3 p$; see the textbooks by Rybicky and Lightman~\cite{Rybicki:2004hfl} and Dermer~\cite{dermer2009high}, for instance.
\end{itemize}
Overall what is happening in the blazar is not quite as spectacular as what we see, observing boosted energies and luminosities over contracted times relative what is actually the reality in the frame of the blazar.

\begin{figure}[ht!]
\begin{center}
\includegraphics[width=0.8\columnwidth]{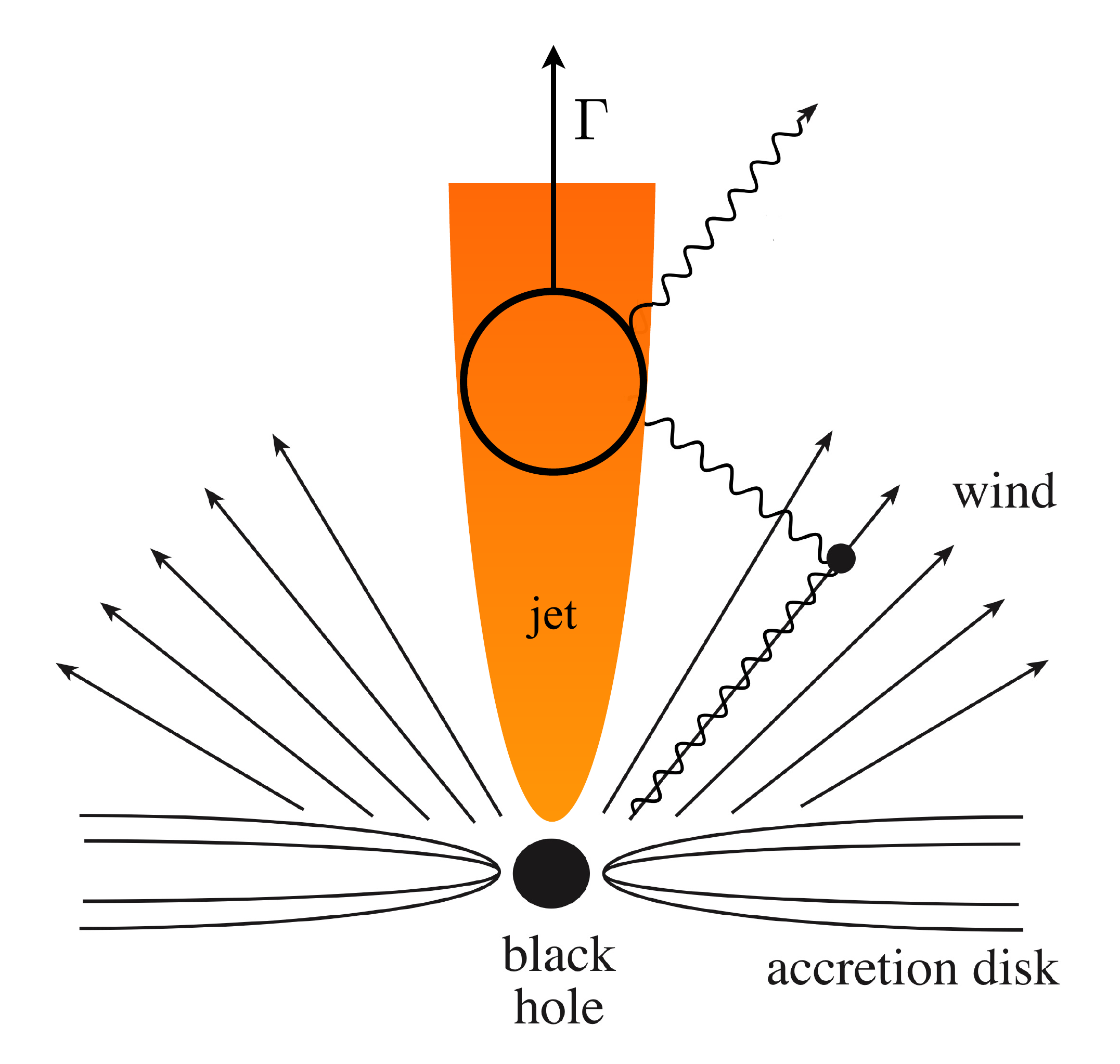}
\caption{Blueprint for the production of high-energy photons and neutrinos near the super-massive black hole powering an AGN. Particles, electrons and protons, accelerated in sheets or blobs moving along the jet, interact with photons radiated by the accretion disk or produced by the interaction of the accelerated particles with magnetic fields. The jet is pointing at the Earth in the subclass of AGN dubbed blazars.}\label{old_bh}
\end{center}
\end{figure}

Blazars are the brightest sources of high-energy gamma rays in the Universe making up most of the diffuse extragalactic gamma ray flux observed by the {\em Fermi} satellite. Their energy density $E_\gamma^2q_\gamma$ is very similar to the one measured by IceCube in cosmic neutrinos, suggesting a possible common origin. Their engines must not only be powerful, but also extremely compact because their luminosities are observed to flare by over an order of magnitude over time periods that are occasionally as short as minutes. Blazars are AGN with the jet directly pointing at our telescopes thus further boosting the energy of the gamma rays and contracting the duration of the episodes of emission. The drawings of a blazar, shown in Figs.~\ref{fig:windingB} and \ref{old_bh}, display its most prominent features: an accretion disk of stars and gas falling onto the spinning supermassive black hole as well as a pair of jets aligned with the rotation axis. Large magnetic fields originating from the inflow of particles on the black hole are wound up along the rotation axis and launch a pair of jets that are the site of the acceleration of electron and proton beams as conceptualized in Fig.~\ref{fig:windingB}.

\begin{figure}[ht!]
  \centering
   \includegraphics[width=0.95\linewidth]{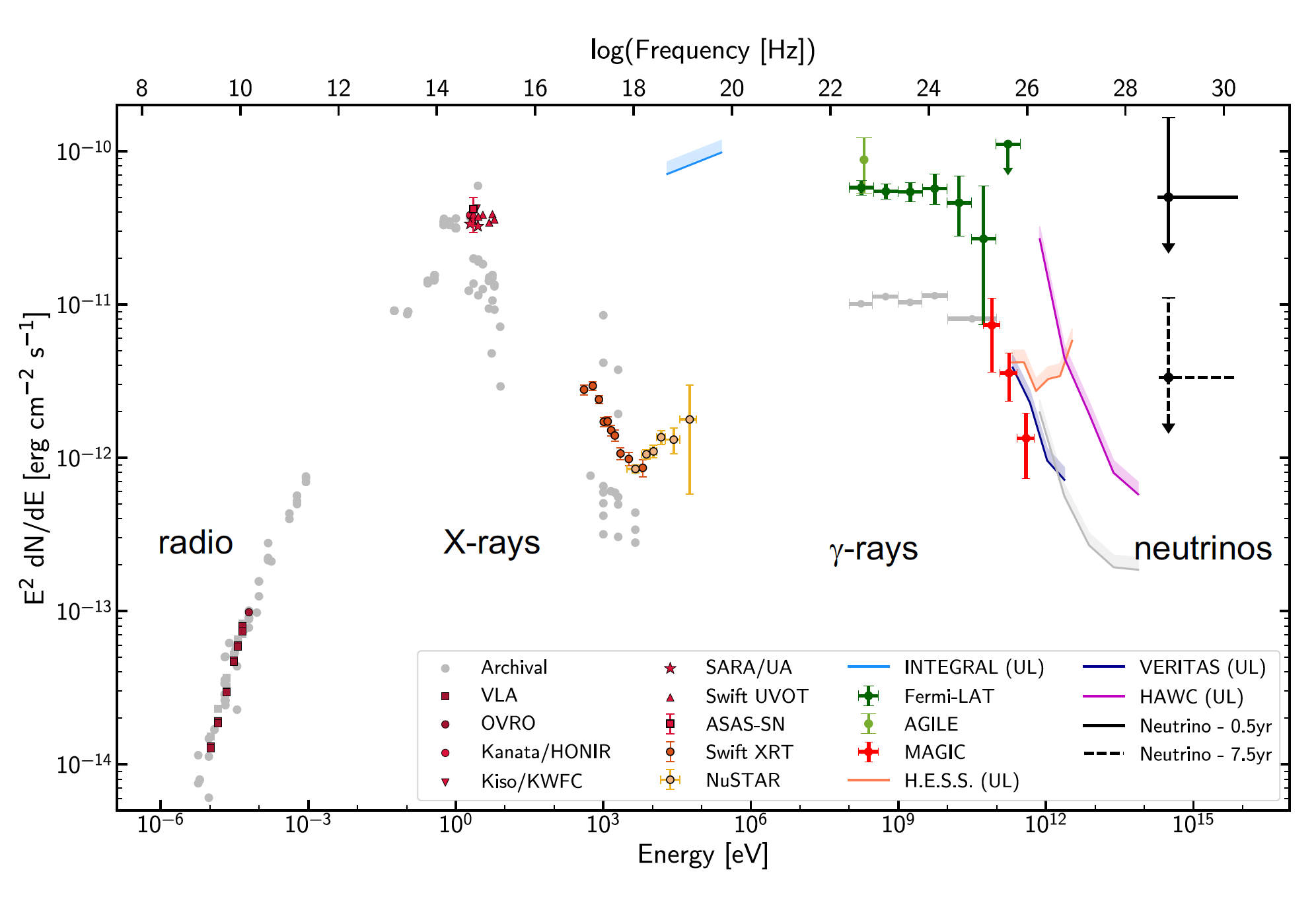}
  \caption{Multiwavelength spectrum of TXS 05060+056 from the multimessenger observations following the IceCube neutrino alert IC170922~\cite{IceCube:2018dnn}}.
  \label{nuFnu}
\end{figure}

The energy spectrum of gamma rays radiated by a blazar jet exhibits the classic double-hump structure with lower energy gamma rays originating from synchrotron radiation by a beam of accelerated electrons, and a higher energy component resulting from inverse Compton scattering of the same photons by the electrons. Other sources of photons may also be upscattered to high energy. In Fig.~\ref{nuFnu} we show the energy spectrum of TXS 0506+056 measured in the multimessenger campaign following the neutrino IC170922. What is special about this spectrum is that the production of neutrinos reveals the co-acceleration of protons along with electrons. While the relative merits of the electron and proton blazar have been hotly debated, the opportunities for studying the first identified cosmic ray accelerator are wonderfully obvious.

\begin{figure}[ht!]
  \centering
   \includegraphics[width=0.8\linewidth]{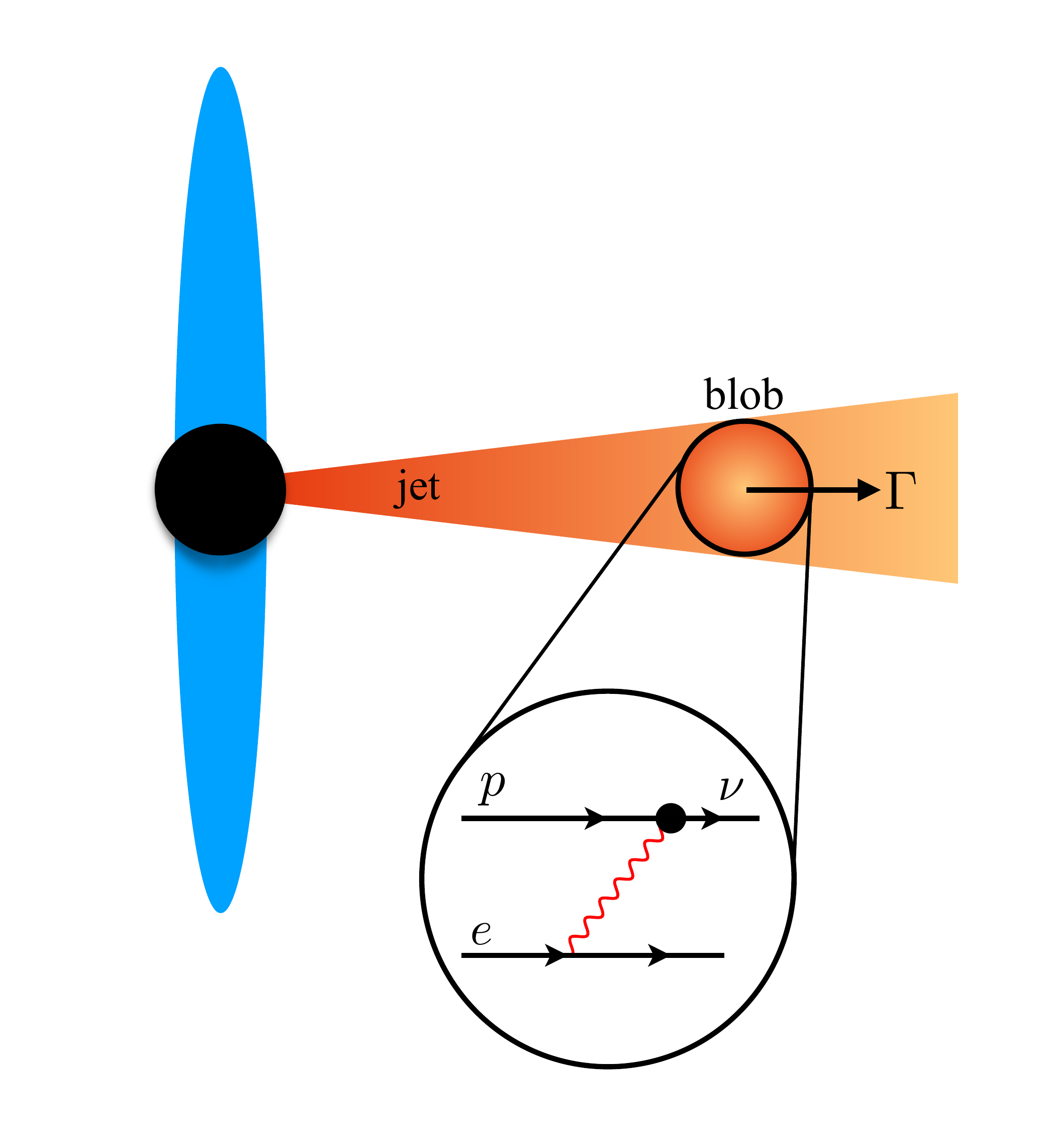}
  \caption{A ``blob" of accelerated protons and electrons moves towards the observer with a boost factor $\Gamma$. Neutrinos are produced in the interactions of protons with synchrotron photons radiated by the electrons. [Note that this is not a Feynman diagram, the photon is real.]}
 \label{bh}
\end{figure}

Because of their reduced synchrotron radiation, protons, unlike electrons, efficiently transfer energy in the presence of the magnetic field in the jet. They provide a mechanism for the energy transfer from the central engine over distances of parsecs as well as for the observed heating of the dusty disk over distances of several hundred parsecs. When protons are accelerated along with the electrons in blobs of plasma opportunities exist for the production of neutrinos by pion photoproduction, e.g., in interactions of the plasma with photons radiated off the accretion disk and the dusty torus in the AGN; see Fig.~\ref{old_bh}. Alternatively, the neutrinos may be produced inside the plasma in interactions of protons with photons radiated by the co-accelerated electrons; see Fig.~\ref{bh}.

Confronted with the challenge of explaining a $\sim E^{-2}$ high-energy photon emission spectrum reaching TeV energies, often radiated in bursts of a duration of less than one day, models have converged on the blazar blueprint shown in Fig.~\ref{bh}. Particles are accelerated by shocks in blobs of matter traveling along the jet with a bulk Lorentz factor of $\Gamma \sim 10$ and higher. The duration of the burst is associated with the size of the blob. The ${\Gamma}$ factor combines the effects of special relativity and the geometry of the moving source; we will routinely assume that maximal Doppler factor because the value of $\Gamma$ is often treated as a free parameter in any case. In the following, primes will refer to a reference frame attached to the blob, which is moving with a Doppler factor $\Gamma$ relative to the observer. In general, the transformation between blob and observer frame is $R' = \Gamma R$ and $E' = E / \Gamma$ for distances and energies, respectively. Blazar bursts appear more spectacular to the observer than they actually are because, in the frame of the blob, distances are larger and times longer.

Episodic high-energy emission is associated with the periodic formation of these blobs. The blobs are clearly identified in X-ray images of AGN jets, e.g., in M87. In order to accommodate bursts lasting a day, or less, in the observer's frame, the size of the blob must be of order $\Gamma\, c\, \Delta t \sim 10^{-2}$\,parsecs or less. The blobs are actually more like sheets, thinner than the jet's size of order parsec. The observed radiation at all wavelengths is produced by the interaction of the accelerated electrons and protons in the blob with the multiple radiation fields in the complex AGN structure, for instance synchrotron photons produced by electrons or the ambient radiation in the AGN which often has a significant prominent component concentrated in the so-called ``UV-bump" of $\sim 10$\,eV photons.

In order for a proton accelerated in the jet to produce pions on target photons of energy $E_\gamma$, a process dominated by the photoproduction of the $\Delta$ resonance, it must exceed the threshold energy:
\begin{equation}
E'_p > \frac{m_{\Delta}^2 - m_p^2}{4E'_\gamma} 
\end{equation}
or,
\begin{equation}
E_p > \Gamma^2\,\frac{m_\Delta^2 - m_p^2}{4E_\gamma}\, ,
\end{equation}
in the observer's frame. Furthermore, the blob must be transparent to gamma rays that are actually detected at Earth from the burst. Therefore the center-of-mass energy $s$ must be below the threshold for $\gamma+\gamma$ interactions in the blob, or
\begin{equation}
s < (2m_e)^2\,,
\end{equation} 
or,
\begin{equation}
E_\gamma E_{\gamma,obs } < \Gamma^2 m_e^2\,.
\end{equation}

The kinematic constraints imply that $\Gamma$ must exceed a few ($\sim30$) for target photon energies 10\,eV (1\,KeV) typical for UV (synchrotron) photons. These values of $\Gamma$ render the jet transparent to 50\,GeV (500\,GeV) photons, for example.

From the observed luminosity $L_\gamma$ we deduce the energy density of target photons in the shocked region of size $R'$: 
\begin{equation}
u'_\gamma = \frac{L'_\gamma \Delta t}{\frac{4}{3}\pi R'^3} = \frac{L_\gamma \Delta t}{\Gamma} \frac{1}{\frac{4}{3} \pi (\Gamma c\Delta t)^3} = \frac{3}{4\pi c^3} \; \frac{L_\gamma}{\Gamma^4 \Delta t^2} \,,
\end{equation}
where $\Delta t$ is the duration of the flare. Here we just boosted energies and contracted times by a factor $\Gamma$ from the primed blob frame to the observer frame.

The fraction of energy $f_\pi$ lost by protons to pion production when traveling a distance $R'$ through a photon field of density $n'_\gamma = u'_\gamma/E'_\gamma$ is given by the number of energy loss lengths of the proton
\begin{equation}
ct_{p\gamma}=c\frac{E_p}{-dE_p/dt}=c\frac{\lambda_{p\gamma}}{\kappa}       
\end{equation}
that fit inside the photon target of size $R'$, or
\begin{equation}
f_\pi=\frac{R^\prime}{c t_{p\gamma}}=\frac{R^\prime}{\lambda_{p\gamma}}\,\kappa
\end{equation}
where $\lambda_{p\gamma}$ is the proton interaction length, $\sigma_{p\gamma\to\Delta\to n\pi^+} \simeq 10^{-28}\rm\,cm^2$ and the inelasticity $\kappa \simeq 0.2$ the energy transferred from the proton to the pion in each interaction. To be clear, we are not asking how many times the proton interacts inside the target region, but after how many interactions it has lost its energy and no longer produces pions. The final result is given by
\begin{eqnarray}\label{pionef}
f_{\pi} \simeq \frac{R^\prime}{\lambda_{p \gamma}} \simeq
\frac{L_{\gamma}}{E_{\rm{ph}}}\frac{1}{\Gamma^2 (c \Delta t)} \frac{3\kappa
\sigma_\Delta}{4\pi c}\,.
\end{eqnarray}
If $f_\pi$ approaches unity, pions will be absorbed before decaying into neutrinos, requiring the substitution of $f_\pi$ by $1-e^{-f_\pi}$.

For a total injection rate in high-energy protons $E_N^2Q_N$, the total energy in neutrinos is $1/2f_\pi t_H E_N^2Q_N$, where $t_H {\sim}10^{10}$\,Gyr is the Hubble time. The secondary $\nu_\mu$ have energy $E_\nu = x_\nu E_p$, with $x_\nu \simeq 0.05$, the fraction of energy transferred, on average, from the proton to the neutrino via the $\Delta$-resonance. The factor 1/2 accounts for the fact that 1/2 of the energy in charged pions is transferred to $\nu_\mu + \bar\nu_\mu$. The neutrino flux is
\begin{equation}
\Phi_\nu = E_\nu\frac{dN}{dE_\nu} = \frac{c}{4\pi} {\frac{1}{E_\nu}\,[\frac{1}{2} f_\pi\, t_H\, E_N^2Q_N]}\,,
\end{equation}

We presented arguments, centered on the 2014-15 neutrino flare of TXS 0506+056 with a duration $\Delta t \simeq$ 110 days, that the observed cosmic neutrino flux required an efficiency for producing pions, $f_\pi$, close to unity. The requirement can be accommodated when protons in neutrino-producing jets interact with the 10\,eV photons in the galaxy via the $\Delta$-resonance; $p \gamma \rightarrow \Delta \rightarrow \pi N$. The pion efficiency of the jet depends on the Lorentz factor of the jet ($\Gamma$), the target photon energy ($E_\gamma$), luminosity of the target photons ($L_\gamma$), and the duration of the flare ($\Delta t$), for instance,
\begin{eqnarray}\label{pionef}
f_{\pi} \simeq
\frac{L_{\gamma}}{E_{\gamma}}\frac{1}{\Gamma^2 (c \Delta t)} \frac{3 \kappa \sigma_\Delta}{4\pi c}\,,
\label{blobresult}
\end{eqnarray}
and
\begin{eqnarray}\label{pionef}
1-e^{-f_\pi} \geq 1 \simeq
\bigg(\frac{L_{\gamma}}{(4-6)\times 10^{46}\, \rm{erg/s}}\bigg)
\bigg(\frac{10\, \rm eV}{E_{\gamma}}\bigg)
\bigg(\frac{1}{\Gamma^2} \bigg)
\bigg(\frac{110 \, \rm d}{\Delta t} \bigg) \nonumber\\ \times
\bigg(\frac{3 \kappa \sigma_\Delta}{4\pi c^2} \bigg)
\end{eqnarray}

This suggests that sources with small Lorentz factors and a UV luminosity exceeding $\mathcal O(10^{46})$ are required for the production of high-energy cosmic neutrinos. This level of UV luminosity has been reported in reference~\cite{Krauss:2014tna}.

The above estimate was performed assuming a monochromatic target photon spectrum, i.e., a line spectrum. A more realistic scenario would take into account the energy distribution of the target photon, ${dn_{\gamma}}/{d\epsilon_\gamma}$, introduced in Eq. \ref{lambdapg}. In this case,~\cite{Stecker:1968uc}
\begin{equation}
t^{-1}_{p\gamma}(E _{p})=\frac{c}{2{\Gamma}^{2}_{p}}\int_{\epsilon^\prime_{\rm{th}}}^{\infty} \!\!\! d\epsilon^\prime \, {\sigma}_{p\gamma}(\epsilon^\prime) {\kappa}_{p}(\epsilon^\prime)\epsilon^\prime \int_{\epsilon^\prime/2{\Gamma}_{p}}^{\infty} \!\!\! d\epsilon \, {\epsilon}^{-2} {dn_{\gamma}}/{d\epsilon_\gamma}, \label{tpgamma}
\end{equation}
where $\epsilon^\prime$ is the photon energy in the rest frame of proton, $\Gamma _{p}$ is the proton Lorentz factor in the comoving frame, $\kappa _{p}$ is the proton inelasticity, and $\epsilon^\prime _{\rm{th}}=145$ MeV is the threshold photon energy for photomeson production. 

The target photon can be modeled as a power-law, $\propto \epsilon^{-\alpha_\gamma}$ with a normalization that is fixed bolometrically by the photon luminosity of the source. Equation \ref{tpgamma} can be simplified when introducing the $\Delta$-resonance approximation:
\begin{equation}
f_{p\gamma}(E_p)\approx\frac{t_{\rm esc}}{t_{p \gamma}} \simeq \frac{2 \kappa \sigma_\Delta}{1+\alpha_\gamma} \frac{\Delta \bar{\epsilon}_{\Delta}}{\bar{\epsilon}_{\Delta}} 
\frac{3L_{\gamma}}{4 \pi c R \Gamma^2  E_{\gamma}} {\left(\frac{E_p}{{E}_p^b} \right)}^{\alpha_\gamma-1}, 
\end{equation}
where $\bar{\epsilon}_{\Delta}\sim0.34$~GeV is the energy value where the cross section peaks, $\Delta\bar{\epsilon}_{\Delta}\sim0.2$~GeV is the width of the resonance peak, and ${E}_p^b\approx0.5\,\Gamma^2\,m_pc^2\,\bar{\epsilon}_\Delta/E_\gamma$. One can verify that the pion efficiency is equal to the one obtained by Eq. \ref{blobresult} when $\alpha_\gamma=1$.

\subsection{An Alternative Approach for the Modeling of Neutrino Production by Proton Beams on External Photons}

\begin{figure}[ht!]
\centering
\includegraphics[width=0.8\linewidth]{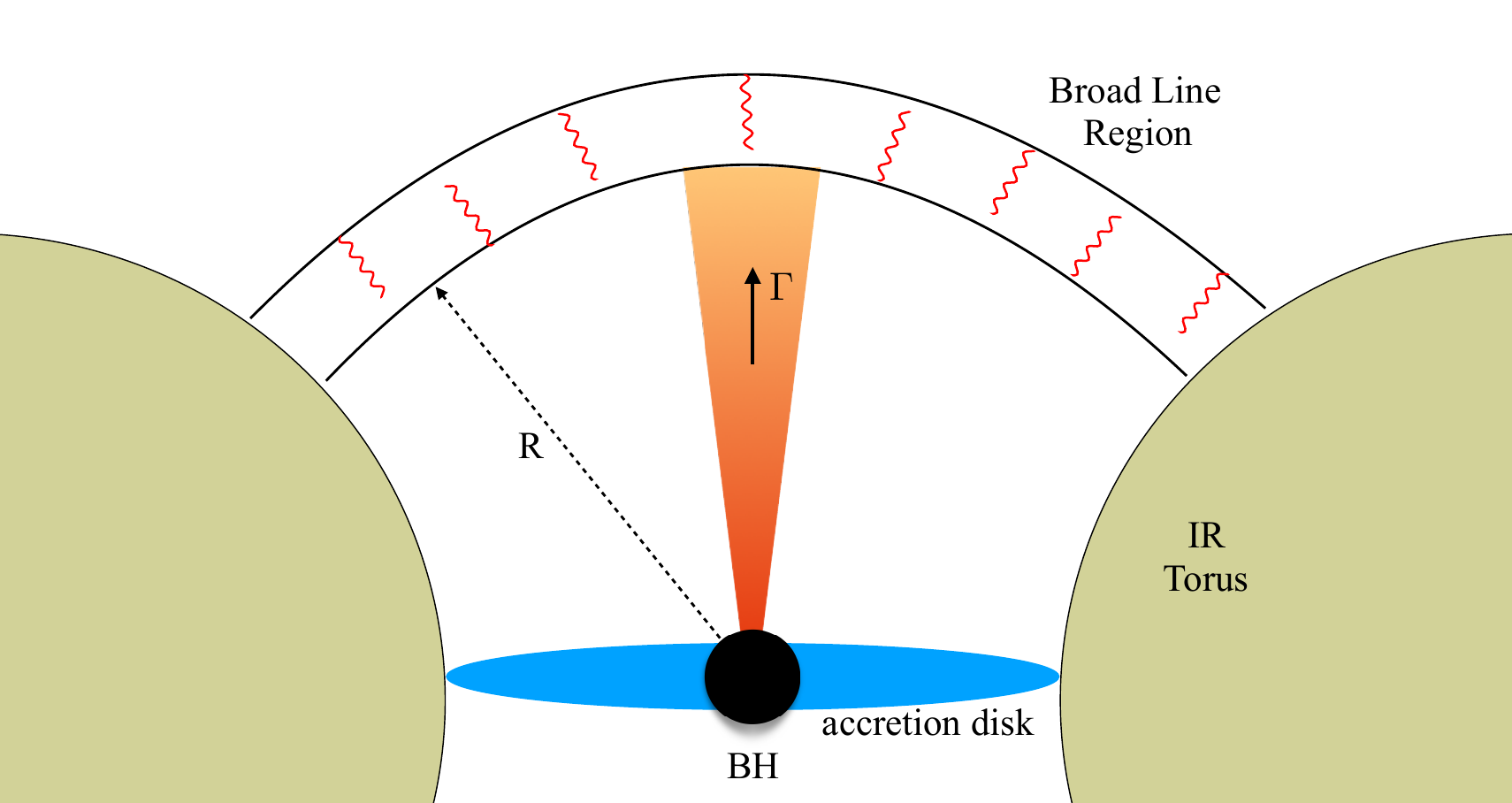}
\caption{A jet of accelerated protons and electrons move towards the observer with a boost factor $\Gamma$. Neutrinos are produced in the interactions of protons with dense photons fields in the galaxy, e.g., photons in the broad line region close to the central black hole.}
\label{blr}
\end{figure}

There is an alternative approach for modeling the production of neutrinos when a jet of protons accelerated near the black hole interacts with dense photon fields surrounding it. In this model the duration of the emission reflects the size of the radiation field rather than that of the blob; see Fig.~\ref{blr}. There are multiple opportunities for the protons to find dense photon targets, from photons associated with pair plasma to the broad-line photons. The framework is illustrated in Fig.~\ref{blr}. The key feature of this model is that in the frame of the protons the target photons acquire boosted energies to produce neutrinos.

The calculation is similar, but in this case a continuous jet of protons interacts with a shell of photons $\pi R'^2 \Delta R'$. As before, the opacity of the photon target is given by:
\begin{equation}
1-e^{-f_\pi} = \kappa\, \frac{\Delta R'}{\lambda_{p\gamma}} =\kappa\, \sigma_{p\gamma}\, n'_\gamma\, \Delta R'\,.
\end{equation}
Here $n'_\gamma$ is the density of photons in the jet frame:
\begin{equation}
n'_\gamma(E') = \frac{1}{V'E'}\,\int{dE'\,E'\frac{dN}{dE'}} = \frac{L'\Delta t'}{V'E'}\,,
\end{equation}
or, with $V' = 4 \pi R'^2 \Delta R'$ and $R' = c \Delta t'$,
\begin{equation}
1-e^{-f_\pi} = \kappa\, \sigma_{p\gamma}\, \frac{1}{4\pi c^2 \Delta t'}\, \frac{L'_\gamma}{E'}\,.
\end{equation}
In the observer frame:
\begin{equation}
1-e^{-f_\pi} = \kappa\, \sigma_{p\gamma}\, \frac{1}{4\pi c}\, \frac{1}{\Gamma^2}\,\frac{1}{c \Delta t}\, \frac{L_\gamma}{E}\,.
\end{equation}
In the last transformation the luminosity transforms as $\Gamma^2$; this was discussed earlier in the section when we introduced superluminal motion. The observed luminosity of the Galaxy $L_\gamma$ should be reduced to only include the fraction $f_{BLR}$ associated with the broad line region. Also $c \Delta t$ is identified with $R_{BLR}$, the distance over which the jet interacts with broad line photons. Therefore:
\begin{equation}
1-e^{-f_\pi} = \kappa\, \sigma_{p\gamma}\, \frac{1}{4\pi c}\, \frac{1}{\Gamma^2}\,\frac{1}{R_{BLR}}\, \frac{f_{BLR} L_\gamma}{E}\,.
\end{equation}

Guided by the 2014-15 burst of TXS 0506+056, one may ask the question at what point the source becomes opaque to photons but, at the same time, provides a sufficiently dense photon target to produce neutrinos. Identifying $R_{BLR}$ with the $110$\,day duration of the burst yeilds the reasonable requirement that $f_{BLR} L_\gamma \geq 10^{47} \rm{erg/s}$ for $10$\,eV photon energy. Also, the value of $R_{BLR}$ is consistent with the common parametrization~\cite{2009MNRAS.397..985G} that  
\begin{equation}
R_{BLR}= 10^{17} \rm cm \sqrt{L_{45}}\,,
\end{equation}
where $L_{45}$ is that $L_\gamma$ in units of $10^{45} \rm erg/s$.

We note that the same phenomenology can be reproduced with the blob model (Eq.~\ref{blobresult}) with a luminosity that is reduced by a factor $f_{BLR}/3$ relative to the luminosity in the broad line model.

However, for the BLR parameters above $f_\pi \leq 1$, and, in any case the neutrino energies too low to accommodate the neutrino energies observed in the 2014-15 flare. A phenomenology that yields appropriate values of $f_\pi$ requires $\Gamma \geq 5.6$ and $L_{BLR}=10^{47} \rm{erg/s}$.

The so-called structured jets model represent a variant of the previous model where gamma rays and neutrinos are produced when protons with boost factor $\Gamma$ catch us with slower moving plasma with boost factor $\Gamma'$. In the BLR formalism one must substitute $\Gamma^2$ by $\Gamma \Gamma'$.

\subsection{Fireballs and GRB}
\label{sec-XX:GRB}

While we have confidence that some of the gamma rays produced in our Galaxy by the decay of neutral pions, there is no straightforward gamma-ray path to the neutrino flux expected from extragalactic cosmic-ray accelerators as already illustrated by our discussion of active galaxies. We presented model calculations to show that AGNs can plausibly accommodate the observations of TXS 0506+056, the multimessenger connection between photons and neutrinos is indirect with the flare indicating the presence of a powerful jet seen in gamma rays, but with the neutrinos produced in periods where gamma rays are absent. Their energy is spread over the electromagnetic spectrum by absorption in the source when the neutrino-producing target is present. Although we showed how different blueprints for the beam dump can plausibly fit the TXS 05060+056 data as well as the diffuse neutrino flux observed, it is certainly premature to conclude that AGN are the sources of the extragalactic cosmic rays.

As discussed in the introduction, there is attractive alternative. Massive stars collapsing to black holes and observed by astronomers as GRBs have the potential to accelerate protons up to 100\,EeV energy. Neutrinos of $100\,\rm TeV - \rm PeV$ energy should be produced by pion photoproduction when protons and photons coexist in the GRB fireball~\cite{Waxman:1997ti}. The model is promising because the observed cosmic-ray flux can be accommodated with the assumption that roughly equal energy is shared by electrons, observed as synchrotron photons, protons and neutrinos.  This is indeed what IceCube has observed.

The phenomenology that successfully accommodates the astronomical observations assumes the creation of a hot fireball of electrons, photons, and protons that is initially opaque to radiation. The hot plasma expands by radiation pressure, and particles are accelerated to a Lorentz factor $\Gamma$ that grows with the expansion until the plasma becomes
optically thin and produces the GRB display. From this point on, the fireball coasts with a Lorentz factor that is constant and depends on its baryonic load. The baryonic component carries the bulk of the fireball's kinetic energy. The rapid variations with time observed in the burst spectrum can be successfully associated with successive shocks (shells), of width $\Delta R$, that develop in the expanding fireball. The rapid temporal variation of the gamma-ray burst, $t_v$, is on the order of milliseconds and can be interpreted as the collision of shells with a varying baryonic load leading to differences in the bulk Lorentz factor. Electrons, accelerated by first-order Fermi acceleration, radiate synchrotron gamma rays in the strong internal magnetic field and thus produce the spikes observed in the burst spectra.

The interpretation~\cite{Guetta:2003wi} of the IceCube observations of GRB is based on the fireball model. The basic assumption is that the energy is dissipated in fireball protons and electrons, with the latter observed via their synchrotron radiation. The neutrino flux is given by
\begin{eqnarray}
E_\nu^2 \frac {dN_{\nu}}{dE_\nu} = \left(\frac{\epsilon_p}{\epsilon_e} \right)\; \frac{1}{2} \;x_{\nu} \;\left[E_{\gamma}^2 \; \frac {dN_{\gamma}}{dE_{\gamma}} \left( E_{\gamma} \right)\right]_{\rm syn}\,,
\end{eqnarray}
where $\epsilon_p$ and $\epsilon_e$ are the energy fractions in the fireball in protons and electrons~\cite{Guetta:2003wi}, respectively. The neutrino observations determine the baryon loading in the GRB fireball, $\epsilon_p\,/\,\epsilon_e$; no prediction is made. The abundance of protons in the fireball is determined by, or, at present, limited by the neutrino observations~\cite{Aartsen:2016qcr}. A prediction for the diffuse flux for GRB emerges when assuming that they produce the observed cosmic rays.
 
Although simulations of GRB fireballs have reached a level of sophistication~\cite{Hummer:2011ms}, a simple energy estimate is sufficient to predict the neutrino flux associated with GRB fireballs assuming that they are the sources of the extragalactic cosmic rays.
 
\begin{figure}[h]
\centering\leavevmode
\includegraphics[trim =1 130 1 130, clip,width=1.0\columnwidth]{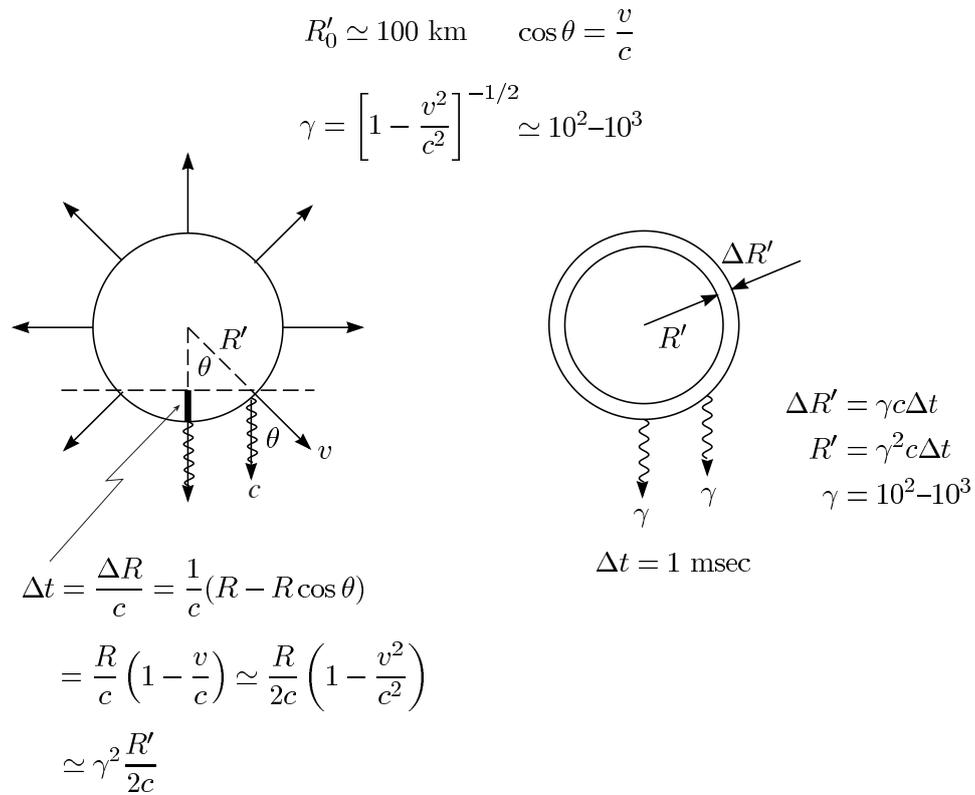}
\caption{Kinematics of a relativistically expanding fireball on the left, and the resulting shell(s) expanding under radiation pressure on the right.}
\label{fig:grbkin}
\end{figure}

The relativistic kinematics summarized in Fig.~\ref{fig:grbkin} relates the radius $R'$ and width $\Delta R'$ of the expanding fireball to the observed duration of the photon burst $c\Delta t$:
\begin{eqnarray}
R' &=& \Gamma^2 (c\Delta t)\\
\Delta R' &=& \Gamma c\Delta t
\end{eqnarray}
From the observed GRB luminosity $L_\gamma$, we compute the photon energy density in the shell:
\begin{equation}
u'_\gamma = { \left( L_\gamma \Delta t \right)/ \Gamma\over 4\pi
R'^2 \Delta R' } = {L_\gamma\over 4\pi \Delta t^2 c^3 \gamma^6} 
\end{equation}
The pion production by shocked protons in this photon field is, as before, calculated from the interaction length:
%
\begin{equation}
{1\over\lambda_{p\gamma}} = N_\gamma \sigma_\Delta  =
{u'_\gamma\over E'_\gamma} \sigma_\Delta  \qquad 
\left(E'_\gamma = {1\over\Gamma}E_\gamma\right).
\end{equation}
Also as before, $\sigma_\Delta$ is the cross section for $p\gamma\to\Delta\to n\pi^+$. As in the calculation of AGN, the fraction of energy going into $\pi$-production is
\begin{eqnarray}
    f_\pi &\cong& {\kappa\,{\Delta R'\over\lambda_{p\gamma}}}\\
    f_\pi &\simeq& \frac{1}{4 \pi c^2}{L_\gamma\over E_\gamma} {1\over\Gamma^4\Delta t}
    {\sigma_\Delta}\,\kappa\\
    f_\pi &\simeq& 0.14 \left[ \frac{L_\gamma}{10^{51}{\rm\,ergs}^{-1}} \right]
    \left[ \frac{1{\rm\ MeV}}{E_\gamma}\right]
    \left[ \frac{300}{\Gamma}\right]^4 \left[ \frac{1{\rm\ msec}}{\Delta t}\right]
\nonumber\\
   && \hskip1.5em\times \left[ \frac{\sigma_\Delta}{10^{-28}{\rm\,cm^2}} \right]
     \left[ \frac{\kappa}{0.2}\right].
\end{eqnarray}
The characteristic photon energy in the problem is 1\,MeV, the energy where the typical GRB spectrum exhibits a break. The contribution of higher energy photons is suppressed by the falling spectrum, and lower energy photons are less efficient at producing pions. Given the large uncertainties associated with the astrophysics, it is an adequate approximation to neglect the explicit integration over the GRB photon spectrum. The proton energy required for the production of pions via the $\Delta$-resonance is   
\begin{eqnarray}
E'_p &=& { m_\Delta^2 - m_p^2\over 4 E'_\gamma}.
 \end{eqnarray}
Therefore,
\begin{eqnarray}
E_p &=& 1.4\times10^{16}\rm\, eV \left(\Gamma\over300\right)^2
\left(1\ MeV\over E_\gamma\right)\\
E_\nu &=& {1\over4}  E_p \simeq 7\times10^{14} \rm\, eV.
\end{eqnarray}
We are now ready to calculate the neutrino flux:
\begin{equation}
\frac {dN_{\nu}}{dE_{\nu}} = {c\over4\pi} {u'_\nu\over E'_\nu} = {c\over4\pi}
{u_\nu\over E_\nu} = {c\over 4\pi} {1\over E_\nu} \left[ {1\over2} f_\pi
t_H \dot E\right],
\end{equation}
where the factor 1/2 accounts for the fact that only 1/2 of the energy in charged pions is transferred to $\nu_\mu + \bar\nu_\mu$. As before, $\dot E$ is the injection rate in cosmic rays beyond the ankle (${\sim} 4\, \times
10^{44}\rm\,erg\,Mpc^{-3}\,yr^{-1}$) and $t_H$ is the Hubble time of ${\sim}\, 10^{10}$\,Gyr. Numerically,
\begin{eqnarray}
\frac {dN_{\nu}}{dE_{\nu}} &=& 2\times10^{-14}{\rm\, cm^{-2}\, s^{-1}\, sr^{-1}}
     \left[ 7\times10^{14}\rm\, eV\over E_\nu \right] \left[ f_\pi\over
0.14\right] \left[ t_H\over 10\rm\ Gyr\right]\nonumber\\
 && \hspace{1.75in} {}\times \left[ \dot E\over10^{44} \rm\, erg\,Mpc^{-3}\,yr^{-1} \right]
\end{eqnarray}

The only subtlety here is the $\Gamma^2$ dependence of the shell radius $R'$; for
a simple derivation, see reference~\citen{Halzen:2002pg}. The result for the diffuse neutrino flux is insensitive to weather the GRBs are beamed or not. Beaming yields more energy per burst, but fewer bursts are actually observed. The predicted rate is also insensitive to the neutrino energy $E_\nu$ because higher average energy yields fewer $\nu$s, but more are detected. Both effects are approximately linear. Neutrino telescopes are essentially background free for such high-energy events and should be able to identify neutrinos at all zenith angles.

For typical choices of the parameters, $\Gamma \sim 300$ and $ t_v \sim 10^{-2}
s$, about 100 events per year are predicted in IceCube, a flux that was already
challenged~\cite{Ahlers:2011jj} by the limit on a diffuse flux of cosmic
neutrinos obtained with one-half of the instrument completed in one year~\cite{Abbasi:2011ji}. The energy density of extragalactic cosmic rays of $\sim10^{44}\,\rm\,TeV\ Mpc^{-3}\ yr^{-1}$ depends on the unknown transition energy between the Galactic and extragalactic components of the spectrum.  In the end, the predictions can be stretched by reducing that number, but having failed by now to observe high-energy neutrinos in spatial and temporal coincidence with over 1500 GRB observations, IceCube has set a limit that is less than 1\% of the PeV cosmic neutrino flux that is actually observed.

One lesson learned from TXS 05060+056 is that sources that emit high-energy gamma rays may be cosmic accelerators, however, at the time of neutrino flares the source is opaque to gamma rays. The failure of IceCube to observe neutrinos from GRB may be the consequence of using GRB alerts based on gamma ray observations. Only a few GRB have been identified that are underluminous in gamma rays and where the jet ``choked" in the dense remnant producing high-energy neutrinos while the photons lose energy and are shifted to low energy. Also, GRB may produce neutrinos even if they are not the sources of the highest energy cosmic rays. For instance, they may produce neutrinos of lower energy in events where the boost factor is limited, $\Gamma \leq 10$. Candidate events have been observed and are referred to as ``low luminosity GRBs.''~\cite{Senno:2015tsn}.

There is also the possibility that high-energy gamma rays and neutrinos are produced when the shock expands further into the interstellar medium. This mechanism has been invoked as the origin of the delayed high-energy gamma rays. Adapting the previous calculation to the external shock is straightforward~\cite{Bottcher:1998qi}. The timescale is seconds rather than milliseconds, and the break in the spectrum shifts from 1 to 0.1\,MeV. Although $f_{\pi}$ is reduced by two orders of magnitude, in the external shocks higher energies can be reached, and this increases the detection efficiency. In the end, the observed rates are an order of magnitude smaller than in internal shocks, but the inherent ambiguities of the estimates are such that it is difficult to establish with confidence their relative neutrino yields. 

\subsection{The ``Guaranteed" Cosmogenic Neutrinos}
\label{sec:cosmogenic}

The production of neutrinos in the sources that accelerate high-energy cosmic rays depends on the source and on its environment. In order to efficiently accelerate cosmic rays, any loss mechanism, including pion production in $p\gamma$ and $pp$ interactions that produce neutrinos, must be suppressed as it reduces the acceleration time. Efficient accelerators are likely to be inefficient beam dumps for producing neutrinos. High-efficiency neutrino production can however be achieved by separating the sites of acceleration and neutrino production, as in TXS 0506+056. Similarly, after acceleration, extragalactic cosmic rays propagate over cosmological distances of more than 10~Mpc and can efficiently produce neutrinos on the EBL, predominantly microwave photons.

In this section, we will briefly discuss the production of neutrinos in the interactions of ultra-high-energy cosmic rays with EBL photons. Soon after the discovery of the cosmic microwave background, Greisen, Zatsepin and Kuzmin~\cite{Greisen:1966jv,Zatsepin:1966jv} (GZK) realized that extragalactic cosmic rays are attenuated by interactions with background photons. Protons interact resonantly via $p\gamma\to\Delta^+\to\pi^+ n$ with background photons with mean energy $\epsilon \simeq 0.33$~meV at energies $E_p \simeq (m^2_\Delta-m_p^2)/4\epsilon \simeq 500$~EeV. The density and width of the Planck spectrum leads to a significant attenuation of proton fluxes after propagation over distances on the order of 150~Mpc above an energy $E_{\rm GZK}\simeq 50$~EeV. This is known as the GZK suppression of the high energy cosmic ray spectrum. Also heavier nuclei are attenuated at a similar energy by photodisintegration of the nucleus by CMB photons via the giant dipole resonance~\cite{Khan:2004nd}. 

The pions produced in GZK interactions decay, resulting in a detectable flux of {\it cosmogenic} neutrinos first estimated by Berezinsky and Zatsepin~\cite{Beresinsky:1969qj} in 1969. This {\it guaranteed} flux of neutrinos became one of the benchmarks for high--energy neutrino astronomy leading early on to the concept of kilometer-scale detectors. The flux of cosmogenic neutrinos peaks at EeV energy. It can be calculated from the observed cosmic ray spectrum and composition, modulo parameters:
\begin{itemize}
    \item the slope of the (power-law) spectrum of the cosmic rays sources that is injected in the EBL prior to absorption,
    \item the minimum and maximum energy of the spectrum,
    \item the cosmological evolution of the sources, i.e., the value of $\xi$.
\end{itemize}
Ahlers {\it{$et al.$}}~\cite{Ahlers:2010fw,Ahlers:2012rz} calculated the range of predictions for the GZK neutrino flux by varying all parameters within reasonable limits. They also determined the most likely flux. The exercise was performed assuming that the highest energy cosmic rays are protons which yields the largest neutrino fluxes~\cite{Yoshida:1993pt,Protheroe:1995ft,Engel:2001hd}. An observable flux is predicted under these assumptions, particularly if the extragalactic component of the proton spectrum extends below the ankle.

An example of models satisfying these assumptions are referred to as ``dip models,'' the ankle results from the absorption of protons by Bethe--Heitler pair production on CMB photons. A fit to the observed cosmic-ray spectrum requires relative strong source evolution with redshift~\cite{Berezinsky:2002nc,Fodor:2003ph,Yuksel:2006qb,Takami:2007pp} that further enhances pion production. However, the corresponding electromagnetic emission via neutral pions as well as $e^\pm$ pairs is constrained by the isotropic gamma-ray background (IGRB) observed by the {\em Fermi}-LAT satellite~\cite{Abdo:2010nz,Ackermann:2014usa} and limits the neutrino intensity of these proton--dominated scenarios~\cite{Berezinsky:2010xa,Ahlers:2010fw,Gelmini:2011kg,Decerprit:2011qe,Heinze:2015hhp,Supanitsky:2016gke}.
\begin{figure}[t]
\centering
\includegraphics[width=0.8\columnwidth]{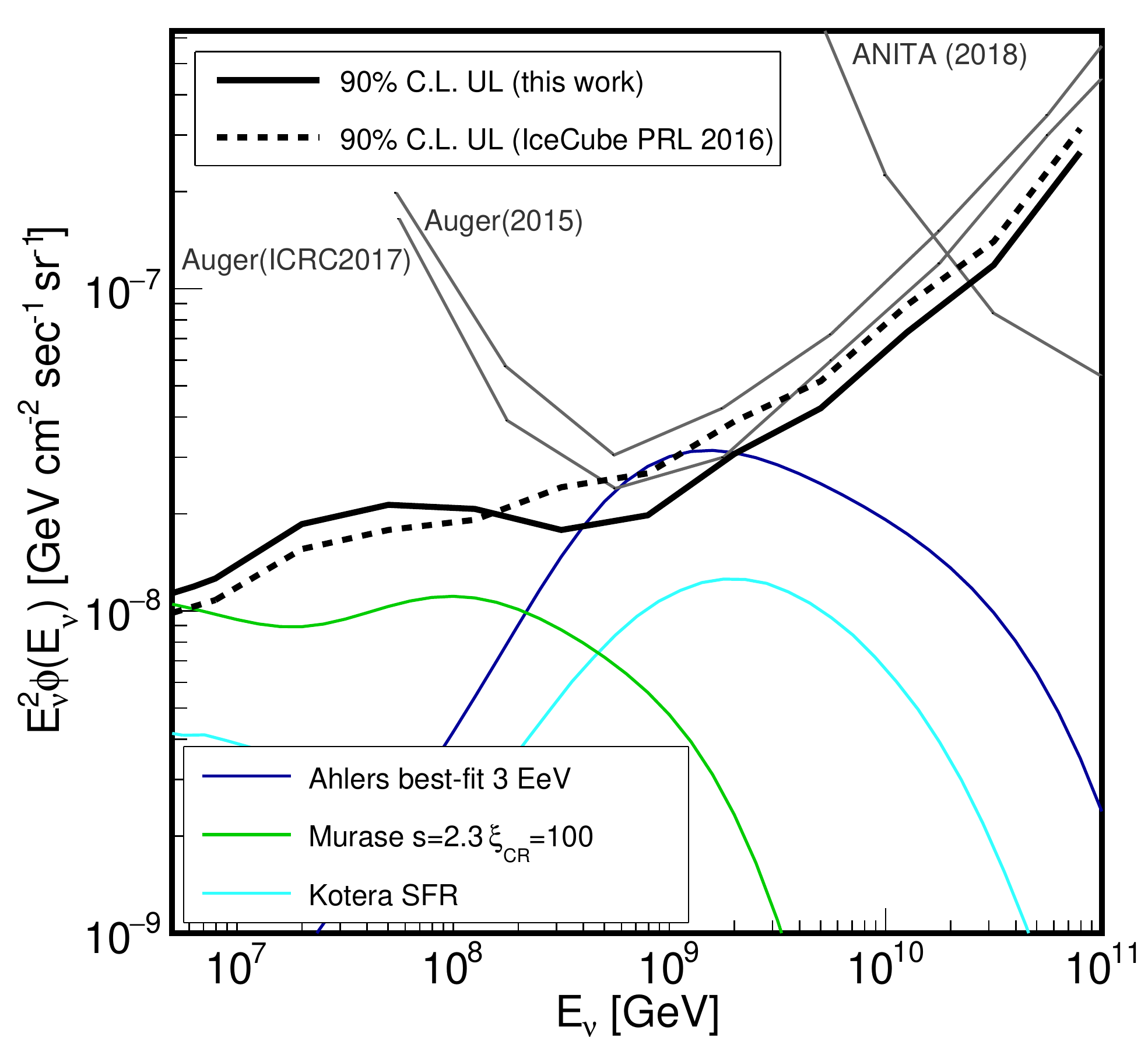}
\caption{ All-flavor upper limit (90\% C.L.) from IceCube nine-year analysis (solid black line) \cite{} and IceCube seven-year analysis (dashed line). The limits are derived using a log-likelihood ratio method. Differential upper limits for  per energy-decade $E^-1$ cosmogenic neutirno flux are also shown for Auger \cite{Aab:2015kma, Zas:2017xdj}, and ANITA \cite{Allison:2018cxu} after proper normalization considering the energy bin width and neutrino flavor. The predicted cosmogenic neutrino flux by Kotera et al. \cite{Kotera:2010yn} (cyan) and by Ahlers et al. \cite{Ahlers:2010fw} (dark blue) and a cosmic neutrino model by Murase et al. \cite{Murase:2014foa} (green) are shown for comparison. All model expectations are shown assuming primary protons for cosmic ray composition.
}
\label{GZK_2016}
\end{figure}

Recent upper limits on the cosmogenic neutrino flux have resulted from the failure of IceCube to observe EeV neutrinos~\cite{Aartsen:2016ngq}; see Fig.~\ref{GZK_2016}. The upper limit on the GZK flux can be accommodated by introducing a heavy nuclear composition into the spectrum of the highest energy cosmic rays. For nuclei, resonant neutrino production still proceeds via the interaction of individual nucleons in the nucleus with background photons, but the threshold of the production is increased to $E_{\rm CR} \gtrsim AE_{\rm GZK}$ for nuclei with mass number $A$. Therefore, efficient cosmogenic neutrino production now requires an injected cosmic-ray flux that extends well above $E_{\rm GZK}$. Especially for a heavier nuclear composition of the primary flux, the production of neutrinos on photons of lower energies in the EBL becomes relatively more important~\cite{Hooper:2004jc,Ave:2004uj,Hooper:2006tn,Allard:2006mv,Anchordoqui:2007fi,Aloisio:2009sj,Kotera:2010yn,Decerprit:2011qe,Ahlers:2011sd,Ahlers:2012rz}. The interaction with optical photons produces neutrino fluxes in the PeV energy range, the overall level is however reduced because of the lower density of the EBL photons. The PeV neutrino flux observed by IceCube cannot be accommodated by the production of neutrinos in the EBL~\cite{Roulet:2012rv,Yacobi:2015kga}. However, it has been recently pointed out~\cite{Safa:2019ege} that the cascading of tau neutrinos of GZK origin through the Earth creates the opportunity to contribute events to the PeV flux at a subdominant level.

\section{Galactic Sources}

The rationale for building kilometer-scale neutrino detectors is that their sensitivity
is sufficient to reveal generic cosmic-ray sources with an energy density in
neutrinos comparable to their energy density in cosmic
rays~\cite{Gaisser1997} and gamma
rays. The expectation was that this relation would also be satisfied by the sources of Galactic cosmic rays~\cite{AlvarezMuniz:2002yr}.

Galactic sources constitute one of the main components of the high-energy emission observed at Earth. The most manifest component of the high-energy radiation from the Galaxy is the measured gamma-ray emission from sources in the Galactic plane. Today, the observed high-energy gamma-rays from Galactic sources reaches 100 TeV, indicating that Galactic sources are capable of accelerating particles to very high energies. While the cosmic rays spectrum extend to energies far beyond these energies, Galactic sources has been understood as the principal contributors to the cosmic ray spectrum up to the ``knee" in the cosmic ray spectrum.

The rationale for searching for Galactic sources of high-energy cosmic neutrinos rely upon the search for PeVatrons. Galactic cosmic rays are thought to reach energies of at least several PeV, the {\em knee} in cosmic ray spectrum. The interaction of cosmic rays with energies greater than PeV with the dense environments in the Galaxy leads to production of charged and neutral meson that would eventually decay to high-energy neutrinos, as well as gamma-rays, and should be seen in neutrino detectors.

In the context of multimessenger connection, identification of the source of high-energy neutrinos and very high energy cosmic rays can benefit from the information brought by other channels of observation. For Galactic source, since they are in our neighborhood, the very high energy gamma rays that should accompany the high-energy neutrinos would not be affected by the extragalactic background light (EBL) and therefore, they can be seen in similar energy ranges that neutrino telescopes operate effectively.

Possible neutrino emission from the Galactic plane appears as a subdominant component to the observed cosmic neutrino flux. However, recent studies imply that identification of the Galactic source of high-energy neutrinos is likely in near future.

As discussed earlier, the necessary condition for a source to be considered as a particle accelerator up to a certain energy is to satisfy the Hillas criterion \citep{1984ARA&A..22..425H}. That is, the gyroradius of the particle shall not excess the size of the astrophysical object. Otherwise, it cannot contain the particle. The other relevant parameters determining the maximum energy an accelerator can achieve are the acceleration mechanism and energy losses at the source. Galactic sources such as supernova remnants can meet Hillas' condition and reach energies beyond the cosmic ray {\em knee} if the magnetic filed is amplified compared to the average interstellar medium's average value. Possible Galactic wind termination shocks can also accelerate cosmic rays beyond the {\em knee} in Milky Way; see reference~\citen{Bustard:2016swa} for details. 

\subsection{Galactic Neutrino-Producing Beam Dumps: Supernova Remnants}

Based on energy considerations, the suggestion that supernova remnants are the sources of the cosmic rays in our own Galaxy originated in the 1930s~\citep{BaadeAndZwicky}. The observed energy density of the cosmic rays in our Galaxy is $\rho_{E} \sim
10^{-12}$\,erg\,cm$^{-3}$. Galactic cosmic rays do not exist forever; they diffuse
within the microgauss fields of our Galaxy and remain trapped for an average containment time
of $\sim 3\times10^{6}$\,years, depending on their energy. The power needed to maintain a steady energy density therefore requires accelerators delivering $10^{41}$\,erg/s. This happens to be 10\% of
the power produced by supernovae releasing $10 ^{51}\,$erg every 30 years ($10
^{51}\,$erg correspond to 1\% of the binding energy of a neutron star after 99\%
is initially lost to neutrinos). This coincidence is the basis for the idea that
shocks produced by supernovae exploding into the interstellar medium are the
accelerators of Galactic cosmic rays.
These sources should be visible to the present generation of gamma ray instruments. A generic supernova remnant releasing an energy of $W\sim10^{50}\,$erg into the
acceleration of cosmic rays will inevitably generate TeV gamma rays by
interacting with the hydrogen in the Galactic disk. 

The emissivity in these pionic gamma rays can be calculated from neutral pion production rate \cite{Stecker:1978ah}, with the substitution $q_{\pi^0}=\frac{1}{2}q_{\pi^\pm}$ in Eq.~\ref{dump}. We first define the production rate $Q_{\pi^0}$, per unit volume and time, by converting $dl$ into $cdt$ in the $\tau'$ integration in Eq.~\ref{dump}. Within the $\Delta$ resonance approximation with $n_\pi = 1$ and $\kappa = 0.8$:
\begin{equation}
Q_{\pi^0}=c\,n\,\sigma_{pp}\,\int dE_p\,n_p\left(E_p\right)\,\delta \left(E_{\pi^0}-\kappa E_p\right)\,,
\end{equation}
or
\begin{equation}
Q_{\pi^0}\left(E_{\pi^0}\right)=c\,\sigma_{pp}\,\frac{1}{\kappa}\,n\,n_p\left(\frac{E_{\pi^0}}{\kappa}\right)\,.
\end{equation}
The emissivity $Q_{\pi^0}$ is proportional to the number density of hydrogen targets and the number of cosmic rays which is determined by the spectrum
\begin{equation}
n_p = \frac{4\pi}{c} \int dE \frac{dN_p}{dE}\,,
\end{equation}
with $n_p \simeq 4\times10^{-14}\,\left(E_p/\rm TeV\right)^{-1.7}\,{\rm cm}^{-3}$.
The emissivity of photons is obtained from the fact that every neutral pion produces two photons with half its energy, therefore
\begin{equation}
Q_\gamma\left(E_\gamma\right)= 2\times2\,Q_{\pi^0}\left(2E_\gamma\right)\simeq Q_{\pi^0}\,,
\end{equation}
where the latter equality holds for an $E^{-2}$ spectrum.
Our final result is
\begin{equation}
Q_\gamma\left(E_\gamma\right) \simeq \frac{c}{\kappa}\,\sigma_{pp}\,n\,n_{cr}\left(E_p>\frac{E_\gamma}{\kappa}\right)\,,
\label{eq:qgam}
\end{equation}
or, assuming an $E^{-2}$ spectrum,
\begin{equation}
Q_\gamma (> 1\,{\rm TeV}) \simeq 10^{-29} \, \left({n \over \rm {cm^3}}\right)\,{\rm cm^{-3}\,s^{-1}}\,.
\end{equation}
Notice the transparency of this result. The proportionality factor in Eq.~\ref{eq:qgam} is determined by particle
physics, where $\lambda_{pp}= (n\sigma_{pp})^{-1}$ is the proton
interaction length ($\sigma_{pp} \simeq 40$\,mb) in a density $n$ of hydrogen
atoms. The corresponding luminosity is
\begin{equation}
L_{\gamma} ({>} 1\,{\rm TeV}) \simeq Q_{\gamma}\, {W \over \rho_E}
\end{equation}
where $W/\rho_E$ is the volume occupied by the supernova remnant; given the ambient density $\rho_E \sim 10^{-12}$\,erg\,cm$^{-3}$
of Galactic cosmic rays~\cite{Gaisser1995}, a supernova with energy $W\sim10^{50}\,$erg in cosmic rays occupies the volume $W/\rho_E$. We here made
the approximation that the density of particles in the remnant is not very
different from the ambient energy density.

We thus predict~\cite{GonzalezGarcia:2009jc,Ahlers:2009ae} a rate of TeV photons from a supernova remnant at a nominal
distance $d$ of
\begin{eqnarray}
&&\int_{E>1{\rm TeV}} \frac{dN_\gamma}{dE_\gamma} dE_\gamma
= {L_\gamma (>{\rm 1 TeV}) \over 4\pi d^2} \nonumber \\
&&\simeq 10^{-12}-10^{-11} \left({\rm TeV\over \rm cm^2\,s}\right)
\left({W\over \rm 10^{50}\,erg}\right) \left({n\over \rm
  1\,cm^{-3}}\right) \left({d\over \rm 1\,kpc}\right)^{-2}.
\label{eq:galactic1}
\end{eqnarray}

As discussed in the introduction, the position of the {\em knee} in the cosmic ray
spectrum indicates that some Galactic sources accelerate cosmic rays to energies of
several PeV. These PeVatrons therefore produce pionic gamma rays whose spectrum
can extend to several hundred TeV without cutoff. For such sources the
gamma-ray flux in the TeV energy range can be parametrized in terms of a
spectral slope $\alpha_{\gamma}$, an energy $E_{cut,\gamma}$ where the
accelerator cuts off, and a normalization $k_{\gamma}$:
\begin{equation}
\frac{dN_{\gamma}(E_\gamma)}{dE_\gamma}
=k_{\gamma}
 \left(\frac{E_\gamma}{\rm TeV}\right)^{-\alpha_{\gamma}} 
\exp\left(-\sqrt{\frac{E_\gamma}{E_{cut,\gamma}}}\right).
\label{eq:cutoff}
\end{equation}
The estimate in Eq.~\ref{eq:galactic1} indicates that fluxes as large as
$dN_{\gamma}/dE_\gamma \sim 10^{-12}$--$10^{-14}$ (TeV$^{-1}$ cm$^{-2}$
s$^{-1}$) can be expected at energies of ${\cal O}$ (10 TeV).

Such sources are well within the sensitivity of existing high-energy gamma ray detectors such air Cherenkov telescopes and large acceptance ground-based detectors like Milagro and HAWC. They may have been first revealed by the highest energy all-sky survey
in $\sim 20$\,TeV gamma rays using the Milagro detector~\cite{Abdo:2006fq}. The survey revealed a subset of sources, located within nearby star-forming regions in Cygnus and in the vicinity of Galactic latitude $l=40^\circ$; some of them cannot
be associated with known supernova remnants or with nonthermal sources
observed at other wavelengths. Subsequently, directional air Cherenkov
telescopes were pointed at three of the sources, revealing them as PeVatron
candidates with an approximate $E^{-2}$ energy spectrum that extends to tens of
TeV without evidence for a cutoff~\cite{Halzen:2016seh}, in contrast with the best
studied supernova remnants RX J1713-3946 and RX J0852.0-4622 (Vela Junior).

Speculations emerged that some Milagro sources may be molecular clouds illuminated by the
cosmic-ray beam accelerated in young remnants located within $\sim\,100$\,pc. Indeed, one
expects that multi-PeV cosmic rays are accelerated only for the short time
period when the shock velocity is high, i.e., towards the end of the free expansion of the remnant. The high-energy particles
can produce photons and neutrinos over much longer periods when they diffuse
through the interstellar medium to interact with nearby molecular
clouds~\cite{Gabici:2007qb}. An association of molecular clouds and supernova
remnants is definitely expected in star-forming regions.

Ground-based and satellite-borne instruments with improved sensitivity
are able to conclusively pinpoint supernova remnants
as the sources of cosmic-ray acceleration by identifying accompanying
gamma rays of pion origin.
The Fermi Large Area Telescope has detected the pion-decay feature in the
gamma-ray spectra of two supernova
remnants, IC 443 and W44~\cite{Ackermann:2013wqa}.
In contrast, GeV gamma-ray data from Fermi LAT have challenged the hadronic
interpretation of the GeV-TeV radiation from one of the best-studied
candidates, RX J1713-3946~\cite{Abdo:2011pb}. On the basis of new observations~\cite{pevatron} the HESS Collaboration has recently argued that the center of the Galaxy hosts a promising PeVatron candidate. Detecting the accompanying neutrinos from
supernova remnants or the Galactic center would provide
incontrovertible evidence for cosmic-ray acceleration.

\begin{figure}[htb]
\begin{center}
\includegraphics[width=12cm,height=6cm,angle=0]{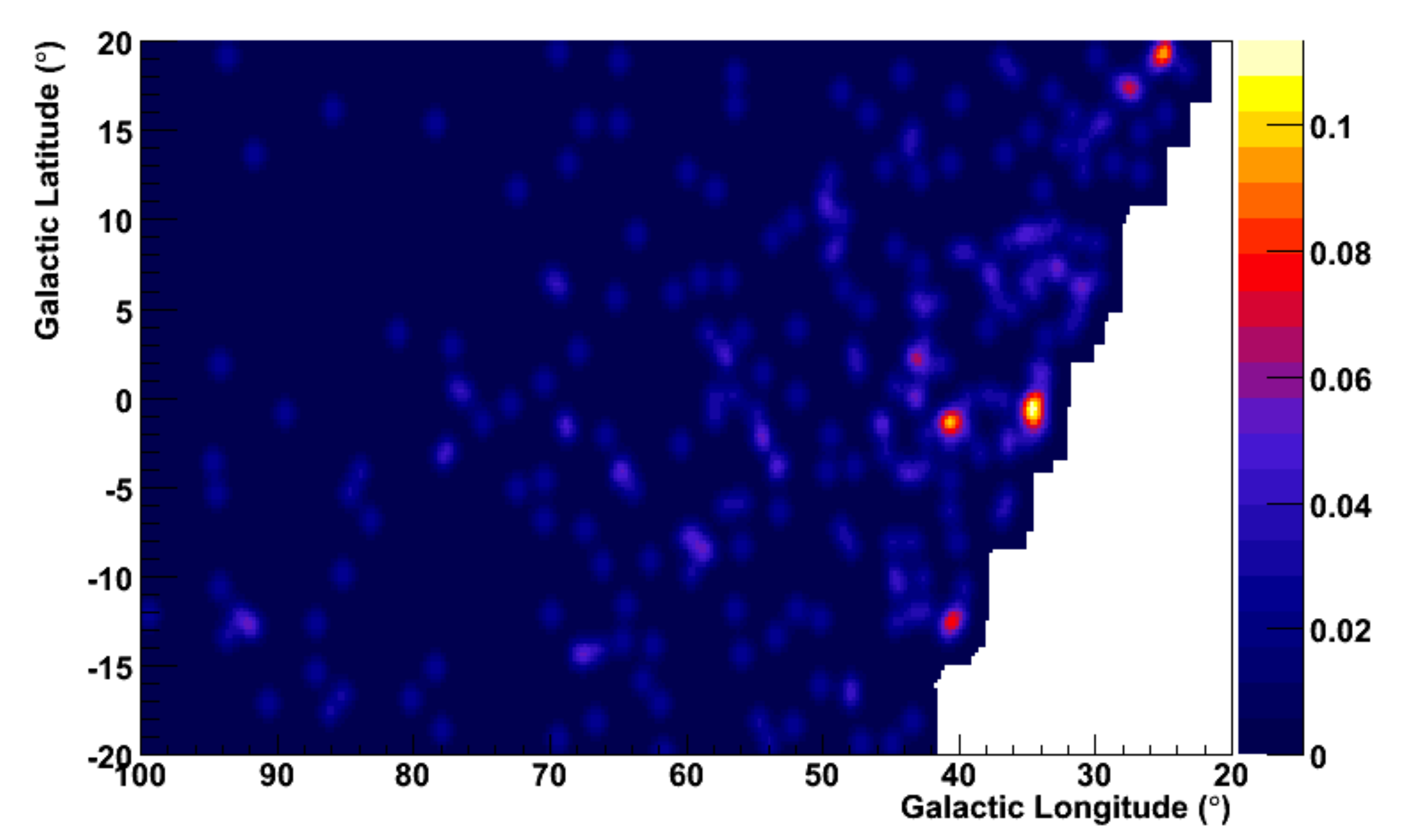}
\caption{Simulated sky map of IceCube in Galactic coordinates after five
years of operation of the completed detector.  Two Milagro
sources are visible with four neutrino events for MGRO J1852+01 and three for MGRO J1908+06 with energy in excess of 40\,TeV. These, as
well as the background events, have been randomly distributed according
to the resolution of the detector and the size of the sources.}
\label{fig-XX.3}
\end{center}
\end{figure}

Particle physics is sufficient to calculate the neutrino flux from potential PeVatrons with the production of a $\nu_\mu+\bar\nu_\mu$ pair for every two gamma rays seen by Milagro. The neutrino flux can also be calculated using the formalism discussed at the beginning of this section, with qualitatively the same outcome. Operating the complete IceCube detector for several
years should yield confirmation that some of the Milagro sources produce pionic gamma rays; see Fig.~\ref{fig-XX.3} and the next section.

The quantitative statistics can be summarized as follows, for average values of the parameters describing the flux, we find that the IceCube detector has the potential to confirm sources in the Milagro sky map as sites of cosmic-ray acceleration
at the $3\sigma$ level in less than one year and at the $5\sigma$ level in three
years~\cite{GonzalezGarcia:2009jc}. We here optimistically assumed a $E^{-2}$ spectrum, and that the source extends to
300\,TeV, or 10\% of the energy of the cosmic rays near the knee in the
spectrum. These results agree with previous estimates~\cite{Halzen:2008zj,Kappes:2009zza}. 
There were intrinsic ambiguities in these early estimate of an astrophysical nature that
may reduce or extend the time required for a $5\sigma$
observation~\cite{GonzalezGarcia:2009jc}. In particular, the poorly known
extended nature of some of the Milagro sources and the value of the cutoff represent a challenge for
IceCube analyses that are optimized for point sources. The absence of any 
observation of an accumulation of high-energy neutrinos in the direction of these sources will seriously challenge the concept 
that gamma-ray telescopes are seeing actual sources of cosmic rays.

These predictions have been stable over the years and have recently been updated in the context of new gamma-ray information, in particular from the HAWC experiment; see reference~\citen{Halzen:2016seh}. The predicted fluxes for Galactic sources in the Northern Hemisphere are close to IceCube's current upper limits on a point source flux of $10^{-12} \,\rm TeV^{-1} cm^{-2} s^{-1}$\cite{Aartsen:2016oji}; see also reference~\citen{Adrian-Martinez:2016fei}.

\subsection{Identifying Galactic Sources of High-Energy Neutrinos}\label{sec:sources}
The search for Galactic neutrino sources is centered on the search for PeVatrons, sources with the required energetics to produce cosmic rays, at least up to the {\em knee} in cosmic ray spectrum. Supernova remnants have been the primary candidates. Accelerating cosmic rays beyond PeV energies, generic PeVatrons should emit very high energy gamma rays whose spectrum extends to several hundred TeV without a cut off. 

Cosmic rays interacting with matter or radiation at the source, or while propagating in the Galactic environment, would result in secondary particles that can be used to pinpoint the origin of the cosmic rays. The threshold for the production of pions in hadronuclear interaction leads to a feature in gamma-ray flux known as the {\em pion bump}, appearing at $\sim$ 70 MeV. Such a feature provides direct evidence for hadronic interactions. Interestingly, the Fermi Large Area Telescope ({\em Fermi}-LAT) identified this feature in gamma-ray spectra of supernova remnants IC 443, W 44, and W 51 \citep{Ackermann:2013wqa, Jogler:2015ddc}. Although identification of the {\em pion bump} is an intriguing evidence for the presence of hadrons in supernova remnants, the particular sources are unfortunately not Pevatron candidates.

As discussed in the previous section, sources of very high energy can be used to identify Pevatron candidates with a neutrino telescope; observation of neutrino emission would provide the incontrovertible evidence of a cosmic ray source. The first very high energy survey of the Galaxy was performed by the Milagro collaboration \citep{Abdo:2006fq}. This survey revealed the brightest gamma-ray emitters in the Galactic plane and pointed towards possible sites of cosmic ray acceleration. Milagro was a large volume water Cherenkov telescope with a wide field of view. The sources identified by the Milagro survey were subsequently scrutinized by imaging air Cherenkov telescopes (IACTs) such as HESS\footnote{High Energy Stereoscopic System}, VERITAS\footnote{Very Energetic Radiation Imaging Telescope Array System}, and MAGIC\footnote{Major Atmospheric Gamma-Ray Imaging Cherenkov} which provided a wealth of data on the gamma-ray emission from the Milky Way. Today, with the commissioning of the HAWC\footnote{High Altitude Water Cherenkov} gamma-ray observatory, a next-generation instrument following Milagro, we have reached an unprecedented sensitivity to very high energy gamma ray emitters in the Milky Way. 

The initial Milagro survey of the Galaxy presented the first view of the Galaxy at 10 TeV and identified a handful of TeV emitters. Today, there are more than 100 sources identified with TeV emission in the Galactic plane, among them pulsar wind nebula (PWN), supernova remnants (SNR), binaries, molecular clouds, and shell SNR. However, the majority of the detected sources are classified as {\em unidentified}. The TeVCat\footnote{tevcat.uchicago.edu} database provides a listing of all TeV sources, including a bibliography of the observations \citep{Wakely:2007qpa}.

The Milagro survey revealed six sources \citep{Abdo:2006fq} as potential Pevatrons because of their hard spectrum that extended to very high energies without evidence for an energy cut-off. Besides the Crab, three of these sources---MGRO J1908+06, MGRO J2019+37, and MGRO J2031+4---stood out. The other two, MGRO J2043+36 and MGRO J2032+37, were candidate sources with relatively high level of gamma-ray emission. Finally, despite its large flux, MGRO J1852+01 did not reach the $5 \sigma$ detection threshold.

Four of these sources were located in the Cygnus region, a star forming zone which contains giant molecular cloud complexes and populous associates of massive young stars. It provides the substantial requirements for particle acceleration and interaction and, as a result, the production of neutrinos. The other two sources were close to the inner galaxy and were characterized by a large flux at TeV energies. They were located at low declination, which is the area in the sky IceCube is most sensitive to. Therefore, provided that the emission was hadronic, they had the highest likelihood of observation in IceCube; see Fig.~\ref{fig-XX.3}.

The early predictions based on the measured fluxes and extensions reported by the Milagro collaboration determined that IceCube should be able to see neutrinos from these sources after five years of operation \citep{Beacom:2007yu, Halzen:2008zj, GonzalezGarcia:2009jc}. Follow-up observations by Milagro reported a low-energy cut-off in the spectrum of these sources \citep{Abdo:2012jg} which reduced their chance of observation in IceCube \citep{Gonzalez-Garcia:2013iha}. The prospect for observing these sources in IceCube data is indeed entangled with the uncertainties in the spectrum and the extension of the sources. Sources with a hard spectrum, close to $E^{-2}$, are easier to separate from the soft spectrum of the atmospheric background. Also, the number of atmospheric background events increase for extended sources. Understanding the precise extension of the source is necessary to obtain the optimal sensitivity \citep{Aartsen:2014cva}.

When IACT and ARGO-YBJ experiments performed follow up observations of these sources discrepancies emerged in the spectral measurements and tensions in the determination of their extension. A clear picture is hard to assess because each experiment faces systematic limitations. The wide-field water Cherenkov telescopes like Milagro are more suited to measure gamma-ray flux from extended sources, while IACTs are pointing observatories and are therefore limited in exploring extended regions. Their small field of view prevents them from separating signal from background in off-source regions. On the other hand, IACTs benefit from good energy resolution and are likely to provide more reliable spectral measurements. However, their sensitivity declines for energies beyond 10 TeV. Here, we revisit the prospects for observation of MGRO J1908+06 as one of the preeminent candidate sources; for a detailed discussion on the rest of sources, see reference~\citen{Halzen:2016seh}.

\subsubsection{MGRO J1908+06}

\begin{figure}[t]
 \centering
 \includegraphics[width=0.7\columnwidth]{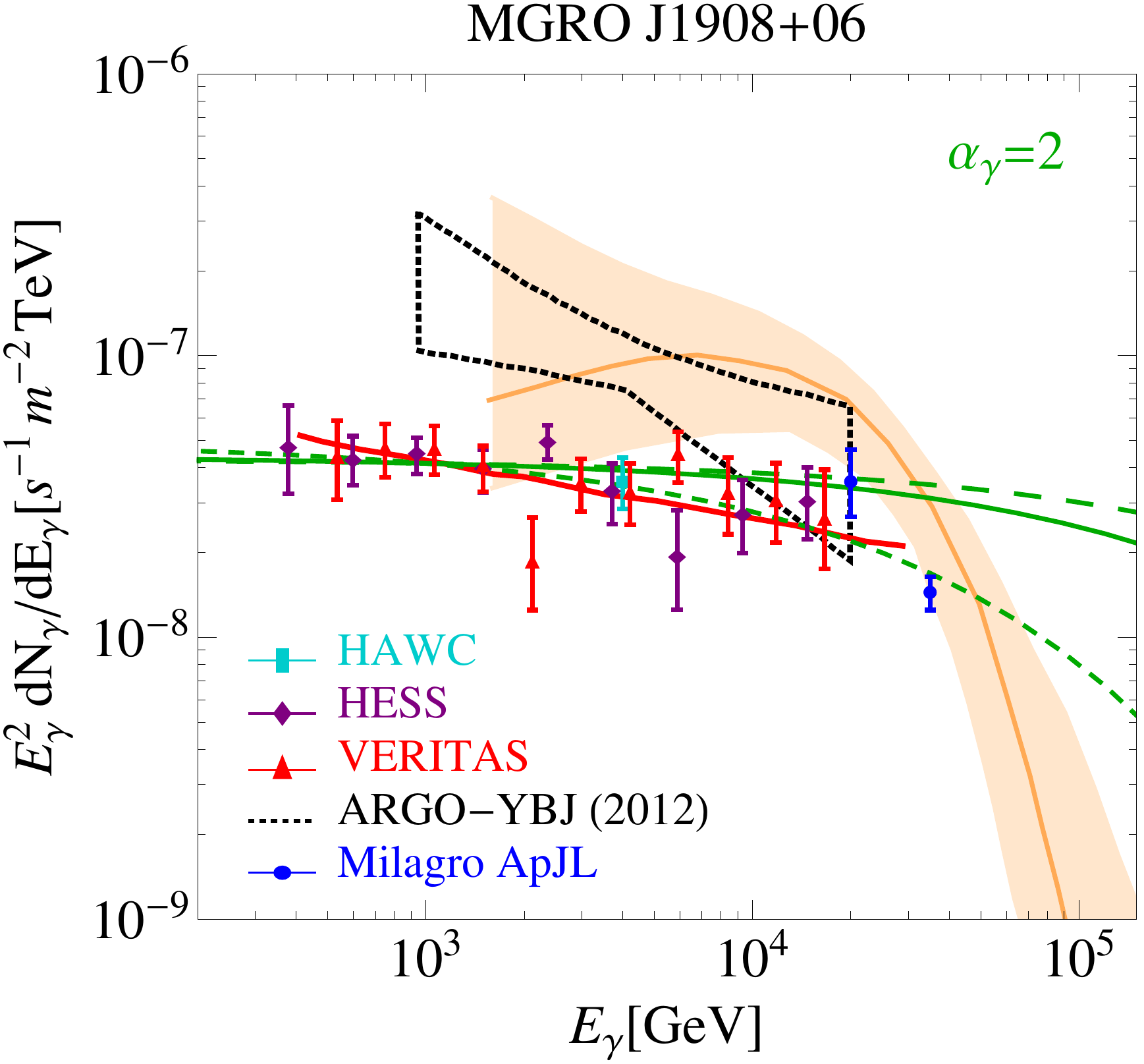}
 \caption{Gamma-ray flux of MGRO J1908+06 from different experiments: the purple data points are measured by HESS~\citep{Aharonian:2009je}, in red the flux from VERITAS~\citep{Aliu:2014rha}, and in cyan the flux from HAWC~\citep{Abeysekara:2015qba}. The blue data points show the original measurements by Milagro~\citep{Abdo:2007ad,Abdo:2009ku}, the orange line and the shaded orange region show the best fit and the $1\sigma$ band \citep{Smith:2010yn} . The dotted area identifies the ARGO-YBJ $1\sigma$ band~\citep{ARGO-YBJ:2012goa}. The green lines show the spectra obtained for a $\alpha_\gamma=2$ with a normalization fixed to the best fit reported by HESS. The result is shown for 3 cut-off enrgies: $E_{\rm cut, \gamma} =30,~300, {\rm and}~800$~TeV (short-dashed, solid, and long-dashed lines, in green). Figure from reference~\citen{Halzen:2016seh}.} 
 \label{fig:1908}
 \end{figure}

Figure \ref{fig:1908} compares the high-energy gamma ray spectra from MGRO J1908+06 reported by the air-shower detectors Milagro~\citep{Abdo:2007ad,Abdo:2009ku,Smith:2010yn} and ARGO-YBJ \citep{ARGO-YBJ:2012goa}, as well as by the IACTs.  The HESS collaboration measures a flux with a hard spectrum and no evidence of a cut-off for energies $<20$~TeV. It is systematically below the Milagro and ARGO-YBJ data \citep{Aharonian:2009je}. With superior angular resolution, it may be that HESS detects the flux from a point source that cannot be resolved by the Milagro and ARGO-YBJ observation. MGRO J1908+06 has been observed by VERITAS \citep{Aliu:2014xra} with a flux that is consistent with the one reported by HESS. Finally, the flux recently reported by HAWC points towards a similar normalization \citep{Abeysekara:2015qba}. 

 \begin{figure}[ht]
 \centering
 \includegraphics[width=0.7\columnwidth]{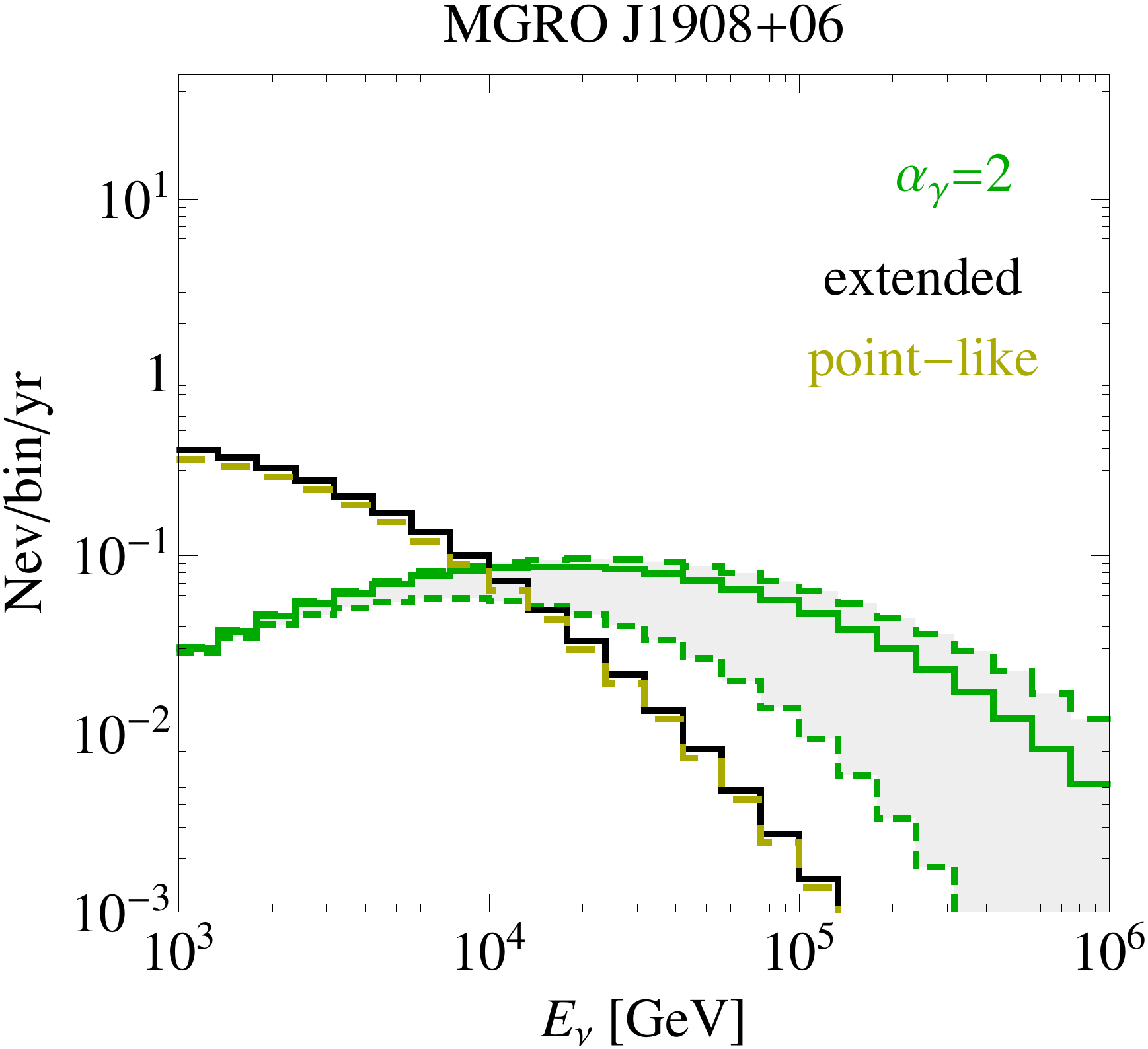}
 \caption{ Event distribution from MGRO 1908+06 for the spectra parametrized by HESS measurement.The grey band encodes the uncertainty on the cut-off energy. The black (gold dashed) line shows the background from atmospheric neutrinos for extended (point-like) sources. Figure from reference~\citen{Halzen:2016seh}.} 
 \label{fig:1908events}
 \end{figure}

The experiments disagree on the extension of the source. ARGO-YBJ finds an extension of $0.5^\circ$ \citep{Bartoli:2012tj}, while HESS measures $0.34^\circ$ \citep{Aharonian:2009je}. VERITAS reports $0.44^\circ$ \citep{Aliu:2014xra}.

MGRO J1908+06 is currently classified as an unidentified source. {\em Fermi}-LAT observes the pulsar PSR J1907+0602 within the extension reported by the Milagro collaboration \citep{Abdo:2010ht}. On the other hand, the large size and the hard spectrum of TeV photons that persist away from the pulsar location, are not characteristic of a PWN scenario \citep{Aliu:2014xra}; MGRO J1908+06 is perhaps a SNR. 

\begin{figure}[t]
\centering
 \includegraphics[width=0.7\columnwidth]{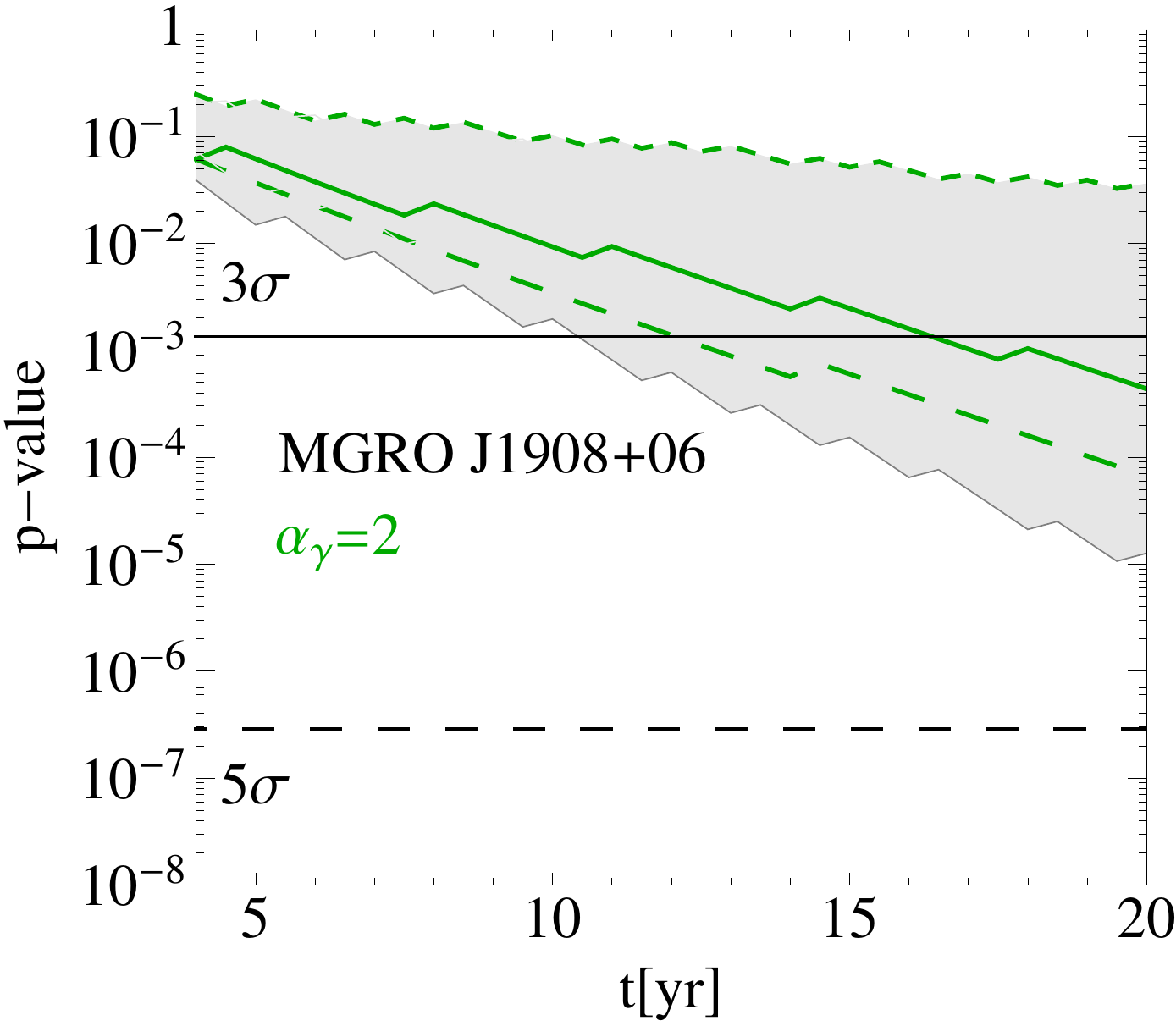}
 \caption{The calculated p-value for MGRO J1908+06 as a function of time for different spectral assumptions compatible with the observations reported by HESS \citep{Aharonian:2009je}; see Fig.~\ref{fig:1908events}. The grey band encodes the uncertainty resulting from different values of $E_{cut,\gamma}$ and the source morphology. Green lines assume an extended source.  Figure from reference~\citen{Halzen:2016seh}.} 
 \label{fig:1908pvals}
 \end{figure}

We use the multimessenger relation between gamma-ray and neutrino production rates in conjunction with the parametrization of Eq.~\ref{eq:cutoff} to estimate the flux of high-energy neutrinos from MGRO J1908+06. One is tempted to associate the cutoff with the knee in the cosmic spectrum. There is however some question whether supernova remnants can reach several PeV particle, and $\sim 300$\,TeV photon energies. On the other hand, it is also believed that the the Galactic cosmic ray spectrum extend well beyond the knee. The green lines in Fig.~\ref{fig:1908} show the result of the calculation assuming $\alpha_\gamma=2$ and fixing the normalization to the best fit reported by HESS. We allow the cut-off energy to vary: $E_{\rm cut, \gamma} =30,~300, {\rm and}~800$~TeV. In Fig.~\ref{fig:1908events} we show the corresponding number of events as a function of the energy cut-off which we varied between 30 and 800 TeV. Also shown is the number of background events for both a point source and an extended source with a size of $0.34^\circ$ extension as reported by HESS. The resulting p-value for observation of neutrino emission in the direction of MGRO J1908+06 is shown in Fig.~\ref{fig:1908pvals}. Different assumptions on the high-energy cut-off of the spectrum and the extension of the source result in different likelihoods for the identification of MGRO J1908+06. Overall, provided that the gamma-ray flux from MGRO J1908+06 extends to energies beyond 100 TeV, we anticipate a $3\sigma$ observation in 10 to 15 years of IceCube data assuming the photon flux reported by HESS.

 \begin{figure}[ht]
 \centering
 \includegraphics[width=0.65\columnwidth]{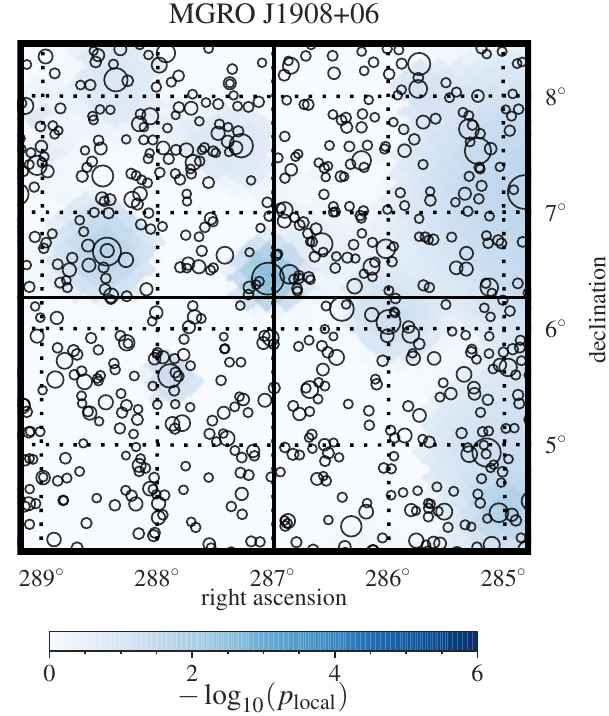}
 \caption{P-value map of neutrino excesses near MGRO 1908+06 in point source search with eight years of IceCube upgoing muons. Figure from reference~\citen{Aartsen:2018ywr}.} 
 \label{fig:8r1908}
 \end{figure}

For all the reasons mentioned above the MGRO 1908+06 source has been included from the beginning in IceCube's triggered search for point-like and extended sources \citep{Aartsen:2014cva, Aartsen:2016oji, Aartsen:2018ywr, Aartsen:2019fau}. A recent search for steady sources in the muon neutrino flux with 8 years of IceCube of data \citep{Aartsen:2018ywr} found this source as the second ``warmest" source in the catalog, with a pre-trial p-value of 0.0088. The fitted flux upper limit is consistent with the expectations discussed above. Figure \ref{fig:8r1908} shows the p-value of the neutrino excess in the vicinity of MGRO J1908+06. This search optimistically assumed no extension of the source. 

A recent search for PeVatron candidates in the HESS Galactic plane survey data \citep{H.E.S.S.:2018zkf} establishes a lower bound on the cut-off energy of MGRO J1908+06 that has further supported the idea that the source may be a PeVatron \citep{Spengler:2019sde}.

Besides monitoring also MGRO J1908+06 and MGRO 2019+37, IceCube performed a stacking analysis that searched for the collective neutrino emission from the six sources initially identified. The stacking likelihood \citep{Achterberg:2006ik} matches the neutrino arrival directions with a collection of sources or a catalog \citep{Aartsen:2017ujz}.

Progress in the search for Galactic sources has been made possible by the commissioning of HAWC as we will discuss further on.

\subsubsection{Pulsar Wind Nebulae}

Progress in instrumentation has revealed more than 150 TeV sources in the Galactic plane, the majority are classified as pulsar wind nebula. Given that they are collectively responsible for a significant fraction of the high-energy emission from the Galaxy, their contribution to the neutrino flux may be important. 

Confined inside supernova remnants, pulsar wind nebulae are bright diffuse nebulae whose emission is powered by pulsar winds emerging from the rapidly spinning magnetized pulsars in their center. Relativistic electron-positron pairs are considered to be the primary nonthermal component of the pulsar winds. Synchrotron emission by relativistic pairs is the dominant process for the low-frequency emission in radio, optical, and X-ray, while inverse Compton scattering of synchrotron photons is responsible for the TeV emission \citep{Kargaltsev:2010jy}.

The possible role of hadrons in PWN was first suggested as an alternative mechanism for generating the very high energy gamma ray emission from the Crab Nebula.  Early proposals included the possibility that protons, accelerated in the outer gap of the pulsar, interact with the nebula \citep{Cheng:1990au} or, alternatively, that heavy nuclei accelerated in the pulsar magnetosphere interact with soft photons \citep{Bednarek:1997cn}. More recent modeling focuses on particle acceleration in the termination shock followed by neutrino production in cosmic ray interactions with the dense environment in the nebula, see e.g., references~\citen{ Guetta:2002hv,Amato:2003kw,Amato:2006ts,Bednarek:2003cv,Lemoine:2014ala,DiPalma:2016yfy}. Minor contamination of ions at the termination shock would result in a significant release of energy in hadrons \citep{Amato:2013fua}. This is particularly important in the context of what is known as the $\sigma$ problem. The parameter $\sigma$ is defined as the ratio of the wind's Poynting to kinetic energy flux. Theoretical modeling of the pulsar magnetosphere and wind point at large $\sigma$ values. However, the magnetohydrodynamic simulations cannot match the shock size and expansion speed at same time. This conflict can be resolved if the majority of the pulsar wind energy is carried by hadrons and, in addition, results in the efficient acceleration of particles in termination shocks \citep{DiPalma:2016yfy}.
 
Searches for neutrino emission from pulsar wind nebulae have targeted individual sources \citep{Aartsen:2013uuv, Aartsen:2014cva, Aartsen:2016oji} as well as their cumulative emission in a stacking search \citep{Aartsen:2017ujz, Liu:2019iga}. So far, neither have yielded evidence for a correlation of the arrival directions of cosmic neutrinos to pulsar wind nebulae and have thus constrained the hadronic component of the termination shock. The results are relevant for evaluating the prospects for observing PWN with KM3NeT; many of these sources are located in the Southern Hemisphere; see reference~\citen{DiPalma:2016yfy}.

\begin{figure}[t]  \label{fig:pwn}
 \centering
 \includegraphics[width=0.7\columnwidth]{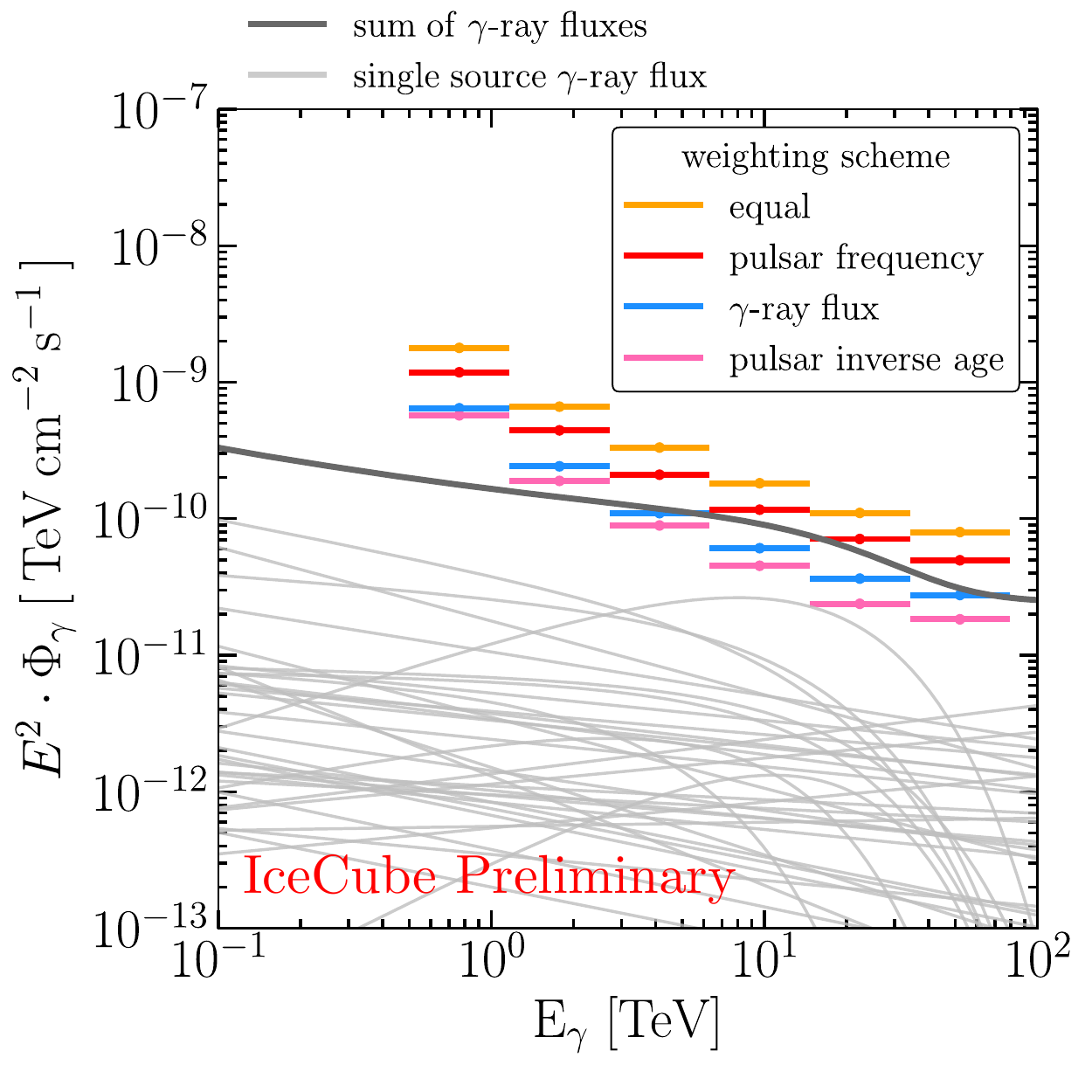}
 \caption{Differential upper limits on the hadronic component of the gamma-ray emission from TeV pulsar wind nebulae. The light grey lines show the observed gamma-ray spectra of the sources while the dark grey line presents the sum. Red, orange, magenta and blue steps show the differential upper limit on the hadronic gamma-ray emission. The upper limits are obtained by converting 90\% CL differential upper limit on the neutrino flux, and each color corresponds to a given weighting method. 
 Figure from reference~\citen{Liu:2019iga,Aartsen:2020eof}.} 
 \end{figure}

The IceCube stacking analysis presented in \citep{Aartsen:2017ujz} targeted a set of 9 supernova remnants with associated pulsar wind nebulae that are gamma ray emitters. In a recent study, IceCube has examined the correlation of high-energy muon neutrinos with a larger set of 35 TeV pulsar wind nebulae with gamma-ray emission above 1 TeV \citep{Liu:2019iga,Aartsen:2020eof}. The selection of the sources were based on HAWC, HESS, MAGIC, and VERITAS observations and the associated pulsars listed in the Australia Telescope National Facility (ATNF) catalog \citep{Manchester:2004bp}. In the absence of any significant correlation, upper limits were derived on the total diffuse neutrino flux from these sources. Figure \ref{fig:pwn} shows the results for four distinct hypothesis tests of the multimessenger relation between the neutrino and gamma ray fluxes. At the highest energies, the stacking analyses find upper limits at the level of the high-energy gamma-ray emission and thus start to constrain the role of hadrons in pulsar wind nebulae. We will return to this search in Section \ref{sec:summary}.

 \subsubsection{The HAWC Very High Energy Gamma Ray Survey of the Galaxy} 
  
 The characteristic energy scale of gamma rays from PeVatrons reaching cosmic ray energies beyond the knee is 100 TeV. The ground-based, wide-field telescope HAWC gamma-ray observatory has provided an unprecedented view of the Galaxy by reaching this energy. HAWC's survey of the Galactic plane has revealed more than 20 new TeV sources, most of them classified as {\em unidentified}.

  \begin{figure}[ht]
   \centering
 \includegraphics[width=0.7\columnwidth]{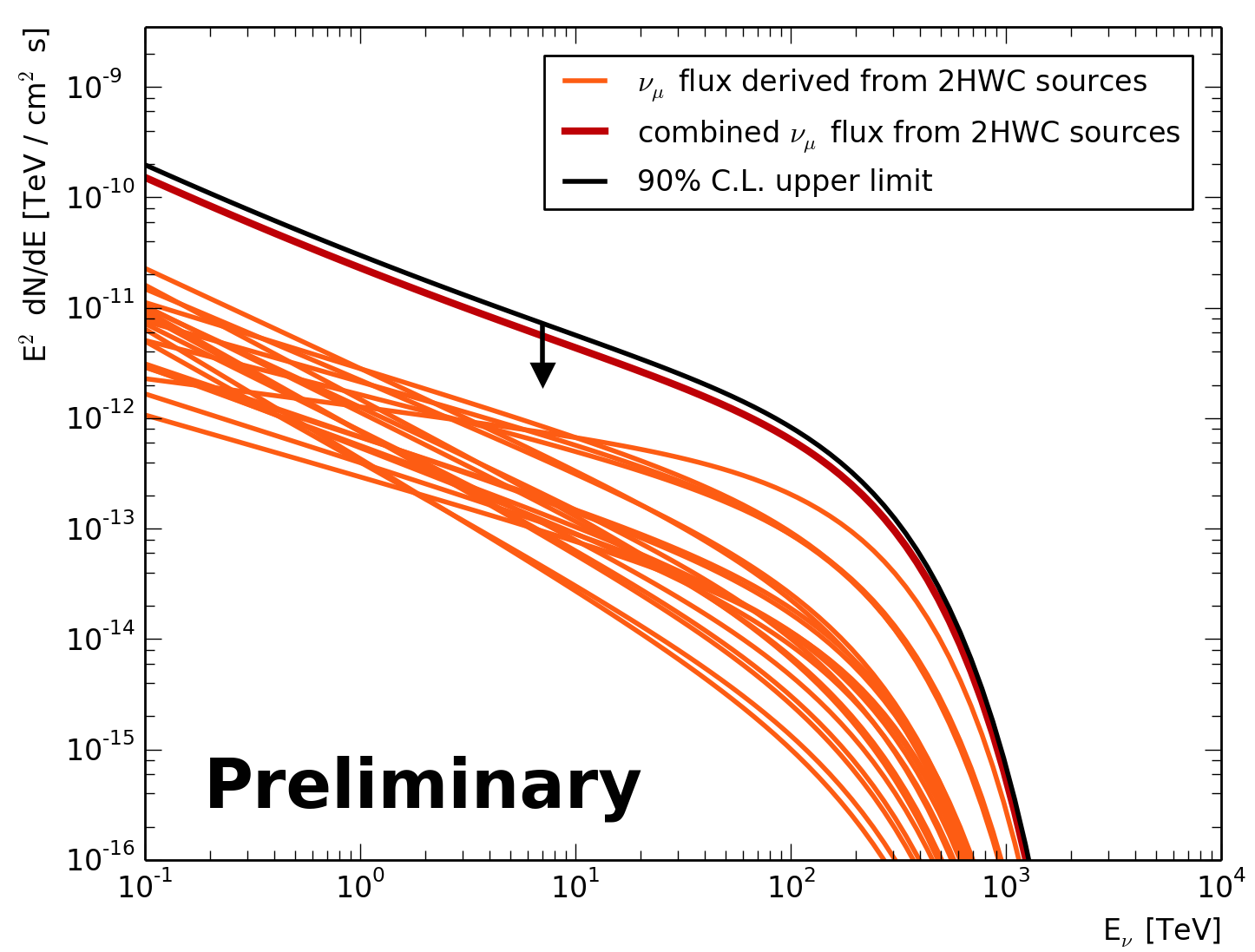}
 \caption{Upper limit (90\% C.L.) of the stacking search for a flux of muon neutrinos (black) associated with sources in the 2HWC catalog (PWN excluded). The thin orange line is the projected muon neutrino flux assuming all HAWC photons are produced in hadronic interactions. The combined flux (red) shows sum of the individual fluxes. Figure from reference~\citen{Kheirandish:2019bke}.}
 \label{fig:hwcstack}
 \end{figure}

In order to find common sources of very high energy gamma rays and neutrinos, the IceCube and HAWC collaborations designed and performed a joint analysis \citep{Kheirandish:2019bke}. Specifically, the analysis examined the correlation of the arrival directions of IceCube muon neutrinos passing through the Earth and sources in the 2HWC catalogue \citep{Abeysekara:2017hyn}, as well as the correlation of the arrival direction of neutrinos with the morphology of the very high energy gamma ray emission. By excluding the sources identified as PWN, the stacking analysis of 2HWC sources focused on unidentified sources. The latter analysis incorporated the full morphology of the gamma-ray emission from the plane visible in the Northern Sky. Finally, special regions were singled out for study: the Cygnus region, MGRO 1908+06, and the inner Galaxy, including MGRO J1852+01 which has been confirmed by HAWC.

Using eight years of IceCube data, constraints are obtained on the contribution of pionic photons to the HAWC gamma-ray flux in the Cygnus region and in the section of the Galactic plane visible to both experiments. The upper limits on the neutrino flux from the other searches are above the flux derived on the basis of the assumption that all HAWC photons are of hadronic origin. The upper limits from the stacking analysis of the 2HWC sources is shown in Fig.~\ref{fig:hwcstack}. More data and improved selections will be required. This is also the case for MGRO J1908+06.

An under-fluctuation of the data for in the Cygnus region led to a stringent constraint on the hadronic component of the high-energy emission.

\subsubsection{The Cygnus Complex}

The Cygnus~X complex is a nearby star-forming region in our Galaxy. It has been extensively studied as a promising nearby source of very high energy cosmic rays and neutrinos \citep{Ackermann:2011lfa, Aharonian:2002ij,Tchernin:2013wfa,Gonzalez-Garcia:2013iha,Nierstenhoefer:2015gta, Guenduez:2017qrw, Aharonian:2018oau}. The complex contains massive molecular clouds, populous associates of massive young stars, and luminous H\textsc{II} regions \citep{1981A&A...101...39B, 1992ApJS...81..267L, 2015ApJ...811...85K}. The observation of a hard gamma-ray spectrum, in combination with dense molecular clouds \citep{2012A&A...541A..79G, 2016A&A...591A..40S} and a large number of young OB stars \citep{2015MNRAS.449..741W} supports the idea that Cygnus~X is a plausible source of cosmic rays and high-energy astrophysical neutrinos.

\setcounter{footnote}{0}

High-energy gamma ray emission from this region was tentatively detected by EGRET\footnote{Energetic Gamma Ray Experiment Telescope} \citep{1996A&AS..120C.423C}. Later, a hard gamma-ray spectrum was established by HEGRA\footnote{High-Energy-Gamma-Ray Astronomy} \citep{Aharonian:2002ij} and confirmed by the detection of TeV emission by the Milagro collaboration \citep{2007ApJ...658L..33A}. More recent observations with the MAGIC \citep{Albert:2008yk}, \textit{Fermi} \citep{Ackermann:2011lfa, FermiLAT:2011lax}, ARGO-YBJ \citep{Bartoli:2014irw}, and VERITAS \citep{Aliu:2014rha} instruments have led to a better understanding of the point-like and extended emission in this region.

Because of its extension, the identification of individual sources has been challenging for the IACT experiments. In contrast, 4 out of 6 sources initially detected by Milagro belonged to this region. MGRO J2031+41 is particularly interesting because its location coincides with the Cygnus cocoon. Because of uncertainties associated with the origin of the flux from the Cygnus cocoon and $\gamma$-Cygni, a complete picture of high-energy emission from MGRO J2031+41 is missing. A combined analysis of ARGO-YBJ and Fermi data finds a hard spectrum \citep{Argo:2014tqa}, which, assuming a point source, implies observation by IceCube after $\sim 15$ years \citep{Halzen:2016seh}.

An important feature of the Cygnus region is the observation of an extended excess of high-energy gamma rays with a hard spectrum referred to as the Cygnus Cocoon. It was first detected by the \textit{Fermi} collaboration \citep{Ackermann:2011lfa} and confirmed by the observation at TeV energies by Milagro, ARGO-YBJ, VERITAS, and HAWC in a region that is partially coincident with the Cygnus Cocoon. The true nature of the Cocoon is still a mystery. 

\begin{figure}[t]
  \centering
 \includegraphics[width=0.95\columnwidth]{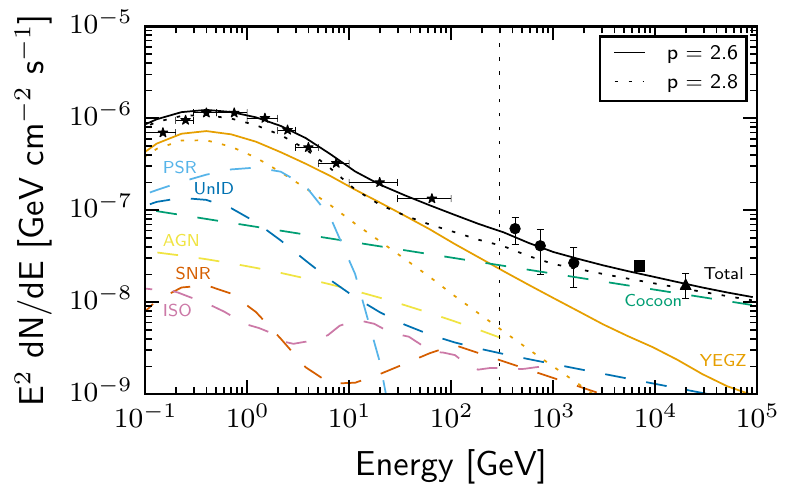}
 \caption{Observed and modeled gamma-ray fluxes from the Cygnus X region. Different components of the spectrum include the gamma-ray emission calculated with the YEGZ model, pulsars (PSRs), active galactic nuclei (AGN), supernova remnants (SNRs), unidentified sources (UnID), the isotropic gamma-ray background (ISO), and the Cygnus Cocoon. Observational points show data from \textit{Fermi} (black stars) \citep{FermiLAT:2011lax}, ARGO-YBJ (black circles) \citep{Bartoli:2015era}, HAWC (black square) \citep{Abeysekara:2017hyn}, and Milagro (black triangle) \citep{2007ApJ...658L..33A}. The vertical dotted black line identifies 300 GeV beyond which the spectrum for Fermi 3FGL sources are extrapolated.  Figure from reference~\citen{Yoast-Hull:2017gaj}.} 
 \label{fig:gammacyg}
 \end{figure}

 We have described the observation in the Cygnux X region with a model of cosmic ray interactions developed for the central molecular zones of starburst galaxies \citep{Yoast-Hull:2013wwa}; see Fig.~\ref{fig:gammacyg}. While at lower energies ($E \sim 10$ MeV) bremsstrahlung by electrons dominates the gamma-ray spectrum, the contribution of pionic photons grows with energy and dominates at $\sim 10\, $--$ 100$ GeV. High-energy gamma ray emission from inverse Compton scattering is negligible at these energies because of the steep spectrum of the cosmic ray electrons. The comparison of the the modeled gamma-ray spectrum with the observations from \textit{Fermi}, Milagro, ARGO-YBJ, and HAWC are shown in Fig.~\ref{fig:gammacyg}. The spectral fits provided in the 3FGL Fermi catalogue only cover the region between 100 MeV and 300 GeV. For the Cocoon, we have extrapolated the spectral fit to higher energies to compare with TeV-energy gamma-ray observations.  Finally, we only included the off-pulse emission for the three brightest pulsars; see reference~\citen{Yoast-Hull:2017gaj} for details. 

\begin{figure}[t]
 \centering
 \includegraphics[width=0.8\columnwidth]{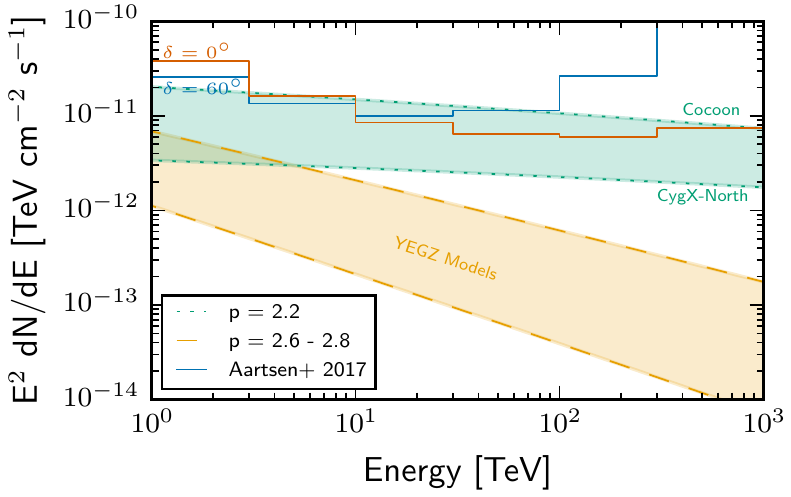}
 \caption{The neutrino energy flux predicted for the diffuse gamma ray flux from a YEGZ model with a soft cosmic ray spectrum with $p = 2.6 - 2.8$, the Cygnus Cocoon, and the CygX-North molecular cloud complex. Also shown is the IceCube differential discovery potential for a point source \citep{Aartsen:2016oji}. Note that the IceCube sensitivity to extended sources is lower, and thus this plot represents the most optimistic case for detection.  Figure from reference~\citen{Yoast-Hull:2017gaj}.} 
 \label{fig:cocoonnu}
 \end{figure}

 Assuming that the Cocoon is a single source and that the spectrum is dominated by gamma-rays from neutral pion decay, the model \citep{Yoast-Hull:2013wwa} predicts a neutrino flux that is slightly larger than the IceCube differential discovery potential at 1 PeV\citep{Aartsen:2016oji}. 
 Therefore, detecting neutrinos from the Cocoon may be possible provided the spectrum extends to PeV energies without steepening.
 
We next turn to a region within the Cocoon that surrounds a large molecular gas cloud complex located left of Cyg OB2 and centered on $(l = 81.5^{\circ},~b = 0.5^{\circ})$,: CygX-North. Its emission is expected to be dominated by hadronic interactions, most likely from an unresolved source, either a supernova remnant or a pulsar wind nebula. Here the model predicts a neutrino flux below IceCube's discovery potential. A summary of the expected neutrino flux from the cocoon and CygX-North is shown in Fig.~\ref{fig:cocoonnu}. Clearly, the prospect for identification of neutrino emission from the Cygnus X region critically depends on the slope of the cosmic ray spectrum.
 
Arguing that diffusion losses play a significant role in modeling Cygnus X, more pessimistic neutrino fluxes are anticipated \citep{Guenduez:2017qrw}. This transport model describes the broad multiwavelength spectrum but predicts a neutrino flux which approaches the sensitivity of IceCube only at very high energies, nominally $> 50$ TeV.

Also of interest was the Milagro source MGRO J1852+01 because of its high gamma ray flux. However, it did not pass the statistical threshold for detection and inclusion in the Milagro survey. HAWC observes a high intensity flux from the source region that is mostly centered on the source 2HWC 1857+027. The large flux reported by Milagro in a $3^{\circ}\times3^{\circ}$ region most likely originated from this source. Given the large flux and the vicinity to the horizon where high-energy neutrinos reach IceCube without absorption by the Earth, this region is a sweet spot in IceCube sensitivity. Interestingly, the largest excess in the joint IceCube-HAWC search is associated with 2HWC 1857+027.

Overall, the joint IceCube-HAWC analysis did not find any significant correlation. However, it may have provided clues about the potential sources of high-energy neutrino in the Galaxy. There are two main outcomes: first, the hadronic component of the emission from Cygnus region has been constrained. Second, in the inner Galaxy the results are consistent with expectations. That is, while IceCube has not reached the sensitivity for identifying emission from those Galactic sources, their observation is within reach as the data accumulates.

In a more recent development, the HAWC collaboration reported the measurement of the gamma-ray flux from the Galaxy above 56 TeV \citep{Abeysekara:2019gov}. HAWC's very high energy survey identifies 9 sources, four of which coincide with Milagro PeVatron candidates. Additionally, this survey resolves two more sources above $-5^\circ$ declination, where IceCube is most sensitive.

\subsubsection{Binaries}

Microquasars are X-ray binaries of a collapsed object with prominent jets rotating around a massive star. They have been identified as potential sites of cosmic ray acceleration and neutrino emission \citep{Torres:2006ub, Levinson:2001as, Guetta:2002hk, Aharonian:2005cx}. High-energy photons are produced in the interaction of the jet with the stellar wind and radiation fields associated with the high-mass companion star. As usual, an alternative possibility is that the high-energy photons result from the inverse Compton scattering of stellar photons by relativistic electrons. In hadronic models, protons in the jet interact with cold protons in the stellar wind \citet{Romero:2003td, Romero:2005fr}. Not only are neutrinos produced, cooling of the other secondary particles produced in decay of the charged pions, i.e., muons and electrons, alter the broadband emission spectrum of gamma-ray binaries. Studies of Cygnus X-1 suggest that protons in the jet could carry a significant fraction of their kinetic energy \citep{Gallo:2005tf,Heinz:2005jc}. Microquasars were actually suggested as the dominant contributors to the Galactic component of the high-energy neutrino flux \citep{Anchordoqui:2014rca}. This was partially motivated by the spatial clustering of the first sample of cosmic neutrino events near LS 5039, a known microquasar which has been historically suggested as a potential cosmic ray accelerator \citep{Aharonian:2005cx}. 

X-ray binaries have periodic outbursts which makes them ideal targets for time-dependent searches. These benefit from lower background rates and can identify the source with fewer signal events. The periodic emission and X-ray flares have been exploited in time-dependent searches for neutrino emission \citep{Abbasi:2011ke, Aartsen:2015wto, Adrian-Martinez:2014ito}.

The discussion presented so far has been centered on potential Galactic sources of high-energy neutrinos in the Northern Sky. In the future, with the commissioning of KM3NeT the sensitivity to neutrino sources in the southern sky will be significantly improved. The HESS collaboration has detected very high energy gamma ray emission near the Galactic center \citep{Abramowski:2016mir} with a spectrum that is difficult to account for by leptonic modeling. Observation of neutrino emission could establish this source as a Galactic PeVatron. The neutrino emission may originate from the Sgr A*, the low-luminosity AGN located at the center of the Milky Way \citep{Fujita:2016yvk, Anchordoqui:2016dcp}. Observations of X-ray echos have revealed a high level of activity in its past \citep{Koyama:1996sj, Ryu:2012ib}. Accelerated cosmic rays will interact with the very dense environment near the center of the Galaxy and generate a flux of high-energy neutrinos.

RX J1713.7-3946, a known supernova remnant, and Vela X \citep{Aiello:2018usb} are promising sources in the field of view of KM3NeT. It is worth mentioning that the IceCube search for point sources using cascades identified RX J1713.7-3946 as the most significant source \citep{Aartsen:2019epb}. With cascades IceCube has a reduced sensitivity, by about a factor of 5, but covers the Southern Hemisphere. HAWC observes high-energy gamma sources in the fraction of the Galactic plane in the Southern Hemisphere that is within its field of view. If hadronic in origin, these sources are within the sensitivity of KM3NeT after 10 years of operation \citep{Niro:2019mzw}.

\subsection{Diffuse Galactic Neutrino Emission }\label{sec:diff}

Over their diffusion time in the Milky Way, very high energy cosmic rays interact with the dense environment of the Galactic plane that provides the target material for producing neutrinos. Therefore, diffuse gamma-ray and neutrino emission from the Galactic plane is considered a guaranteed non-isotropic component in the high-energy sky \citep{Stecker:1978ah, Domokos:1991tt, Berezinsky:1992wr, Ingelman:1996md, Evoli:2007iy}.

The intensity of the Galactic diffuse neutrino emission is expected to vary along the plane tracing the gas density and the distribution of the sources. The intensity and morphology of the diffuse Galactic neutrino flux can be further constrained by modeling the observed diffuse Galactic gamma-ray emission profile using the local measurements of the cosmic ray spectrum \citep{Ahlers:2013xia,Joshi:2013aua,Kachelriess:2014oma}.

\begin{figure}[t]
\centering
 \includegraphics[width=0.85\columnwidth]{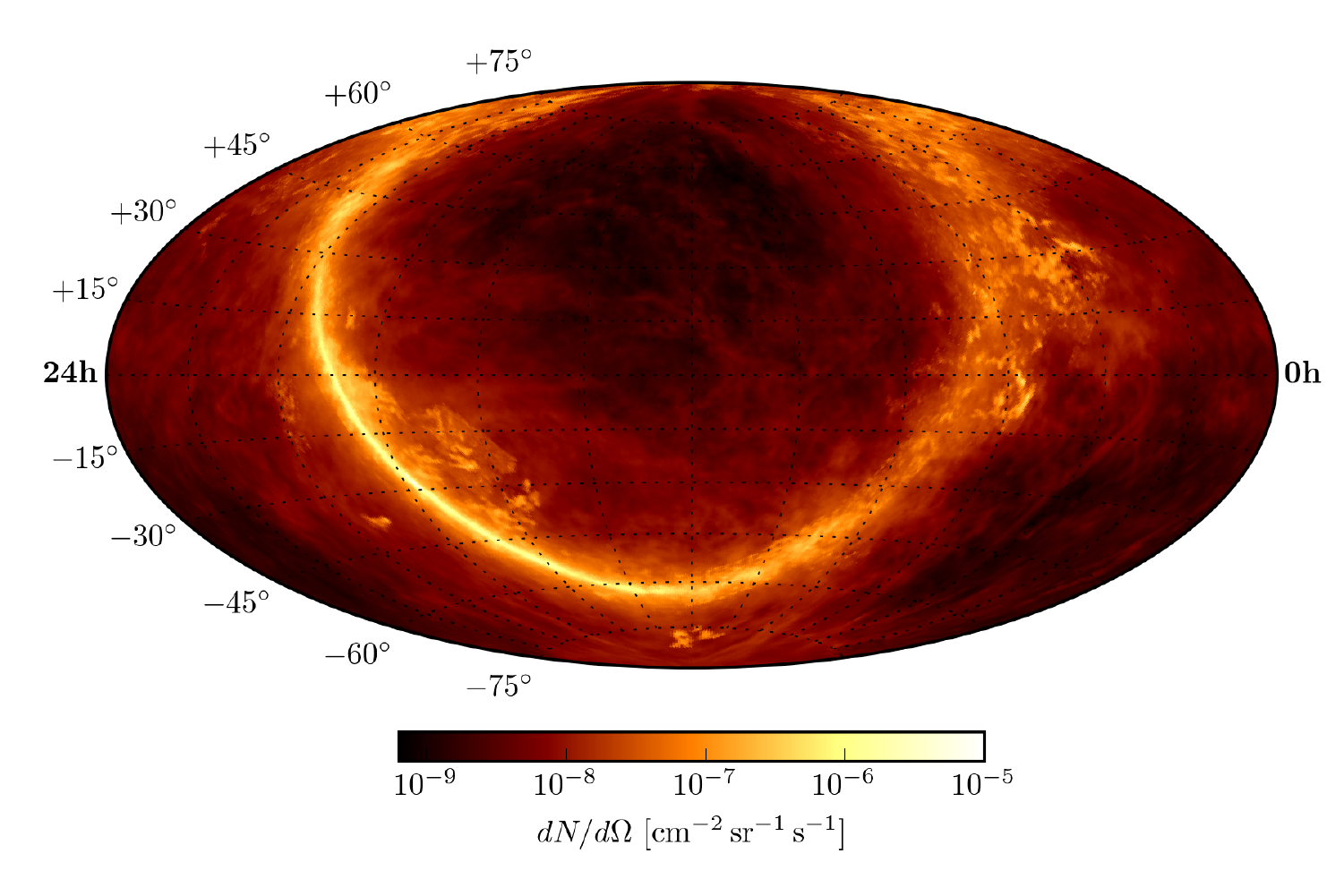}
 \caption{Diffuse gamma-ray emission from the Galactic plane. 
 The color map shows the energy integrated neutrino flux, assuming the KRA$_\gamma^5$ model~\citep{Gaggero:2015xza}, as a function of their arrival direction in equatorial coordinates.
 Figure from reference~\citen{Albert:2018vxw}.}
 \label{fig:diffnu}
 \end{figure}

The diffuse component of gamma-ray emission from Galactic plane has been measured by {\em Fermi}-LAT in the range 100 MeV to tens of GeV \citep{Ackermann:2012pya, Acero:2016qlg}. A tool routinely used to model these observations is the GALPROP \citep{2011CoPhC.182.1156V} code. The diffuse Galactic gamma-ray component is derived by assuming a distribution of potential sources of cosmic rays and by modeling their propagation and diffusion in the interstellar medium. While good agreement can be achieved with the gamma-ray measurements below 10 GeV, the modeling fails to account for the observed flux from the inner Galaxy above that energy \citep{Ackermann:2012pya}. The so-called KRA$_\gamma$ models~\citep{Gaggero:2015xza, Gaggero:2014xla, Gaggero:2017jts} propose to resolve this by modifying the diffusion coefficient as a function of energy. Energy cut-offs at 5 and 50 {PeV} are chosen to accommodate the cosmic ray measurements by KASCADE~\citep{Antoni:2005wq} and KASCADE-Grande~\citep{Apel:2013uni}. Figure \ref{fig:diffnu} shows the anticipated morphology of neutrinos predicted in the KRA$_\gamma^{5}$ model integrated over energy.

IceCube has searched for the Galactic component of the diffuse cosmic neutrino flux \citep{Ahlers:2013xia, Ahlers:2015moa, Albert:2017oba, Aartsen:2017ujz} based on a variety of assumptions for the energy dependence of the Galactic spectrum, including the one predicted by the KRA$_\gamma$ models. 

  \begin{figure}[ht]
  \centering
 \includegraphics[width=0.7\columnwidth]{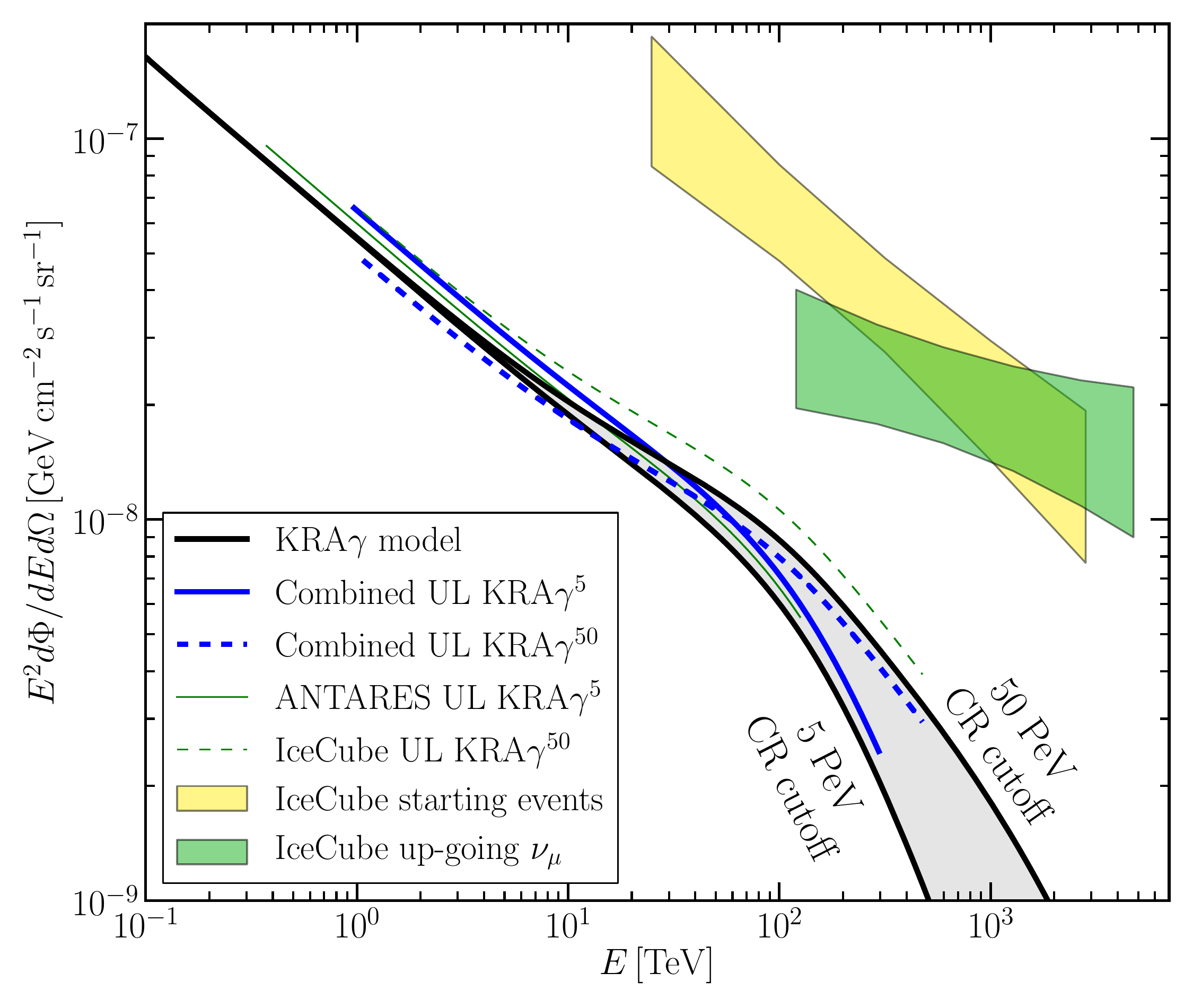}
 \caption{Combined upper limits at 90\% confidence level (blue lines) on the three-flavor neutrino flux for the KRA$_\gamma$ with the 5 and 50 PeV cutoffs (black lines). The boxes represent the diffuse astrophysical neutrino fluxes measured by IceCube using
an isotropic flux template with starting events (yellow) and upgoing tracks (green). Figure from reference~\citen{Albert:2018vxw}.}
 \label{fig:difflim}
 \end{figure}

In a recent study, using ten years of ANTARES muon neutrino and cascade data and seven years of IceCube muon neutrino data, a joint search search has been performed for a diffuse Galactic component of the cosmic neutrino spectrum \citep{Albert:2018vxw}. The results of the maximum likelihood analysis finds a non-zero component, however, the excess falls short of statistical significance because the upper limits obtained from the test fall above the actual Galactic flux; see Fig.~\ref{fig:difflim}.

It has been suggested \citep{Crocker:2010dg} that the extended region of the {\em Fermi} bubbles may be a site of cosmic ray diffusion and high-energy neutrino emission \citep{Lunardini:2013gva, Ahlers:2013xia}. Gamma ray emission with a hard spectrum up to $\sim$ 100 GeV has been observed \citep{Fermi-LAT:2014sfa}. IceCube and ANTARES have searched for a diffuse Galactic flux from this region but only established upper limits \citep{Aartsen:2019epb, Albert:2017bdv}. 

\subsection{Closing In on Galactic Sources of High-Energy Neutrinos}\label{sec:summary}

The arrival directions of high-energy neutrinos indicate an extragalactic origin. However, a subdominant contribution from a Galactic component is expected, as discussed above. 

Galactic sources are in fact a {\em guaranteed} contributor to the flux of cosmic neutrinos. Cosmic rays in the Galaxy are believed reach energies of few PeV. Their interaction with dense stellar and interstellar gas and radiation results in the production of neutrinos and gamma rays.

In this section, we have discussed the prospects for identifying neutrino emission from Galactic sources and reviewed the status of past searches. The discussion was centered around major gamma-ray emitters in the Galaxy because they provide, if not a measurement, upper limits on the production of pionic gamma rays.

The subdominant Galactic component of the high-energy neutrino flux is estimated to contribute at most 10\% of the total measured cosmic neutrino flux. However, identification of Galactic sources with strong neutrino emission is expected with upgraded and next-generation detectors. While a significant contribution to the flux at very high energies is not expected \citep{Ahlers:2015moa}, features in the cosmic neutrino spectrum and the excess flux observed in IceCube below $\sim 10$\,TeV may already indicate a Galactic contribution from the southern sky with a harder spectrum.

In addition, according to our best estimates, the searches for point-like and extended sources in the Galaxy will soon reach the required sensitivity to be successful. In this context, continued improvement of the gamma ray data will be helpful.

HAWC observations of the gamma ray flux at energies exceeding 50 TeV continue to support the idea that sources like MGRO J1908+06 may be sources of high-energy neutrinos. The projected Southern Wide-field Gamma-ray Observatory (SWGO) \citep{Albert:2019afb} will provide a deeper survey of the Galactic plane in the Southern Hemisphere. Soon the Cherenkov Telescope Array (CTA) \citep{Acharya:2013sxa, Acharya:2017ttl} and LHAASO \citep{DiSciascio:2016rgi} will provide enhanced sensitivity reaching into the PeV regime, essential to identify the PeVatrons in the Galaxy. Based on current estimates, CTA is expected to resolve $\sim$ 100 supernova remnants \citep{Zanin:2017rao}. It is also worth mentioning that proposed MeV telescopes like eAstrogram and AMEGO have the capability of identifying the {\em pion bump} in the broadband spectrum of Galactic sources. This will contribute to the superior modeling of the diffuse emission because unresolved sources with harder intrinsic cosmic ray spectra can have a significant contribution to diffuse gamma-ray and neutrino fluxes \citep{Ahlers:2013xia, Lipari:2018gzn}.
 
Anticipated improvements in angular resolution and, specifically, in identifying more starting muon tracks, will improve IceCube's sensitivity to Galactic sources in the southern sky \citep{Silva:2019fnq, Mancina:2019hsp}. Addition of cascade data sets, despite their reduced angular resolution, has proved to add substantial capabilities in this search. 

Future neutrino detectors will considerably improve the likelihood of finding Galactic sources of high-energy neutrinos. Development of neutrino detectors in the Northern Hemisphere such as KM3NeT and Baikal  \citep{Avrorin:2019vfc} will boost the sensitivity towards Galactic center and other regions of interest in the southern sky. The next-generation IceCube detector is expected to improve the sensitivity to sources by at least a factor of 5 \citep{Aartsen:2014njl, Ahlers:2014ioa, Aartsen:2019swn}. Increased sensitivity to point-like and extended sources will be achieved improved angular resolution, better energy resolution, higher statistics, and reduction of the background. 

A potential more exotic Galactic component contributing to the high-energy neutrino flux is dark matter. The decay and annihilation of dark matter particles create distinctive signatures in the observed neutrino spectrum; see e.g., references~ \citen{Beacom:2006tt, Murase:2012xs, Bai:2013nga, Murase:2015gea, Bhattacharya:2019ucd, Arguelles:2019ouk}. An enhancement in the direction of the Galactic center could result from the higher local concentration of dark matter. In contrast, the possible interaction of cosmic neutrinos with dark matter in the Galaxy would create a deficit in the neutrino flux in the same direction \citep{Arguelles:2017atb}.

More important, IceCube has a unique capability of identifying the annihilation of dark matter particle trapped in the sun. We will discuss this in the next section.

\section{Beyond Astronomy}

We now turn to the study of the neutrinos themselves. Over more than a decade of operations IceCube has focused on three traditional methods to study neutrino physics: the study of neutrino oscillations exploiting the atmospheric neutrino beam, the search for MeV neutrinos from supernovae explosions in our Galaxy with their extraordinary potential to reveal neutrino physics, and the search for dark matter particles that have accumulated in the sun to annihilate in Standard Model particles that decay into neutrinos.

\subsubsection{Atmospheric Neutrino Oscillations}
\label{subsection1}

Deepcore~\cite{Collaboration:2011ym} is the IceCube tool to study the oscillations of neutrinos produced in the atmosphere. It instrumented a volume with improved photocathode coverage at the bottom center of the detector with eight additional strings that were deployed inside the regular IceCube grid. The PMTs on these strings have higher quantum efficiency and are located in the clearest ice below 2.1\,km. The DeepCore instrumented volume is roughly $10^7$\,m$^3$ with a module density about five times that of the surrounding IceCube array that is used as a veto removing with high efficiency the cosmic ray muon background.

Oscillation measurements~\cite{Aartsen:2013jza,Aartsen:2014yll,Aartsen:2017nmd} use samples of neutrinos of more than 100,000 events per year from the decay of pions and kaons produced by cosmic ray air showers in the Earth's atmosphere.
The initial atmospheric flux is dominated by ${\nu_e}$ and $\nu_\mu$ and contains negligible numbers of ${\nu_\tau}$~\cite{Bulmahn:2010pg}.
The atmospheric neutrinos interacting in the DeepCore subarray of IceCube travel distances ranging from $L\approx20$\,km (vertically downward-going) to $L\sim 1.3\times10^4$\,km (vertically upward-going or the full diameter of the Earth).
For vertically upward-going neutrinos, the first peak of the $\nu_\mu \rightarrow \nu_\tau$ oscillation probability occurs at 25\,GeV. This is comfortably above the $E_\nu \sim 5$\,GeV threshold of DeepCore~\cite{Leuermann:2018}.
The energy corresponding to the oscillation maximum is also well above the kinematic threshold of $E_{\nu_\tau} = 3.5$\,GeV for charged current $\nu_\tau$-nucleon interactions.

\begin{figure}[htbp]
    \includegraphics[width=\linewidth]{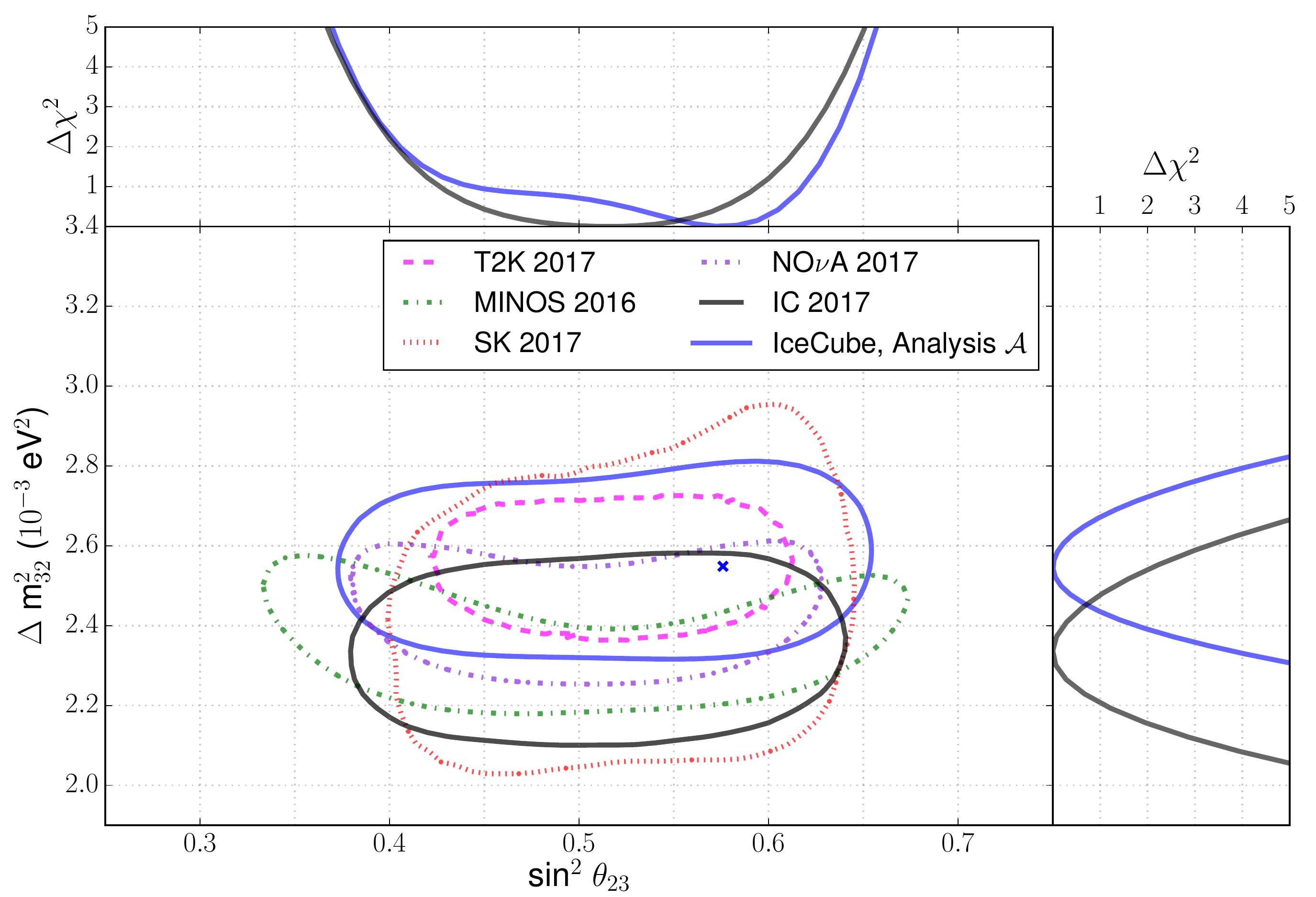}
    \caption{The 90\% allowed region using the data sample from Analysis A in blue compared to other experiments\cite{Singh:2017thk,Haegel:2017ofz,Whitehead:2016xud,Wendell:2014dka,Aartsen:2017nmd}. The best fit point from Analysis A is shown as the blue cross mark. The IceCube 2017 result~\cite{Aartsen:2017nmd}, represented in black, uses the data sample from Analysis B. The top and right plots are the 1-d $\Delta \chi^2$ profiles of the measured oscillation parameters.}
    \label{greco_numu}
\end{figure}

Results from two separate analyses with different strategies for event selection and background estimation are shown in Fig.~\ref{greco_numu}. The main analysis A targets a high statistics sample of neutrino events of all flavors resulting in 13462 events in three years. The background estimation is simulation-driven. In contrast, the confirmatory analysis B is optimized for higher rejection of background events and its atmospheric muon background estimation is data-driven.

Fig.~\ref{greco_numu} shows the 90\% allowed region of atmospheric neutrino oscillation parameters for the analyses based on Analysis A, and the allowed regions reported by other experiments. Overall, the 90\% allowed regions for the two analyses are statistically consistent, and both results compare favorably with the latest published 90\% contours from other neutrino experiments~\cite{Singh:2017thk,Haegel:2017ofz,Whitehead:2016xud,Wendell:2014dka}.

The shift between the two IceCube contours in both $\Delta$m$^2_{23}$ and $\sin^2 \theta_{23}$ can be explained by statistical fluctuations alone, as 65\% of events in Analysis A are unique with respect to Analysis B.
Additionally, detailed investigation show that differences in the analyses, specifically differences in the parametrization of detector effects, in the event selection and in reconstruction, can lead to small systematic shifts of $<0.5\sigma$ as well.

While the measured oscillation parameters agree with previous experiments, it is important to keep in mind that they have been measured at a roughly ten times higher energy. The measurement is therefore also sensitive to any new neutrino physics, an important consideration when the precision of the IceCube measurements will be further improved.

An upgrade project is underway that will result in the deployment of 7 new strings that will further increase the light collection in DeepCore ~\cite{Aartsen:2019kpk}. By decreasing the separation between IceCube strings from 125 to 75 meters and the distance between light sensors along a string from 17 to 7 meters, DeepCore reduced the energy threshold from the TeV-and-above scale suitable for neutrino astrophysics to the GeV scale required for measurements of neutrino properties at the highest energies. The seven new strings will be deployed with vertical and horizontal spacings that are three times smaller than DeepCore and will also include advanced calibration devices. 

This new instrumentation will dramatically boost IceCube's performance at the lowest energies, increasing the samples of atmospheric neutrinos by a factor of ten. As a result of this upgrade, IceCube will yield the world's best measurements in neutrino oscillations as well as critical measurements that could provide evidence for new physics in the neutrino sector.  
\subsubsection{Supernova Neutrinos}
\label{subsection2}

A core-collapse supernova radiates the vast majority of the binding energy of the resulting compact remnant in the form of neutrinos of all flavors. Information about the astrophysics of the collapse and subsequent explosion, and about the physics of neutrinos, is encoded in the time, energy, and flavor structure of the neutrino burst. When supernova 1987A exploded just a couple of tens of events provided sufficient information not only to probe the physics of the explosion, but also to shed new light on neutrino properties such as their mass, magnetic moment, and lifetime. Neutrino oscillations produce significant modifications of the neutrino spectra. Matter transitions in the expanding remnant have significant time- and energy-dependent effects on the spectra of either neutrinos or antineutrinos, depending on the neutrino mixing parameters and the neutrino mass hierarchy. Furthermore, more exotic ``collective" effects resulting from neutrino-neutrino interactions can result in significant modulation of the spectra, such as ``spectral swaps,'' which exchange flavor spectra above or below a particular energy threshold.
 
Although designed as a high-energy Cherenkov detector with a nominal threshold of 100 GeV ($\sim 5$\,GeV in DeepCore), IceCube is still capable of detecting MeV-energy neutrinos from core-collapse supernovae by counting them, albeit without recording the direction of individual events. A large burst of below-threshold MeV neutrinos produced by a supernova will nevertheless be detected as a collective increase in photomultiplier counting rates on top of their low dark noise in sterile ice. The origin of this increased rate is Cherenkov light from shower particles produced by supernova electron antineutrinos interacting in the ice, predominantly by the inverse beta decay reaction. Additionally, a new supernova data acquisition that is capable of recording multiple photons from a single neutrino event provides a proxy for measuring their energy.

In ice, positron tracks of about $0.6 {\rm cm} \times (E_\nu/ \rm MeV)$ length radiate $178 \times (E_{e^+}/ \rm MeV)$ Cherenkov photons in the 300-600 nm wavelength range. From the approximate $E^2$ dependence of the neutrino cross section and the linear energy dependence of the track length, the light yield per neutrino scales with $E^3$. With absorption lengths exceeding 100 m, photons travel long distances in the ice so that each DOM effectively monitors several hundred cubic meters of ice. Typically, only a single photon from each interaction reaches one of the photomultipliers, which are vertically separated by roughly 17 m and horizontally by 125 m. The DeepCore subdetector, equipped with a denser array of high efficiency photomultipliers, provides higher detection and coincidence probabilities. Although the rate increase in individual light sensors is not statistically significant, the effect will be clearly seen once the rise is considered collectively over many independent sensors. For a supernova at 10\,kpc, a signal of one million photoelectros will trace a detailed movie of the time evolution of the neutrino signal. The additional measurement of the rate of two-photon correlations is sensitive to the energy of the supernova neutrinos. With a two megaton effective volume for supernova neutrinos IceCube is the most precise detector for analyzing their neutrino light curve~\cite{Abbasi:2011ss}.

While IceCube does not resolve individual neutrino events, it provides a high-statistics and detailed movie of the supernova as a function of time. Rather than just monitoring the noise in the DOMs, IceCube buffers the information from all photomultiplier hits~\cite{HeeremanvonZuydtwyck:2015mbs}. Every photon will be recorded to an accuracy of 2 ns in the case of a supernova. This low-level data provides several advantages: the complete detector information is available and the data-buffered at an early stage of the data acquisition system on the so-called string hubs-will be available in the unlikely case that the data acquisition fails, for instance in the case of an extremely close supernova, which could exhaust the system. The automatized hitspooling has been working reliably for several years, including the automatic transfer of data in case of serious alarms to the North. The data from these triggered events have been carefully studied and compared to the results from the standard supernova data acquisition.

\begin{figure}[htbp]
    \includegraphics[width=\linewidth]{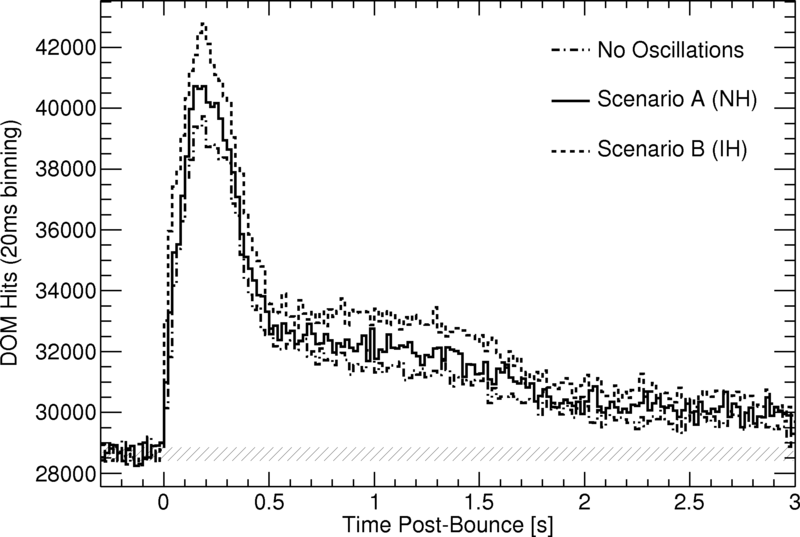}
    \caption{Expected supernova neutrino events as a function of their arrival times for a supernova at 10 kpc  distance. Scenarios A (normal hierarchy) and B (inverted hierarchy) are contrasted while the case for no oscillations is shown as a reference.}
    \label{supernova}
\end{figure}

The excellent sensitivity to neutrino properties such as the neutrino hierarchy as well as the possibility of detecting the neutronization burst, a short outbreak of anti-\,$\nu_e$ released by electron capture on protons soon after collapse are discussed in~\cite{Abbasi:2011ss}. Also, the formation of a black hole or a quark star as well as the characteristics of shock waves are investigated to illustrate IceCube's capability for supernova detection. In Fig.~\ref{supernova} we illustrate the sensitivity of IceCube to the hierarchy for a supernova at the most likely distance of 10\,kpc.

\subsubsection{Search for Dark Matter}

A neutrino ``telescope" is a powerful tool to search for the particle nature of dark matter. Although IceCube, was primarily designed as a discovery instrument, searching for dark matter was from inception one of its priority missions. Its DeepCore infill that lowers IceCube's threshold over a significant part of the volume was initially proposed by Per Olof Hulth as a way to enhance IceCube's capabilities for detecting dark matter particles with lower masses. In this context it is worthy of note that the AMANDA detector, the forerunner and proof of concept for IceCube, received a significant fraction of its initial funding from the Berkeley Center for Particle Astrophysics to search for dark matter.

IceCube searches indirectly for dark matter by looking for neutrinos from concentrations of weakly interacting massive particles (WIMPs)
in the Sun~\cite{Aartsen:2012kia,Aartsen:2016exj,Aartsen:2016zhm}, in the Milky Way~\cite{Aartsen:2017ulx,Aartsen:2015xej,Aartsen:2016pfc,Abbasi:2011eq,Aartsen:2014hva}
and in nearby galaxies~\cite{Aartsen:2013dxa}.  The neutrinos are secondary products of the annihilation of pairs of WIMPs into standard Model particles which includes decays into neutrinos, either directly or via their leptonic decays. 

If ``weakly" interacting massive particles (WIMPs) make up the dark matter, they have been swept up by the Sun for billions of years as the Solar System moves about the galactic halo. Though interacting weakly, they will occasionally scatter elastically with nuclei in the Sun and lose enough momentum to become gravitationally bound. Over the lifetime of the Sun, a sufficient density of WIMPs may accumulate in its center so that an equilibrium is established between their capture and annihilation. The annihilation products of these WIMPs represent an indirect signature of halo dark matter, their presence revealed by neutrinos which escape the Sun with minimal absorption. The neutrinos are, for instance, the decay products of heavy quarks and weak bosons resulting from the annihilation of  WIMPs into $\chi\chi\rightarrow \tau^{+}\tau^{-}$ , $b\bar{b}$ or $W^+ W^-$. Neutrino ``telescopes'' are sensitive to such neutrinos because of their relatively large neutrino energy reflecting the mass of the parent WIMP.

The beauty of the indirect detection technique using neutrinos originating in the sun is that the astrophysics of the problem is understood. The source in the sun has built up over solar time sampling the dark matter throughout the galaxy; therefore, any unanticipated structure in the halo has been averaged out over time. Other astrophysical uncertainties, such as the dark matter velocity distribution, have little or no impact on the capture rate of dark matter in the sun~\cite{Choi:2013eda,Danninger:2014xza}. Given a WIMP mass and cross section (and the assumption that the dark matter is not exclusively matter or antimatter), one can unambiguously predict the signal in a neutrino telescope. If not observed, the model will be ruled out. This is in contrast to indirect searches for photons from WIMP annihilation, whose sensitivity depends critically on the structure of halo dark matter; observation requires cuspy structure near the galactic center or clustering on appropriate scales elsewhere; observation not only necessitates appropriate WIMP properties, but also favorable astrophysical circumstances. 

Extensions of the Standard Model of quarks and leptons, required to solve the hierarchy problem, naturally yield dark matter candidates. For instance, the neutralino, the lightest stable particle in most supersymmetric models, has been intensively studied as a possible dark matter candidate. Detecting it has become the benchmark by which experiments are evaluated and mutually compared. We will follow this tradition here. Supersymmetric models allow for a large number of free parameters and, unfortunately, for a variety of parameter sets that are able to generate the observed dark matter density in the context of standard big bang cosmology. The neutralino interacts with ordinary matter by spin-independent (e.g., Higgs exchange) and by spin-dependent (e.g., Z-boson exchange) interactions. The first mechanism favors direct detection experiments~\cite{Sadoulet:2007pk} because the WIMP interacts coherently, resulting in an increase in sensitivity proportional to the square of the atomic number of the detector material. Given the rapid improvement of the sensitivity of direct experiments, indirect detection experiments are therefore not competitive for models where the dark matter has a large spin-independent coupling. This is not the case for spin-dependent models for which IceCube achieves world-best limits. Within the context of supersymmetry, direct and indirect experiments are complementary.

Although IceCube detects neutrinos of all flavors, sensitivity to neutrinos produced by WIMPs in the sun is \textit{primarily} achieved by exploiting the degree accuracy with which muon neutrinos can be pointed back to the Sun. Searches optimized to probe WIMP masses below about 50~GeV benefit from extending the detection channel to all neutrino flavors.

IceCube is most sensitive to WIMPs with significant spin-dependent interactions with protons because they result in large concentrations of WIMPS in the Sun, a nearby and readily identifiable source of protons.
The signal would simply be an excess of neutrinos of GeV energy and above, depending on the WIMP mass, from the direction of the Sun over the atmospheric neutrino background in the same angular window. 
There would be no alternative astrophysical explanation of such a signal that represents a smoking gun for dark matter particles.
In most WIMP scenarios, the cross section for WIMP capture ($\sigma_{\chi,p}$) is large enough so that an equilibrium between capture and annihilation would have been achieved within the age of the solar system~\cite{Gondolo:2004sc}.  In this case, limits on neutrinos from the Sun can be expressed
in terms of the capture cross section, $\sigma_{\chi,p}$.  If equilibrium is not reached, 
weaker limits can still be derived.

\begin{figure}
 \begin{center}
\includegraphics[width=0.95\linewidth,angle=0]{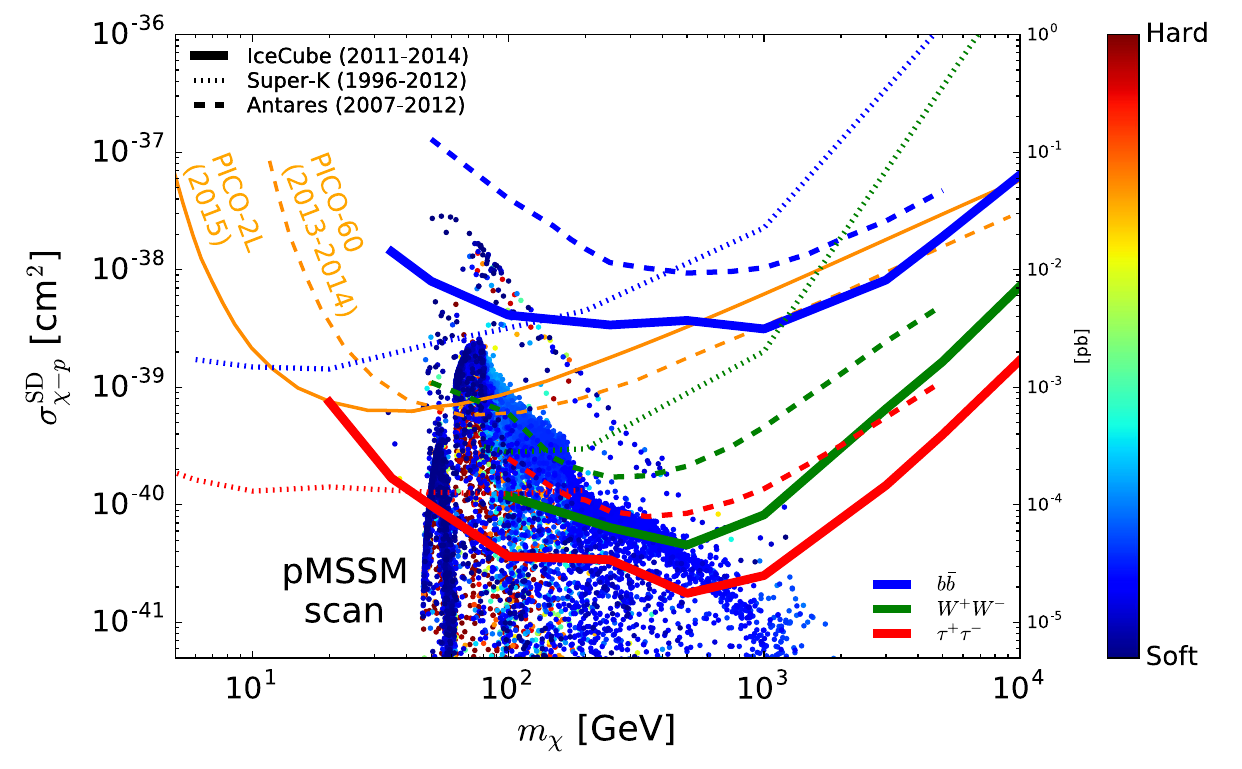}
\end{center}
\caption{Limits on the spin-dependent WIMP-proton cross section ($\sigma^{\mathrm{SD}}_{\chi\mathrm{-}p}$) from IceCube 3-year study \cite{Aartsen:2016zhm}   compared to results from other neutrino detectors and direct detection experiments. Various points corresponding to neutralinos from a scan of the phenomenological minimally supersymmetric standard model (pMSSM) are also shown, color-coded by their leading annihilation channel. Points close to the red end of the spectrum annihilate into harder channels such as $\tau^+ + \tau^-$ and can be excluded by the red line from IceCube.}
\label{IC79dmlimits}
\end{figure}

Quantitative interpretation of the IceCube 
limits depends on detailed analysis
of capture rates and annihilation channels.  
The analysis traditionally uses DarkSUSY~\cite{Gondolo:2004sc},
which builds on the classic calculations of Gould~\cite{Gould:1991hx}
for capture and annihilation rates. The neutrino spectrum from
annihilation of pairs of WIMPs depends on the dominant channel for coupling to standard
model particles.  The neutrino spectrum at production is also modified by subsequent
interactions and oscillations of neutrinos as they propagate out from the solar core.
Three flavor oscillations with matter effects are included as in reference~\citen{Blennow:2007tw}.
The range of possibilities is bracketed by calculating two extremes,
\begin{equation}
\chi\,\chi\rightarrow b\bar{b}\,{\rm (soft)}\,\,{\rm and}\,\,\chi\,\chi\rightarrow \tau^{+}\tau^{-}\,{\rm (hard)}.
\label{channels}
\end{equation}
In recent works $\tau^{+}\tau^{-}$ serves typically as hard channel, in older works often a combination of $W^\pm$ and $\tau^\pm$ are used, where $\tau^\pm$ is replaced with $W^\pm$ for $m_{\chi} > m_{W}$. The world-best limits on the spin-dependent interaction cross section of WIMPs with normal matter are shown in Fig.~\ref{IC79dmlimits}. Even stronger limits, or a discovery, are forthcoming using an 8-year data set and improved reconstructions.

\subsection{IceCube, the Facility}

During its construction phase, IceCube demonstrated a significant potential for
facilitating a range of other research.  For example, a dust logger provided
measurements with millimeter precision that are valuable for event reconstruction
in IceCube but that also provide a record of surface winds over more
than 100,000 years~\cite{dustlogger}. 
 
Already during construction of AMANDA, antennas
forming the RICE detector were deployed in deep holes to expand the target volume in the
search for cosmogenic neutrinos~\cite{Kravchenko:2011im}.  
An acoustic test setup of receivers in the upper portion of four IceCube holes
was deployed in 2007 to explore the acoustic technique for
detecting ultra-high-energy neutrinos. 
A retrievable transmitter (pinger) was submerged briefly in several
newly prepared holes at various depths and distances from the receivers
to measure the attenuation of sound in ice.   The attenuation length of $300$~m
is significantly less than had been expected~\cite{Abbasi:2010vt}. In contrast,
the Askaryan Radio Array (ARA)~\cite{Allison:2011wk} is taking advantage of 
the kilometer-scale attenuation for radio signals in ice
to operate a detector that performs R\&D for a much larger array that will be colocated with the next-generation IceCube detector.
The ARA detectors send data to computers housed in the IceCube Lab (ICL)
 for staging and transmission to the north.

The DM-Ice experiment~\cite{Cherwinka:2011ij} proposes to repeat
the DAMA experiment in the quiet and sterile environment of the
Antarctic ice.  An interesting feature of the measurement
is the fact that the seasonal modulation of the muon
rate has the opposite phase relative to the motion of the Earth
 through the halo of dark matter compared to 
a detector in the Northern Hemisphere.
A test detector to explore the noise environment for DM-Ice
was deployed at the bottom of an IceCube string in December 2010.
Its computers and data transmission are also hosted in the
ICL. The experiment will be deployed in the 2022-23 Antarctic summer as part of the Upgrade project, which we will discuss next.

The enhancement of the low-energy capabilities of IceCube
provided by the DeepCore subarray had led to the PINGU proposal~\cite{PINGU} to deploy an additional 40 strings within the
existing DeepCore detector.  This aimed to lower the threshold below $5$~GeV ($<25$~m muon track length in ice).
In this energy range, matter effects in the Earth lead to resonant
oscillations of $\nu_\mu\leftrightarrow\nu_e$ 
($\bar{\nu}_\mu\leftrightarrow\bar{\nu}_e$) for normal (inverted)
hierarchy~\cite{Akhmedov:2012ah} that depend on zenith angle.
By taking advantage of the fact that the neutrino cross section
is larger than that for antineutrinos, coupled with the excess
of $\nu_\mu$ compared to $\bar{\nu}_\mu$, a measurement sensitive
to the neutrino mass hierarchy is possible on a relatively short
timescale.  PINGU would also have had sensitivity to $\nu_\mu$ disappearance,
$\nu_\tau$ appearance, and maximal mixing.
The lower energy threshold would enhance the
indirect searches for dark matter with IceCube as well as
the sensitivity to neutrinos from supernova explosions.  
In addition, PINGU presented the potential for neutrino tomography of the Earth. The science goals of PINGU are now pursued with the IceCube Upgrade project. Although the construction of a PINGU-style detector can be launched in the future, the next-generation IceCube detector will target high-energy neutrino astronomy instead. The ORCA detector, with the same science goals and performance, is now under construction in the Mediterranean Sea~\cite{Adrian-Martinez:2016fdl}. 

The so-called IceCube Upgrade is also under construction. It will deploy seven new strings at the bottom of the detector array that have been designed as an incremental extension of the DeepCore detector and as a test bed for the technology of the next-generation detector.

The new strings will increase the light collection in what is already a denser region of the gigaton neutrino detector. By decreasing the separation between IceCube strings from 125 to 75 meters and the distance between light sensors along a string from 17 to 7 meters, DeepCore reduced the energy threshold from the TeV-and-above scale suitable for neutrino astrophysics to the GeV scale required for measurements of neutrino properties at the highest energies. The seven new strings will be deployed with vertical and horizontal spacings that are three times smaller than DeepCore and will also include advanced calibration devices. 

This new instrumentation will dramatically boost IceCube's performance at the lowest energies, increasing the samples of atmospheric neutrinos by a factor of ten, and will enhance the pointing resolution of astrophysical neutrinos. As a result of this upgrade, IceCube will yield the world's best measurements in neutrino oscillations as well as critical measurements that could provide evidence for new physics in the neutrino sector.  

The new calibration devices will advance our understanding of the response variability of the light sensors in both current and new strings, resulting in enhanced reconstruction of cascade events and better identification of tau neutrinos. Cascade signatures are the light patterns of more than 75\% of astrophysical neutrinos and currently have a poorer pointing resolution than the so-called tracks--cascades allow reconstructing the direction of the incoming neutrino with a typical angular resolution of 5-10 degrees, while tracks point to the neutrino's origin to within less than 1 degree. The refined calibration of the existing sensors will also enable a reanalysis of more than ten years of archival data and increase the discovery potential of neutrino sources.

\subsection{From Discovery to Astronomy, and More...}

Accelerators of cosmic rays produce neutrino fluxes limited in energy to roughly 5\% of the maximal energy of the protons or nuclei. For Galactic neutrino sources, we expect neutrino spectra with a cutoff of a few hundred TeV. Detection of these neutrinos requires optimized sensitivities in the TeV range. At these energies, the atmospheric muon background limits the field of view of neutrino telescopes to the downward hemisphere. With IceCube focusing on high energies, a second kilometer-scale neutrino telescope in the Northern Hemisphere would ideally be optimized to observe the Galactic center and the largest part of the Galactic plane.

Following the pioneering work of DUMAND~\cite{Babson:1989yy}, several neutrino telescope projects were initiated in the Mediterranean in the 1990s~\cite{Aggouras:2005bg,Aguilar:2006rm,Migneco:2008zz}. In 2008, the construction of the ANTARES detector off the coast of France was completed. With an instrumented volume at a few percent of a cubic kilometer, ANTARES reaches roughly the same sensitivity as AMANDA and is currently the most sensitive observatory for high-energy neutrinos in the Northern Hemisphere. It has demonstrated the feasibility of neutrino detection in the deep sea and has provided a wealth of technical experience and design solutions for deep-sea components. 

While less sensitive than IceCube to a diffuse extragalactic neutrino flux, ANTARES has demonstrated its competitive sensitivity to neutrino emission from the Galactic center~\cite{Adrian-Martinez:2015wey,Adrian-Martinez:2014wzf} and extragalactic neutrino sources in the Southern Hemisphere~\cite{AdrianMartinez:2012rp}. The important synergy between Mediterranean and Antarctic neutrino telescopes has been demonstrated by the first joint study of continuous neutrino sources~\cite{Adrian-Martinez:2015ver} as well as neutrino follow-up campaigns of gravitational waves~\cite{ANTARES:2017bia}.

\begin{figure}[t]
\centering
\includegraphics[width=0.6\linewidth]{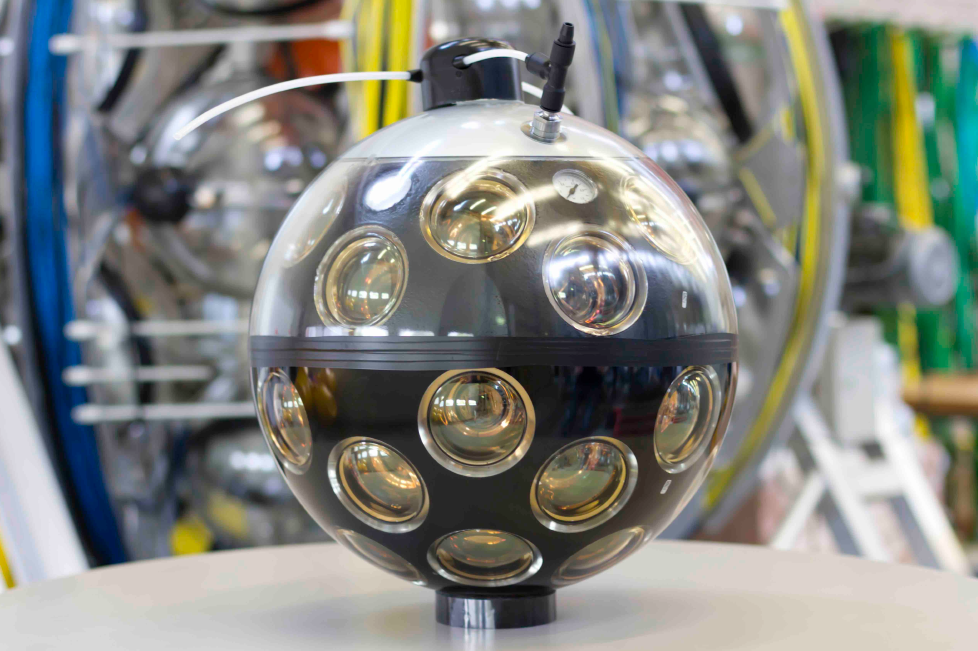}
\caption[]{The KM3NeT optical module (from reference~\citen{Adrian-Martinez:2016fdl}). The optical module consists of a glass sphere with a diameter of 42~cm, housing 31 photosensors (yellowish disks). The glass sphere can withstand the pressure of the water and is transparent to the faint light that must be detected to see neutrinos.\label{fig:mdom}}
\end{figure}

An international collaboration has started construction of a multi-cubic-kilometer neutrino telescope in the Mediterranean Sea, KM3NeT~\cite{Adrian-Martinez:2016fdl}. Major progress has been made in establishing the reliability and the cost-effectiveness of the design. Most importantly, this includes the development of a digital optical module that incorporates 31 3-inch photomultipliers instead of one large photomultiplier tube; see Fig.~\ref{fig:mdom}. The advantages are a tripling of the photocathode area per optical module, a segmentation of the photocathode allowing for a clean identification of coincident Cherenkov photons, some directional sensitivity, and a reduction of the overall number of penetrators and connectors, which are expensive and failure-prone. For all photomultiplier signals exceeding the noise level, time-over-threshold information is digitized and time-stamped by electronic modules housed inside the optical modules. This information is sent via optical fibers to shore, where the data stream will be filtered online for event candidates.

KM3NeT in its second phase~\cite{Adrian-Martinez:2016fdl} will consist of two units for astrophysical neutrino observations, each consisting of 115 strings (detection units) carrying more than 2,000 optical modules, and one ORCA detector studying fundamental neutrino physics with atmospheric neutrinos. The detection units are anchored to the seabed with deadweights and kept vertical by submerged buoys. The vertical distances between optical modules will be 36\,meters, with horizontal distances between detection units at about 90\,meters. Construction is now ongoing near Capo Passero (east of Sicily).

A parallel effort is underway in Lake Baikal with the construction of the deep underwater neutrino telescope Baikal-GVD (Gigaton Volume Detector)~\cite{Avrorin:2015wba}. The first GVD cluster was upgraded in spring 2016 to its final size (288 optical modules, a geometry of 120 meters in diameter, 525 meters high, and instrumented volume of 6 Mton). Each of the eight strings consists of three sections with 12 optical modules. At present 7 of the 14 clusters have been deployed reaching a sensitivity close to the diffuse cosmic neutrinos flux observed by IceCube.

Further progress requires larger instruments. IceCube therefore proposes as a next step to capitalize on its large absorption length to instrument $10\rm\,km^3$ of glacial ice at the South Pole and thereby improving on IceCube's sensitive volume by an order of magnitude~\cite{Aartsen:2015dkp,Aartsen:2019swn,Aartsen:2020fgd}. This large gain is made possible by the unique optical properties of the Antarctic glacier revealed by the construction of IceCube. As a consequence of the extremely long photon absorption lengths in the deep Antarctic ice, the spacing between strings of light sensors can be increased from 125 to close to 250 meters without significant loss of performance of the instrument. The instrumented volume can therefore grow by one order of magnitude while keeping the construction budget of a next-generation instrument at the level of the cost of the current IceCube detector. The new facility will increase the event rates of cosmic events from hundreds to thousands over several years. The superior angular resolution of the longer muon tracks will allow for the discovery of the sources seen at the $3 \sigma$-level in the 10-year sky map. Construction of a next-generation instrument with at least five times higher sensitivity would likely result in the observation of cosmogenic neutrinos~\cite{Aartsen:2014njl}. Obviously, higher sensitivity would also benefit the wide range of measurements performed with the present detector, from the search for dark matter to the precision limits on any violation of Lorentz invariance.

\section{Acknowledgements}

Discussion with collaborators inside and outside the IceCube Collaboration, too many to be listed, have greatly shaped this presentation. Thanks. 
This research was supported in part by the U.S. National Science Foundation under grants~PLR-1600823 and PHY-1913607 and by the University of Wisconsin Research Committee with funds granted by the Wisconsin Alumni Research Foundation.

\bibliographystyle{ws-rv-van}
\bibliography{bib,gc,frontiers}
\end{document}